\documentclass[twoside,12pt]{Classes/myThesis}  

\graphicspath{{./Chapters/Chapter_1/Chapter_1_Fig/}     
              {./Chapters/Chapter_2/Chapter_2_Fig/}
              {./Chapters/Chapter_3/Chapter_3_Fig/}
              {./Chapters/Chapter_4/Chapter_4_Fig/}
              {./Chapters/Chapter_5/Chapter_5_Fig/}
              {./Chapters/Chapter_6/Chapter_6_Fig/}
              {./Chapters/Chapter_7/Chapter_7_Fig/}
              {./Chapters/Chapter_8/Chapter_8_Fig/}
              {./ThesisFigs/}}

\newcommand{\theAuthor}{Jintao Shuai}
\newcommand{\authorEmail}{j.shuai@leeds.ac.uk}
\newcommand{\myTitle}{Interaction of surface acoustic waves and magnetic thin films}
		\pdfcatalog { /PageMode (/UseOutlines)
                  /OpenAction (fitbh)  }

\title{\myTitle}
\author{\href{https://scholar.google.com/citations?hl=en&user=sVp6dlcAAAAJ}{\theAuthor}}
\crest{\includegraphics[width=40mm]{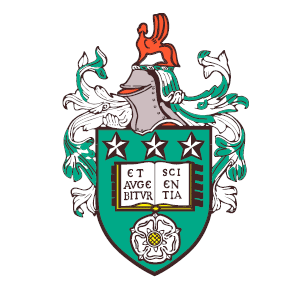}}
\logo{\includegraphics[width=50mm]{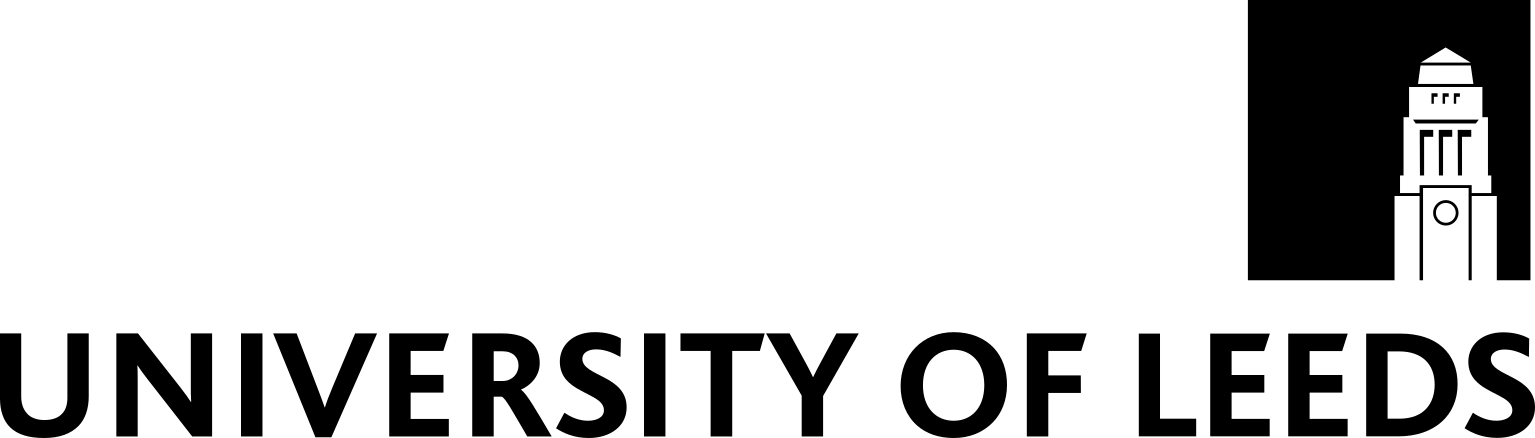}} 
\university{\href{http://www.leeds.ac.uk}{University of Leeds}}
\collegeordept{\href{https://eps.leeds.ac.uk/physics}{School of Physics and Astronomy}}

\deg{Doctor of Philosophy}
\degreedate{September 2023}

\patchcmd{\bibliography}{\bibname}{References}{}{}

\begin{document}
\frontmatter
\pagenumbering{roman}
\maketitle
\cleardoublepage
\setcounter{secnumdepth}{2}
\setcounter{tocdepth}{2}



\begin{dedication} 

\Large{To my parents}

\end{dedication}



\cleardoublepage

\begin{ipstatement} 

\setcounter{page}{1}
The candidate confirms that the work submitted is his own, except where work which has formed part of jointly authored publications has been included. The contribution of the candidate and the other authors to this work has been explicitly indicated below. The candidate confirms that appropriate credit has been given within the thesis where reference has been made to the work of others.\par
This copy has been supplied on the understanding that it is copyright material and that no quotation from the thesis may be published without proper acknowledgement.\par
The right of \theAuthor\ to be identified as Author of this work has been asserted by him in accordance with the Copyright, Designs and Patents Act 1988.
\\
\\
\copyright \yeardate\today\ The University of Leeds and \theAuthor.
\\
\\
This thesis incorporates work from the jointly authored publications:
\\
\\
\textbf{Chapter~\ref{Chapter4_anisotropy_control}}\\
\textbf{Jintao Shuai}, Mannan Ali, Luis Lopez-Diaz, John E. Cunningham, and Thomas A. Moore. Local anisotropy control of Pt/Co/Ir thin film with perpendicular magnetic anisotropy by surface acoustic waves. \href{https://doi.org/10.1063/5.0097172}{\textit{Applied Physics Letters}, 120, 252402 (2022)}.\\
\textbf{Contributions by candidate:}
\DOE \EOE \DOS \EOS \DA \MP\\
\textbf{Contributions by Others:}
Mannan Ali: \MI \ME
Luis Lopez-Diaz: \DOS \CR \RD \ME
John E. Cunningham and Thomas A. Moore: \SV \DOE \MI \RD \ME
\\
\\
\textbf{Chapter~\ref{Chapter5_heating_effect}}\\
\textbf{Jintao Shuai}, Robbie G. Hunt, Thomas A. Moore, and John E. Cunningham. Separation of heating and magnetoelastic coupling effects in surface-acoustic-wave-enhanced creep of magnetic domain walls. \href{https://doi.org/10.1103/PhysRevApplied.20.014002}{\textit{Physical Review Applied}, 20, 014002 (2023)}.\\
\textbf{Contributions by candidate:}
\DOE \EOE \DA \MP \\
\textbf{Contributions by Others:}
Robbie G. Hunt: \MI \ME
Thomas A. Moore and John E. Cunningham: \SV \DOE \MI \RD \ME
\\
\\
\textbf{Chapter~\ref{Chapter6_DW_dynamics}}\\
\textbf{Jintao Shuai}, Luis Lopez-Diaz, John E. Cunningham, and Thomas A. Moore. {Surface acoustic wave effect on magnetic domain wall dynamics}, \href{https://doi.org/10.1103/PhysRevB.108.104420}{\textit{Physical Review B}, 108, 104420 (2023)}.\\
\textbf{Contributions by candidate:}
\DOS \EOS \DA \MP \\
\textbf{Contributions by Others:}
Luis Lopez-Diaz: \DOS \CR \RD \ME
Thomas A. Moore and John E. Cunningham: \SV \DOS \MI \RD \ME
\\
\\
\textbf{Chapter~\ref{Chapter7_Skyrmion_motion}}\\
\textbf{Jintao Shuai}, Luis Lopez-Diaz, John E. Cunningham, and Thomas A. Moore. Precise transport of skyrmions by surface acoustic waves. \href{https://arxiv.org/abs/2305.16006}{\textit{arXiv:2305.16006} (2023)}.\\
\textbf{Contributions by candidate:}
\DOS \EOS \DA \MP \\
\textbf{Contributions by Others:}
Luis Lopez-Diaz: \DOS \CR \RD \ME
John E. Cunningham and Thomas A. Moore: \SV \DOS \MI \RD \ME

\end{ipstatement}


\cleardoublepagewithnumberheaderipstatement

\begin{acknowledgements}      

First and foremost, my heartfelt gratitude goes to my supervisors, Dr Thomas Moore and Prof John Cunningham. Their consistent guidance, coupled with their deep expertise, played a pivotal role in shaping my research. Their constructive feedback, encouragement, and patience have been instrumental, ensuring I navigated challenges effectively. I truly valued our in-depth discussions and the broad spectrum of knowledge they imparted during our time together.

I am deeply thankful to Prof Luis Lopez-Diaz, Prof Rudolf Sch\"{a}fer, Dr Ivan Soldatov, Prof Bert Koopmans, Dr Reinoud Lavrijen, Mouad, Roc\'{i}o, Felipe, Pingzhi, and Adrien for their support during my secondments in Salamanca, Dresden, and Eindhoven. Dr Liza Herrera Diez, Clare Desplats and PIs from MagnEFi deserve special mention for their incredible guidance, management of the project, and for uniting everyone. My journey was further enriched by my fellow ESRs from MagnEFi: Golam, Rohit, Adithya, Giovanni, Mandy, Gyan, Cristina, Sreeveni, Beatrice, Adriano, and Subhajit.

My deepest appreciation goes to the respected academics in our group. Their insights and encouragement were invaluable throughout my research journey. Further, I want to express my heartfelt thanks to Mannan, Philippa, Matt, Nathan, and Craig for their invaluable training and support. I extremely appreciate Mark and Li for their foundational cleanroom training and technical assistance, with additional thanks to Li for his assistance for the cleanroom fabrication during the Covid lockdown.

Cheers to the ``Lunch Club'' -- Chris, Dan, and Robbie. Your company transformed our daily meals into delightful gatherings; lunches just weren not the same without you. My thanks also go out to Ahmet, Robert, Emily, Lin, Sean, Ioannis, Callum, Thomas, Servet, Hari, and Eloi for the shared moments and memories. My years at Royal Park Flat were made memorable by friends such as Ayhan, Ayo, Nidhin, Srinath, Smruthi, Chrissy, Drea, Manmeet, Prateek, Manan, Xiao, Harry, David, Oscar, and Dylan.

A special acknowledgement for Rutvij -- thank you for being an incredible pillar of support. My time in York was made unforgettable thanks to Ding, Vincenzo, Shashaank, Alistair, Cordelia, Jenny, Beth, Peiyun, Jack, Jade, Hamid, Ambroise, Simen, Ayan, Shefali, Nekhel, and Anjali. To Weng Wei, Chen Wen, Chenyang, Zhou Jia, Beibei, and Ruru -- your presence was a comforting reminder of home in a foreign land.

Above all, my deepest appreciation goes out to my parents and family. Their boundless love, unwavering support, and sacrifices have been the driving force behind my achievements. This accomplishment is as much theirs as it is mine.

I am grateful to the European Union's Horizon 2020 research and innovation programme for their support through the Marie Sk\l{}odowska-Curie grant agreement No. 860060 \href{https://doi.org/10.3030/860060}{``Magnetism and the effect of Electric Field'' (MagnEFi)}. This funding not only facilitated my research but also elevated my entire PhD journey to a unique and enriching experience.

Finally, I would like to thank Dr Satoshi Sasaki and Dr Tom Hayward for their comprehensive examination of this thesis.

\end{acknowledgements}



\cleardoublepagewithnumberheaderacknowledgement
\begin{abstracts}       
Surface acoustic waves (SAWs) have emerged as innovative and energy-efficient means to manipulate domain walls (DWs) and skyrmions in thin films with perpendicular magnetic anisotropy (PMA) owing to the magnetoelastic coupling effect. 
This thesis focuses on the complex interplay between SAWs and magnetic thin films with PMA through both experimental and micromagnetic studies. 
The effects of the standing SAWs on the magnetisation dynamics in a Ta(5.0 nm)/Pt(2.5 nm)/Co(1.1 nm)/Ir(1.5 nm)/Ta(5.0 nm) thin film with PMA were first investigated. SAWs with frequency of 93.35 MHz significantly reduced the coercivity of the thin film by 21\% and enhanced the magnetisation reversal speed by 11-fold. Standing SAWs introduce a dynamic energy landscape with a unique spatial distribution, forming striped domain patterns in the thin film. 
The use of radio-frequency signals for generating SAWs inevitably causes a heating effect in the device. This heating effect was carefully examined \textit{in situ} in a SAW device featuring a Ta(5.0 nm)/Pt(2.5 nm)/Co(0.9 nm)/Ta(5.0 nm) thin film with PMA using an on-chip platinum thermometer. It was shown that the temperature increased by 10 K in the presence of SAWs at centre freqneucy of 48 MHz. The DW velocity was significantly enhanced in the presence of the standing SAWs by a factor of 4 compared to that with temperature change alone owing to the magnetoelastic coupling effect. 
To understand the SAW-enhanced DW motion, comprehensive micromagnetic simulations were performed on thin films in the presence of travelling SAWs with frequencies from 50 to 200 MHz. The findings highlighted that SAW-induced vertical Bloch lines within DWs can simultaneously boost DW depinning and dissipate energy at the DW via spin rotation. The SAW-induced strain gradient can be exploited to control skyrmion motion in a current-free manner. Micromagnetic simulations revealed that the use of orthogonal SAWs, combining horizontal travelling and vertical standing waves, offered a promising approach to direct skyrmion motion along desired trajectories, avoiding undesirable Hall-like motion. In conclusion, this thesis offers a significant contribution to the understanding of how SAWs influence magnetisation dynamics.

\end{abstracts}

\cleardoublepagewithnumberheaderabstract
\tableofcontents
\cleardoublepagewithnumberheader
\listoffigures
\cleardoublepagewithnumberheader
\listoftables
\cleardoublepagewithnumberheader
\printglossary[type=\acronymtype, title=LIST OF ABBREVIATIONS]
\cleardoublepagewithnumberheader
\begin{publications}

{\large \textbf {Journal Publications}}
\begin{itemize}[leftmargin=*]
    \item \textbf{J. Shuai}, L. Lopez-Diaz, J. E. Cunningham, T. A. Moore, {Influence of surface acoustic wave frequency on domain wall dynamics}, \href{https://doi.org/10.1103/PhysRevB.108.104420}{{\emph{Physical Review B}, 108, 104420 (2023)}}.
    
    \item \textbf{J. Shuai}, R. G. Hunt, T. A. Moore, J. E. Cunningham, {Separation of heating and magneto-elastic coupling effects in surface acoustic wave-enhanced magnetic domain wall creep motion}, \href{https://doi.org/10.1103/PhysRevApplied.20.014002}{{\emph{Physical Review Applied}, 20, 014002 (2023)}}.

    \item \textbf{J. Shuai}, M. Ali, L. Lopez-Diaz, J. E. Cunningham, T. A. Moore, {Local anisotropy control of Pt/Co/Ir thin film with perpendicular magnetic anisotropy by surface acoustic waves}, \href{https://doi.org/10.1063/5.0097172}{{\emph{Applied Physics Letters}, 120, 252402 (2022)}}.

    \item \textbf{J. Shuai}, L. Lopez-Diaz, J. E. Cunningham, T. A. Moore, {Precise transport of skyrmions by surface acoustic waves}, \href{https://arxiv.org/abs/2305.16006}{\textit{arXiv:2305.16006} (2023)}.
\end{itemize}

\vspace{5 mm}

{\large \textbf{Conference presentations}}\par
Note: ``\textbf{$\bullet$}'' denotes oral presentation and ``\textbullet'' denotes poster presentation.

\begin{itemize}[leftmargin=*]
    \item[\textbullet] Surface acoustic wave effect on magnetic domain wall dynamics. Spin Electronics and Nanomagnetism Colloquium 2023, Nancy, France, August 2023.
    
    \item[\textbf{$\bullet$}] Surface acoustic wave effect on magnetisation dynamics in thin films. Royce Student Summit 2023, Leeds, United Kingdom, July 2023.
    
    \item[\textbf{$\bullet$}] Interaction of surface acoustic waves and magnetic thin films. IOP Condensed Matter and Quantum Materials, Birmingham, United Kingdom, June 2023.
    
    \item[\textbullet] Precise transport of skyrmions by surface acoustic waves. IEEE Intermag, Sendai, Japan, May 2023.
    
    \item[\textbf{$\bullet$}] Separation of heating and magneto-elastic coupling effects in surface acoustic wave-enhanced magnetic domain wall creep motion. IOP Magnetism, Manchester, United Kingdom, April 2023.
    
    \item[\textbullet] Precise transport of skyrmions by surface acoustic waves. IOP Magnetism, Manchester, United Kingdom, April 2023.
    
    \item[\textbf{$\bullet$}] Surface acoustic waves effect on domain wall motion. IEEE MMM 2022, online, November 2022.
    
    \item[\textbullet] Precise trajectory control of skyrmion motion by surface acoustic waves. IEEE MMM 2022, online, November 2022.
    
    \item[\textbf{$\bullet$}] Enhanced domain wall motion by surface acoustic waves. IEEE AtC-AtG 2022, online, August 2022.
    
    \item[\textbf{$\bullet$}] Interaction of surface acoustic waves and magnetic thin films. Home workshop, Universidad de Salamanca, Salamanca, Spain, July 2022.
    
    \item[\textbf{$\bullet$}] Interaction of surface acoustic waves and magnetic thin films. Joint meeting of the BeMAGIC and MagnEFi ITN networks, Aalto University, Espoo, Finland, June 2022.
    
    \item[\textbf{$\bullet$}] Local anisotropy control of Pt/Co/Ir thin films with perpendicular magnetic anisotropy by surface acoustic waves. Bragg PhD Colloquium 2022, University of Leeds, Leeds, United Kingdom, June 2022.
    
    \item[\textbf{$\bullet$}] Local anisotropy control of Pt/Co/Ir thin films with perpendicular magnetic anisotropy by surface acoustic waves. IOP Magnetism 2022, University of York, York, United Kingdom, March 2022.
    
    \item[\textbf{$\bullet$}] Coercivity control of thin films with perpendicular magnetic anisotropy by surface acoustic waves. Bragg Exchange Conference, University of Leeds, Leeds, United Kingdom, January 2022.
    
    \item[\textbullet] Coercivity control of thin films with perpendicular magnetic anisotropy by surface acoustic waves. 2022 Joint MMM-INTERMAG, online, January 2022.
    
\end{itemize}

\end{publications}

\cleardoublepagewithnumberheaderpublication

\mainmatter

\pagenumbering{arabic}

\chapter{Introduction}
\label{Chapter1_introduction}
\section{Background}
Spintronic devices have drawn wide attention for their significant potential in the next generation of data storage and logic processing applications~\cite{vzutic2004spintronics,fert2008nobel,bader2010spintronics}. 
A promising strategy is to encode information using magnetic spin structures, such as \glspl{DW} and skyrmions~\cite{parkin2008magnetic,fert2017magnetic}.
Thin films with \gls{PMA} are useful in this context because the \gls{PMA} guarantees stability of the stored information as well as a high storage density~\cite{dieny2017perpendicular}.
One of the challenges in thin magnetic films with \gls{PMA} is to move \glspl{DW} and skyrmions efficiently. 
\Glspl{SAW}, which are elastic waves travelling on the surface of piezoelectric materials, offer a potential solution to this. \Glspl{SAW} can introduce dynamic strain waves into magnetic thin films~\cite{yang2021acoustic}. Through the magnetoelastic coupling effect, these dynamic strain waves generate a dynamic energy landscape, thereby triggering magnetisation precession~\cite{thevenard2013irreversible, thevenard2016precessional, camara2019field}, facilitating magnetisation switching~\cite{shuai2022local,li2014acoustically}, and enhancing the \gls{DW} and skyrmion motion~\cite{dean2015sound,edrington2018saw, vilkov2022magnetic,wei2020surface,cao2021surface,shuai2023separation,nepal2018magnetic}.

\section{Magnetisation dynamics control}
\subsection{By magnetic field}
Magnetic \glspl{DW} are transitional regions separating adjacent magnetic domains with opposing magnetisation directions~\cite{hubert2008magnetic}.
When an external magnetic field is applied to a magnetic material, its intrinsic magnetic moments tend to align with this field, in order to reduce the overall energy of the system. 
Consequently, this magnetic field can favour the expansion of these domains, which in turn drives the \gls{DW} motion~\cite{chauve2000creep}.
At low fields, domains expand due to a slightly distorted wall structure associated with internal torque, which leads to creep regime. 
Above a critical field, which is called the Walker breakdown, the \glspl{DW} motion decreases with the field. 
When the magnetic field is sufficient higher than the Walker breakdown, \gls{DW} enters the flow regime in which the magnetisation within the \gls{DW} precesses. 
Notably, Metaxas \et studied the \gls{DW} velocity in Pt/Co/Pt thin films with \gls{PMA} over a large range of applied magnetic fields showing a complete velocity-field characteristics with motion regimes predicted from general theories for driven elastic interfaces in weakly disordered media~\cite{metaxas2007creep}. \par
\subsubsection{Limitations}
While the motion of \glspl{DW} driven by magnetic fields has been extensively studied and is readily attainable in laboratory settings, challenges arise when aiming for high-speed \gls{DW} motion. Achieving such rapid motion usually necessitates the application of a substantial external magnetic field, thereby limiting practical device applications~\cite{beach2005dynamics}. Furthermore, achieving local control over individual domains through a magnetic field poses challenges, since it inevitably affects neighbouring domains.

\subsection{Via electric current}

\subsubsection{Spin--transfer torque}
\begin{figure}[ht!]
    \centering
    \includegraphics[width=0.8\textwidth]{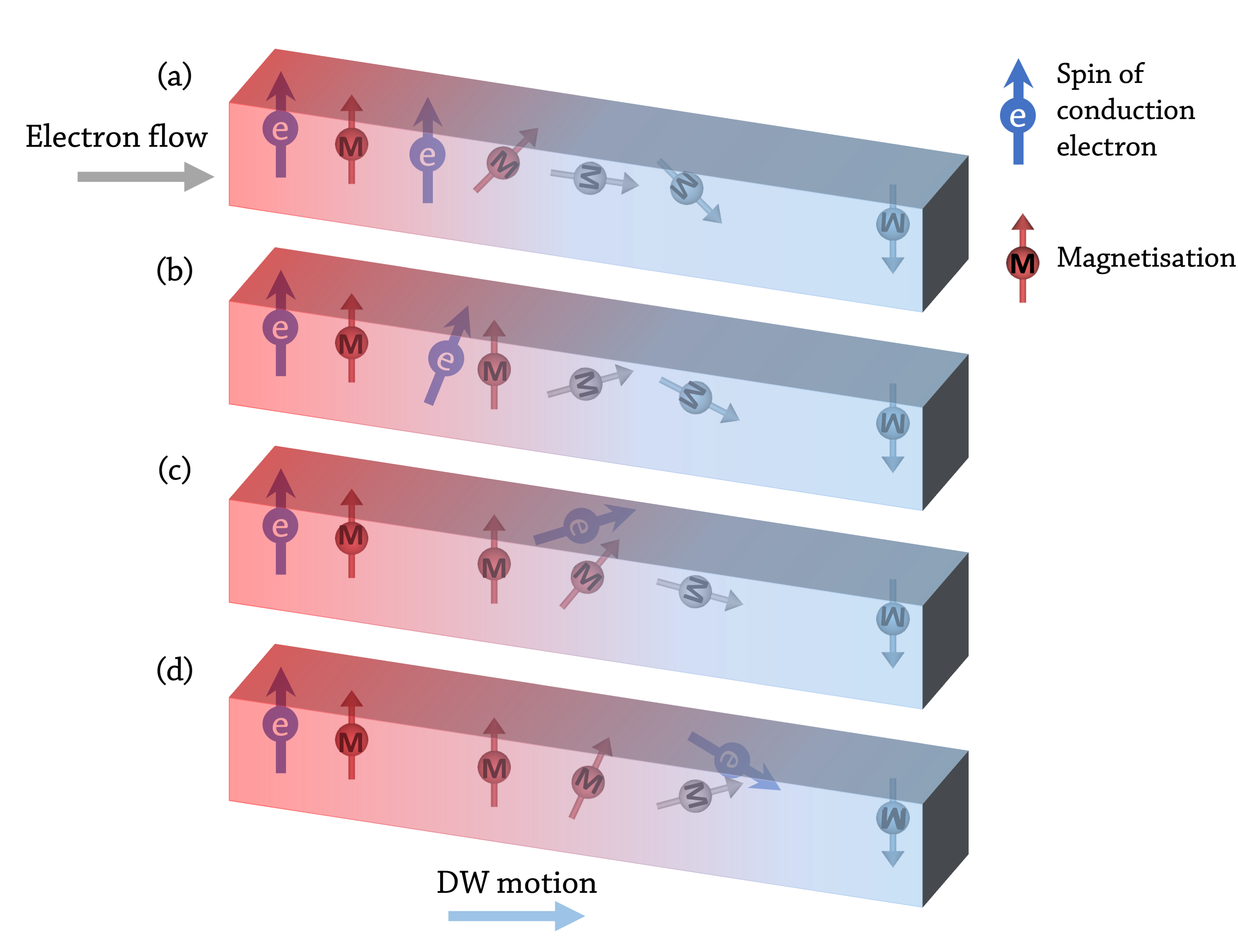}
    \caption[Diagram of domain wall motion induced by spin--transfer torque.]{\revisiontwo{Diagram of \gls{STT}-induced \gls{DW} motion. (a) Conduction electrons are injected from one side of the thin film containing a \gls{DW}. The local magnetisation attempts to align the spins of these conduction electrons. (b--c) Aligned spins exert an equal and opposite torque on the local magnetisation, leading to its rotation. (d) This reciprocal action can switch the magnetisation or induce the motion of the \gls{DW}.} \figref{\cite{kumar2022domain}}}
    \label{fig:STT}
\end{figure}

An electric current can lead to \gls{DW} motion in thin films with \gls{PMA} via \gls{STT}~\cite{li2004domain,li2004domain2}.
\gls{STT} describes the transfer of angular momentum from spin-polarised electrons to a magnetic material, subsequently exerting a torque on magnetic moments of the magnetic material~\cite{ralph2008spin}.
\revisiontwo{Figure~\ref{fig:STT} shows the schematic diagram of \gls{STT}-induced \gls{DW} motion. When conduction electrons (injected from left of the magnetic layer in Figure~\ref{fig:STT}), whose spins are not aligned with the magnetisation of the ferromagnetic layer, flow through a magnetic layer, the local magnetisation attempts to align these electron spins by exerting a torque on them (see  Figure~\ref{fig:STT}b).
Owing to the conservation of spin angular momentum, these conduction electrons also exert an equal and opposite torque on the local magnetisation (see Figure~\ref{fig:STT}c). 
This reciprocal action can be used to switch magnetisation or drive \gls{DW} motion (as shown in Figure~\ref{fig:STT}d)~\cite{kumar2022domain,chureemart2011dynamics,thiaville2004domain}.}

\subsubsection{Spin--orbital torque}
\begin{figure}
    \centering
    \includegraphics[width=0.8\textwidth]{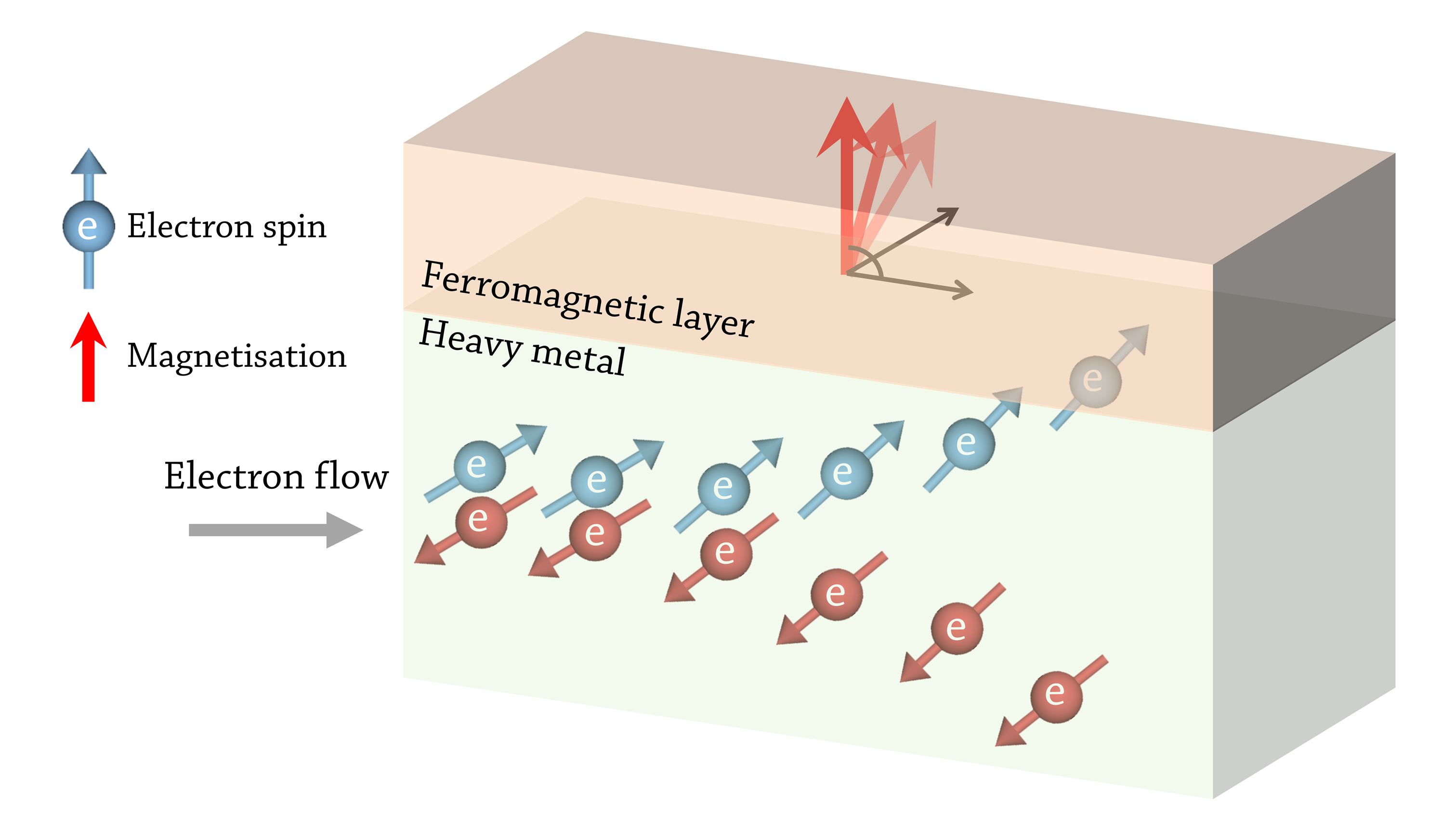}
    \caption[Diagram of domain wall motion induced by spin--orbit torque.]{Diagram of \gls{SOT}-induced \gls{DW} motion. \revisiontwo{In heavy metal/ferromagnetic heterostructures, the pronounced \gls{SOC} within the heavy metal converts the charge current into a spin current when an electric current passes through. This spin current is injected into the adjacent ferromagnetic layer, exerting a torque on the magnetisation. This torque ressults in the magnetisation switching and \gls{DW} motion.} \figref{\cite{ryu2020current}}} 
    \label{fig:SOT}
\end{figure}

Electric currents can induce \gls{SOT}, a phenomenon also capable of driving \gls{DW} motion. This torque arises from the pronounced \gls{SOC}. 
As shown in Figure~\ref{fig:SOT}, in heavy metal/ferromagnetic heterostructures, the pronounced \gls{SOC} within the heavy metal converts the charge current into a spin current when an electric current passes through. 
Subsequently, this spin current is injected into the adjacent ferromagnetic layer, exerting a torque on the magnetisation. 
\revisiontwo{The presence of this torque induces the rotation of magnetisation. Depending on the polarity of the interfacial \gls{DMI} (arises between the heavy metal and ferromagnetic layer) strength, the motion of the \gls{DW} is either parallel or anti-parallel to the direction of electron flow (indicated by the grey arrow in Figure~\ref{fig:SOT})~\cite{ryu2013chiral}.}
\gls{SOT} can facilitate magnetisation switching~\cite{li2019manipulation,garello2014ultrafast,safeer2016spin} and prompt rapid \gls{DW} motion within these heavy metal/ferromagnet systems~\cite{murray2019field,shiino2016antiferromagnetic,bhowmik2015deterministic,barker2023domain}. 

\revision{\gls{SOT}-driven \gls{DW} motion is more energy-efficient than that driven by \gls{STT}~\cite{roy2011hybrid}. \gls{STT} achieves charge-to-spin conversion through spin filtering in a ferromagnetic layer, capping efficiency at one. In contrast, \gls{SOT} uses multiple spin-orbit scatterings in a heavy metal layer, allowing for greater spin angular momentum transfer from each electron, with additional contributions from the atomic lattice of the heavy metal. Therefore, there is no fundamental limit to the charge-to-spin conversion efficiency~\cite{ryu2020current}.}\par

\subsubsection{Limitations}
\revisiontwo{Material with strong \gls{PMA} have a high energy barrier for reorientation of the magnetisation.} Therefore, a large current is required to reverse the magnetisation or to move \glspl{DW} in order to write or transfer data for both \gls{STT} and \gls{SOT}.
This high current can lead to energy wastage and Joule heating, imposing limitations on the packing density of practical devices~\cite{miron2011fast, yamaguchi2005effect,moore2008high}. For example, Miron \et demonstrated \gls{DW} motion reaching velocities of approximately 400 m/s in a Pt/Co/AlO\textsubscript{x} nano-wire when a current with a density of around 3.2$\times$10\textsuperscript{12} A/m\textsuperscript{2} was applied~\cite{miron2011fast}.
Therefore, there is much interest in manipulating spin structures without external magnetic field or electric current, for example, using strain.

\subsection{With static strain}
\begin{figure}[ht!]
    \centering
    \includegraphics[width=0.9\textwidth]{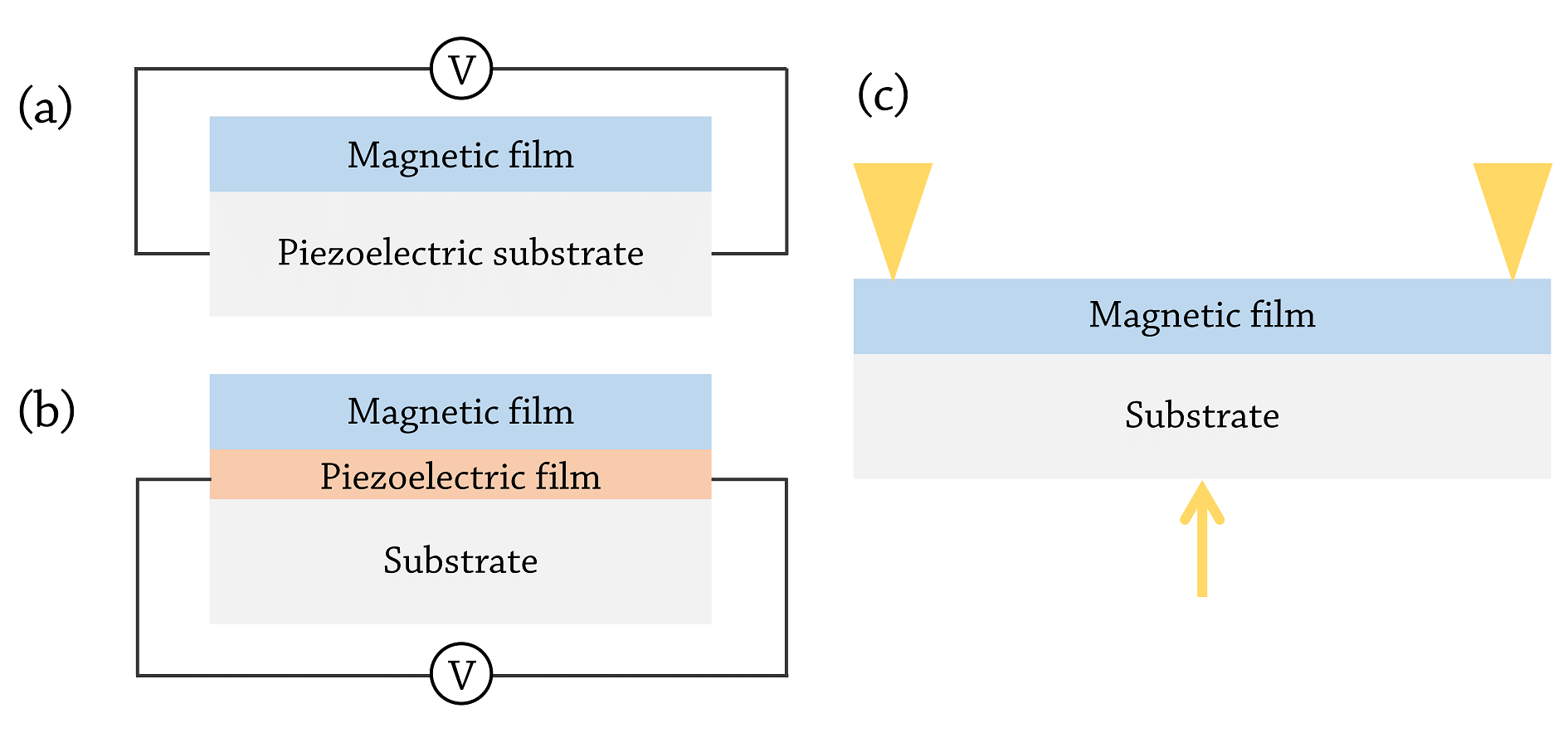}
    \caption[Methods of introducing static strain into magnetic thin films.] {Methods of introducing static strain into magnetic thin films: (a) Deposition of thin films onto a piezoelectric substrate followed by voltage application to the substrate. (b) Deposition of thin films onto a piezoelectric film and subsequent voltage application to the piezoelectric layer. (c) Application of mechanical strain via bending.}
    \label{fig:apply_strain}
\end{figure}

Owing to the magnetoelastic coupling effect, strain also serves as a method of manipulating to control thin film magnetism.
Both theoretical insights and experimental findings underscore that strain-driven \gls{DW} motion is more energy-efficient than that driven by electric current~\cite{roy2011hybrid,roy2011switching}. 
Various approaches can introduce static strain into magnetic thin films. One approach is by depositing the magnetic thin films onto a piezoelectric substrate or onto a piezoelectric thin film, and applying of voltage to the piezoelectric substrate or film facilitates strain generation (as shown in Figure~\ref{fig:apply_strain}a and b)~\cite{ba2021electric,shepley2015effect,masciocchi2023generation,fattouhi2022absence}. Another approach is mechanical bending, such as three-point bending as shown in Figure~\ref{fig:apply_strain}c~\cite{masciocchi2021strain,masciocchi2022control}.\par

Strain-induced modifications in magnetisation, magnetic anisotropy, and \gls{FMR} have been the focus of a wide range of studies spanning various magnetic systems~\cite{hunt2023strain,hunt2023Temperature,liu2009giant,shepley2015modification,ba2021electric}. 
To illustrate, Shepley \et modified Pt/Co/Pt thin films with \gls{PMA} by applying static strain using \gls{dc} voltage applied to a piezoelectric transducer. The static strain reduced the coercivity, and increased the \gls{DW} creep velocity by up to 100\%~\cite{shepley2015modification}.
De Ranieri \et utilised a piezoelectric stressor to induce a 5\% reduction in anisotropy in a perpendicularly magnetised GaMnAsP/GaAs ferromagnetic semiconductor, yielding a 500\% fluctuation in \gls{DW} mobility~\cite{de2013piezoelectric}. 
Gopman \et achieved over a 30\% reduction in the coercivity of Co/Ni multilayers by inducing a 0.1\% expansion in Pb(Zr\textsubscript{0.52}Ti\textsubscript{0.48})O\textsubscript{3}~\cite{gopman2016strain}.\par

Skyrmions are chiral magnetic spin structures, which can be found in thin films with \gls{PMA} that lack spatial inversion symmetry. 
\revision{Skyrmions can be stabilised by a competition between \gls{PMA} and the \gls{DMI}. Thus, the modification of \gls{PMA} can create or annihilate skyrmions.} 
For example, Ba \et demonstrated skyrmion creation, deformation and annihilation by modifying \gls{PMA} via strain~\cite{ba2021electric}.
Feng \et modified the anisotropy in Pt/Co/Ta multilayers to manipulate domain patterns~\cite{feng2021field}. By utilising the shape memory effect of a TiNiNb substrate, they successfully achieved strain-induced, field-free skyrmion nucleation and annihilation.\par

Significant advancements have been achieved in understanding the magnetoelastic coupling mechanism in the case of static strain-controlled magnetism. However, dynamic strain-controlled magnetism remains intricate and comparatively understudied. \par

\subsection{Through dynamic strain}
\begin{figure}[ht!]
    \centering
    \includegraphics[width=0.9\textwidth]{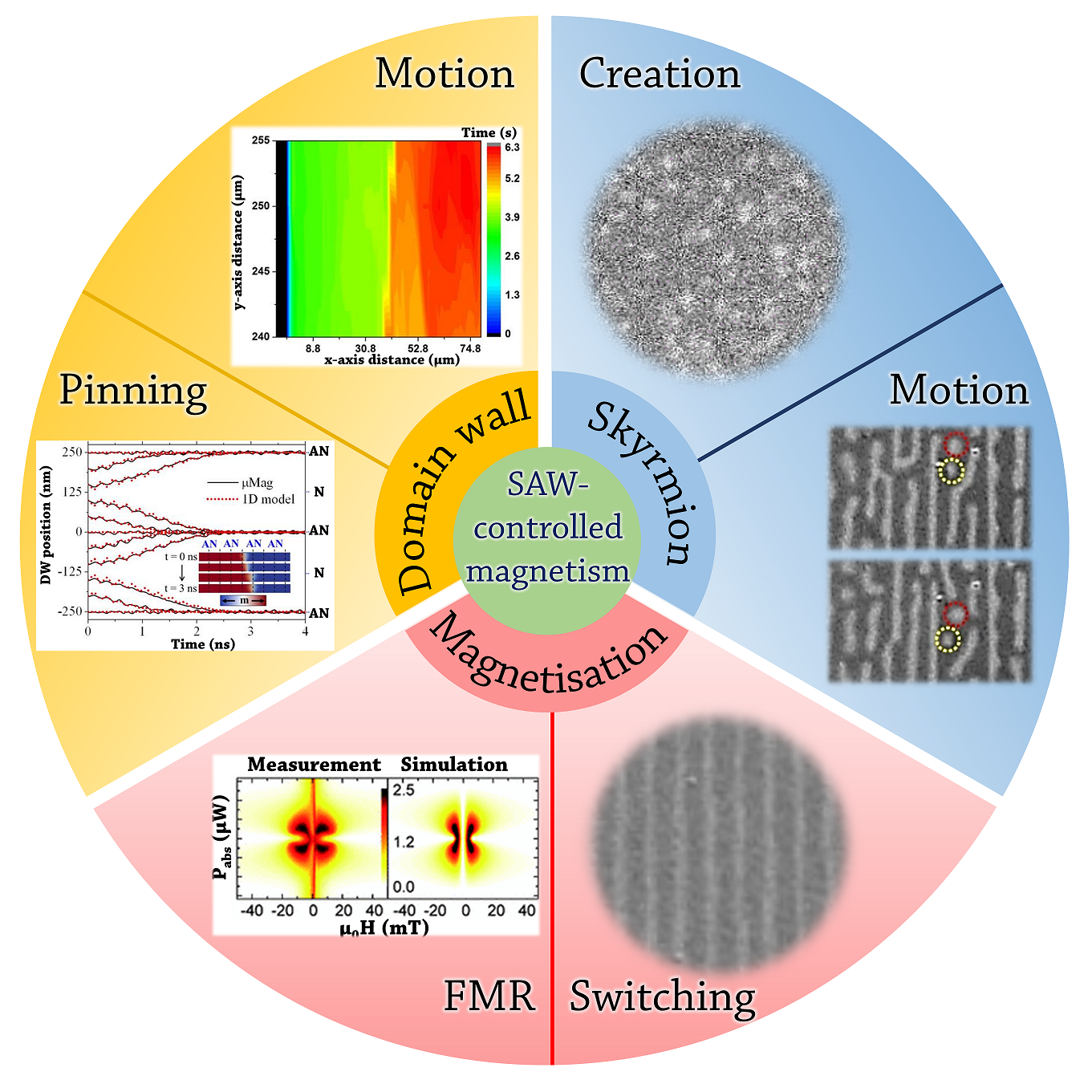}
    \caption[Applications of surface acoustic wave-controlled magnetism.]{Applications of \gls{SAW}-controlled magnetism. \figrefs{\cite{yokouchi2020creation,chen2023ordered,li2014writing,dreher2012surface,dean2015sound,edrington2018saw}}}
    \label{fig:SAW_control_magnetism}
\end{figure}

\glspl{SAW} are dynamic strain waves that travel along the surface of piezoelectric materials and are typically generated through the application of \gls{rf} signals to \glspl{IDT} that convert these signals to mechanical waves on a piezoelectric surface. 
Although the application of this technology is extensive in areas that include wireless communications~\cite{hays1976surface}, microfluidics~\cite{dung2010surface}, sensing~\cite{zida2021current,liu2016surface,mujahid2017surface} and television~\cite{weigel2002microwave}, the utilisation of \glspl{SAW} to introduce dynamic strain into magnetic thin films and thus to manipulate magnetism remains a relatively new and under-explored field~\cite{yang2021acoustic}.
\glspl{SAW} demonstrate promise in efficiently controlling magnetism, offering a broad spectrum of applications. These encompass magnetisation dynamics, as well as the manipulation of \glspl{DW} and skyrmions (as shown in Figure~\ref{fig:SAW_control_magnetism}).
For example, Davis \et showcased \gls{SAW}-induced magnetisation dynamics in 10-nm Co bars. Their findings aligned with the magnetoelastic free energy introduced by strain~\cite{davis2010magnetization}. Their research provided a proof-of-concept that magnetisation rotation triggered by dynamic strain from \glspl{SAW} can substantially alter the magnetisation direction.
Li \et presented experimental evidence for \gls{SAW}-assisted magnetic recording (as shown in Figure~\ref{fig:SAW_control_magnetism} ``Switching'' section). They utilised standing and focused \glspl{SAW} to reduce the coercivity in specific regions of 57-nm FeGa thin films~\cite{li2014acoustically,li2014writing}.\par

The periodic strain in both time and space introduced by \glspl{SAW} can change the anisotropy, which in turn exerts a torque on the magnetisation, triggering \gls{FMR}. For example, Weiler and Dreher with their colleagues demonstrated both experimentally and numerically that \gls{FMR} can be purely driven by \glspl{SAW} without an external \gls{rf} magnetic field being applied to a nickel thin film (see Figure~\ref{fig:SAW_control_magnetism} ``\gls{FMR}'' section)~\cite{weiler2011elastically,dreher2012surface}. 
Thevenard \et reported numerical prediction and experimental observation of \gls{SAW}-driven \gls{FMR} in (Ga,Mn)(As,P) epilayers between a temperature range of 5 to 85 K~\cite{thevenard2013irreversible,thevenard2014surface,thevenard2016precessional}. They also observed a strong coercivity reduction in these thin films~\cite{thevenard2016strong}.\par

\gls{SAW}-induced anisotropy changes also contribute to \gls{DW} motion~\cite{vilkov2022magnetic}. For instance, Dean \et presented theoretical evidence that standing \glspl{SAW} can solely drive \gls{DW} motion and create pinning sites remotely due to the \gls{SAW}-induced strain gradient (as shown in Figure~\ref{fig:SAW_control_magnetism} ``Pinning'' section). They demonstrated that, by adjusting the \gls{SAW} frequency, multiple \glspl{DW} can be synchronised and moved simultaneously~\cite{dean2015sound}. Edrington \et conducted experimental investigations showing that standing \glspl{SAW} effectively drive \gls{DW} motion from the creep regime to the flow regime in Co/Pt multilayers (see Figure~\ref{fig:SAW_control_magnetism} ``Domain wall motion'' section)~\cite{edrington2018saw}. Adhikari \et explored the impact of \glspl{SAW} on enhancing \gls{DW} motion within the creep regime by the \gls{SAW}-induced effective field and increasing the likelihood of \gls{DW} depinning~\cite{adhikari2021surface, adhikari2021surface2}.
Cao \et utilised \glspl{SAW} to assist \gls{SOT} magnetisation switching in a Pt/Co/Ta thin films resulting in a reduced critical current density and a higher \gls{DW} velocity~\cite{cao2021surface}.\par

Owing to the magnetoelastic coupling effect, \glspl{SAW} can also be used to create or annihilate skyrmions. Yokouchi \et experimentally observed the creation of skyrmions by \glspl{SAW} in a Pt/Co/Ir thin film owing to the inhomogeneous effective torque arising from both \glspl{SAW} and thermal fluctuations via magnetoelastic coupling (as shown in Figure~\ref{fig:SAW_control_magnetism} ``Skyrmion creation'' section)~\cite{yokouchi2020creation}. However, due to the high pinning energy of the thin film used in their study, \glspl{SAW} could not move skyrmions.
Nepal \et theoretically studied the dynamical pinning of skyrmion bubbles at the anti-nodes of standing \glspl{SAW} in a FePt nano-wire, revealing the mechanism of the \gls{SAW}-driven skyrmion motion~\cite{nepal2018magnetic}. 
Chen \et experimentally presented the ordered generation of magnetic skyrmions~\cite{chen2023ordered}. With the application of both electric current and \glspl{SAW}, they demonstrate the skyrmions motion with negligible skyrmion Hall-like motion (as shown in Figure~\ref{fig:SAW_control_magnetism} ``Skyrmion motion'' section). 
Using a similar concept, Chen \et theoretically demonstrated the suppression of skyrmion Hall-like motion via standing \glspl{SAW}~\cite{chen2023suppression}.
Using the energy potential created by two standing \glspl{SAW}, 
Miyazaki \et theoretically demonstrated the precise trapping of skyrmions. They showed the possibility to move skyrmions in a straight line by tuning the frequencies of the standing \glspl{SAW}~\cite{miyazaki2023trapping}.\par

\subsubsection{Advantages}
Utilising \glspl{SAW} to manipulate thin film magnetism brings distinct advantages compared to using magnetic field, electric current and static strain:
\begin{itemize}
    \item Technological maturity: \gls{SAW} technology is well-established and can be easily integrated with \gls{CMOS} systems. \Gls{IDT} design and fabrication are relatively straightforward.
    \item Energy efficiency: Generating \glspl{SAW} on specific powerful piezoelectric materials using transducers is efficient. These waves can propagate a few centimetres with minimal energy dissipation.
    \item Design simplification and precision: One pair of \glspl{IDT} can remotely establish pinning site arrays. This enables the precise positioning of numerous magnetic spin structures simultaneously within the \gls{SAW} beam path.
    \item Enhanced magnetisation dynamics: The dynamic strain introduced by \glspl{SAW} alters the magnetisation dynamics, offering a novel approach to control the magnetisation behaviour.
\end{itemize}

\subsubsection{Limitations}
\revision{However, controlling thin film magnetism with \glspl{SAW} also presents certain limitations, including:}
\begin{itemize}
    \item \revision{Cost and processing complexity: Although \gls{SAW} technology itself is mature, the piezoelectric substrates needed for the most efficient \gls{SAW} generation and propagation -- like lithium niobate and lithium tantalate -- are more expensive and can require more complex processing than silicon substrates.}
    \item \revision{Wake magnetoelastic coupling: While \glspl{SAW} can propagate over large areas with minimal attenuation, this implies that the energy transfer between the \gls{SAW} and the magnetic film is not particularly strong. Effective manipulation of magnetism with \glspl{SAW} requires the magnetic film to be sensitive to the relatively small strains produced by the \glspl{SAW}, which limited the choice of the magnetic materials. This weak coupling means that it is possible to control many devices simultaneously, the control over each might not be very strong or efficient.}
    \item \revision{Long wavelength: Typical \gls{SAW} frequencies range from tens of MHz to a few GHz, resulting in \gls{SAW} wavelengths ranging from hundreds of microns to hundreds of nanometres. Consequently, \glspl{SAW} are not suitable for controlling devices at the nanometres scale.}
    \item \revision{Limited material choices: The need for strong piezoelectric materials limits the choice of substrates for \gls{SAW} devices. Not all piezoelectric materials are equally effective, and some of the most efficient ones are not compatible with standard semiconductor processing, limiting their integration with other electronics.}
    
\end{itemize}

\subsubsection{Open questions}
Moreover, there are still scientific and engineering questions need to be addressed:
\begin{itemize}
    \item Magnetisation dynamics: The interplay between \glspl{SAW} and magnetisation dynamics, particularly in thin films with \gls{PMA}, needs further explanation. Specifically, the contributions of nodes and anti-nodes within standing \glspl{SAW} during the magnetisation reversal process \revision{needs a deeper understanding}.
    \item \gls{SAW} properties and magnetisation correlation: There exists an unclear correlation between the \gls{SAW} characteristics -- namely frequency, wavelength, and amplitude -- and the consequent influence on the magnetisation from these parameters needs to be elucidated.
    \item \gls{SAW} and \gls{DW} dynamics: Knowledge of the precise nature of the interaction between \glspl{SAW} and \glspl{DW} is incomplete. The potential perturbations introduced by intrinsic defects and disorders in thin films on \gls{SAW}-assisted \gls{DW} motion remains under-investigated for example. The influence of \gls{SAW} frequency on \gls{DW} dynamics, especially across films with different levels of disorder, also needs further investigation. 
    \item Thermal impact from \gls{rf} power: Potential thermal implications on devices pose another layer of complexity. The introduction of \gls{rf} power may increase internal device temperatures, thereby affecting their operational efficacy. Thus, quantifying thermal effects and understanding their influence on device performance are essential for ensuring robust and efficient device applications, and to ensure that the correct mechanism for interaction is confirmed.
\end{itemize}

\section{Thesis overview}
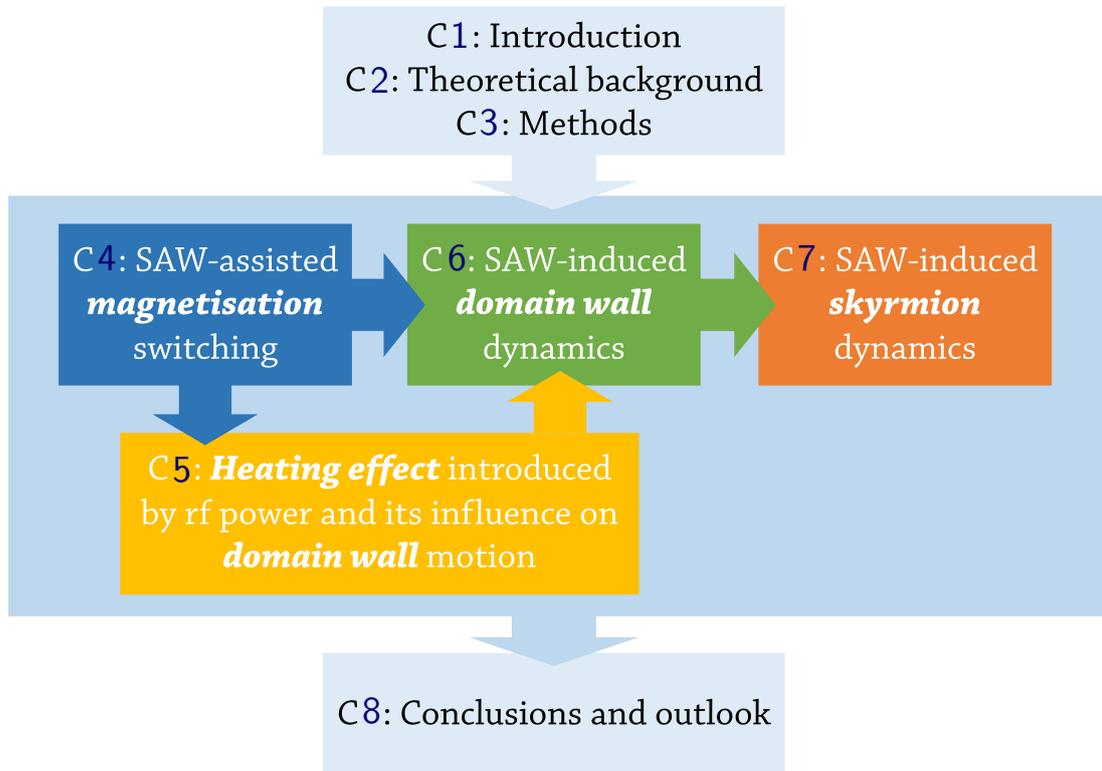
\begin{figure}[ht!]
    \centering
    \includegraphics[width=1\textwidth]{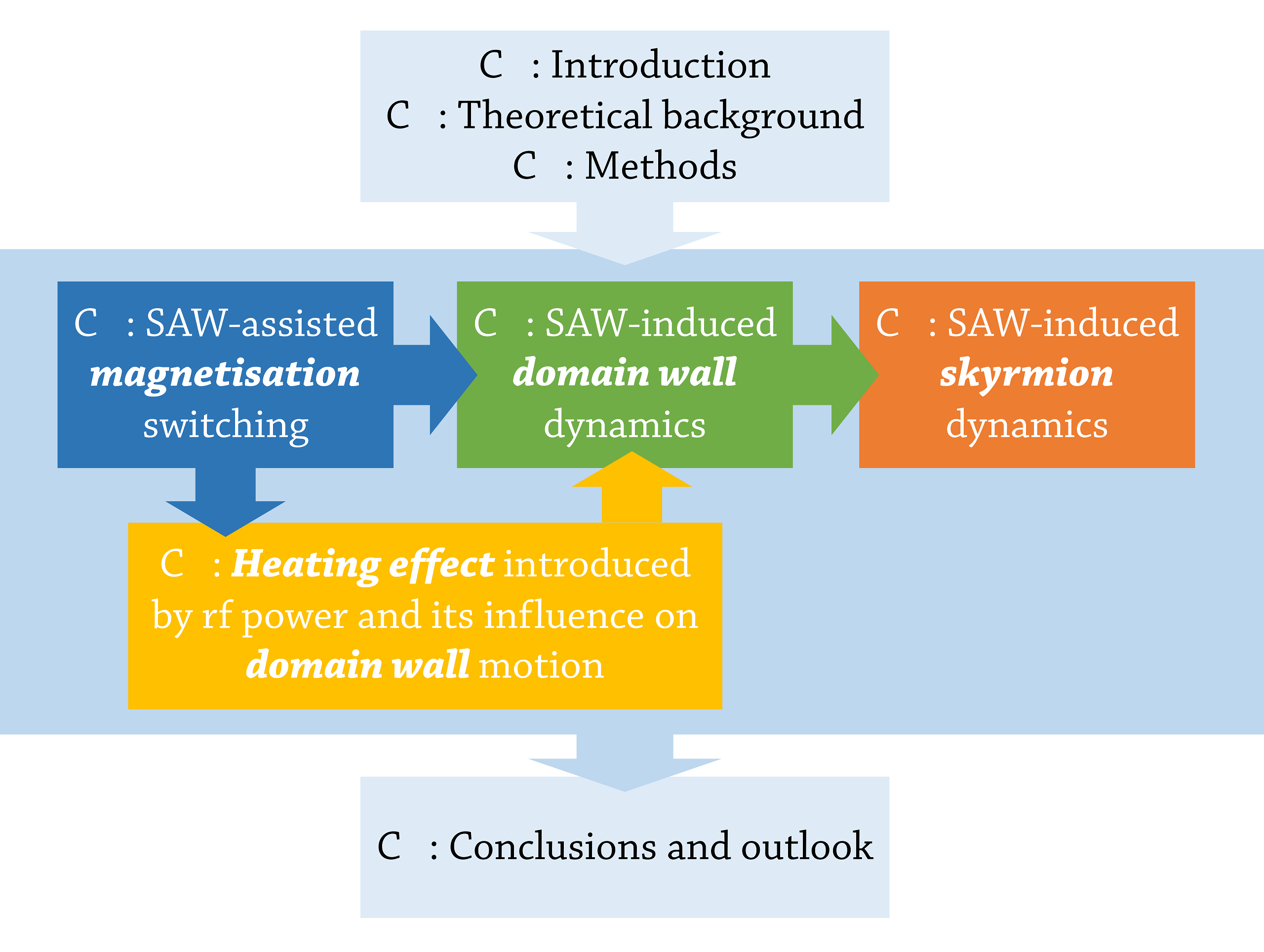}
\begin{picture}(0,0)
\put(-45,300){ \large \ref{Chapter1_introduction}}
\put(-75,283){ \large \ref{chapter2_theory}}
\put(-34,267){ \large \ref{chapter3_methods}}
\put(-177,216){ \large \ref{Chapter4_anisotropy_control}}
\put(-149,136){ \large \ref{Chapter5_heating_effect}}
\put(-46,216){ \large \ref{Chapter6_DW_dynamics}}
\put(85,216){ \large \ref{Chapter7_Skyrmion_motion}}
\put(-78,45){ \large \ref{chapter8_conclusion_and_outlook}}
\end{picture}
    \caption{Thesis overview.}
    \label{fig:thesis_overview}
\end{figure}

In this thesis, a comprehensive exploration of the interplay between \glspl{SAW} and magnetic thin films with \gls{PMA} is presented, encompassing both experimental and theoretical perspectives. A schematic outline of the thesis structure is shown in Figure~\ref{fig:thesis_overview}, along with the interactions between its constituent chapters.\par
\begin{itemize}[leftmargin=*]
    \item Chapter~\ref{Chapter1_introduction} sets the stage by introducing the background and motivations behind this project.
    \item Chapter~\ref{chapter2_theory} elucidates the foundational theoretical frameworks underpinning this study.
    \item Chapter~\ref{chapter3_methods} explains the experimental methodologies and computational method employed.
    \item Chapter~\ref{Chapter4_anisotropy_control} describes local magnetic anisotropy modification in Ta/Pt/Co/Ir/Ta thin films via standing \glspl{SAW} through both experiments and micromagnetic simulations.
    \item Chapter~\ref{Chapter5_heating_effect} provides the experimental findings on the thermal implications due to \gls{rf} power, revealing its origin and influence on magnetic properties.
    \item Chapter~\ref{Chapter6_DW_dynamics} shows a detailed numerical investigation of \gls{SAW} frequency impact on \gls{DW} motion, particularly within thin films with varied anisotropy disorder using micromagnetic simulations.
    \item Chapter~\ref{Chapter7_Skyrmion_motion} introduces an innovative approach for the precise transport of skyrmions along predefined trajectories via orthogonal \glspl{SAW} using micromagnetic simulations.
    \item Chapter~\ref{chapter8_conclusion_and_outlook} provides concluding remarks and offers ideas for the potential extension of this work.
\end{itemize}
\cleardoublepagewithnumberheader
\chapter{Theoretical background}
\label{chapter2_theory}
\section{Introduction}
This chapter focuses on the basic physical concepts that provide the theoretical foundations for this thesis, drawing inspiration from various textbooks and papers~\cite{coey2010magnetism,blundell2001magnetism,johnson1996magnetic,chikazumi1997physics,ballantine1996acoustic,campbell2012surface}. 
It provides a detailed discussion on the underlying theories and principles essential to understand the results presented in subsequent chapters.
We firstly discuss magnetisation dynamics in terms of free energy and the effective field in thin films with \gls{PMA}. 
We introduce the external magnetic field, the exchange interaction, the anisotropy, and the \gls{DMI} all of which are key contributors to the behaviour of magnetisation in this project.
Two spin structures, namely \glspl{DW} and skyrmions, are then discussed in detail. 
\glspl{DW} are the boundaries that separate two uniformly magnetised magnetic domains, while skyrmions are topologically protected particle-like spin structures that emerge in materials lacking inversion symmetry.
Finally, the excitation and propagation of \glspl{SAW} are described. 
\glspl{SAW} are strain waves that travel along the surface of a piezoelectric material, with an amplitude typically decaying exponentially into the substrate.
\glspl{SAW} can induce dynamic strain in magnetic thin films, resulting in the magnetoelastic coupling effect that influences magnetisation dynamics.
These concepts of magnetisation dynamics, spin structures, and \glspl{SAW} form the foundational pillars of this project, providing the necessary theoretical framework to comprehend and interpret the research findings. \par

In this study, we use a concise notation for representing vectors and their magnitudes. Bold fonts, such as $\B{V}$, indicate vectors. For the magnitude of vector $\B{V}$, we employ $V$ as opposed to the conventional $|\B{V}|$. This approach is particularly useful when graphing figures in relation to the magnitude of magnetic fields.

\section{Magnetisation dynamics}
\begin{figure}[ht!]
    \centering
    \includegraphics[width=0.8\textwidth]{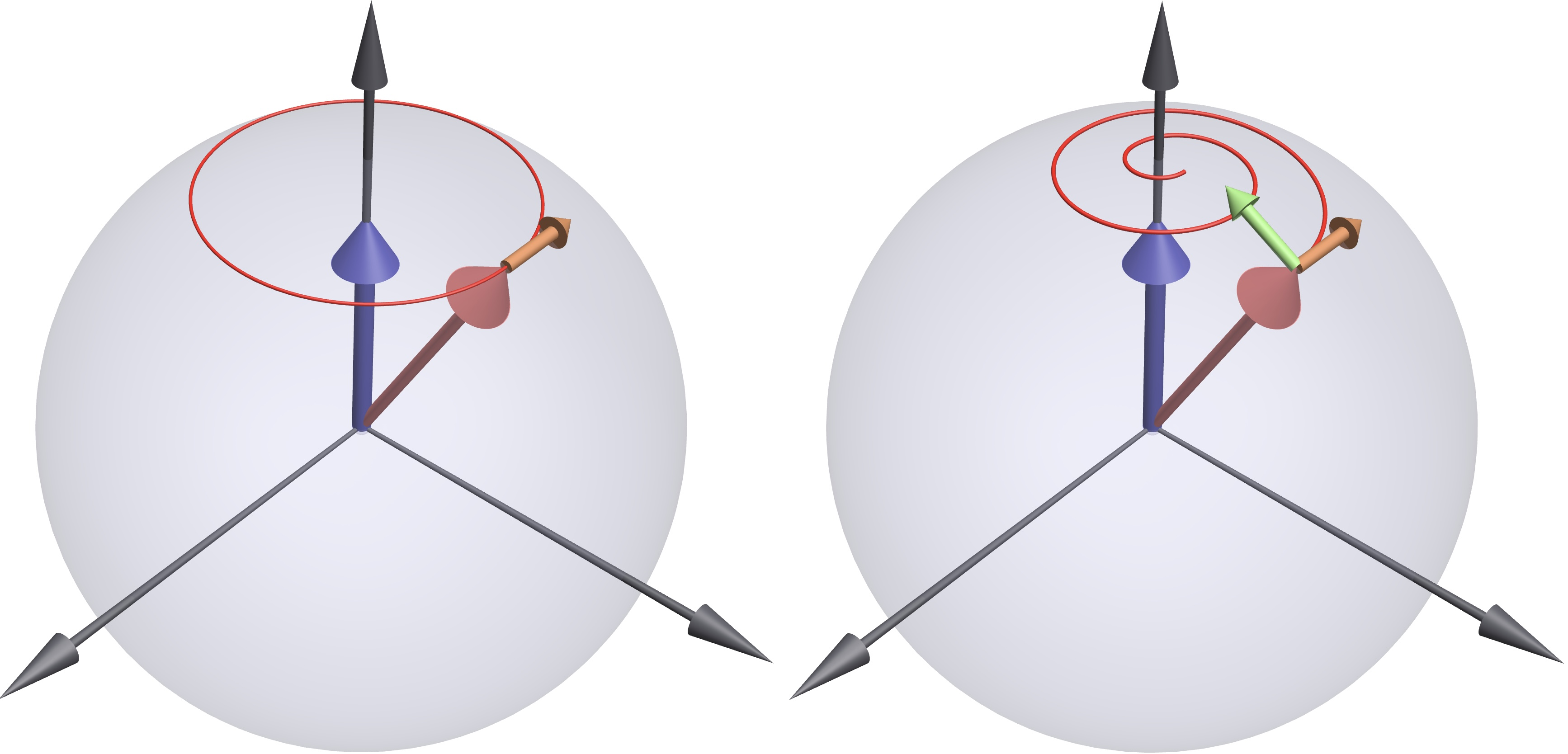}

\begin{picture}(0,0)
\put(-180,163){\small (a)}
\put(0,163){\small (b)}

\put(-167,18){\large $x$}
\put(-9,25){\large $y$}
\put(-83,168){\large $z$}
\put(-120,120){\textcolor[HTML]{00008B}{\large $\B{H}_\RM{eff}$}}
\put(-57,103){\textcolor[HTML]{840000}{\large $\B{M}$}}
\put(-45,123){\textcolor[HTML]{cc5400}{\large $\frac{\RM{d}\B{M}}{\RM{d}t}$}}

\put(0,18){\large $x$}
\put(158,25){\large $y$}
\put(85,168){\large $z$}
\put(48,120){\textcolor[HTML]{00008B}{\large $\B{H}_\RM{eff}$}}
\put(110,103){\textcolor[HTML]{840000}{\large $\B{M}$}}
\put(123,123){\textcolor[HTML]{cc5400}{\large $\frac{\RM{d}\B{M}}{\RM{d}t}$}}
\put(80,138){\textcolor[HTML]{09cc07}{\small $\B{M}\times\frac{\RM{d}\B{M}}{\RM{d}t}$}}

\end{picture}

    \caption[Magnetic moment precesses around the effective field.] {Magnetic moment precesses around the effective field $\B{H}_\RM{eff}$: (a) with a constant polar angle, and (b) with a spiral trajectory due to the damping.}
    \label{fig:llg}

\end{figure}
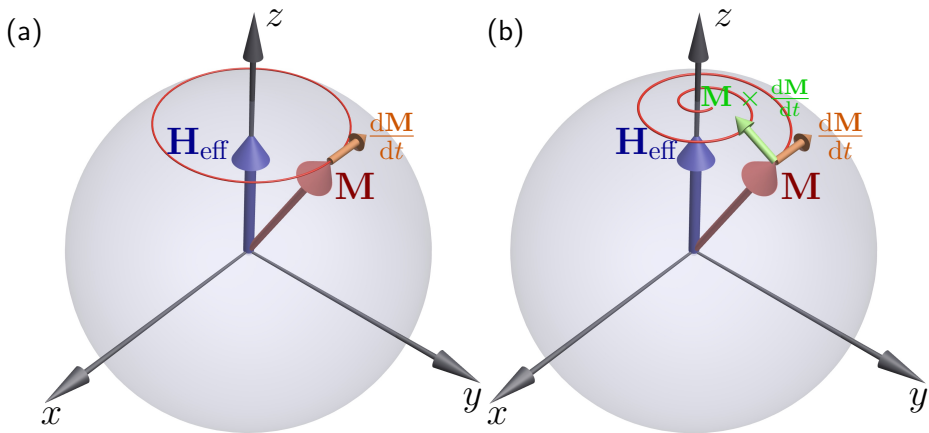

\define{Magnetisation}, $\B{M}$, in ferromagnetic materials are inclined to align with a local field to minimise the energy of the system. 
This local field, known as the \define{effective field} $\B{H}_\RM{eff}$, governs the dynamics of the magnetic moments at any position $\B{r}(x,y,z)$. 
The magnetisation dynamics can be described by
\begin{equation}
\frac{\RM{d}\B{M}(\B{r})}{\RM{d}t} = -\gamma \B{M}(\B{r}) \times \B{H}_\RM{eff}(\B{r}) ,
\label{eqn:LL}
\end{equation}
where $\gamma$ represents the gyromagnetic ratio. 
As a result, the magnetic moments undergo perpetual precession at a constant angle around the effective field (as shown in Figure~\ref{fig:llg}a). \par

However, experimental observations demonstrate that in the presence of an external magnetic field, the magnetisation tends to align predominantly parallel to the direction of that external field. 
To reconcile this discrepancy, a phenomenological dissipative term is introduced into the $\B{H}_\RM{eff}$ in Equation~\ref{eqn:LL}. This dissipative term is given by
\begin{equation}
    -\frac{\alpha}{\gamma M_\RM{s}}\frac{{\RM{d}\B{M}}}{{\RM{d}t}} ,
    \label{eqn:dissipative_term}
\end{equation}
where $\alpha$ represents the Gilbert damping constant, and $M_\RM{s}$ is the saturation magnetisation. 
We have dropped the explicit dependence of the vector $\B{r}$ from $\B M(\B{r})$ to keep the notation more compact. We will use this shorthand throughout the thesis.
By incorporating this dissipative term (Equation~\ref{eqn:dissipative_term}) to $\B{H}_\RM{eff}$ in Equation~\ref{eqn:LL}, we obtain the well-known \gls{LLG} equation
\begin{equation}
    \frac{{\RM{d}\B{M}}}{{\RM{d}t}} = -\gamma \B{M} \times \B{H}_\RM{eff} + \frac{\alpha}{M_\RM{s}}\B{M}  \times \frac{{\RM{d}\B{M}}}{{\RM{d}t}}.
    \label{eqn:LLG}
\end{equation}\par
The \gls{LLG} equation accounts for the damping effect, causing the magnetic moments to precess in a spiral path towards their equilibrium position, as shown in Figure~\ref{fig:llg}b. This phenomenological adjustment delivers a more precise depiction of the magnetisation dynamics, aligning better with experimental observations.\par

The \gls{LLG} equation describes the dynamics of magnetic moments in a ferromagnetic material. 
We define \define{reduced moment} by $\B{m} = \B{M}/M_\RM{s}$, which follows the direction of the spin. 
The effective field ($\B{H}_\RM{eff}$) can be expressed as the first derivative of the total energy density with respect to the magnetisation, given by
\begin{equation}
    \B{H}_\RM{eff} = -\frac{1}{\mu_0 M_\RM{s}} \frac{\delta E_\RM{total}}{\delta \B{m}} ,
    \label{eqn:effecttive_field}
\end{equation}
where $\mu_0$ is the permeability of vacuum and $E_\RM{total}$ represents the total internal free energy density of the system. This free energy density, $E_\RM{total}$, encompasses different energy contributions, depending on the material properties, magnetisation configurations, and external conditions. 
In most systems, $E_\RM{total}$ can be obtained by
\begin{equation}
    E_\RM{total} = E_\RM{exch} + E_\RM{ani} + E_\RM{me} + E_\RM{DMI} + E_\RM{Zeeman}.
\end{equation}
The elements in this equation represent different energy contributions: exchange, anisotropy, magnetoelastic coupling, \gls{DMI}, and external field. Together, these energy terms govern the behaviour of the magnetic moments within the ferromagnetic medium. We will examine each of these energy terms in the following sections.
\section{Free energy and effective field}
\subsection{Exchange}
In ferromagnetic materials, both the magnetic moment and spontaneous magnetisation are consequences of the exchange interaction among electrons, which is a quantum mechanical phenomenon. Magnetic moments inherently tend to align with each other, resulting in spin ordering that extends over macroscopic distances.\par

Consider an infinite lattice composes of spins located at positions $\{\B{r}_i\}_{i}$. Each site located at $\B{r}_i$ possesses a spin represented by $\B{S}_i$. Adhering to the Heisenberg model, the interaction Hamiltonian for the site $\B{r}_i$ is given by
\begin{equation}
\hat{\mathcal{H}}_i = -2 \sum_{j} J_{i,j} \B{S}_i \cdot \B{S}_j,
\label{eqn:HH}
\end{equation}
where the above sum is over the entire lattice and $J_{i,j}$ is the exchange constant. \par

While Equation~\ref{eqn:HH} represents the most general form of the interaction Hamiltonian, the Heisenberg model introduces two additional key assumptions. Firstly, it is assumed that all lattice points are identical in terms of their physical properties. Secondly, the interactions between spins are considered to be short-ranged. The interactions in the model are, in fact, restricted to only nearest neighbours. Let $\mathcal{N}(i)$ be the set of all the nearest neighbours of the site $i$, then the assumptions taken lead us to the following form for $J_{i,j}$
\begin{equation}
    J_{i,j} =
    \begin{cases}
        J & \text{if} \ j \in \mathcal{N}(i), \\
        0 & \text{otherwise.}
    \end{cases}
\end{equation}
It is further reasonable to assume that the two spins that are nearest neighbours are almost parallel to each other. Thus, of the the angle $\phi_{ij}$ between two adjacent spins $S_{i}$ and $S_{j}$ can be taken to be small. Incorporating this assumption gives
\begin{equation}
    E_i = -2 J S^2 \sum_{j \in \mathcal{N}(i)} \cos{\phi_{ij}} =  \RM{constant} + J S^2 \sum_{j \in \mathcal{N}(i)} {\phi_{ij}}^2 ,
\end{equation}
where the last equality in holds using $\cos \phi_{ij} \approx 1- \phi_{ij}^2/2$ for $\phi_{ij} \ll 1$.\par

\begin{figure}[ht!]
    \centering
    \includegraphics[width=0.5\textwidth]{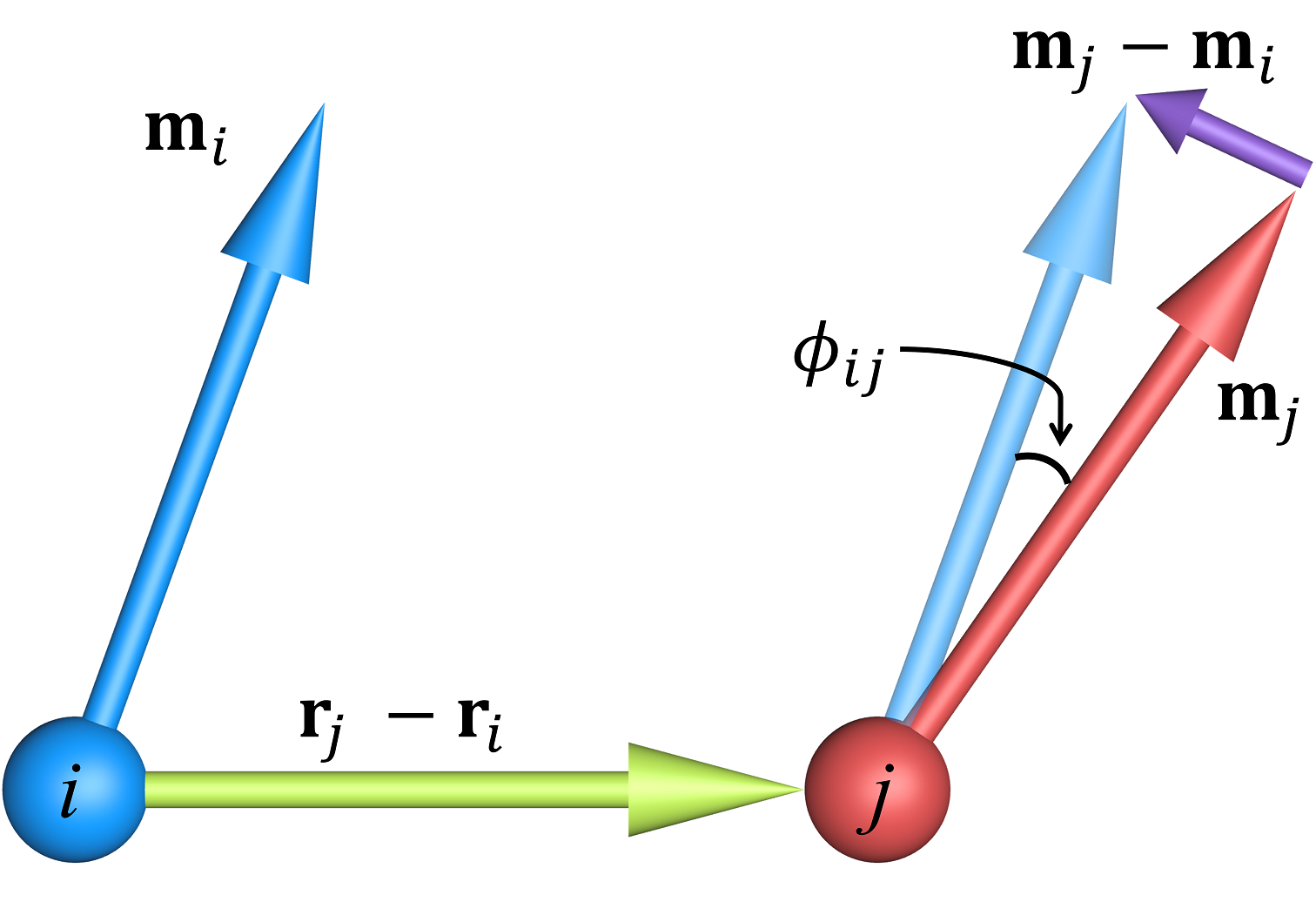}   
    \caption[Magnetic moments at the nearest neighbours $i$ and $j$.]{Magnetic moments at the nearest neighbours $i$ and $j$ represented by the reduced moments $\B{m}_i$ and $\B{m}_j$. The angle between the two moments is $\phi_{ij}$. \figref{\cite{blundell2001magnetism}}}
    \label{fig:SiSj}
\end{figure}

The value of $|\phi_{ij}|$ can be approximated in terms of the reduced moments $\B{m}_i$ and $\B{m}_j$ (see Figure~\ref{fig:SiSj}) as
\begin{equation}
     \left| \B{m}_i - \B{m}_j \right| \approx \left| \big( \left( \B{r}_i - \B{r}_j\right) \cdot  \nabla \big)\B{m}(\B{r}_i)\right| ,
\end{equation}
thus the energy can be rewritten as
\begin{equation}
    E_i = \RM{constant} + JS^2 \sum_{j \in \mathcal{N}(i)}\bigg( \big( \left( \B{r}_{i} - \B{r}_{j} \right) \cdot  \nabla \big)\B{m}(\B{r}_i) \bigg)^2 .
\end{equation}

At the scales we are working with, the discrete lattice is can be effectively treated as a continuum. Taking this continuum approximation, we can write the exchange energy of the system (upto an constant term, which represents the energy of the system when all the spins in the material are fully aligned) as
\begin{equation}
    E_\RM{exch} = A_\RM{exch} \int_V \left( \left(\nabla m_x\right)^2 + \left(\nabla m_y\right)^2 + \left(\nabla m_z\right)^2 \right)\RM{d}^3r , 
    \label{eqn:ED_exch}
\end{equation}
where $A_\RM{exch}$ is the exchange stiffness constant given by
\begin{equation}
    A_\RM{exch} = 2 J S^2 z a_0 .
\end{equation}
In this equation, $z$ is the number of sites in the unit cell and $a_0$ is the distance between two nearest neighbours (lattice parameter). 
Plug Equation~\ref{eqn:ED_exch} into Equation~\ref{eqn:effecttive_field}, the exchange contribution to the effective field thus is 
\begin{equation}
    \B{H}_\RM{exch} = \frac{2A_\RM{exch}}{\mu_0M_\RM{s}} \nabla^2 \B{m} .
    \label{eqn:H_exch}
\end{equation}

\subsection{Magnetic Anisotropy}

Magnetic materials exhibit specific directions known as the ``easy'' and ``hard'' directions of magnetisation. The magnetic moments naturally prefer to align in the easy direction, and to maintain a configuration where they do not align in this direction, additional energy is required. This energy variation required to magnetise a material depending on the direction of the applied field is referred to as \define{magnetic anisotropy}. The presence of uniaxial magnetic anisotropy gives rise to an extra energy term, which only relies on the angle $\theta_0$ between the magnetic moments and the easy axis $\hat{\B{u}}$.  The anisotropy energy density, $E_\RM{ani}$, up to the leading order in $\theta_0$ can be expressed as follows
\begin{equation}
    E_\RM{ani} = K \sin^2{\theta_0} =  K\left( 1- \left( \B{m} \cdot \hat {\B{u}} \right)^2 \right) ,
    \label{eqn:anisotropy_energy_high}
\end{equation}
where $K$ is the first order anisotropy constant, and $\B{m} \cdot \hat {\B{u}} = \cos \theta_0$.
The effective field that associates with anisotropy thus can be expressed as
\begin{equation}
    \B{H}_\RM{ani} = \frac{2K}{\mu_0 M_\RM{s}} \left( \B{m} \cdot \hat{\B{u}}\right)\hat{\B{u}}.
\end{equation}

\subsubsection{Magnetocrystalline and shape anisotropy}
Two primary sources contribute to magnetic anisotropy: spin--orbit interaction and magnetic dipolar coupling. 
Spin--orbit interaction is characterised by a coupling between the magnetic moment and the orbital momentum, leading to an energy dependence on the orientation of magnetisation relative to the crystal axes. This phenomenon is referred to as \define{magnetocrystalline anisotropy}, as it reflects the crystal inherent symmetry. By applying strain to the material, the magnetocrystalline anisotropy can be modified, resulting in what is known as \define{magnetoelastic anisotropy}. \par

The dipolar interaction, or \define{shape anisotropy}, is influenced by the shape of the sample because of its long-range character~\cite{coey2010magnetism}. If the magnetisation is uniformly aligned within a material, magnetic charges will appear at the surfaces, creating dipolar energy. This is especially significant for in-plane magnetisation in thin films. The energy density of shape anisotropy \revision{in thin films $E_\RM{sh}$} can be expressed as follows
\begin{equation}
E_\RM{sh} = \frac{1}{2} \mu_0 M_\RM{s}^2 \cos^2{\theta_0} ,
\end{equation}
\revision{where recall that $\theta_0$ represents the angle between the magnetic moment and the easy axis. Shape anisotropy is associated with the demagnetising field, which is the magnetic field generated by the magnetic material.}
\subsubsection{Magnetoelastic anisotropy}
When a ferromagnetic material is magnetised, it undergoes a phenomenon known as \define{magnetostriction}, where its size experiences a slight change due to crystal deformation. This change in size reduces the anisotropy energy of the material. On the contrary, any deformation occurring in the crystal structure can also impact the anisotropy energy of the material. This connection between magnetic and elastic behaviour is commonly referred to as \define{magnetoelastic coupling}.
The energy density of the magnetoelastic anisotropy $E_\RM{me}$ associated with this effect can be written as
\begin{equation} \label{eqn:E_me}
    E_\RM{me} = B_1\sum_{i = x, y, z}m_i^2\epsilon_{ii} + B_2\sum_{i \neq j}m_im_j\epsilon_{ij},
\end{equation}
where $B_1$ and $B_2$ represent the magnetoelastic coefficients, $\epsilon_{ij}$ denotes the strain tensor, $m$ is the normalised magnetisation, $i$ and $j$ are the Cartesian components ($x,y,z$).
The effective field introduced by the magnetoelastic interaction thus can be expressed as
\begin{equation}
\B{H}^{(i)}_\RM{me} = -\frac{2}{\mu_0 M_\RM{s}} \left(B_{1} m_{i} \epsilon_{ii} + B_{2} \sum_{j: j \neq i} m_{j} \epsilon_{ij} \right),
\end{equation}
where $\B{H}^{(i)}_\RM{me}$ is the component of the effective field along the axis labeled by $i$.

\subsubsection{Perpendicular magnetic anisotropy}
\define{Perpendicular magnetic anisotropy} refers to the magnetic material having an easy axis perpendicular to its surface. The \gls{PMA} results from the magnetic anisotropy at an interface, which is different from the magnetic anisotropy in the bulk.
\Gls{PMA} is typically observed in thin films or multilayers that consist of magnetic materials with specific structural properties and interface conditions. 
It is commonly found in materials composed of transition metals combined with heavy elements or alloys with strong \gls{SOC}. 
The magnetic anisotropy of the thin film $K$ can be described by the volume and surface contributions $K_\RM{v}$ and $K_\RM{s}$ as
\begin{equation}\label{K_eff}
    K= K_\RM{v}+\frac{2K_\RM{s}}{d} ,
\end{equation}
where $d$ is the thickness of the magnetic layer in a thin film. \revision{$K_\RM{v}$ includes shape anisotropy, magnetocrystalline anisotropy, and magnetoelastic anisotropy contributions, with shape anisotropy being the dominant factor in this contribution.} 

\subsection{Interfacial Dzyaloshinskii--Moriya interaction}
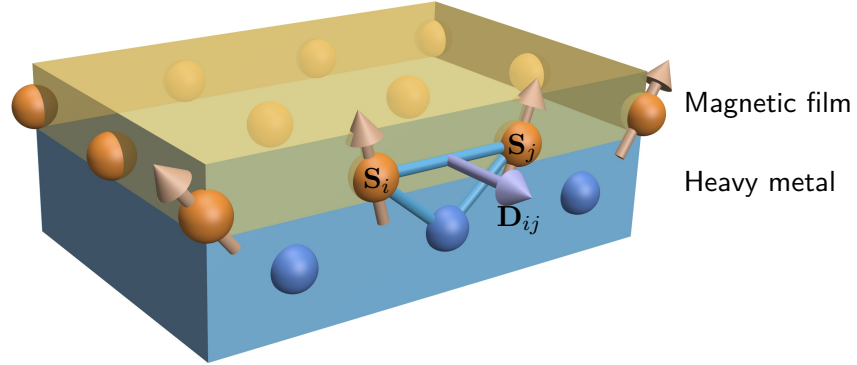
\begin{figure}[ht!]
    \centering
    \includegraphics[width=.6\textwidth]{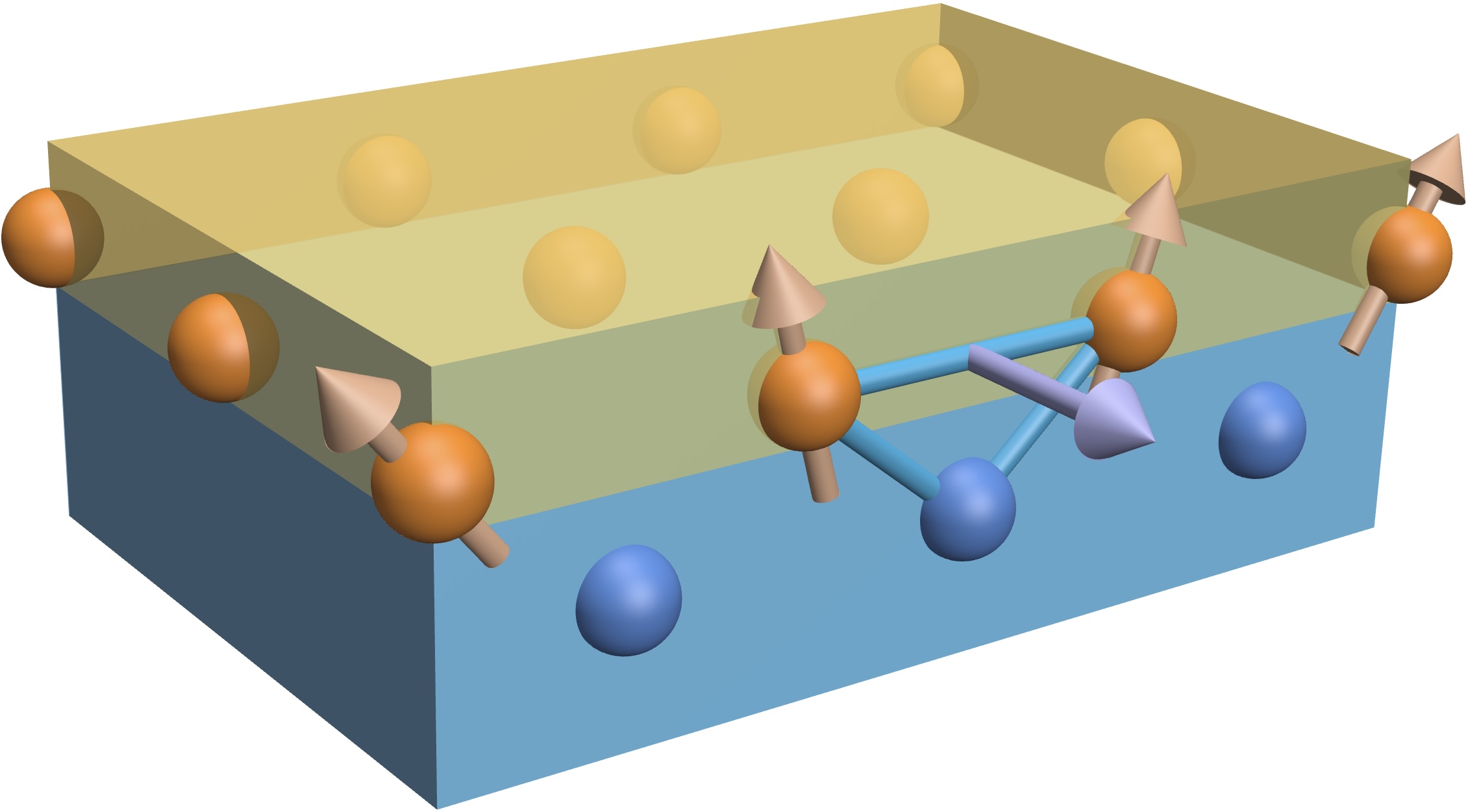}
    \caption[Schematic diagram illustrating the interfacial Dzyaloshinskii--Moriya interaction at the interface between a magnetic film and a heavy metal.]{Schematic diagram illustrating the interfacial \gls{DMI} at the interface between a magnetic film and a heavy metal. $\B{S}_i$ and $\B{S}_j$ are the spins of neighbouring atoms, and $\B{D}_{ij}$ is the corresponding Dzyaloshinskii--Moriya vector. \figref{\cite{fert2017magnetic}}}
\begin{picture}(0,0)
\put(10,186){\small $\B{S}_{i}$}
\put(64,199){\small $\B{S}_{j}$}
\put(60,172){\small $\B{D}_{ij}$}
\put(130,215){\small Magnetic film}
\put(130,185){\small Heavy metal}
\end{picture}
    \label{fig:DMI}
\end{figure}

The \revision{interfacial} \acrfull{DMI} arises in magnetic thin films that lack inversion symmetry. The Hamiltonian term acting between two adjacent spins $\B{S}_i$ and $\B{S}_j$ can be expressed as
\begin{equation}
    \mathcal{\hat H}_\RM{DMI}=(\B{S}_i \times \B{S}_j)\cdot \B{D}_{ij} ,
\end{equation}
where $\B{D}_{ij}$ denotes the Dzyaloshinskii--Moriya vector (as illustrated in Figure~\ref{fig:DMI}). The Dzyaloshinskii--Moriya vector can be written as
\begin{equation}
    \B{D}_{ij} = D_{ij}(\hat{\B{z}}_0 \times \hat{\B{r}}_{ij}) ,
\end{equation}
where $D_{ij}$ is the \revision{interfacial} \gls{DMI} strength, and $\hat{\B{z}}_0$ and $\hat{\B{r}}_{ij}$ are unit vectors perpendicular to the interface and pointing in the \revision{plane} of the magnetic layer and from site $i$ to site $j$ (i.e., unit vector in the direction of the vector $\B{r}_{j} - \B{r}_{i} $), respectively.

The sign of the $D_{ij}$ value determines whether the \gls{DMI} favours anticlockwise or clockwise rotations from $\B{S}_i$ to $\B{S}_j$. The rotation direction, in turn, influences whether the \gls{DMI} lowers or increases the spin energy. A strong \gls{DMI} induces a spin tilt around the $\B{D}_{ij}$ vector. At the interface between the magnetic material and heavy metal, the Dzyaloshinskii--Moriya vector lies within the film plane as shown in Figure~\ref{fig:DMI}. The energy density term due to the presence of \gls{DMI} is
\begin{equation}
    E_\RM{DMI} = D \big( \left( \B{m} \cdot \nabla\right)m_z - \left(\nabla \cdot \B{m} \right)m_z \big),
\end{equation}
where $D$ is the average \gls{DMI} strength. The \gls{DMI}-induced effective field can be expressed by
\begin{equation}
    \B{H}_\RM{DMI} = \frac{2D}{\mu_0 M_\RM{s}} \big( \left( \nabla \cdot \B{m} \right)\hat{\B{z}}_0 - \nabla m_z\big) .
\end{equation}

\subsection{Zeeman energy}
Ferromagnetic materials respond to an external magnetic field $\mathbf{H}_\mathrm{ext}$, defining the magnetisation process and hysteresis loop. The energy associated with the external magnetic field is known as the \define{Zeeman Energy}. The Zeeman energy density is given by the equation
\begin{equation}
  E_\RM{Zeeman} = -\mu_0 \B{M} \cdot \B{H}_\RM{ext} .
\end{equation}
The effective field associated with Zeeman energy is simply $\B{H}_\RM{eff} = \B{H}_\RM{ext}$.

\section{Magnetisation reversal}
\begin{figure}[ht!]
    \centering
    \includegraphics[width=0.9\textwidth]{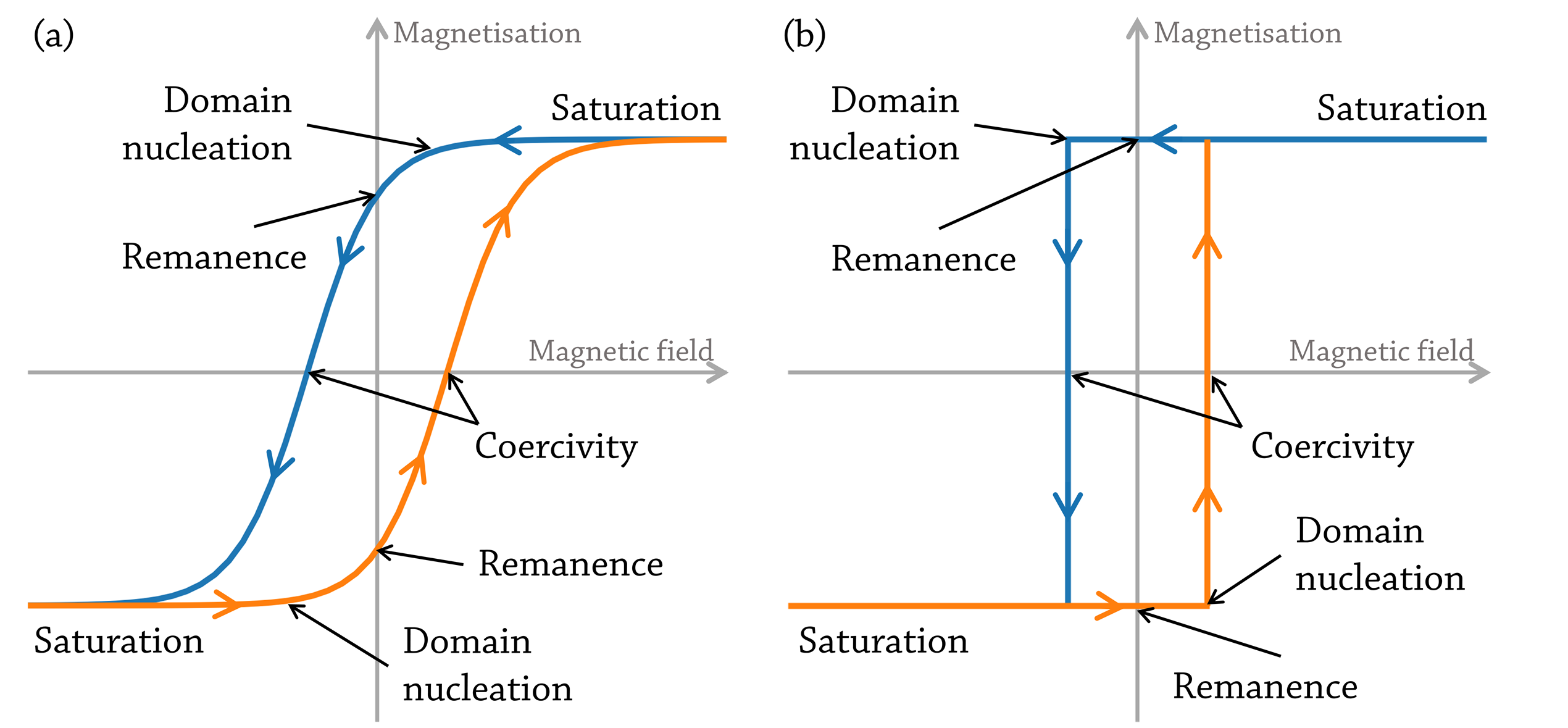}
    \caption[Typical hysteresis loops of a ferromagnetic material and a thin film with perpendicular magnetic anisotropy.]{\revision{(a) Typical hysteresis loop of a ferromagnetic material. (b) Typical hysteresis loop of a thin film with \gls{PMA}. The direction of the arrows shows the increasing or decreasing trend of the applied magnetic field.}}
    \label{fig:typical_HL}
\end{figure}

\revision{The process of magnetisation reversal often involves the nucleation and propagation of magnetic domains, where the magnetisation aligns uniformly.}
The process of magnetisation reversal in thin films exhibits several distinct characteristics. 
The nucleation field marks the point at which domains nucleate and the magnetisation begins to reverse. 
Remanence refers to the remaining proportion of the saturation magnetisation when no external field is applied. The coercivity is the strength of the field required for the average magnetisation in the film to reach zero. In the case of thin films with PMA, after the nucleation of domains, a propagation field is necessary to sustain the motion of DWs. This sequential process contributes to the complete magnetisation reversal in the thin film.
\revision{A hysteresis loop is a graphical representation that illustrates the relationship between the magnetisation in a ferromagnetic material and the magnetic field strength applied to it.}
\revision{Figure~\ref{fig:typical_HL}a shows a typical hysteresis loop for a ferromagnetic material. If a material is magnetised to its saturation magnetisation by an external magnetic field, then when the applied field is reduced to zero, the magnetisation of the material reduces to remanence. A magnetic field with the strength of the coercivity is required to switch the magnetisation into the opposite direction. During this process, the magnetisation starts to reduce at a particular field strength, marking the nucleation of domains.
In thin films with \gls{PMA} shown in Figure~\ref{fig:typical_HL}b, magnetisation remains constant as the external field is reduced to zero, indicating a remanence that is similar as the saturation magnetisation.
The magnetisation retains its value until the magnetic field is decreased further, resulting in a sharp switching at the domain nucleation field, which is comparable in magnitude to the coercivity.}

\subsection{Domain and domain wall}
\begin{figure}
    \centering
    \includegraphics[width=1\textwidth]{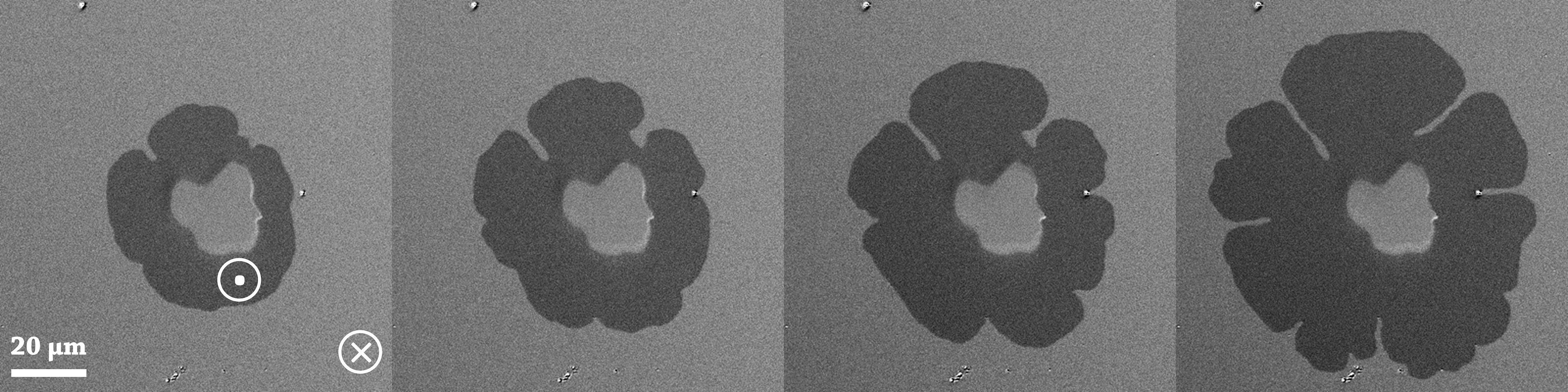}
    \caption[Domain and domain wall motion with an applied out-of-plane field in a thin film with perpendicular magnetic anisotropy.]{Domain and \gls{DW} motion with an applied out-of-plane field in a Ta(5.0 nm)/Pt(2.5 nm)/Co(0.6 nm)/Pt(2.5 nm)/Ta(5.0 nm) thin film with \gls{PMA}. \revisiontwo{The light grey area in the centre is the region where the thin film peeled off}. The darker grey ring is magnetisation pointing out of the plane, and the light grey pointing into the plane.}
    \label{fig:DWmotion}
\end{figure}

\begin{figure}
    \centering
    \includegraphics[width=.7\textwidth]{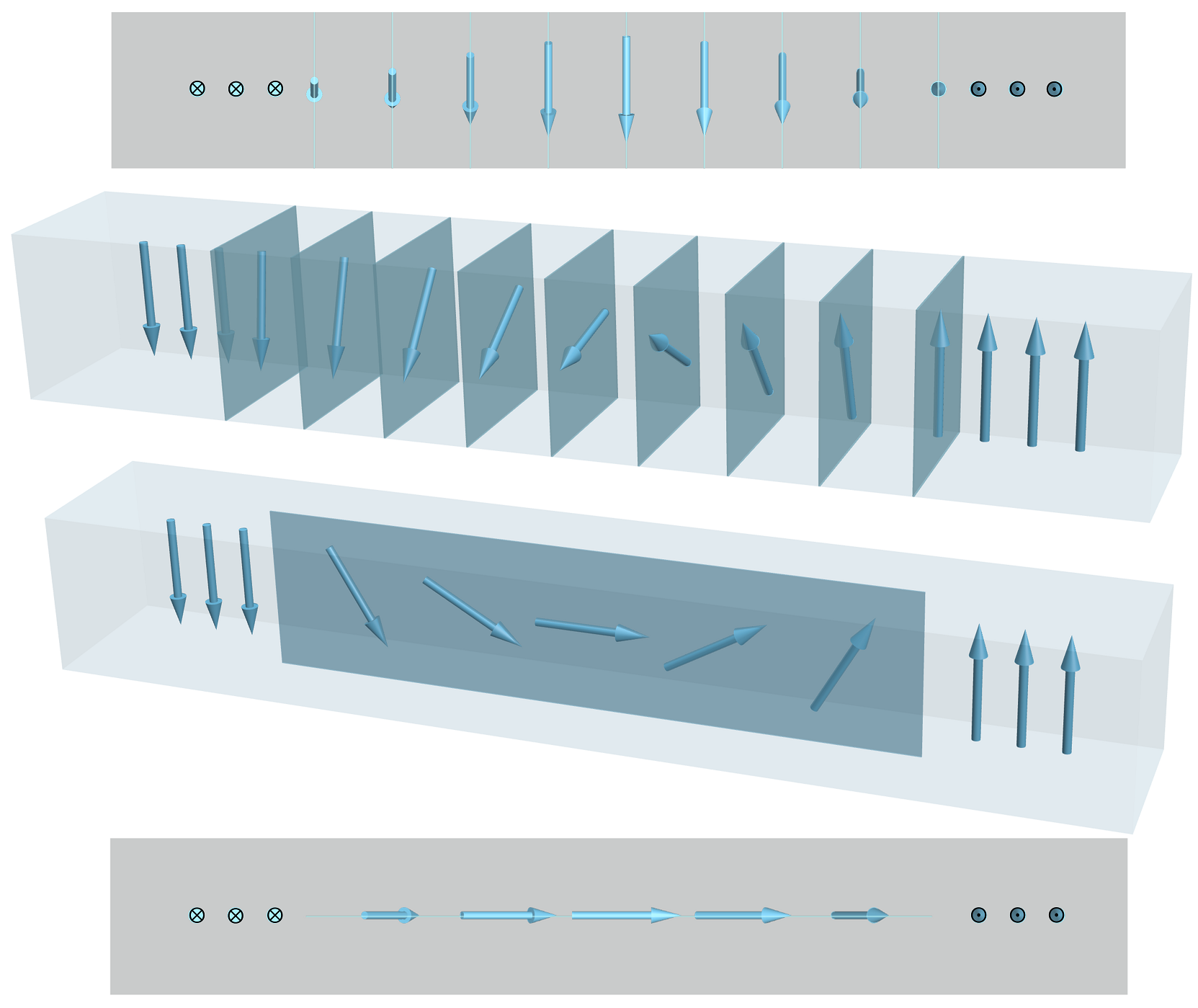}
\begin{picture}(0,0)
\put(-310,220){\small (a)}
\put(-310,170){\small (b)}
\put(-310,100){\small (c)}
\put(-310,20){\small (d)}

\end{picture}
    \caption[Diagram of the domain wall structures.]{Diagram of the \gls{DW} structures: (a) top view of the Bloch wall, (b) isometric view of Bloch wall, (c) isometric view of the N\'eel wall, and (d) top view of the N\'eel wall.}
    \label{fig:DW_structure}
\end{figure}
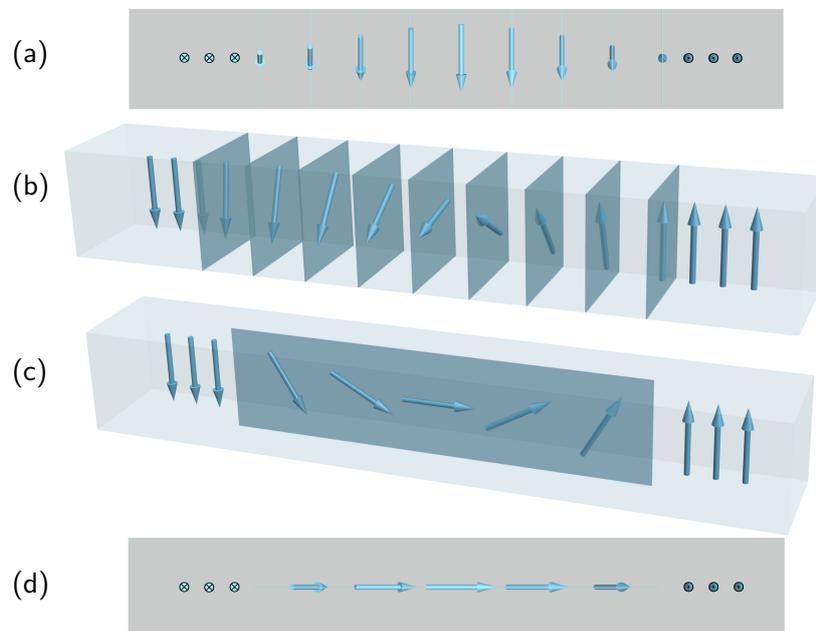

\define{Magnetic domains} correspond to regions within a material where the magnetisation is aligned in the same direction~\cite{hubert2008magnetic}. 
Each magnetic domain acts like a small magnet with its own magnetic field. 
A \acrfull{DW} is a narrow region separating two magnetic domains with different magnetisation directions. 
\glspl{DW} can be moved by an external field, as shown in Figure~\ref{fig:DWmotion}. 
Two different types of \glspl{DW}, Bloch and N\'eel walls, are demonstrated in Figure~\ref{fig:DW_structure}~\cite{shepley2015effect}. \revision{In a Bloch wall, magnetisation rotates uniformly in a plane perpendicular to the domain wall, while in a N\'eel wall, magnetisation rotates within the plane of the wall itself.}
In thin films with \gls{PMA}, the magnetisation direction tends to be ``up'' and ``down''. The \glspl{DW} exist between two domains, gradually and smoothly reorienting the magnetic moments from ``up'' to ``down'' or from ``down'' to ``up''.

\subsection{Domain wall motion}
\begin{figure}[ht!]
    \centering
    \includegraphics[width=1\textwidth]{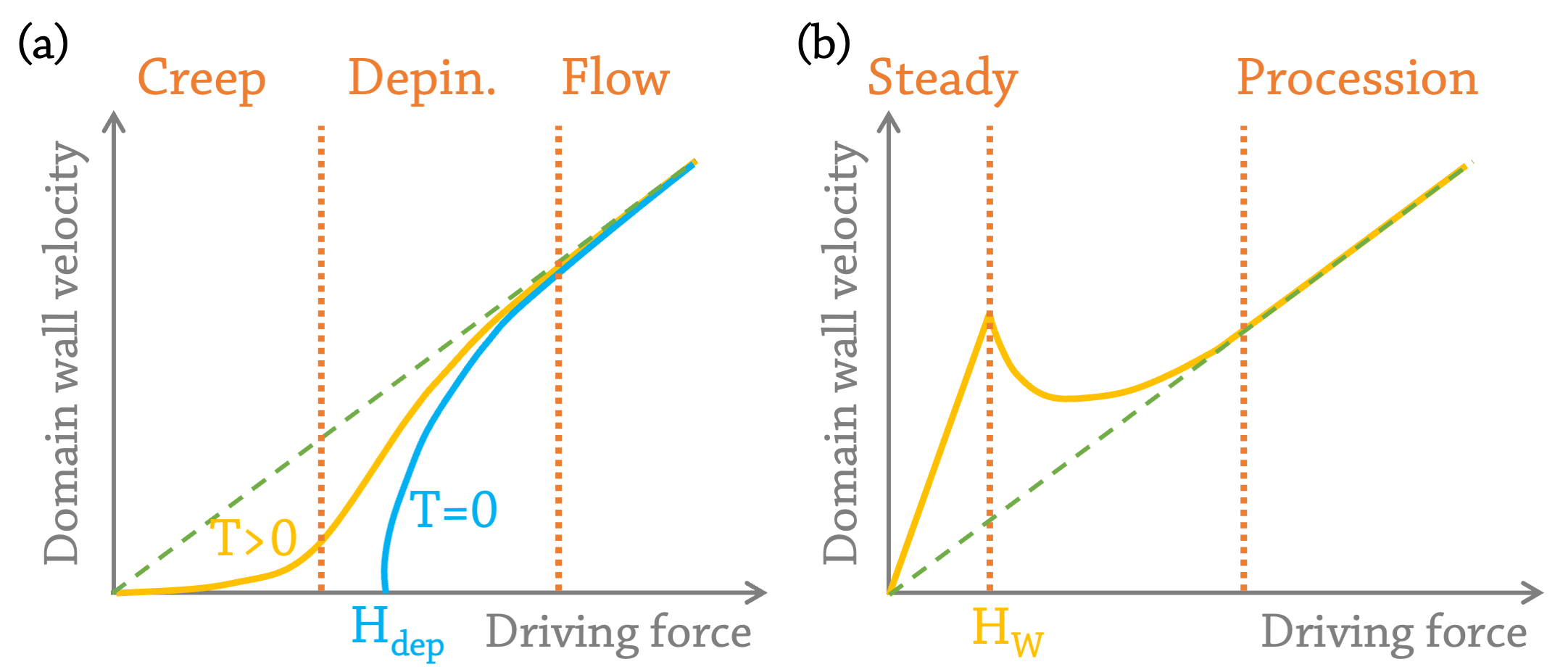}
    \caption[Domain wall velocity regimes in relation to driving force.]{\gls{DW} velocity regimes in relation to driving force. (a) \gls{DW} motion is within the creep and thermally activated regimes at a low driving force and a finite temperature. \gls{DW} motion shows less dependent on thermal energy in the depinning regime at a higher driving force and eventually shows a linear dependence on the driving force in the flow regime. (b) Steady and precessional regimes are separated by the Walker breakdown ($\RM{H}_\RM{W}$), where the \gls{DW} velocity reduces with the driving force increases. }
    \label{fig:DW_vs_driving_force}
\end{figure}

\gls{DW} motion can be induced by an external magnetic field through two distinct mechanisms (Figure~\ref{fig:DW_vs_driving_force}). In an ideal thin film without any defects or disorder, when subjected to low fields, the \gls{DW} velocity increases with the field strength along with a slight distortion in the wall structure. Above a threshold $\RM{H}_\RM{W}$, \gls{DW} velocity experiences an abrupt decrease with the field (Figure~\ref{fig:DW_vs_driving_force}b). This phenomenon is widely known as Walker breakdown, which is characterised by an alternating \gls{DW} structure and the precession of magnetisation inside the \gls{DW} around the applied field~\cite{schryer1974motion}.\par

Defects and disorders within thin films significantly influence the \gls{DW} motion. In a thin film containing such imperfections, at zero temperature, the \gls{DW} becomes strongly entangled with the defects and remains pinned for all fields below a critical value known as the \define{depinning field} $\RM{H}_\RM{dep}$ (see Figure~\ref{fig:DW_vs_driving_force}a). The depinning field is defined as the minimum external magnetic field required to surpass the local energy barrier that hinders the \gls{DW} motion~\cite{jeudy2016universal,ferre2013universal}.
At finite temperatures $\RM{T}$, \gls{DW} motion can be activated by thermal fluctuations, which initiate localised depinning process. This regime is commonly referred to as the creep regime, where \gls{DW} velocity increases with the external magnetic field and follows the creep law, which can be expressed as 
\begin{equation}\label{eqn:creep_law}
    v = v_0 \exp\left( -\frac{U_c}{k_B \RM{T}} \left(\frac{\RM{H}_{\RM{dep}}}{\RM{H}_\RM{ext}}  \right)^{\frac{1}{4}}\right),
\end{equation}
where ${U_c}$ is the pinning energy barrier induced by disordered energy landscape, $k_B$ is the Boltzmann constant, and $v_0$ is a numerical prefactor~\cite{ferrero2013numerical,metaxas2007creep}. This behaviour holds even when the $\RM{H}_\RM{ext}$ is smaller than the $\RM{H}_\RM{dep}$. 

\begin{figure}[ht!]
    \centering
    \includegraphics[width=0.6\textwidth]{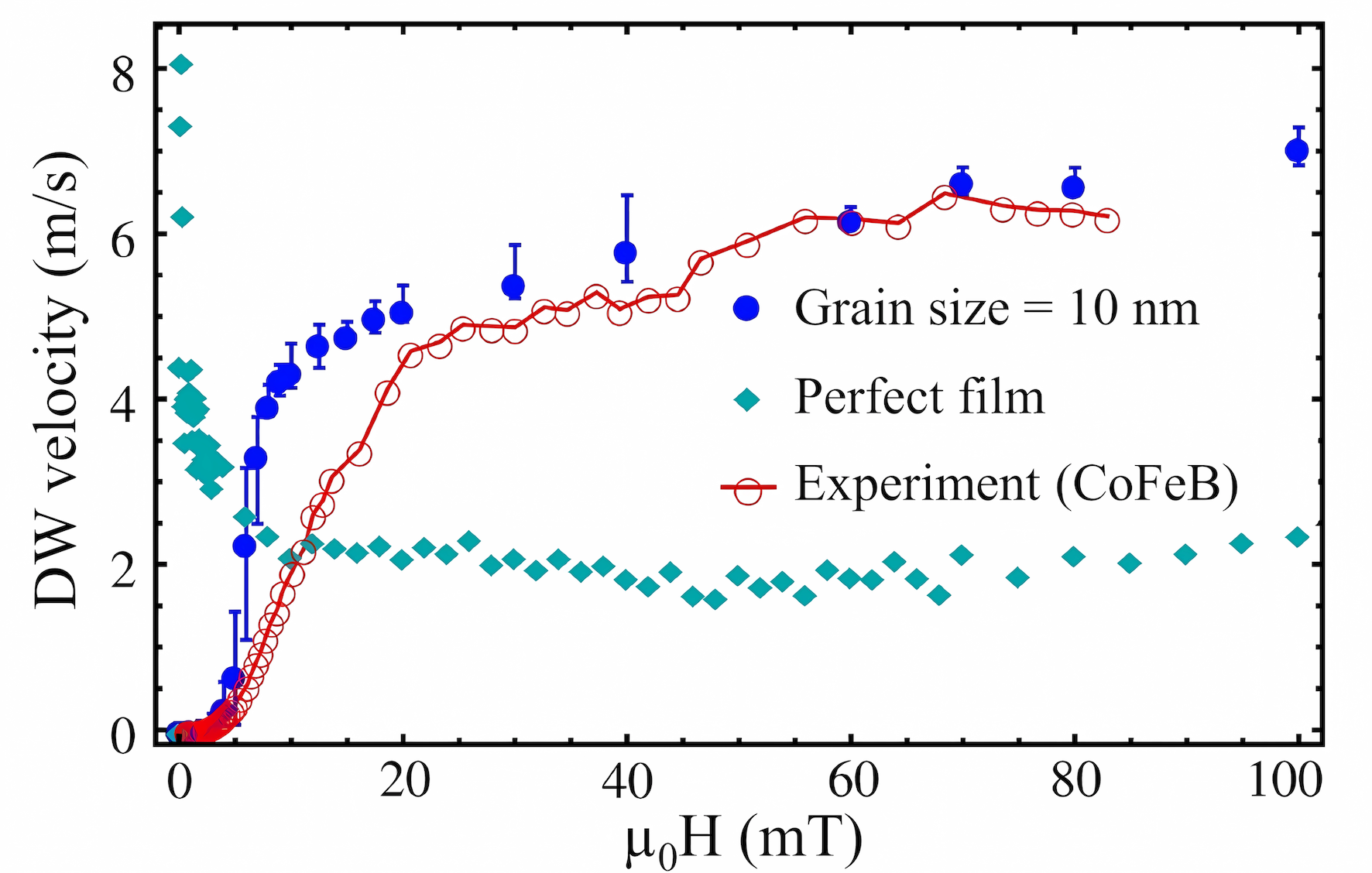}
    \caption[Domain wall velocity as a function of the magnetic field in thin films with disorder.]{\revision{\gls{DW} velocity as a function of the magnetic field in thin films with disorder. Simulated data for a grain size of 10 nm with 5\% anisotropy dispersion shows a velocity plateau indicating incoherent precession. Experimental result from a Ta/CoFeB/MgO film is without the Walker breakdown. A perfect film response includes the Walker breakdown. \figref{\cite{voto2016effects}}}}
    \label{fig:DW_motion_with_plateau}
\end{figure}

For fields exceeding $\RM{H}_\RM{dep}$ at both temperature conditions (zero and finite temperatures), the \gls{DW} velocity increases with the field until it enters the viscous regime where the \gls{DW} velocity exhibits a wide plateau~\cite{burrowes2013low,mougin2007domain,voto2016effects,herranen2015domain} instead of a Walker breakdown \revision{(as shown in Figure~\ref{fig:DW_motion_with_plateau}~\cite{voto2016effects})}. This \gls{DW} velocity saturation effect can be attributed to an enhanced energy dissipation at defects in the thin film~\cite{voto2016effects}, as well as the blocking~\cite{yamada2015excitation} and annihilation~\cite{yoshimura2016soliton} of \glspl{VBL}. \glspl{VBL} are curling magnetic structures that appear inside \glspl{DW} and introduce complexity to the static and dynamic properties of the \glspl{DW}~\cite{lian1985observation,herranen2017bloch}. Defects within the thin films give rise to variations in the energy landscape along the \gls{DW}, which in turn affect the wall structure and the local mobility of the \gls{DW} as it traverses its path. The \gls{DW} structure thus directly influences the local mobility. As a result, the nucleation, propagation, and annihilation of \glspl{VBL} have a significant impact on the overall \gls{DW} dynamics.\par

\subsection{Skyrmions}

\define{Magnetic skyrmions} are topologically protected particle-like magnetic spin structures. Within a skyrmion, the spins undergo a gradual rotation with a consistent chirality, transitioning from an upward direction at one edge, through the center, and finally to a downward direction at the other edge. The symmetry of the system plays a crucial role in determining the energetically preferred spin configuration of the formed skyrmions. 

\begin{figure}[ht!]
    \centering
    \includegraphics[width=.8\textwidth]{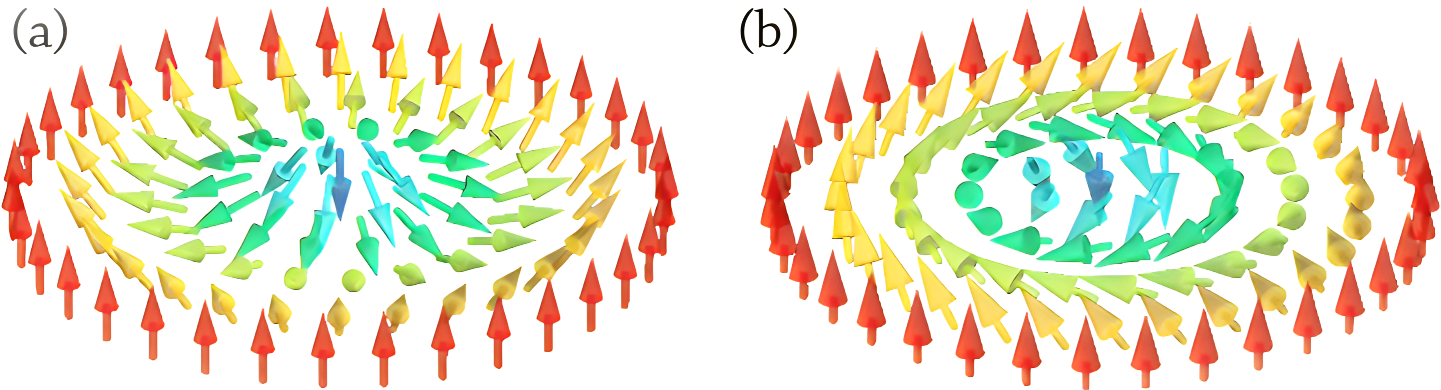}
    \caption[Magnetic texture of skyrmions.]{Magnetic texture of skyrmions: (a) N\'eel skyrmion and (b) Bloch skyrmion. \figrefs{\cite{fert2017magnetic}}}
    \label{fig:skyrmion_dmi}
\end{figure}

\revision{As shown in Figure~\ref{fig:skyrmion_dmi}, there are two typical types of magnetic skyrmions according to the spin configuration: N\'eel ~\cite{kezsmarki2015neel,boulle2016room} and Bloch skyrmions~\cite{muhlbauer2009skyrmion,yu2010real}.} The distinct spin configuration arises from the inherent symmetries in the spin interactions, which can be influenced by factors such as the crystal lattice or interface properties.
\revision{For instance, Bloch skyrmions are typically observed in bulk materials, facilitated by the absence of crystal inversion symmetry and the presence of high \gls{SOC} atoms, as seen in certain ferromagnetic alloys like B20 materials~\cite{finocchio2016magnetic}. In contrast, N\'eel skyrmions are commonly associated with interfacial \gls{DMI} in multilayer thin films~\cite{kuepferling2023measuring}.}

The spin direction at a given spatial position $\B{r}=(x,y)$ can be represented by $\B{m}(\B{r})$. The spin configuration of a skyrmion is characterised by its topological skyrmion number ($N_\RM{sk}$), defined as
\begin{equation}
N_\RM{sk} = \frac{1}{4\pi}\int\B{m}\cdot \left(\partial_x \B{m} \times \partial_y \B{m} \right) \RM{d}x\RM{d}y = 1 .
\end{equation}
Similarly, a spin configuration with a skyrmion number of $-$1 is known as an anti-skyrmion, which can be observed in magnets exhibiting $D_\RM{2d}$ symmetry~\cite{nayak2017magnetic, nagaosa2013topological}. The finite size of skyrmions enables them to exhibit particle-like behaviour, allowing for movement, interaction, and excitation at specific dynamic modes.

\section{Surface acoustic waves}
\subsection{Piezoelectric effect}
\begin{figure}[ht!]
    \centering
    \includegraphics[width=0.7\textwidth]{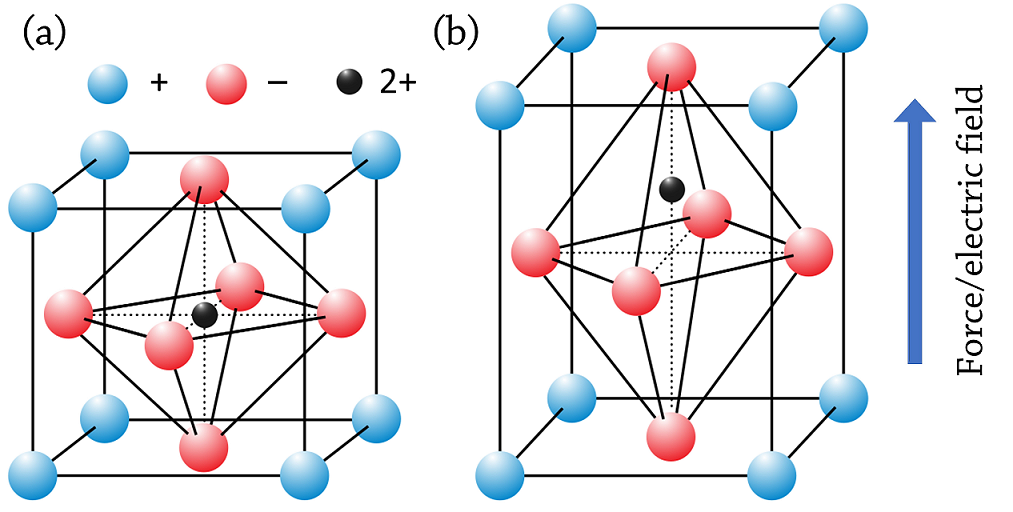}
    \caption[Schematic of a piezoelectric unit cell and a piezoelectric unit cell with force or electric field applied.]{Schematic of (a) a piezoelectric unit cell and (b) a piezoelectric unit cell with force or electric field applied. Charges of molecules are arbitrary and for illustration only. \figrefs{\cite{wiki:pzt}}}
    \label{fig:piezoeletric_material_cell}
\end{figure}

The \define{piezoelectric effect} refers to the ability of certain materials to generate an electric charge when subjected to mechanical strain, known as the direct piezoelectric effect. This effect is reversible, meaning that materials exhibiting piezoelectricity can also generate mechanical strain in response to an electric field, known as the converse piezoelectric effect~\cite{o2010spatial}.
The origin of the piezoelectric effect lies in the linear electromechanical interaction between the mechanical and electrical states in crystalline materials lacking inversion symmetry~\cite{campbell2012surface}. When a stress is applied to such a crystal, an imbalance of charge occurs due to the movement of charges within the crystal, as depicted in Figure~\ref{fig:piezoeletric_material_cell}. This charge movement leads to a change in surface charge density and the creation of an electric field between the faces of the crystal. The strength of the electric field increases with the magnitude of the applied stress. On the contrary, by applying an electric voltage, a (compressive/tensile) strain can be induced in the material. The magnitude of the strain is proportional to the applied voltage.

\subsection{SAW excitation and propagation}

\revision{Based on their wave motion characteristics, several types of \glspl{SAW} exist, including Rayleigh waves and Love waves. Rayleigh waves propagate along the free surface of a solid material, and they are characterised by elliptical particle motion. This motion includes both vertical and horizontal components, leading to an overall elliptical trajectory. Love waves exhibit purely horizontal particle motion that occurs in a horizontal plane parallel to the surface. This motion is restricted to the plane of the wavefront. In this thesis, Rayleigh waves were chosen because Rayleigh waves can easily form both travelling waves and standing waves, allowing us to explore different types of waves on the magnetisation dynamics. However, when the Rayleigh waves encounter the thin films, due to the different mechanical properties between the piezoelectric substrate and the thin film, Love wave may occur in the thin film.}\par
\begin{figure}
    \centering
    \includegraphics[width = 0.9\textwidth]{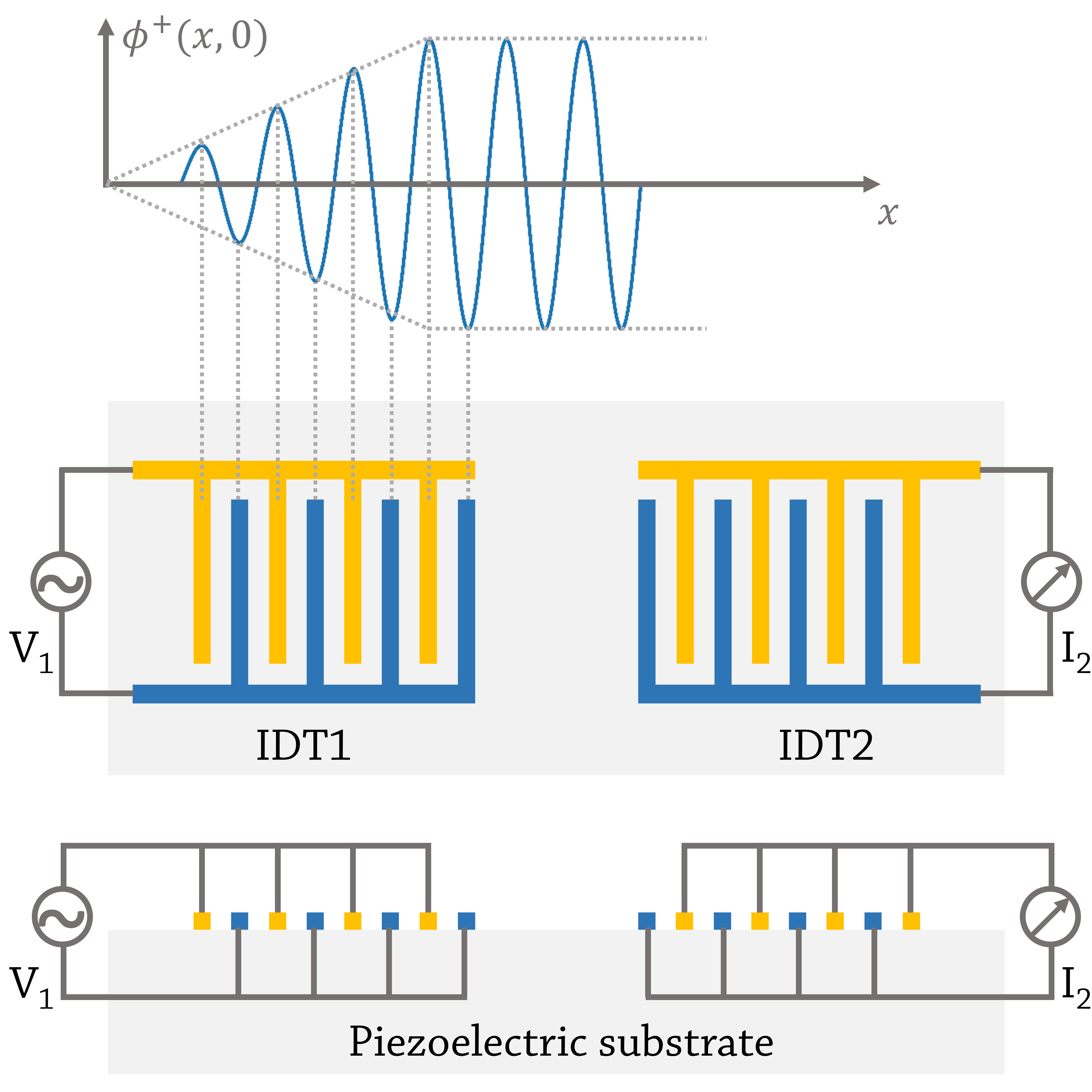}
    \caption[Excitation of surface acoustic waves.]{Excitation of \glspl{SAW}. (a) Wave potential as a function of distance. \gls{SAW} electrical potential can be formed by applying alternating voltage to the \glspl{IDT} that are patterned onto a piezoelectric material. (b) and (c) are the top and side view of the \gls{SAW} device, respectively. The \gls{rf} power is applied to the \gls{IDT}1, generating \glspl{SAW} that move towards \gls{IDT}2, where they are subsequently detected. \figrefs{\cite{ballantine1996acoustic}}}
    \label{fig:SAW_generation}

\begin{picture}(0,0)
\put(-200,520){\small (a)}
\put(-200,385){\small (b)}
\put(-200,240){\small (c)}
\end{picture}

\end{figure}
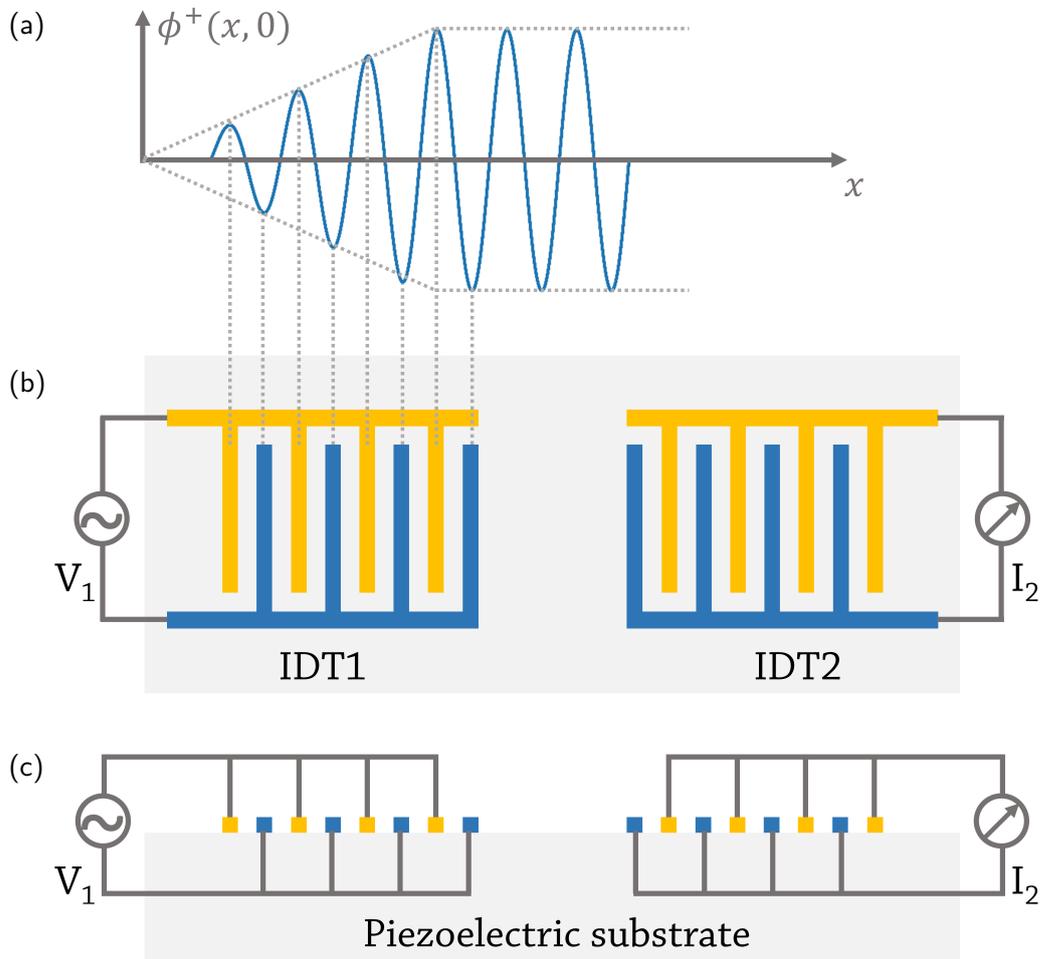

\glspl{SAW} are commonly generated on a piezoelectric material using \glspl{IDT} consisting of interleaved electrodes (as shown in Figure~\ref{fig:SAW_generation}b). By applying an alternating signal delivering \gls{rf} power to either \glspl{IDT} (\gls{IDT}1 in Figure~\ref{fig:SAW_generation}b), a periodic strain field is induced on the surface of the piezoelectric material, resulting in the formation of a standing \gls{SAW} (see Figure~\ref{fig:SAW_generation}a). This standing \gls{SAW} propagates in both directions (left and right in Figure~\ref{fig:SAW_generation}b) away from the \glspl{IDT}. The \glspl{IDT} exhibit optimal \gls{rf} power to mechanical wave efficiency when the \gls{SAW} wavelength $\lambda_\RM{SAW}$ matches the periodicity of the electrode structure $d_\RM{IDT}$. 
The relationship between the centre frequency $f_0$ and the wavelength $\lambda_\RM{SAW}$ of the \glspl{SAW} then can be expressed as
\begin{equation}
f_0=\frac{v_\RM{SAW}}{\lambda_\RM{SAW}} ,
\end{equation}
where $v_\RM{SAW}$ represents the speed of the \gls{SAW}. Therefore, by adjusting the pitch of the \glspl{IDT}, the centre frequency of the \glspl{SAW} can be modified accordingly.\par

The propagation of a mechanical wave in a piezoelectric material gives rise to an associated electrostatic wave potential, denoted as $\phi$ (as depicted in Figure~\ref{fig:SAW_generation}a). The wave potential of an electrode subjected to a continuous wave voltage $V_1$ can be described as
\begin{equation}
\phi^\pm = \mu_\RM{s} V_1 ,
\end{equation}
where $\phi^+$ and $\phi^-$ correspond to the wave propagating towards the right and left directions, respectively, while $\mu_\RM{s}$ represents a substrate-dependent constant that is unaffected by the frequency of the applied voltages.

Each electrode within the \gls{IDT} can be considered as a discrete source for the generation of \glspl{SAW}. When \glspl{IDT} with electrodes spaced periodically with a period of $d_\RM{IDT}$ are excited by alternating voltages $V_n = (-1)^n V_0$ with a frequency of $f$, the resulting wave potential for the rightward propagating wave $\phi^+$ can be calculated as the vector sum of contributions from each electrode. It can be represented as
\begin{equation}
\phi^+(0) = \mu_s V_0 \sum_{n=0}^{N_\RM{f}-1} (-1)^n \exp \left(\frac{i n k d_\RM{IDT}}{2}\right) ,
\end{equation}
where $n$ represents the $n^\RM{th}$ electrode, $N_\RM{f}$ is the total number of electrodes, and \revision{$i^2 = -1$}. The wavenumber $k = \omega/v$, where the angular frequency $\omega = 2\pi f$. This wave potential is a geometric series, where the elements become unity and add constructively when $k d_\RM{IDT}/2 = m \pi$ with $m$ being an odd integer. \revision{As a result, the wave potential reaches its maximum value when $d_{\RM{IDT}} = m\lambda_{\RM{SAW}}$ for some odd integer $m$.} \par

\begin{figure}[ht!]
    \centering
    \includegraphics[width=0.9\textwidth]{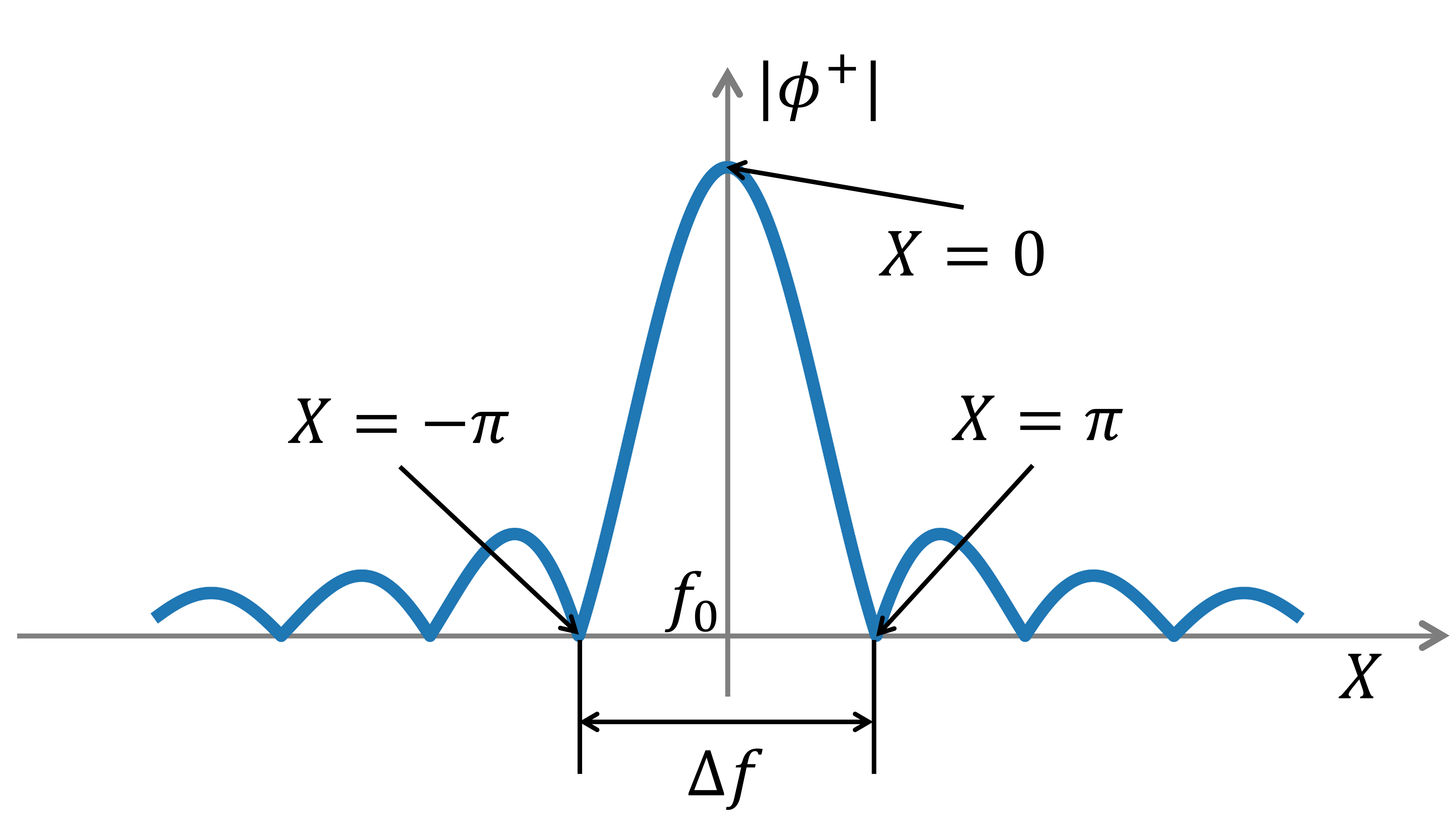}
    \caption[Interdigitated transducer response as a function of the detuning parameter.]{\gls{IDT} response ($|\sin{\left (X \right )}/X|$) as a function of the detuning parameter $X$. \figref{\cite{ballantine1996acoustic}}}
    \label{fig:wave_potential_vs_f}
\end{figure}

As the frequency moves away from the centre frequency ($f_0$), the addition of components from individual electrodes becomes incoherent, resulting in a wave potential given by
\begin{equation}
\left |\phi^+(f) \right| = \left|\frac{\sin{X}}{X}\right| ,
\end{equation}
where $X$ is the detuning parameter defined as
\begin{equation}
X = \frac{N_\RM{p} \pi(f-f_0)}{f_0} .
\label{eqn:detunning}
\end{equation}
In this equation, $N_\RM{p}$ corresponds to the number of \gls{IDT} periods \revision{($N_\RM{p} = N_\RM{f}/2$)}. The variation of the wave potential with respect to the detuning parameter $X$ is illustrated in Figure~\ref{fig:wave_potential_vs_f}. 

\revision{The wave potential $\phi^{+}$ becomes zero when $X$ in Equation~\ref{eqn:detunning} is an integer multiple of $\pi$. This phenomenon occurs due to the complete cancellation between electrode contributions. The frequency interval $\Delta f$ between the first nulls on either side of the centre frequency (see Figure~\ref{fig:wave_potential_vs_f}), therefore, can be written as}
\begin{equation}
    \Delta f = \frac{2}{N_\RM{p}} f_{0}.
\end{equation}
\revision{Thus, the bandwidth of the \gls{IDT} $B$ defined as $\Delta f/f_0$ is given by}
\begin{equation}
    B = \frac{2}{N_\RM{p}}.
    \label{eqn:bandwidth}
\end{equation}

When the acoustic wave reaches the receiving electrodes, the wave potential induced by the incident wave leads to a flow of current in the receiving interdigitated loaded transducer (\gls{IDT}2 in Figure~\ref{fig:SAW_generation}b). This current flow can be detected by an external detection circuit. Similar to the transmitting \gls{IDT}, the receiving \gls{IDT} also exhibits optimal performance when the periodicity of the transducer matches the wavelength of the acoustic wave.

\section{Summary}
This chapter laid the essential theoretical groundwork necessary for interpreting the experimental results presented in the subsequent chapters. The dynamics of magnetisation were analysed in detail, with a specific emphasis on the roles of free energy and the effective field within thin films with \gls{PMA}. The discussion focused on the various energy terms contributing to magnetisation dynamics, including exchange interaction, anisotropy, \gls{DMI}, and Zeeman energy.
Two primary spin structures that form the subject of this thesis, \glspl{DW} and skyrmions, were introduced in detail.
The chapter provided insights into the generation and propagation of \glspl{SAW}. It emphasised the dynamic strain that \glspl{SAW} induce in magnetic thin films, leading to a magnetoelastic coupling effect that, in turn, influences magnetisation dynamics.
In summary, this chapter laid the groundwork for the project, with critical insights into magnetisation dynamics, spin structures, and \glspl{SAW} serving as its cornerstone.
\cleardoublepagewithnumberheader
\chapter{Methods}
\label{chapter3_methods}
\section{Introduction}

This chapter provides an overview of the methods and equipment used for the fabrication and characterisation of the thin films and \gls{SAW} devices investigated in this study.
The fabrication of magnetic thin films was achieved using \gls{dcMS}. 
Layer thickness measurements, determination of \gls{PMA}, and observation of the magnetic domains were carried out using \gls{XRR} and wide-field Kerr microscopy.
The \glspl{IDT} were prepared in a class 100 cleanroom facility using a maskless photolithography technique.
\Glspl{S-parameter} of the \glspl{IDT} were characterised using a \gls{VNA}. 
Micromagnetic simulations were also performed to study the effect of \glspl{SAW} on magnetisation dynamics using Mumax3. Ta/Pt/Co/Ir/Ta and Ta/Pt/Co/Ta thin films with \gls{PMA} were chosen for investigation in Chapter \ref{Chapter4_anisotropy_control} and Chapter \ref{Chapter5_heating_effect}, respectively.\par 
Ta was chosen to be the buffer layer since it tends to form a (111) face-centred-cubic layer and introduce an atomically smooth interface between the Pt/Co layers. This can promote the formation of the smooth and (111) preferred orientation Pt layer and improve the \gls{PMA}. 
To determine the basic magnetic properties such as thickness and \gls{PMA}, thin films were deposited on silicon wafers. 
Lithium niobate (\LNO) wafers were also used as substrates to support the propagation of \glspl{SAW}. 
The \gls{IDT} geometry was designed to excite \glspl{SAW} with frequencies of $\sim$100 MHz and $\sim$50 MHz for Chapter \ref{Chapter4_anisotropy_control} and Chapter \ref{Chapter5_heating_effect}, respectively. Note that the chosen \gls{SAW} frequencies are well-below the \gls{FMR} frequency of the thin films.
The \gls{SAW} device was mounted in a \gls{CPW} made of the copper-covered \gls{PCB} and then was connected to a \gls{VNA}. \par
\clearpage

\section{Magnetic thin film fabrication}
\subsection{Substrate and cleaning process}
\subsubsection{Substrate}

\begin{figure}[ht!]
    \centering
    \includegraphics[width=0.8\textwidth]{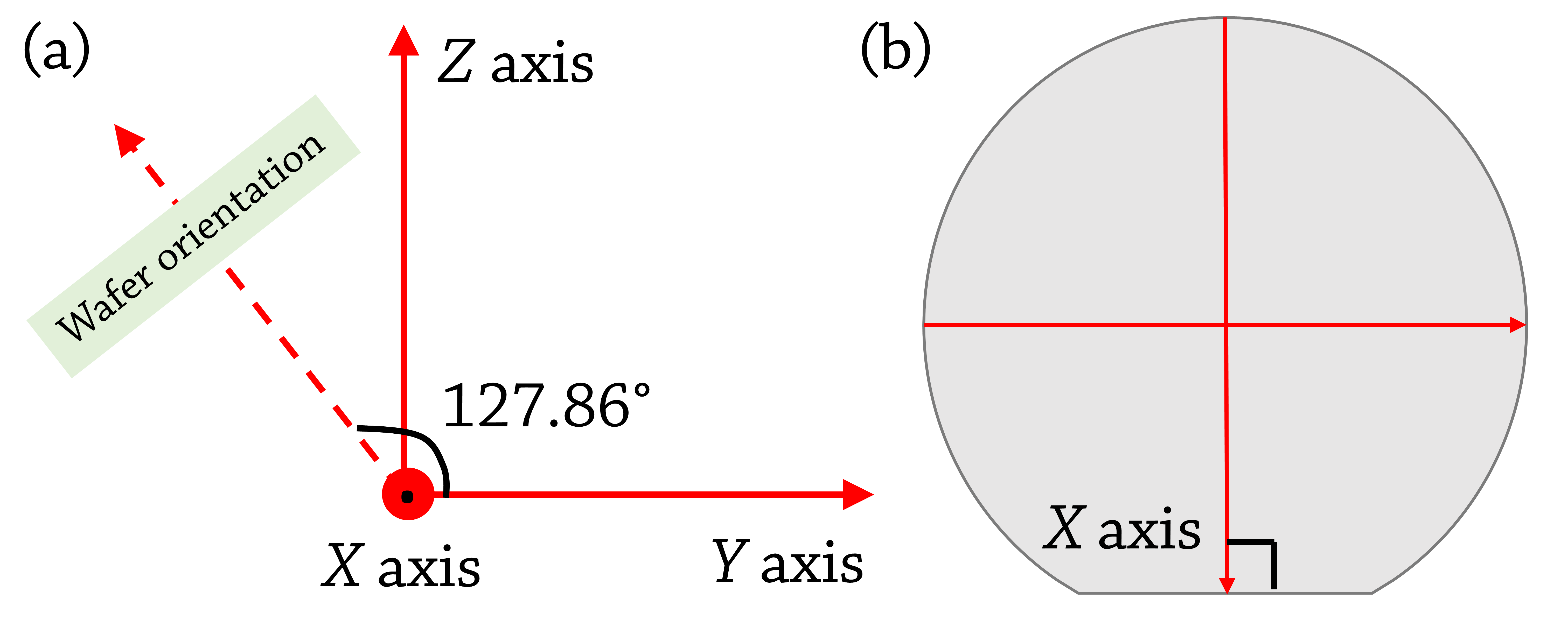}
    \caption[Schematic representation of the 128\textdegree\ Y-cut lithium niobate.]{\revision{Schematic representation of the 128\textdegree\ Y-cut \LNO. (a) Wafer orientation is normal to the $X$ direction, and with rotation of this crystal axis by 128\textdegree\ from the $Y$ direction. (b) $X$ axis and the direction perpendicular to $X$ axis of the \LNO. \figrefs{\cite{o2010spatial}}}}
    \label{fig:LNO}
\end{figure}

One-side polished 128\textdegree\ Y-cut \LNO\ (PI-KEM) substrates with a thickness of 0.5 mm were chosen in this study. This material is known for its strong piezoelectric coupling efficiency. \revision{The substrate is 128\textdegree\ rotated from the $+Y$ axis through the $+Z$ axis about the $X$ axis (see Figure~\ref{fig:LNO}a)}. The \gls{SAW} propagating velocities in $X$ axis and the direction perpendicular to $X$ axis (see Figure~\ref{fig:LNO}b) are 3982 m/s and 3640 m/s, respectively~\cite{paskauskas1995velocity}. For this study, \gls{SAW} propagation along the $X$ direction was preferred.
By applying an electric voltage to the piezoelectric material, the piezoelectric effect induces deformation in the substrates. The chosen \LNO\ substrates provided the necessary piezoelectric properties for efficient \gls{SAW} generation and propagation.

\subsubsection{Cleaning process}
In order to achieve high-quality thin films with low defects and strong adhesion, the substrates underwent a thorough cleaning process. The cleaning procedure involved ultrasonic cleaning using acetone for 5 minutes to remove any dirt and organic residues present on the substrate surface. Subsequently, the substrates were subjected to another 5 minutes of ultrasonic cleaning using isopropanol. After the cleaning process, the substrates were carefully dried using compressed air before being loaded into the deposition chamber. This cleaning process ensured that the substrates were free from contaminants and ready for the deposition of thin films.

\subsection{Thin film deposition}
\label{Chapter3_Section:dcMS}

The magnetic thin films used in this thesis were fabricated by \acrfull{dcMS}. This deposition technique is widely utilised in research and industry due to its advantaged such as high deposition rate and cost-effectiveness, and precise control over thin film structure and thickness, even down to sub-nanometre.\par

\begin{figure}[ht!]
    \centering
    \includegraphics[width = .9\textwidth]{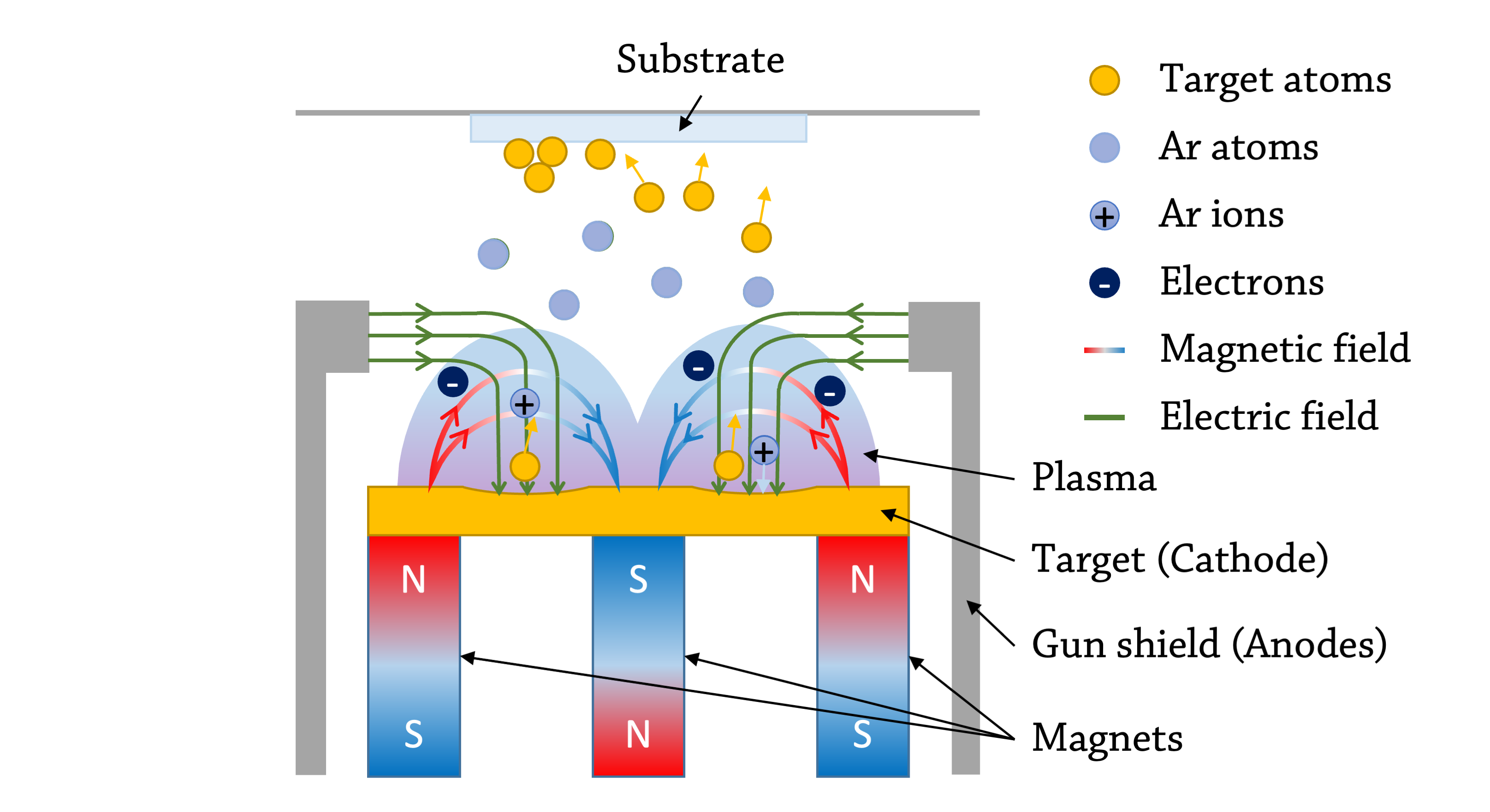}
    \caption[Diagram of the direct current magnetron sputtering.]{\revision{Diagram of the \gls{dcMS}. An electric field is applied between the target (cathode) and gun shield (anode) to ionise Ar. The plasma shape is controlled by permanent magnets behind the target. Material is removed from the target by the Ar ions that repeatedly hit the target. The target material is then impinged onto the substrate surface. \figref{\cite{shepley2015effect}}}}
    \label{fig:dc_sputtering}
\end{figure}

\begin{figure}[ht!]
    \centering
    \includegraphics[width = 1\textwidth]{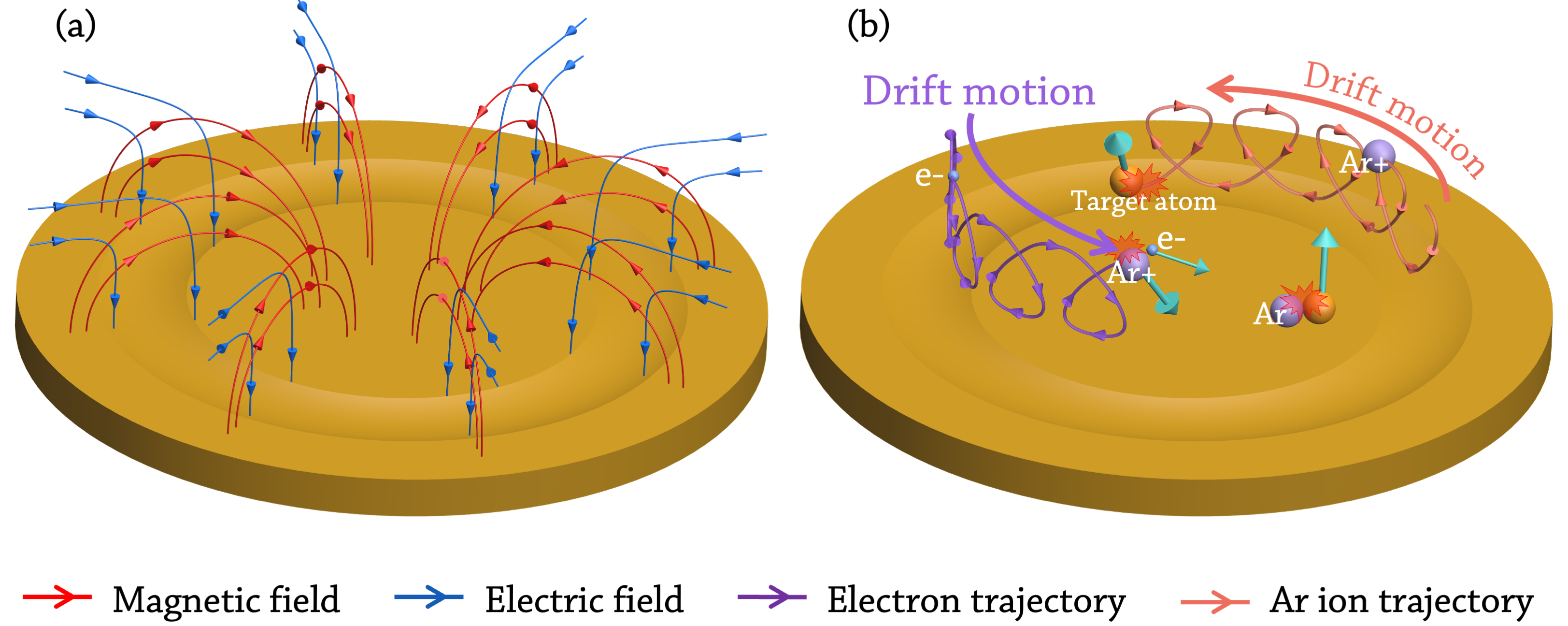}
    \caption[Diagram of the distribution of the magnetic and electric fields in the direct current magnetron sputtering, alongside the particle motion influenced by these fields.]{\revisiontwo{(a) Diagram of the distribution of the magnetic and electric fields in \gls{dcMS}. (b) Diagram of the motion of the particles in \gls{dcMS}. In the presence of the electric and magnetic fields, electrons and Ar ions exhibit cyclotron motion with drift motion within zones where $\B{E} \cdot \B{B} = 0$. The collision of Ar ions with the target material can cause the ejection of target atoms. These target atoms then collide with the plasma, eventually reaching the substrate and forming the thin film. Further collisions between electrons and Ar atoms generate additional Ar ions and electrons, significantly boosting deposition efficiency. The cyclotron motion radii of Ar ions and electrons are similar, with the electron trajectory typically positioned above that of Argon ions (due to the target is negative charged), resulting in partial overlap of their paths.}}
    \label{fig:particle_motion}
\end{figure}

The \gls{dcMS} system used in this work was equipped with eight targets, two of which were specially designed for depositing magnetic materials. The base pressure of the chamber can reach was 3.0 $\times$ 10\textsuperscript{--6} Pa with the application of cryo-pump and liquid nitrogen. During the deposition process, an Argon (Ar) atmosphere was maintained at a pressure of $\sim$3.3 $\times$ 10\textsuperscript{--4} Pa. 
A schematic of a magnetron gun, which is a key component in the \gls{dcMS} system, is shown in Figure~\ref{fig:dc_sputtering}. The electric and magnetic fields are shown in Figure~\ref{fig:particle_motion}a. In the deposition process, Ar gas is ionised into plasma by an electric field applied between the target and gun shield~\cite{swann1988magnetron}. \revisiontwo{The magnetic field generated by the permanent magnets is arranged in such a way that it traps electrons near the target surface. This increases the efficiency of ionisation and confines the plasma close to the target. The motion of the electrons and Ar ions is in cyclotron motion with a drift motion in the presence the magnetic field $\B{B}$ and electric field $\B{H}$ (as shown in Figure~\ref{fig:particle_motion}b). The trajectory of the positively charged Ar ions is lower than that of the electrons as the target is negatively charged due to the electric field. When these Ar ions collide with the target, target atoms are ejected from the target material (see Figure~\ref{fig:particle_motion}b). These target atoms then collide with the plasma, eventually reaching the substrate and forming the thin film. The cyclotron with drift motion of the electrons and ions is the most significant at the region where $\B{E} \cdot \B{B} = 0$. Therefore, target material removal occurs at these regions forming a ring shape on the target (as shown in Figure~\ref{fig:particle_motion}). Subsequent collisions between electrons and Ar atoms generate additional Ar ions and electrons, significantly boosting deposition efficiency.}
The substrates were fixed on a rotatable wheel, allowing them to be positioned in front of different target materials. Magnetic thin films with desired structures can be obtained by moving substrates among different materials. 
The thickness of the deposited thin films was controlled by the deposition rate of the material and deposition time.\par

\section{Magnetic thin film characterisation}

\subsection{Deposition rate calibration}

\acrfull{XRR} is a surface-sensitive analytical technique employed to determine the thickness and roughness of thin films. In this work, \gls{XRR} was utilised to calibrate the deposition rate of the thin films. For our measurements, we detected the intensity of X-rays reflected from a flat surface at low angles ranging from 0.5\textdegree\ to 8\textdegree.\par

\begin{figure}[ht!]
    \centering
    \includegraphics[width = 1\textwidth]{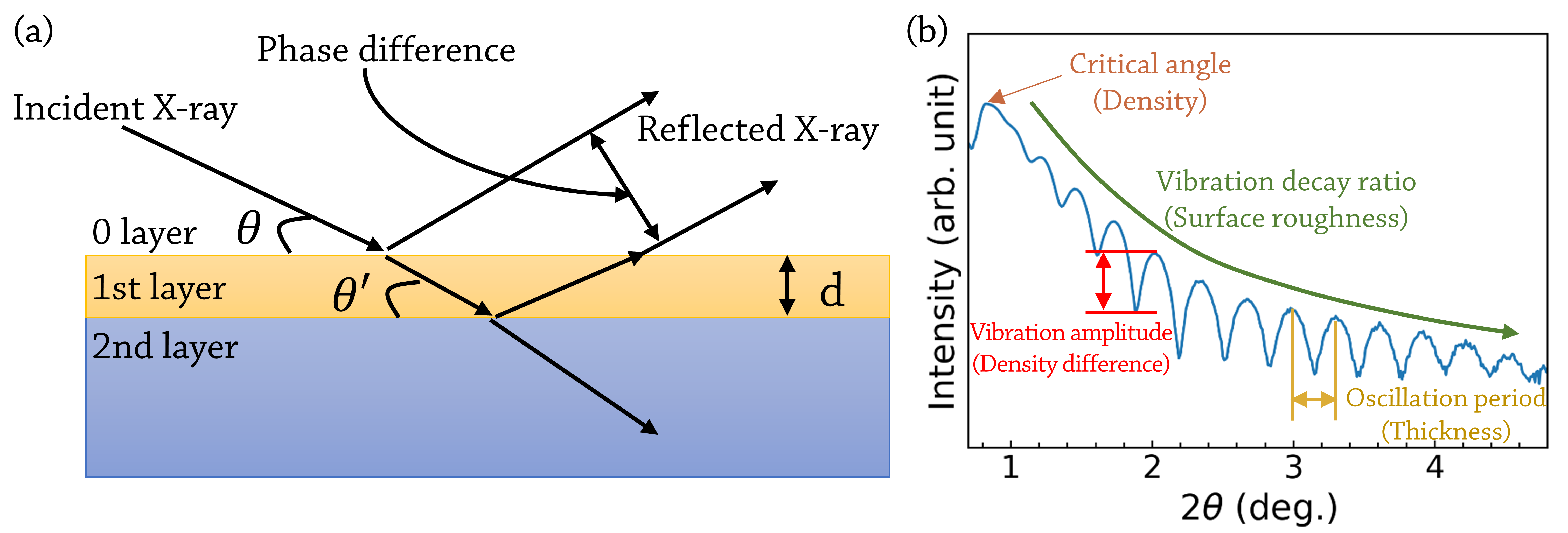}
    \caption[Diagram of X-ray reflectometry principle and X-ray reflectometry patterns.]{(a) Diagram of \gls{XRR} principle and (b) \gls{XRR} patterns for the Co thin film deposited onto a Si substrate.}
    \label{fig:xrr_principle}
\end{figure}

The basic principle of \gls{XRR} is to detect the intensity of a beam of reflected X-ray from a flat surface~\cite{briscoe2007applying}, as illustrated in Figure~\ref{fig:xrr_principle}a.
If the interface is not perfectly sharp or smooth, the intensity of the reflected X-ray deviates from the prediction based on the law of Fresnel reflectivity. 
To obtain the thickness, roughness, and density of each thin film layer, the measured \gls{XRR} data was fitted with a simulated curve using the recursive Parratt formalism combined with the rough interface formula~\cite{parratt1954surface}. The relation of the fringes and the film thickness can be expressed by a modified Bragg equation as
\begin{equation}\label{Bragg_equation}
    n\lambda_\RM{X}=2d\sqrt{\sin^2a_i-\sin^2a_c} \hspace{1mm},
\end{equation}
where the $n$ is integer number of fringes, $\lambda_\RM{X}$ is the wavelength of the X-ray, $d$ is the thickness of the film, $a_i$ is the angle of the fringes, and the $a_c$ is the critical angle. GenX 3 software was used to fit the \gls{XRR} data~\cite{bjorck2007genx}. Figure~\ref{fig:xrr_principle}b shows an example \gls{XRR} pattern for a Co thin film deposited on a Si substrate. 
The deposition rates of the materials used in this work, obtained at specific applied currents, are summarised in Table~\ref{tb:deposition_rate}.

\renewcommand{\arraystretch}{1.2}
\begin{table}[ht!]
\centering
\caption{Deposition current and deposition rate of materials used in this thesis.}
\label{tb:deposition_rate}
\begin{tabular}[t]{ccc}
\midrule[0.4mm]
Material&Current (mA)&Deposition rate (\revision{\AA}/s)\\
\midrule[0.2mm]
Ta&50&1.43 $\pm$ 0.03\\
Pt&25&1.72 $\pm$ 0.03\\
Ir&25&0.91 $\pm$ 0.02\\
Co&50&0.89 $\pm$ 0.02\\
\midrule[0.4mm]

\end{tabular}
\end{table}%

\subsection{Hysteresis loop and domain image}

A wide-field Kerr microscope was used to measure hysteresis loops and image magnetic domains. The \gls{MOKE} refers to when light reflects from a magnetised surface and may change its polarisation and reflected intensity~\cite{hubert2008magnetic}. The \gls{MOKE} measurement is a highly sensitive technique for detecting changes in magnetisation within a region of a sample. It serves as a rapid and reliable method for determining the presence of \gls{PMA} in magnetic thin films. \par

\begin{figure}
    \centering
    \includegraphics[width=0.8\textwidth]{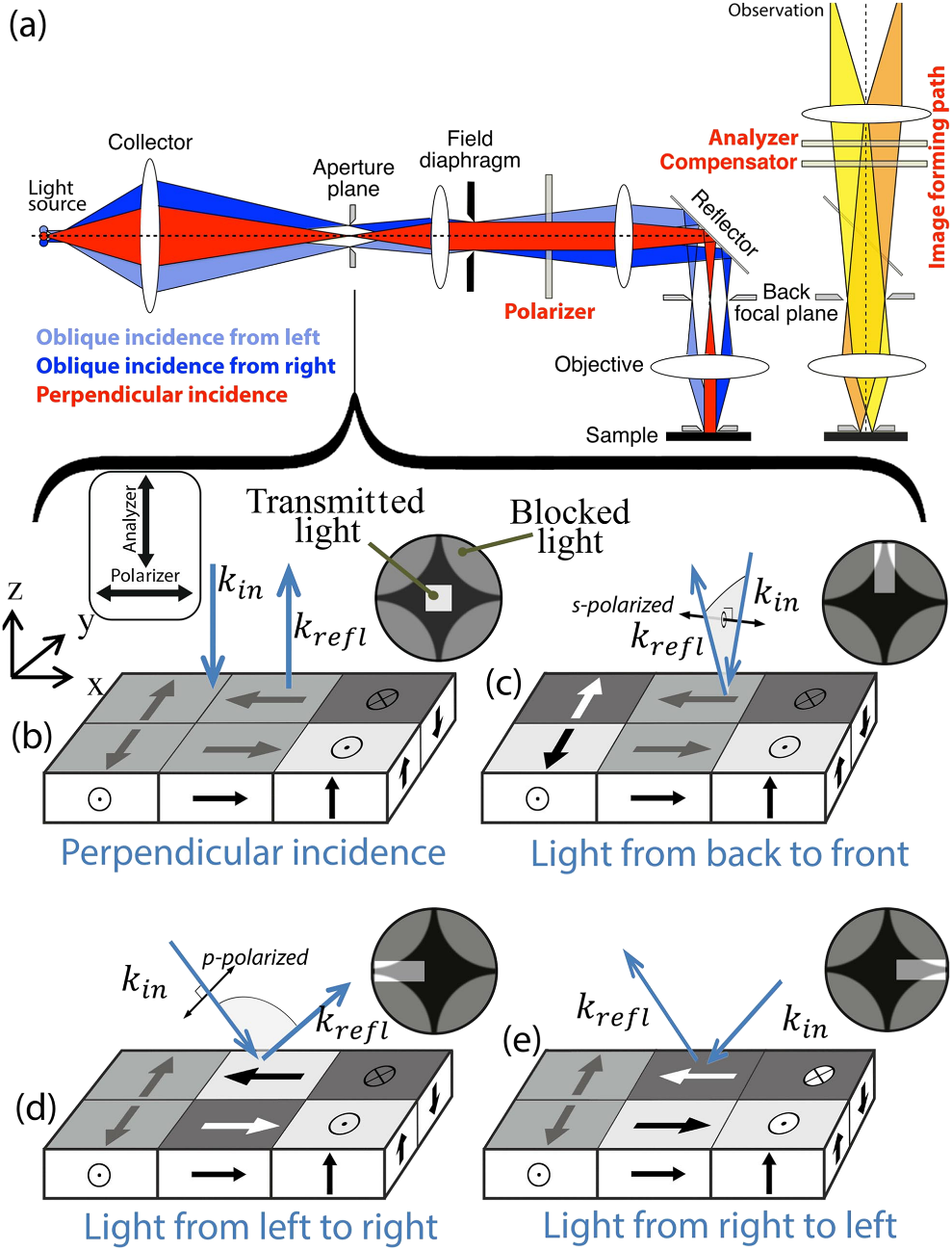}
    \caption[Diagram of the wide-field Kerr microscope.]{Diagram of the wide-field Kerr microscope. (a) The light paths for illumination and the image formation. The movable slit can be used to adjust the sensitive directions: (b) polar contrast, (c) longitudinal with s-polarised light, and (d) and (e) the longitudinal in transverse direction with p-polarised light with direct and inverted contrast, respectively~\cite{soldatov2017selective}. }
    \label{fig:wide_field_keer}
\end{figure}

The magnet coils of the Kerr-microscope were powered by a Kepco power supply and cooled by water. Figure~\ref{fig:wide_field_keer}a shows the light path from the \gls{LED} through the optics to the surface of the samples and then reflected to the camera~\cite{soldatov2017selective}. \gls{MOKE} can be further classified into four categories based on the direction of the magnetisation vector relative to the reflecting surface and the plane of incidence: polar \gls{MOKE}, longitudinal \gls{MOKE}, transversal \gls{MOKE}, and quadratic \gls{MOKE}. In this study, only polar \gls{MOKE} was employed to characterise the thin films. For polar \gls{MOKE}, a slit was used to select the centre of the polarisation cross, ensuring that the reflection magnetisation vector was perpendicular to the reflection surface and parallel to the plane of incidence.\par

Figure~\ref{fig:wide_field_keer}b--e depict the primary magneto-optical effects observed in a Kerr microscope, focusing on the polar and longitudinal Kerr effects. For light incident perpendicularly to the sample, in-plane magnetised domains exhibit no contrast, as no reflected light aligns with the in-plane direction. On the other hand, out-of-plane domains display the most potent vectorial components, yielding maximum contrast.
To obtain a contrast for in-plane domains, one can adjust the light incidence to an oblique angle. Figure~\ref{fig:wide_field_keer}c presents a scenario where light is directed onto the sample from the rear of the figure. This setup produces contrast between domains with magnetisation components parallel to the incidence plane. Yet, transverse domains lack contrast, as their magnetisation component does not align with the reflected light. This phenomenon is evident in Figure~\ref{fig:wide_field_keer}d, where such domains display maximum contrast when light is incident from the left, while vertical in-plane domains remain without contrast.
Figure~\ref{fig:wide_field_keer}e illustrates the effects when the direction of light incidence is inverted along the same axis as Figure~\ref{fig:wide_field_keer}d. Here, while transverse domains retain their lack of contrast, the longitudinal domains exhibit a contrast reversal compared to Figure~\ref{fig:wide_field_keer}d.
In all cases of oblique incidence, the polar Kerr effect is apparent because the vectorial components of polar domains consistently align with the reflected light beam. The orientation of the incident and reflected beams does not influence the polar Kerr effect.

\section{Surface acoustic wave device}

\subsection{Device design}

\subsubsection{Interdigitated transducer geometry}
\begin{figure}[ht!]
    \centering
    \includegraphics[width=.9\textwidth]{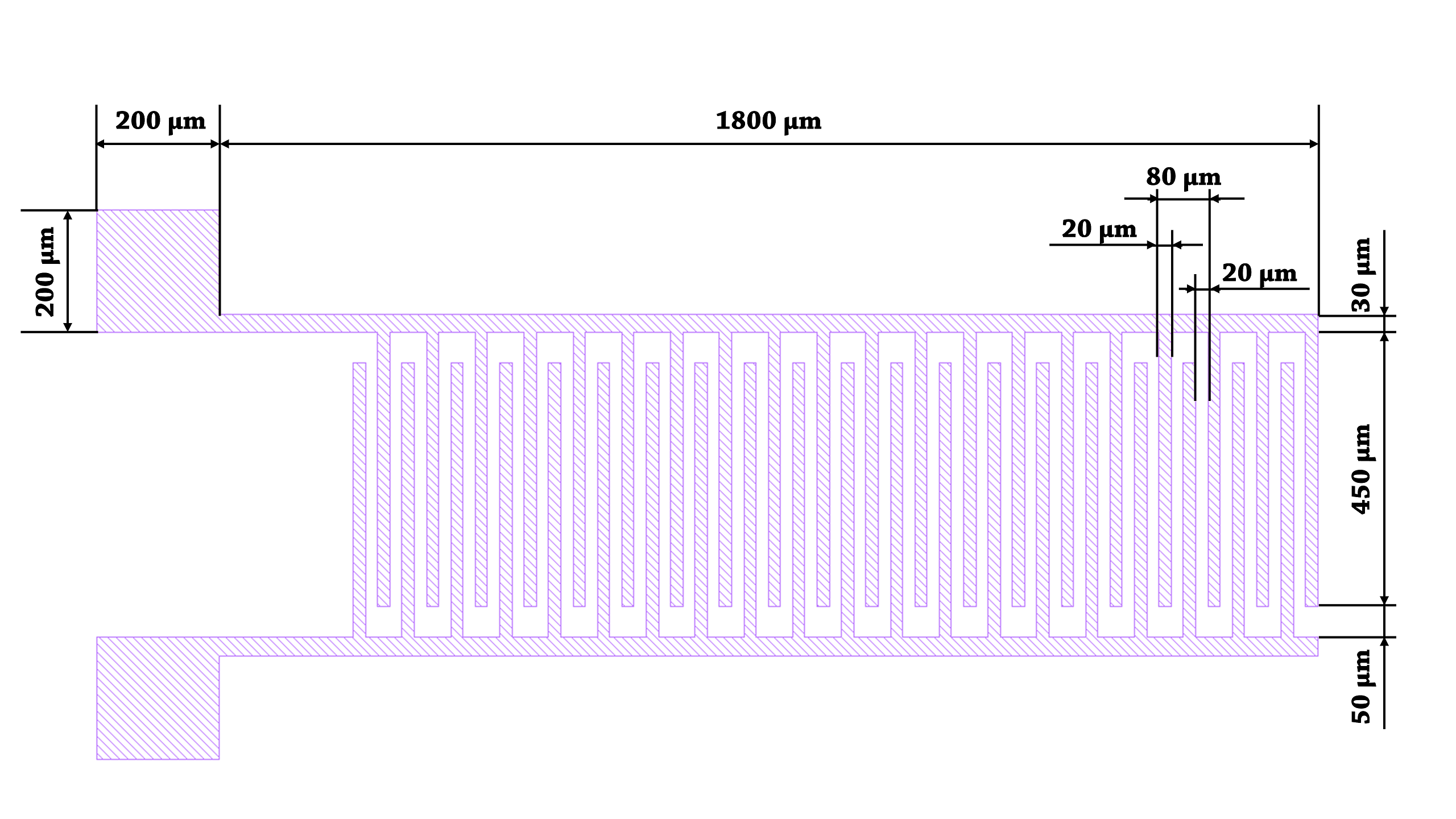}
    \caption[Example of the interdigitated transducer geometry used to generate surface acoustic waves.]{Example of the \gls{IDT} geometry used to generate \glspl{SAW} with a frequency of $\sim$50 MHz. The centre frequency of the \gls{SAW} depends on the widths of the electrodes, spacing between neighbouring electrodes, and the \gls{SAW} propagating velocity. With electrode widths and spacing set to 20 \textmu m, the resulting \gls{SAW} centre frequency is $\sim$50 MHz. The \glspl{IDT} feature an aperture of 500 \textmu m, contact pad sizes of 200$\times$200 \textmu m\textsuperscript{2}, and a total of 20 electrodes.}
    \label{fig:geomotry_IDT}
\end{figure}

\gls{SAW} devices with various centre frequencies were designed and fabricated for this study. An example of the design of \glspl{IDT} that generate \glspl{SAW} with a centre frequency of approximately 50 MHz is illustrated in Figure~\ref{fig:geomotry_IDT}. In all the \glspl{IDT} fabricated in this thesis, the ratio of electrode width to spacing was maintained at 1:1 to maximise the \gls{SAW} power. 
The acoustic aperture (defined by the length between the overlapped region of \gls{IDT} electrodes in the direction perpendicular to the \gls{SAW} propagation direction) was designed to be 500 \textmu m for convenient observation under an optical microscope and to maximise the interaction with our patterned magnetic films.\par

The widths of the electrodes and the separation between them play a crucial role in determining the centre frequency of the \gls{SAW} device. Taking into account the \gls{SAW} propagating velocity of $\sim$4000 m/s, to generate \glspl{SAW} with frequencies of approximately 50 MHz and 100 MHz (devices used in Chapter~\ref{Chapter4_anisotropy_control} and Chapter~\ref{Chapter5_heating_effect}, respectively), the widths of the electrodes were set as 20 \textmu m and 10 \textmu m, respectively, with the same separations as the widths.
The dimensions of the contact pads used for wire bonding were 200$\times$200 \textmu m\textsuperscript{2} to allow several attempts to bond a device, should the first fail.
These design parameters were selected to achieve the desired \gls{SAW} frequencies and ensure the proper functioning of the \gls{SAW} devices in the experimental setup.\par

The bandwidth of the \gls{SAW} device can be determined using Equation~\ref{eqn:bandwidth}. By decreasing the number of electrodes, the bandwidth can be increased, while adding more electrodes results in a reduced bandwidth. However, it is important to note that reducing the number of \gls{IDT} electrodes also leads to a decrease in the wave amplitude. Therefore, a trade-off needs to be considered between \gls{SAW} amplitude and bandwidth. In this study, a configuration with 20 electrodes was chosen, achieving a bandwidth of 10\% of the centre frequency, ensuring a balance between amplitude of the \glspl{SAW} obtained and a sufficiently wide bandwidth.\par

\subsubsection{Device layout}
\begin{figure}[ht!]
    \centering
    \includegraphics[width=.7\textwidth]{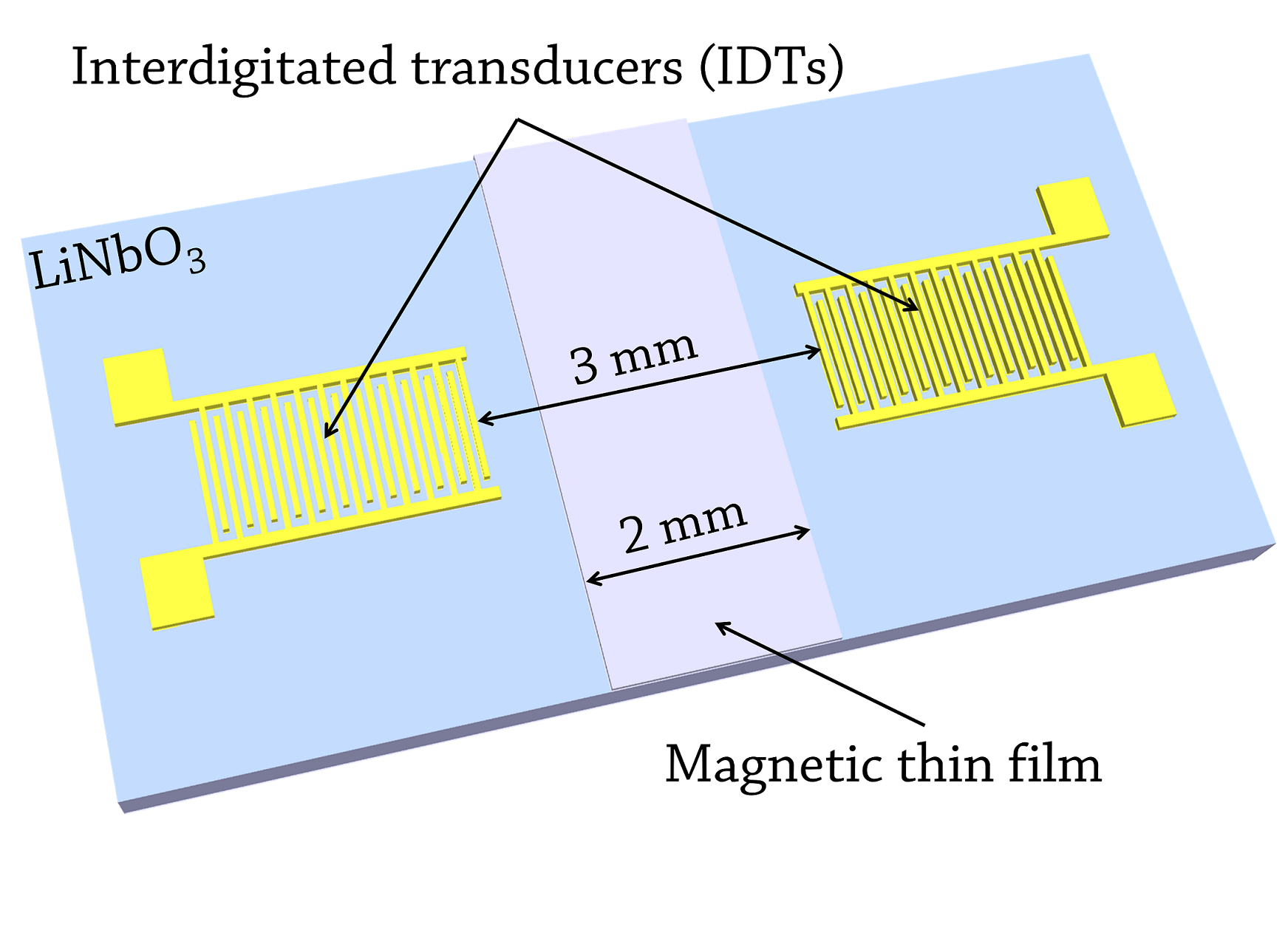}
    \caption[Layout of a surface acoustic wave device.]{Layout of a \gls{SAW} device (not to scale). A 2-mm wide magnetic thin film is deposited onto a \LNO\ substrate. One pair of \glspl{IDT} is fabricated on both ends of the thin film with a spacing of 3 mm.}
    \label{fig:SAW_device}
\end{figure}

The devices described in Chapter~\ref{Chapter4_anisotropy_control} and Chapter~\ref{Chapter5_heating_effect} share a common design layout, as shown in Figure~\ref{fig:SAW_device}.
Both devices feature two \glspl{IDT} positioned opposing each other.
This design was chosen because it allows for the investigation of the effects of travelling and standing \glspl{SAW} on the magnetic properties of thin films. Specifically, travelling \glspl{SAW} can be generated by one set of \glspl{IDT} and subsequently detected by the opposite set. Moreover, the design permits the concurrent excitation of \glspl{SAW} from both \glspl{IDT}, leading to the formation of standing \glspl{SAW} within the delay line.
In this configuration, the gap between the two \glspl{IDT}, which forms the delay line, was designed to be 3 mm. This spacing was chosen to accommodate the deposition of a 2-mm-wide magnetic thin film in between the \glspl{IDT}. Further specifics about the device layout and the experimental setup are provided in the relevant chapters.

\subsection{Device fabrication}
\label{Chapter3_nanofabrication}
High-quality electrodes for the \gls{SAW} devices were fabricated using standard cleanroom maskless photolithography techniques. 
Traditionally, photolithography involves defining a photoresist pattern on a substrate by exposing the photoresist to high-intensity \gls{UV} light through an optical mask with the desired pattern, followed by developing away the exposed photoresist.
In this work, a \gls{MLA} was employed, allowing patterns to be directly written onto the photoresist without the use of an optical mask. This approach eliminates the need for a physical mask while providing more flexibility in pattern design. 
The general procedure used is as follows:
\begin{enumerate}[label=\roman*.]
    \item Design of the patterns for the magnetic thin film and \glspl{IDT} to the required pattern using commercial KLayout software~\cite{klayout}.
    \item Perform photolithography to transfer the designed patterns onto the substrate prior to the deposition of magnetic thin films using \gls{dcMS}.
    \item Deposition of the magnetic thin films.
    \item Use a lift-off process to remove the excess undesired metal on top of the photoresist layers.
    \item Repeat the photolithography process to define the patterns for the electrical connections and \glspl{IDT}.
    \item Pattern the electrical connections and \glspl{IDT} using thermal evaporation technique, followed by the metal lift-off process to remove the excess metal.
\end{enumerate}

\begin{figure}
    \centering
    \includegraphics[width=1\textwidth]{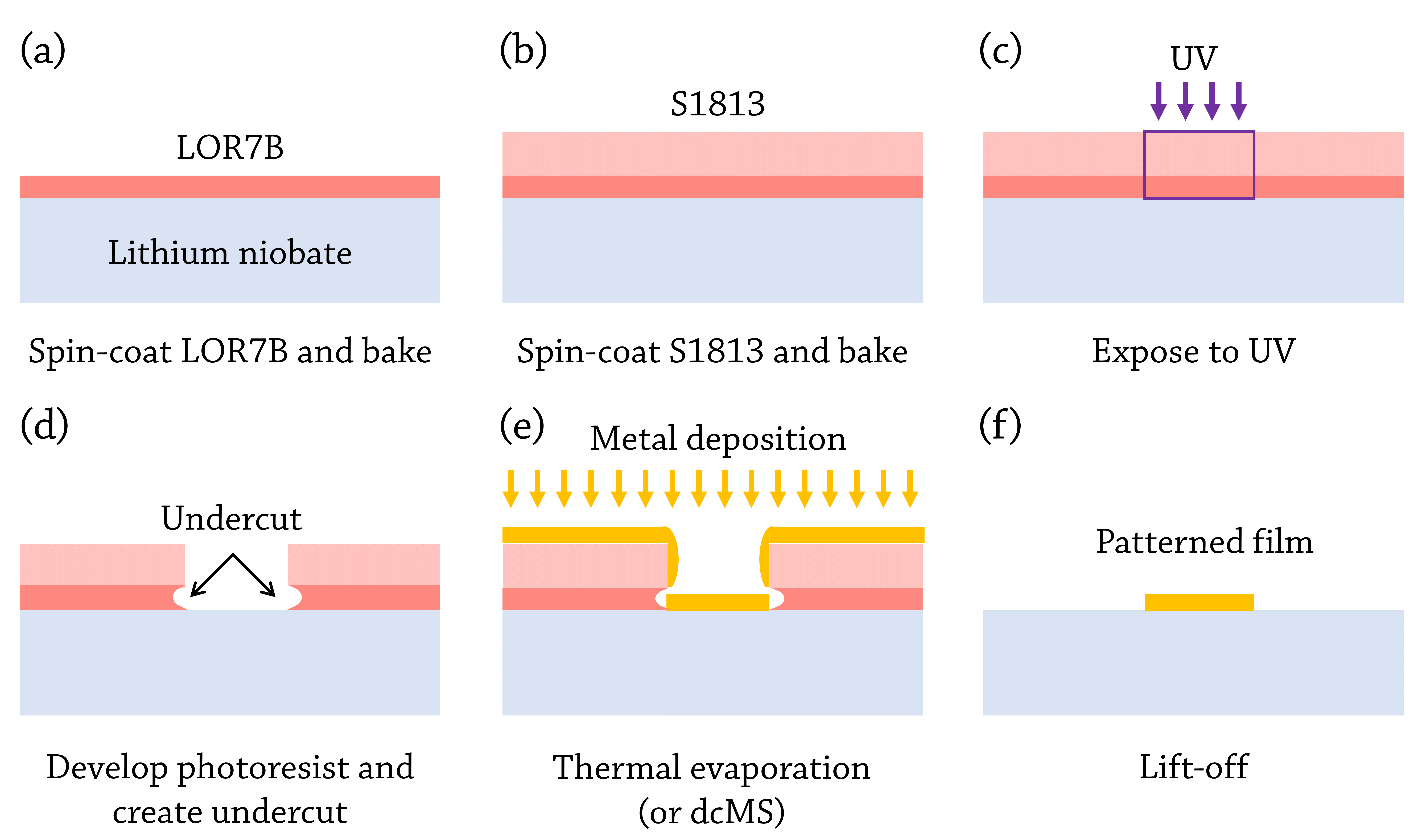}
    \caption[Illustration of the photolithography process.]{Illustration of the photolithography process. (a) A layer of LOR7B is spin-coated and baked onto the \LNO\ substrate. (b) Another layer of S1813 is spin-coated and baked on top of the LOR7B layer. (c) The sample is exposed to \gls{UV} light, forming the desired patterns. \revision{(d) The sample is immersed in a developer solution, removing the exposed photoresist and creating an undercut (LOR7B develops isotropically). (e) Magnetic thin films are deposited using \gls{dcMS}, or electrical connections and \glspl{IDT} are deposited using the thermal evaporation technique.} (f) The bi-layer photoresist and the metal layer above are lifted off by immersion in acetone, which removes all the resist and the metal layer on top of unexposed regions, leaving only the film patterned with the desired geometry in the region of interest.}
    \label{fig:cleanroom_process}
\end{figure}

The workflow of the photolithography process is illustrated in Figure~\ref{fig:cleanroom_process}. Further details of the process are discussed in the following section.

\subsubsection{Sample preparation}
The \LNO\ wafer was firstly diced into the desired sizes using a wafer saw. The diced substrates were then subjected to agitation in acetone and isopropanol for 5 minutes each, followed by rinsing in deionised water to clean them. After rinsing, the samples were dried using a stream of nitrogen gas.
Next, a layer of MicroChem LOR7B was spin-coated onto the substrate at a spin speed of 6000 rpm for 45 seconds. The coated sample was then baked at 165\degreeC for 3 minutes, as shown in Figure~\ref{fig:cleanroom_process}a. 
Subsequently, another layer of Microposit S1813 positive photoresist was spin-coated onto the baked sample at a spin speed of 4000 rpm for 45 seconds. This was followed by a 2-minute baking at 110\degreeC, as depicted in Figure~\ref{fig:cleanroom_process}b. The resulting thicknesses of the LOR7B layer and S1813 layer were approximately 0.5 \textmu m and 1.4 \textmu m, respectively.
This bilayer process was used to improve undercut of the resist and subsequent lift-off of excess metal outside the patterned region.

\subsubsection{Forming patterns on photoresist}
The maskless aligner (Heidelberg \gls{MLA}150) is a lithography tool that allows direct writing of pattern onto a substrate without requiring a traditional optical mask. It utilises a technology based on the \gls{DMD} to project the patterns on to the photoresist layer. The workflow of the \gls{MLA} is as follows:
\begin{enumerate}[label=\roman*.]
    \item Import desired pattern that created by Klayout into \gls{MLA} software. 
    \item The \gls{MLA} is equipped with \gls{DMD} chip, whose mirrors each correspond to one pixel on the pattern. These mirrors can be individually controlled to either reflect or block light.
    \item Load samples into \gls{MLA}.
    \item The \gls{DMD} precisely controls the exposure of specific areas on the photoresist, forming the desired patterns. 
    \item Unload samples for further development and processing. 
\end{enumerate}

The samples were then immersed in Microposit 351 developer (in a ratio of 3 parts H\textsubscript{2}O to 1 part Microposit 351) for 40 seconds, followed by rinsing in deionised water. The bi-layer photoresist was used to create an undercut, resulting in thin films with smooth and uniform edges (Figure~\ref{fig:cleanroom_process}d).

\subsubsection{Deposition of electrical connections and \glspl{IDT}}

After the defining patterns using the \gls{MLA}, the LOR7B and S1813 photoresists together formed a mask on the substrate. Metal deposition then took place on the surface of the substrate through the mask. The metal was deposited over the entire substrate by thermal evaporation for \glspl{IDT} and electrical contacts, or by \gls{dcMS} for magnetic thin films (see Section \ref{Chapter3_Section:dcMS}). Gold electrodes (Au, thickness of $\sim$90 nm), which are suitable for forming electrical connections, were used to pattern the \glspl{IDT} and electrical contacts. A $\sim$10 nm layer of titanium (Ti) was deposited prior to the Au layer to enhance the adhesion between the Au electrodes and the substrate.\par

\begin{figure}[ht!]
    \centering
    \includegraphics[width=0.7\textwidth]{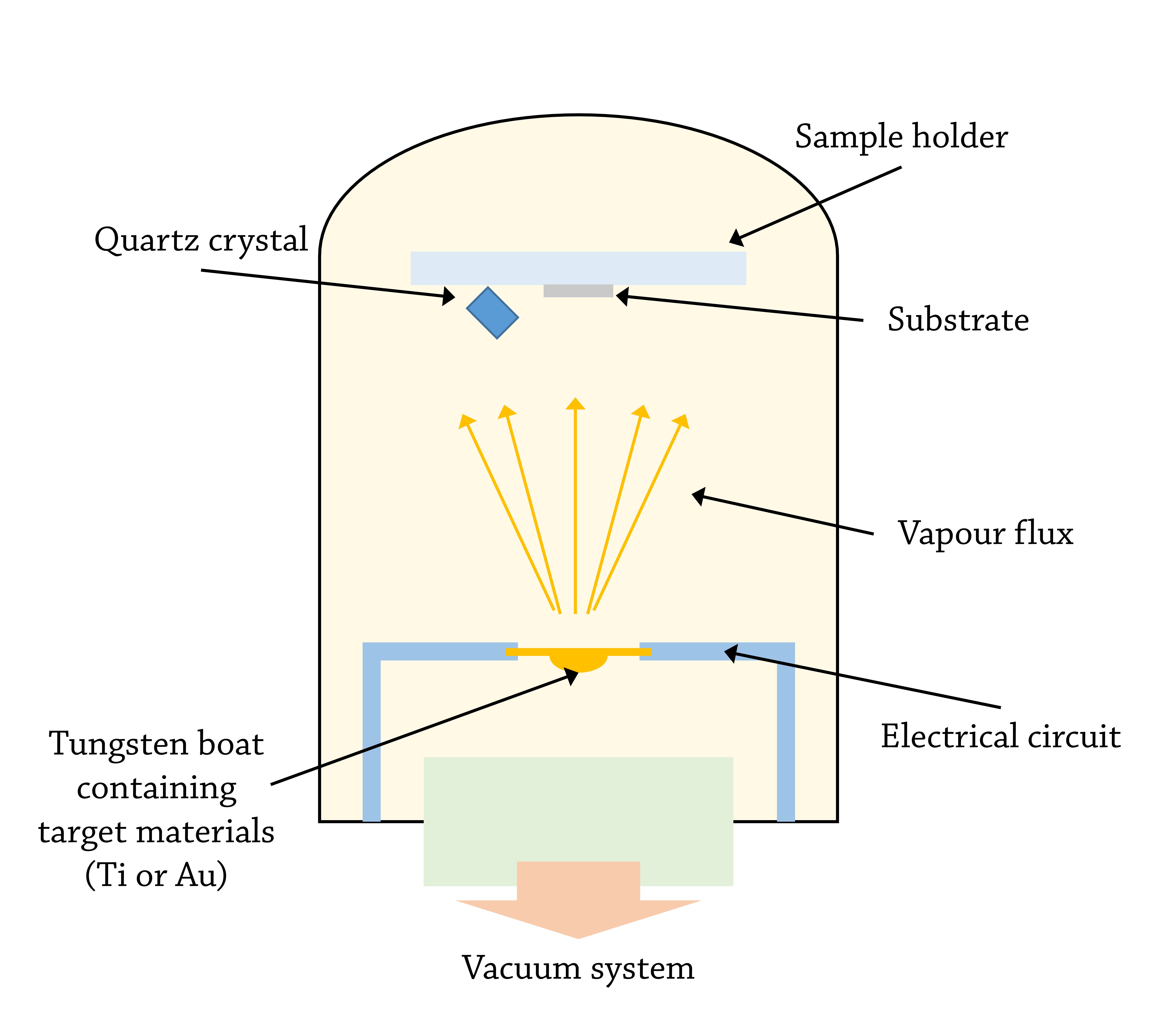}
    \caption[Diagram of the thermal evaporator.]{Diagram of the thermal evaporator. The substrates are mounted on a sample holder that is fixed at the upper part of the chamber. The base pressure of the chamber is pumped down to 1 $\times$ 10\textsuperscript{--6} Pa before performing the thermal evaporation. Electrical current is applied to the tungsten boat containing target material (Ti or Au) to generate vapour flux consisting of atoms and clusters of atoms. The vapour flux condenses on the surface of the substrates. A quartz crystal is used to monitor the thickness of the deposited metal. \figref{\cite{bashir2020chapter}}}
    \label{fig:thermal_evaporator}
\end{figure}

In this work, a thermal evaporator (Edwards Auto 306) was used to deposit Ti/Au for the fabrication of \glspl{IDT} and electrical connections. Figure~\ref{fig:thermal_evaporator} illustrates the diagram of the thermal evaporator. The substrates were mounted on a sample holder, which was positioned facing downwards towards the target materials (Ti and Au). The chamber was initially pumped down to achieve a base pressure of 1 $\times$ 10\textsuperscript{--6} Pa.
The target materials, Ti and Au, were placed in separate tungsten boats within the evaporator chamber. To induce vaporisation, an electric current was passed through the tungsten boat, resulting in the heating of the target material. The required current ranges from 20 A to 30 A for Ti and 30 A to 40 A for Au.
As the target material was heated, it evaporated and formed a vapor flux consisting of both individual atoms and clusters of atoms~\cite{bashir2020chapter}. This vapor flux travelled towards the substrate and condenses on its surface. The deposition process was controlled by a shutter located between the target material and the substrate, which regulated the exposure of the substrates to the vapor flux~\cite{bashir2020chapter}.
The thickness of the deposited materials was monitored by measuring the frequency shift of a quartz crystal located near the substrates. 

\subsubsection{Lift-off}
To form the \glspl{IDT} and electrical connections, a technique called ``lift-off'' was employed after the metal deposition process. This technique involved the removal of undesired metal and the underlying photoresist (LOR7B and S1813) to leave behind the desired patterned metal structures.
The lift-off process began by immersing the samples into acetone for 10 minutes, followed by immersion in developer for 2 minutes. In some cases, if the metal and photoresist were not easily removable after the acetone immersion, the samples were gently agitated for 3 minutes to aid in the removal process.
After the lift-off process, the samples were rinsed in isopropanol and deionised water, respectively, to eliminate any residual traces of solvent or impurities. Finally, the samples were dried using a nitrogen gas stream to ensure a clean and dry surface for further processing or characterisation.

\subsection{Device characterisation}

\subsubsection{Design of the co-planar waveguide}

\begin{figure}[ht!]
    \centering
    \includegraphics[width=.7\textwidth]{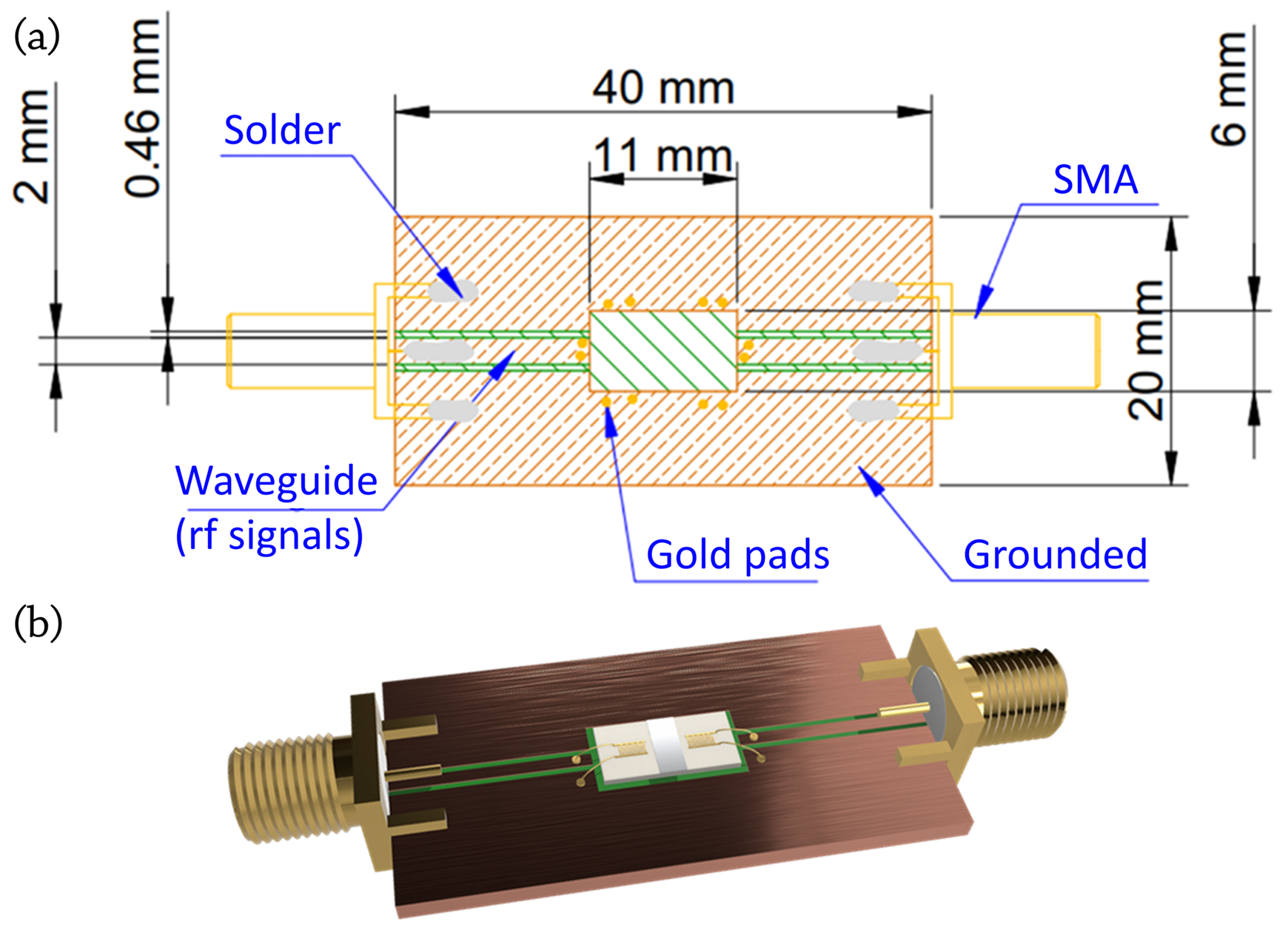}
    \caption[Layout and schematic diagram of the co-planar waveguide used in this thesis.]{(a) Layout and (b) schematic diagram of the \gls{CPW} used in this thesis. The \gls{CPW} was designed with an impedance of 50 $\Omega$ and features two \gls{SMA} connectors for connections to the \gls{VNA} to be made. The \gls{SAW} device was fixed on the \gls{CPW} using a double-sided tape. Gold pads were positioned on the \gls{CPW} using silver paint to establish reliable connections. A wedge wire-bonder was employed to establish connections between the \glspl{IDT} and the gold pads.}
    \label{fig:sample_holder}
\end{figure}

A \acrfull{CPW} is a transmission line configuration commonly used for guiding and propagating electromagnetic waves. It offers advantages over other waveguides including wider bandwidth, lower signal attenuation, and better isolation between the signal line and the ground planes~\cite{gevorgian1995cad}. The design of the \gls{CPW} used in this study is illustrated in Figure~\ref{fig:sample_holder}. The \gls{SAW} device was mounted in the centre of the \gls{CPW} using double-sided tapes. Gold pads were fixed on the \gls{CPW} using silver paint to ensure a reliable connection between the \glspl{IDT} and the \gls{CPW}. Wire bonding was employed to connect the \glspl{IDT} to the ground and conducting planes of the \gls{CPW}. \Gls{SMA} connectors were used to establish the connection between the \gls{CPW} and the \gls{rf} power source. The \gls{CPW} was fabricated from a 2 $\times$ 4 mm\textsuperscript{2} Cu-epoxy \gls{PCB}. The width of the conducting line was set to 2 mm, and the \gls{CPW} designed using an \gls{rf} circuit calculator to have an impedance of 50 $\Omega$, which indicated that the gap between the signal ground and the signal conductor needed to be 0.46 mm to obtain this value.\par

\subsubsection{Scattering parameter measurements}
\label{Chapter3_S-parameters_measurement}
\begin{figure}[ht!]
    \centering
    \includegraphics[width=.7\textwidth]{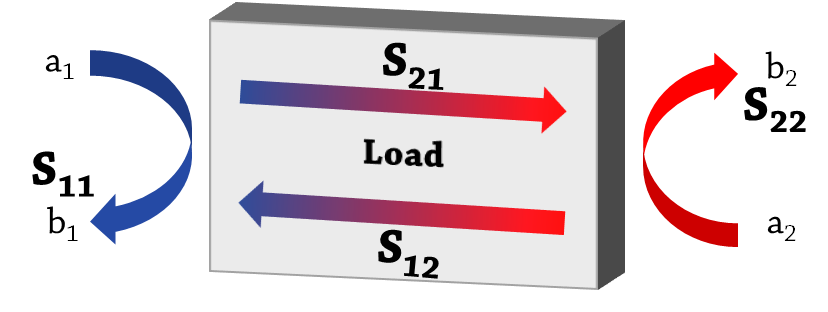}
    \caption[Schematic representation of the scattering parameters in a two-port system.]{Schematic representation of the \glspl{S-parameter} in a two-port system.}
    \label{fig:S_parameter_def}
\end{figure}

\Acrfullpl{S-parameter} were used to characterise the electrical behaviour of linear networks when stimulated by electrical signals~\cite{ballantine1996acoustic}. In this study, the frequency response of the \gls{IDT} was measured using \glspl{S-parameter} obtained from a \gls{VNA}. A two-IDT system, as shown in Figure~\ref{fig:S_parameter_def}, allows the measurement of four \glspl{S-parameter}: $\RM{S}_\RM{11}$, $\RM{S}_\RM{22}$, $\RM{S}_\RM{21}$, and $\RM{S}_\RM{12}$.
$\RM{S}_\RM{11}$ ($\RM{S}_\RM{11}=b_1/a_1$) and $\RM{S}_\RM{22}$ ($\RM{S}_\RM{22}=b_2/a_2$) represent the power reflected at port 1 and port 2, respectively, and are commonly referred to as reflection coefficients. The transmission coefficients, $\RM{S}_\RM{21}$ ($\RM{S}_\RM{21}=b_2/a_1$) and $\RM{S}_\RM{12}$ ($\RM{S}_\RM{12}=b_1/a_2$), measure the power transmitted from port 1 to port 2 and from port 2 to port 1, respectively.
S-parameters are typically expressed in decibels (dB), given by the equation:
\begin{equation}
\mathcal{A}_{ij}=20\log(\RM{S}_{ij}),
\end{equation}
where $i$ and $j$ can be either 1 or 2.\par

\subsubsection{Circuit for \gls{S-parameter} measurements}

\begin{figure}
    \centering
    \includegraphics[width=1\textwidth]{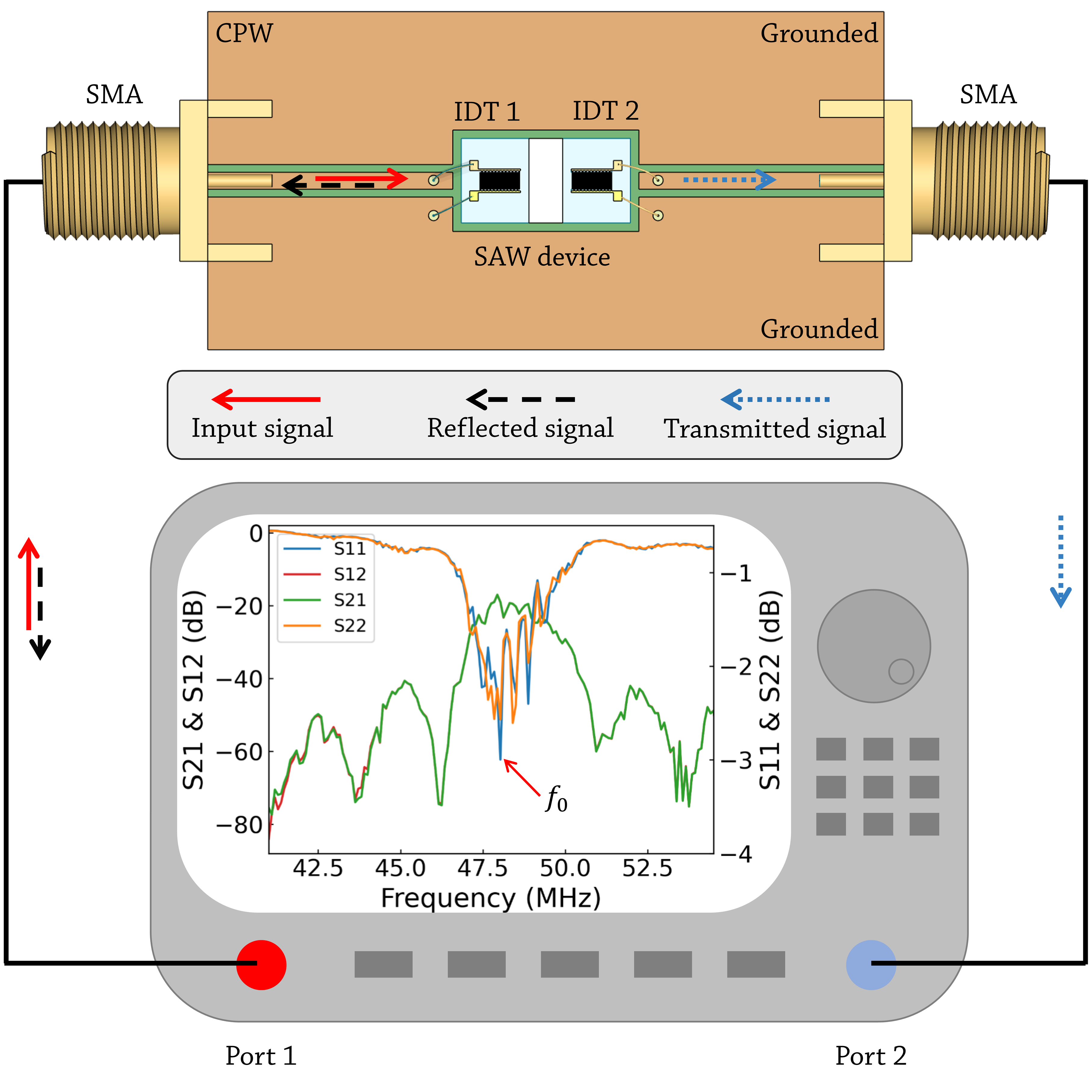}
    \caption[Schematic of the radio-frequency circuit used to characterise scattering parameters.]{\revision{Schematic of the \gls{rf} circuit used to characterise \glspl{S-parameter}. The inset graph shows an example of the typical \glspl{S-parameter} of the \glspl{IDT} used in this study.}}
    \label{fig:S-parameters_circuit}
\end{figure}

Figure~\ref{fig:S-parameters_circuit} illustrates the circuit setup employed to measure the \glspl{S-parameter} of the \glspl{IDT}. Two \glspl{IDT} were connected to the ports of a \gls{VNA} (Agilent E5062A) via a \gls{CPW} using \gls{SMA} connectors. One port of the \gls{VNA} was used to apply \gls{rf} signals, while the other port was used to detect the transmitted signals.
The inset graph in Figure~\ref{fig:S-parameters_circuit} presents an example of the \glspl{S-parameter} obtained from the \glspl{IDT} in this study. The centre frequency is determined as the frequency at which the transmission coefficient is maximal and the reflection coefficient is at minimal.

\section{Micromagnetic simulations}
Micromagnetic simulations were conducted using Mumax3 throughout this thesis to investigate how \glspl{SAW} influence magnetisation dynamics of the thin films~\cite{Vansteenkiste2014}. 
Mumax3 integrates numerically the \gls{LLG} equation (Equation~\ref{eqn:LLG}).
By modelling the interactions between the \glspl{SAW} and the magnetic thin films at the microscopic level, the simulations offer insights into the underlying physical mechanisms and help elucidate the observed experimental phenomena. 
This combined experimental and theoretical approach enhances our overall understanding of the \gls{SAW}-induced magnetisation dynamics and thus contributes to a more comprehensive understanding of the system.\par

\glspl{SAW} were implemented by the Mumax3 built-in strain tensor using spatial and time profiles~\cite{vanderveken2021finite}. 
It should be noted that the film is considered acoustically thin, meaning that it is sufficiently thin and rigid for the chosen \gls{SAW} frequency in this study. \revision{The entire film moves synchronously with the substrate surface, where the in-plane gradients are predominant compared to those normal to the surface, leading to a consistent displacement throughout the film thickness~\cite{ballantine1996acoustic,martin1994dynamics}}. Therefore, only in-plane strain ($\epsilon_{xx}$) was taken into consideration, and $B_2$ in Equation~\ref{eqn:E_me} was set to zero in this thesis. \par

\begin{figure}[ht!]
    \centering
    \includegraphics[width=0.7\textwidth]{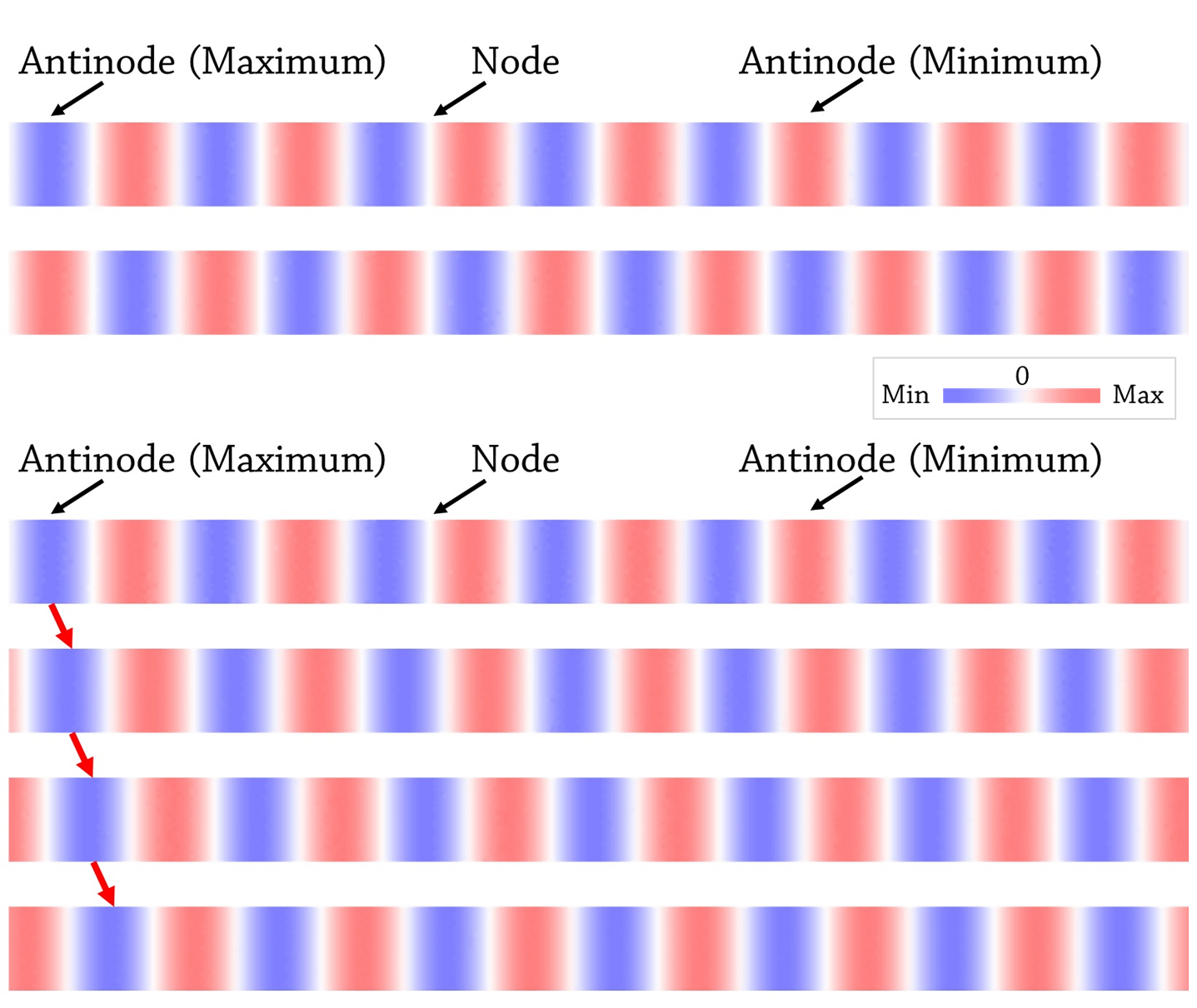}
\begin{picture}(0,0)
    \put(-310,225){\small (a)}
    \put(-310,130){\small (b)}
\end{picture}
    \caption[Example of standing and travelling surface acoustic waves implemented using Mumax3.]{Example of (a) standing and (b) travelling (b) \glspl{SAW} implemented using Mumax3.}
    \label{fig:Implementation_of_SAW}
\end{figure}
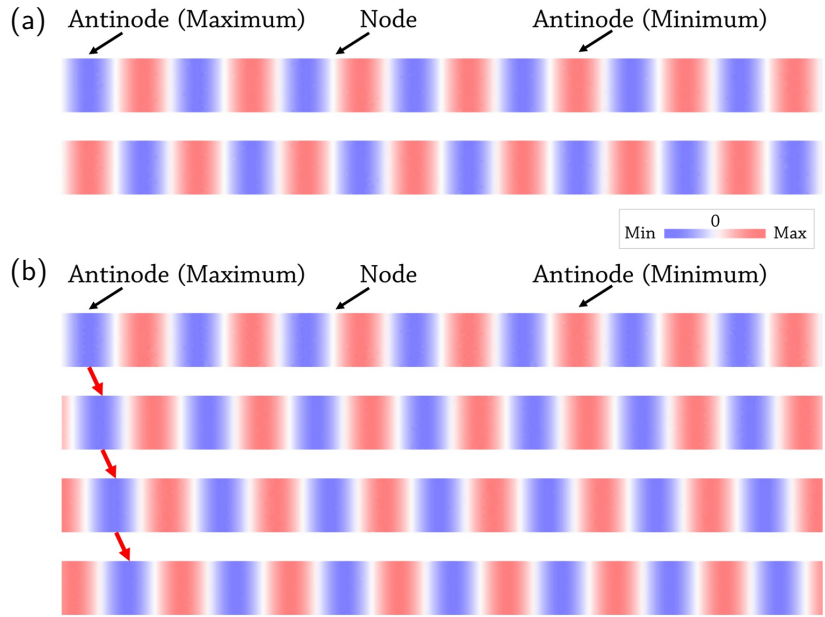

Figure~\ref{fig:Implementation_of_SAW} depicts a typical example of the spatial profile of strain for both standing and travelling \glspl{SAW}. The blue and red stripes in the figure indicate the anti-nodes of the \glspl{SAW}, representing regions with maximum (compressive) and minimum (tensile) strain, respectively. The white regions between the blue and red stripes signify the nodes, where the strain is minimal.\par

The $\epsilon_{xx}$ of standing \glspl{SAW} was implemented as follows
\begin{equation}\label{eqn:standingSAW}
    \epsilon_{xx} = \sin(kx)\cos(\omega t),
\end{equation}
where $k$ and $\omega$ are the wavenumber, angular frequency of the standing \gls{SAW}, respectively, and $t$ is time.\par

Similarly, the $\epsilon_{xx}$ of travelling \gls{SAW} was implemented as follows
\begin{equation}\label{eqn:travellingSAW}
    \epsilon_{xx} = \sin(\omega t)\cos(kx)-\cos(\omega t)\sin(kx),
\end{equation}
where $k$ and $\omega$ are the wavenumber, angular frequency of the travelling \gls{SAW}, respectively, and $t$ is time.
Detailed information regarding the material parameters used and the simulation procedures can be found in the respective chapters.\par

\section{Summary}

This chapter presented an overview of the methods and techniques employed for the design, fabrication, and characterisation of magnetic thin films and \gls{SAW} devices.
The magnetic thin films were deposited using \gls{dcMS}, and their deposition rates calibrated using \gls{XRR}. The magnetic properties of the thin films, such as hysteresis loops and domain patterns, were examined using Kerr microscopy, which allows for the visualisation and analysis of magnetisation behaviour.
\gls{SAW} devices, which consisted of \glspl{IDT} and magnetic thin films, were designed, fabricated, and characterised to investigate the influence of \glspl{SAW} on the magnetic properties. The \glspl{IDT} were prepared through standard photolithography techniques, and their \glspl{S-parameter} were measured using a \gls{VNA}, which provided valuable information about the electrical behaviour and performance of the \glspl{IDT}.
To gain further insights into the magnetisation dynamics induced by \glspl{SAW}, micromagnetic simulations were performed using the Mumax3 software. These simulations enabled a detailed investigation of the \gls{SAW} effects on the magnetisation dynamics of the thin films, providing valuable theoretical predictions that completing the experimental measurements.
The subsequent chapters of this thesis detail specific aspects of these experimental measurements and simulations, together providing an improved understanding of the \gls{SAW}-induced magnetisation dynamics in the magnetic thin films.
\cleardoublepagewithnumberheader
\chapter{Local magnetic anisotropy control}
\label{Chapter4_anisotropy_control}

\section{Introduction}
Materials with strong \gls{PMA} are promising candidates for future generations of data storage and processing devices owing to their stable magnetisation states and narrow \glspl{DW}~\cite{jung2008current,parkin2008magnetic,emori2013current,ryu2013chiral}. These features confer the stability of the stored information while providing a high storage density~\cite{dieny2017perpendicular}. However, unfortunately materials with strong \gls{PMA} typically also require a large current to reverse the magnetisation or to move \glspl{DW} in order to write or transfer data. This high current can lead to energy wastage and Joule heating, imposing limitations on the packing density of practical devices~\cite{miron2011fast, yamaguchi2005effect}. There is thus much interest in reducing the energy required to manipulate \glspl{DW}. \par

\gls{SAW}-induced coercivity \revision{modification} and \gls{DW} motion show significant potential for energy-efficient information storage and data processing devices, since the magnetisation switching is driven by voltage instead of current. However, the \gls{SAW}-induced magnetisation switching mechanism of thin films with \gls{PMA} remains unclear, and especially the role of the nodes and anti-nodes of the standing \gls{SAW} in the magnetisation reversal process. The correlation between the properties of \glspl{SAW} and magnetisation changes also is uncertain. \par

In this chapter, we demonstrate local anisotropy control of a \revision{Ta/Pt/Co/Ir/Ta} thin film with \gls{PMA} by standing \glspl{SAW} at a centre frequency of 93.35 MHz. In the presence of \glspl{SAW}, the coercivity of the thin films decreases significantly, while the magnetisation reversal speed increases. These experimental results along with micromagnetic simulations reveal that the anti-nodes of the standing \glspl{SAW} locally reduce the anisotropy of the Ta/Pt/Co/Ir/Ta thin film, which lowers the coercivity and favours magnetisation reversal. 
\revision{Existing literature, such as the work by Li \et, has explored the effects of \glspl{SAW} on magnetisation switching within thin films that exhibit in-plane magnetic anisotropy~\cite{li2014writing}. In this chapter, we will focus on the impact of the standing \glspl{SAW} on the magnetic properties of thin films with \gls{PMA}.}
The results of this work were published as \textit{``Local anisotropy control of Pt/Co/Ir thin film with perpendicular magnetic anisotropy by surface acoustic waves''}, \href{https://doi.org/10.1063/5.0097172}{Applied Physics Letters, 120, 252402 (2022)}~\cite{shuai2022local}.\par

\section{Experimental details}

\subsection{Magnetic thin film and SAW device}

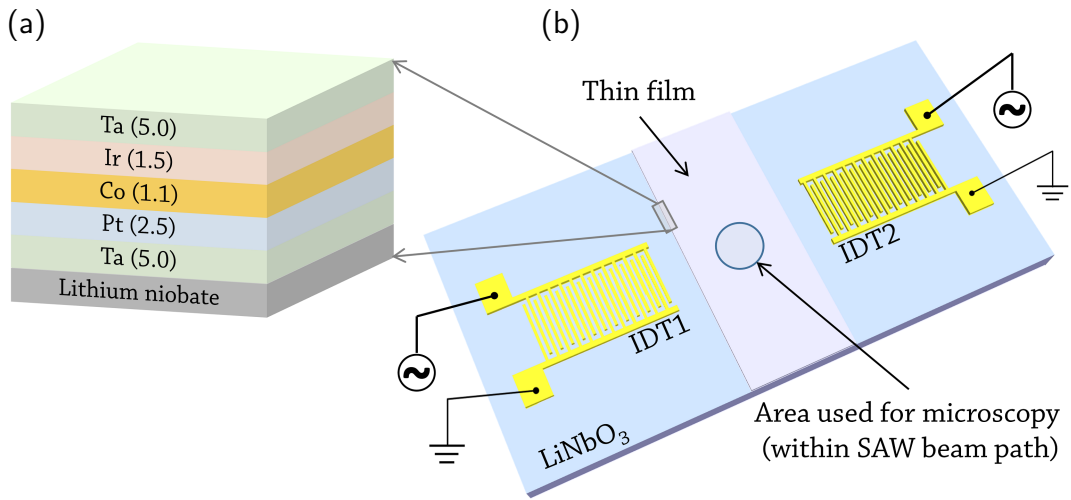
\begin{figure}[ht!]
\centering
\includegraphics[width=1\textwidth]{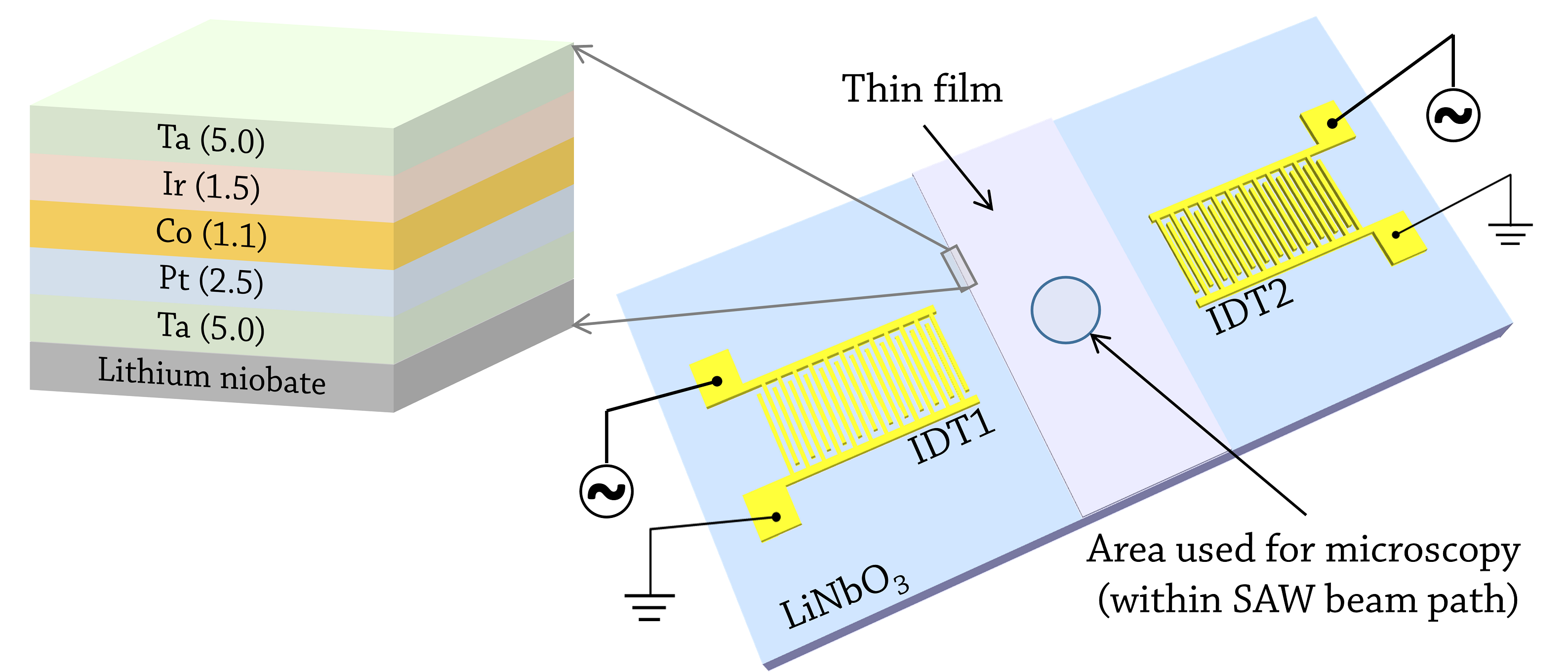}

\begin{picture}(0,0)
    \put(-200,190) {(a)}
    \put(0,190) {(b)}
\end{picture}

\caption[Thin film structure and schematic of experimental arrangement.]{(a) Thin film structure: Ta(5.0 nm)/Pt(2.5 nm)/Co(1.1 nm)/Ir(1.5 nm)/Ta(5.0 nm). (b) Schematic of experimental arrangement (not to scale). The 2-mm-wide thin film was deposited onto the \LNO\ substrate. \glspl{IDT} were patterned on each side of the thin film to launch the \glspl{SAW}.}
\label{fig:setup} 
\end{figure}

Figure~\ref{fig:setup} shows a schematic of the experimental arrangement.
The device comprises a 2-mm-wide stripe of Ta(5.0 nm)/Pt(2.5 nm)/Co(1.1 nm)/Ir(1.5 nm)/Ta(5.0 nm) thin film with \gls{PMA}, which was prepared by \gls{dcMS} on a 128\textdegree\ Y-cut \LNO\ substrate.
As shown in Figure~\ref{fig:setup}b, one pair of Ti(10 nm)/Au(90 nm) \glspl{IDT}, each consisting of 20 pairs of electrodes, was patterned using optical lithography (exposure of the resist was achieved using a \gls{MLA}, with subsequent metal evaporation and lift-off forming the \glspl{IDT} as mentioned in Section~\ref{Chapter3_nanofabrication}). The aperture of the \glspl{IDT} and the \gls{SAW} propagation distance were designed to be 500 \textmu m  and 3 mm, respectively. The electrode width and spacing were both set at 10 \textmu m. \par

\subsection{SAW circuit and S-parameters}
\begin{figure}[htbp!]
\centering
\includegraphics[width=0.6\textwidth]{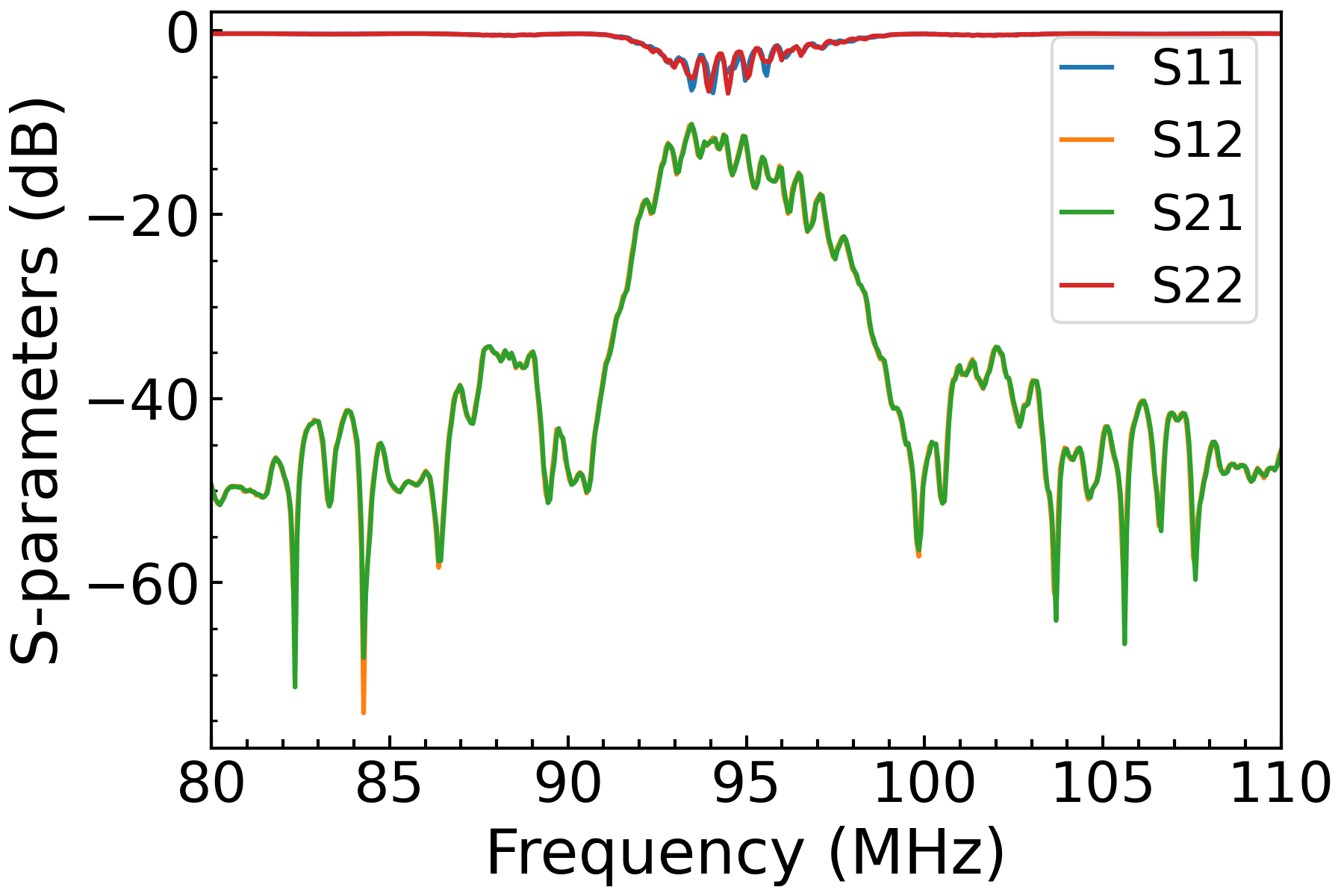}
\caption[Scattering parameters of the interdigitated transducers.]{\label{fig:S-parameters} 
\glspl{S-parameter} of the \glspl{IDT}. The delay line comprising both \glspl{IDT} and the substrate shows a centre frequency of 93.35 MHz, and 3 dB bandwidth of 9.34 MHz.}
\end{figure}

\begin{figure}[ht!]
    \centering
    \includegraphics[width=0.9\textwidth]{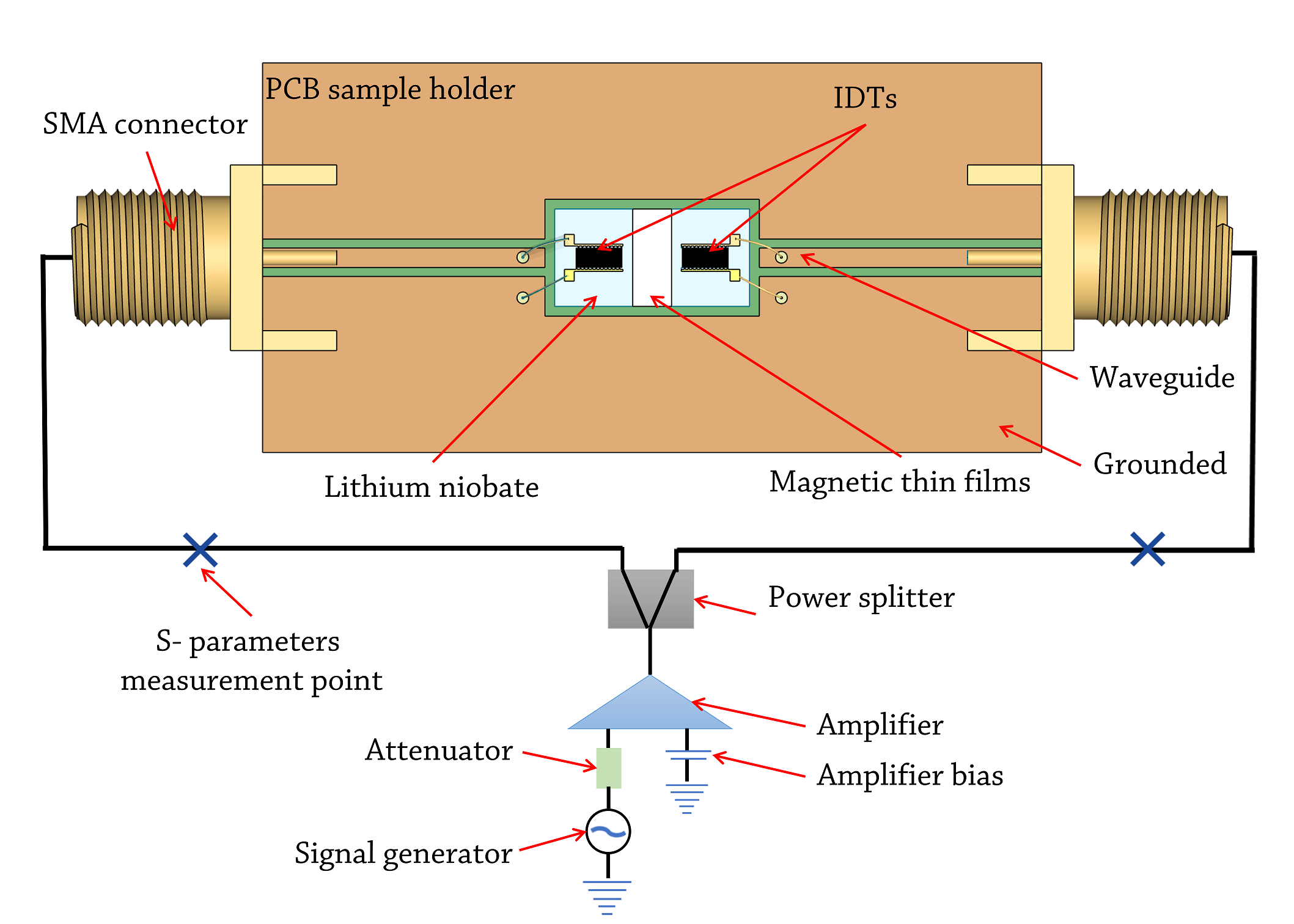}
    \caption[Schematic of the radio-frequency circuit for generating standing surface acoustic waves.]{Schematic of the \gls{rf} circuit for generating standing \glspl{SAW}. For \gls{S-parameter} measurements, the signal generator, attenuator, amplifier, and power splitter were removed, and a \gls{VNA} was connected at the ``X'' points.}
    \label{fig:SAW circuit}
\end{figure}

\glspl{S-parameter} of the \glspl{IDT} were measured using the method mentioned in Section~\ref{Chapter3_S-parameters_measurement}, allowing the peak value of transmission as a function of frequency to be determined, at which point \glspl{SAW} are generated most strongly (this point typically showing 11 dB transmission loss at 93.35 MHz, as shown in Figure~\ref{fig:S-parameters}). 
The total loss associated three factors: 
(i) cabling, 
(ii) transduction at the \gls{IDT} (expected to be the largest factor given no impedance matching was used) and 
(iii) \gls{SAW} propagation to the centre of the device, together introduces 11 dB/2 = 5.5 dB total loss from each port to the centre of the (symmetric) device. 
The \gls{VNA} output was then connected to the device as shown in Figure~\ref{fig:SAW circuit}. 
An attenuator (10 dB, Mini-Circuit BW-S10W2+) was connected between the signal generator and the amplifier (gain 33 dB, Mini-Circuits ZHL-2-8-S+) to ensure the applied \gls{rf} power does not exceed specifications of the amplifier.\par

To generate standing \glspl{SAW}, a power splitter (Mini-Circuit ZAPD-30-S+) was used to divide the signal from the \gls{VNA} output port into two, which each of the resulting signals then sent to one \gls{IDT}. The power from the \gls{VNA} was varied from 0 to 5 dBm. 
This power was reduced down to --10 to --5 dBm by an attenuator to ensure the output was kept inside the linear range of the amplifier. 
The final signal applied \gls{rf} power was thus amplified to between 23 to 28 dBm before being split using the 3 dB splitter (with each \gls{IDT} thus receiving $\sim$20 dBm to $\sim$25 dBm). 
Taking into account the measured losses at peak transmission, the expected total \gls{rf} power transduced into the standing \gls{SAW} mode was in the range $\sim$17.5 to $\sim$22.5 dBm. 
Figure~\ref{fig:S-parameters} shows the reflection (S11 and S22) and transmission (S21 and S12) characteristics of the \gls{SAW} transducers and substrate showing a centre frequency of 93.35 MHz, yielding a wavelength of 42.667 $\pm$ 0.004 \textmu m. The \gls{VNA} used in our study had a frequency accuracy of 5 ppm, and the propagating velocity of the \gls{SAW} on \LNO\ is known to be 3982 m/s~\cite{paskauskas1995velocity}. \par

\subsection{Determination of magnetisation reversal speed}
\begin{figure}[!ht]
    \centering
    \includegraphics[width=0.65\textwidth]{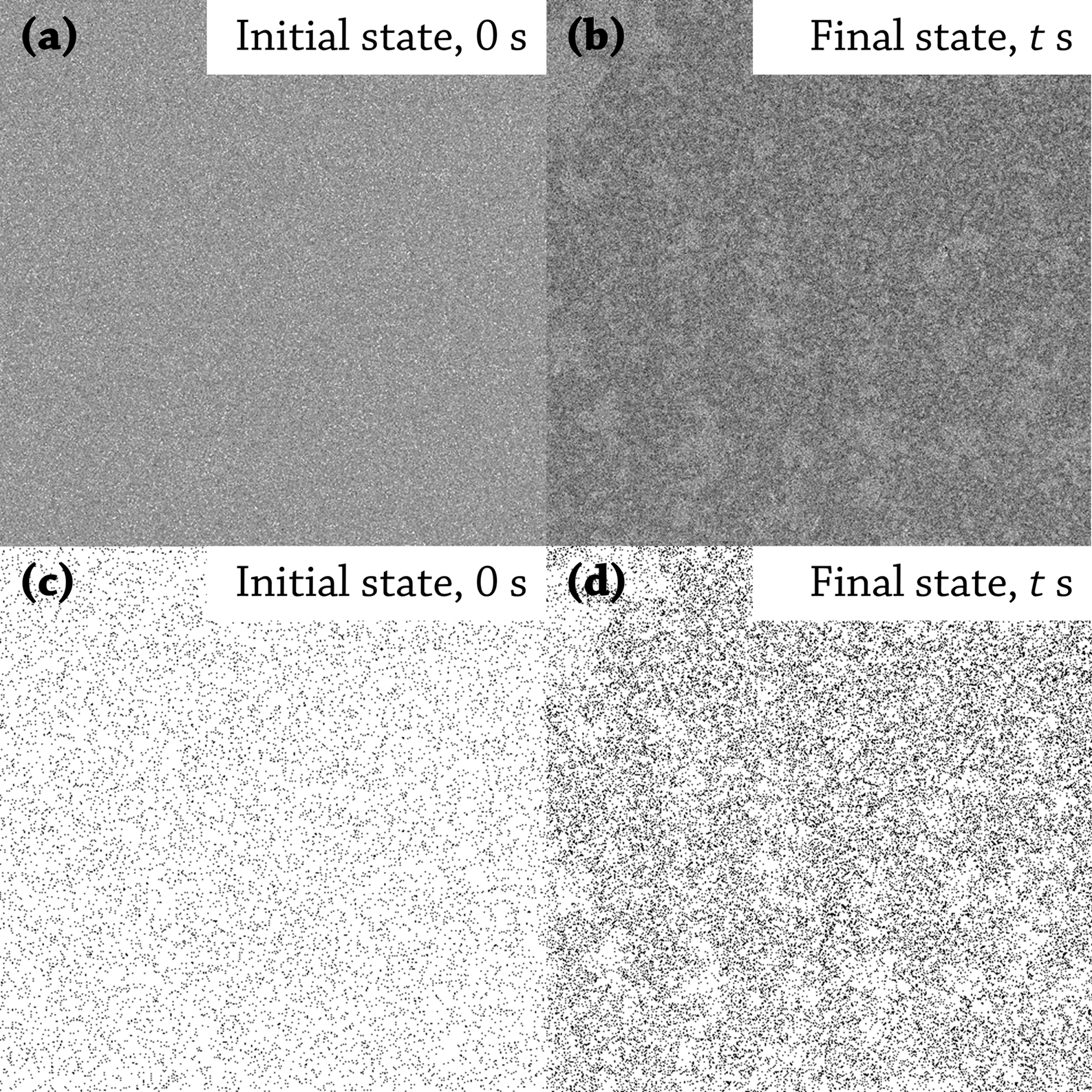}
    \caption[Determining magnetisation reversal speed using Kerr microscopy.]{Determining magnetisation reversal speed \revision{in the Ta/Pt/Co/Ir/Ta thin film} using Kerr microscopy: (a) Original image of the initial state (background noise) at 0 mT; (b) Final state after the application of a pulse magnetic field and \gls{SAW}; (c) Binary representation corresponding to (a); (d) Binary representation corresponding to (b).}
    \label{fig:determinationofmagnetisationreversalspeed}
\end{figure}

To determine the magnetisation reversal speed, we employed Kerr microscopy using the following procedure:
\begin{enumerate}[label=\roman*.]
    \item We first saturated the sample with a 50 mT out-of-plane field. After turning off the field, an initial image was captured to assess background noise, as illustrated in Figure~\ref{fig:determinationofmagnetisationreversalspeed}a.
    \item We then applied a pulsed field in the opposite direction (ranging from --2 to --3.2 mT). This was done to nucleate domains and promote \gls{DW} motion, both with the \gls{SAW} turned on and \gls{SAW} off (\gls{SAW} modal power varied between 17.5 to 22.5 dBm). Subsequently, we captured a second image representing the final state, as seen in Figure~\ref{fig:determinationofmagnetisationreversalspeed}b.
    \item Both images were converted into binary values, as depicted in Figure~\ref{fig:determinationofmagnetisationreversalspeed}c and d. In these binary representations, magnetisation pointing ``up'' was denoted by white pixels, while ``down'' was denoted by black pixels.
    \item We determined the number of black pixels in the initial and final images as $P_0$ and $P_1$, respectively. The difference between $P_0$ and $P_1$ represented reversed magnetisation.
\end{enumerate}

Thus, we expressed the magnetisation reversal speed as
\begin{equation}
    S_\RM{mag} = \frac{(P_1-P_0)\phi_0}{t_\RM{p}},
\end{equation}
where $\phi_0$ signifies the area per pixel (0.07 \textmu m\textsuperscript{2}/pixel for 50$\times$ lens), and $t_\RM{p}$ represents the pulse width of the field, set at 5 s. We used an image size of 500 $\times$ 500 pixel\textsuperscript{2} (equivalent to 132 $\times$ 132 \textmu m\textsuperscript{2}) to determine the magnetisation reversal speed. It is important to note that since we assessed the background noise on a global scale, our estimations for the magnetisation reversal speed might be underestimated.

\subsection{Computational details}
\begin{figure}[ht!]
    \centering
    \includegraphics[width=0.8\textwidth]{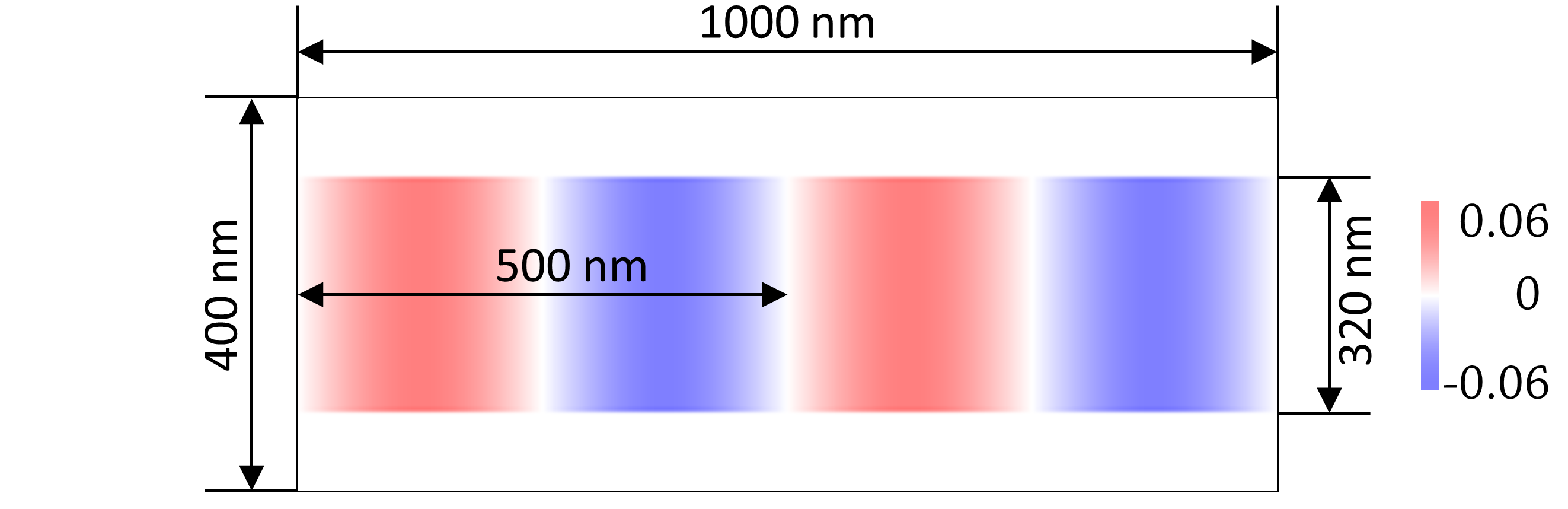}
    \caption[Configuration of the micromagnetic simulations.]{Configuration of the micromagnetic simulations. The entire dimension of the simulation is 400 $\times$ 1000 $\times$ 1 nm\textsuperscript{3}, with a \gls{SAW} beam path with a width of 320 nm. The chosen wavelength of the \gls{SAW} is 500 nm. The amplitude (strain) of the \gls{SAW} is 0.06.}
    \label{fig:SI2}
\end{figure}

Micromagnetic simulations were performed using Mumax3, based on the \gls{LLG} equation (Equation~\ref{eqn:LLG})~\cite{Vansteenkiste2014}. The magnetoelastic energy density of the system can be described as Equation~\ref{eqn:E_me}.
The material parameters were set as follows~\cite{yokouchi2020creation}: 
saturation magnetisation $M_\RM{s}$ = 6.0 $\times$ 10\textsuperscript{5} A/m, exchange constant $A_\RM{exch}$ = 1.0 $\times$ 10\textsuperscript{--11} J/m, 
PMA $K_\RM{u}$ = 8.0 $\times$ 10\textsuperscript{5} J/m\textsuperscript{3},
and magnetoelastic coupling coefficient $B_1$ = 1.0 $\times$ 10\textsuperscript{7} J/m\textsuperscript{3}. \revisiontwo{This $B_1$ value is taken from~\cite{yokouchi2020creation}, which is closed to the experimental observation as reported in~\cite{gutjahr2000magnetoelastic} (1.5 $\times$ 10\textsuperscript{7} J/m\textsuperscript{3} to 3.5 $\times$ 10\textsuperscript{7} J/m\textsuperscript{3}).}
A computational region of dimensions \revision{500 $\times$ 200 $\times$ 1 nm\textsuperscript{3}} with a mesh size of \revision{2 $\times$ 2 $\times$1 nm\textsuperscript{3}} was used (as shown in Figure~\ref{fig:SI2}).\par

\revision{The chosen mesh size is smaller than the exchange length, which defines the scale at which the atomic-level exchange interactions become more prominent than the typical magnetostatic fields. The exchange length, $l_\RM{exch}$, is given by $l_\RM{exch} = \sqrt{A_\RM{exch}/K_\RM{u}}$,  resulting in an exchange length of 3.5 nm in our simulations.}
The velocity of the \gls{SAW} was set to 4000 m/s~\cite{paskauskas1995velocity}. For the simulation, the wavelength of the \gls{SAW} was set as 500 nm owing to the limitations of computing capacity~\cite{dean2015sound,yokouchi2020creation,nepal2018magnetic}.
\revision{It should be noted that the material parameters and \gls{SAW} characteristics differ from those explored in experimental studies, therefore, the findings should be carefully interpreted.}
Standing \glspl{SAW} were implemented using Equation~\ref{eqn:standingSAW} in the central part of the computational region representing the beam path of the \gls{SAW} with a width of 320 nm (the total width is 400 nm).

\section{Results and discussion}
%
\subsection{Reduced coercivity}
\begin{figure}
    \centering
    \includegraphics[width=0.95\textwidth]{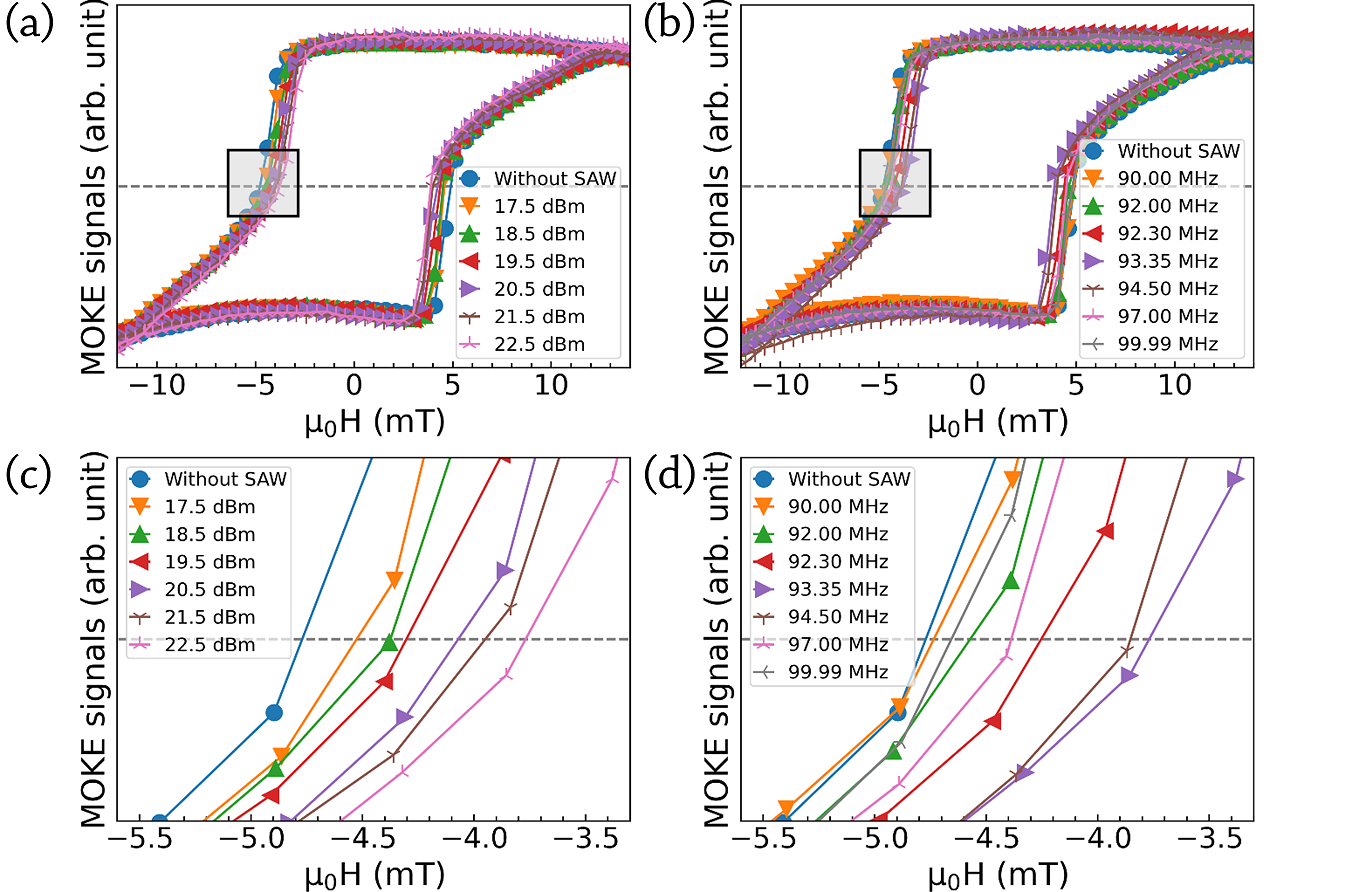}
    \caption[Hysteresis loops of the Ta/Pt/Co/Ir/Ta thin film.]{Hysteresis loops of the \revision{Ta/Pt/Co/Ir/Ta} thin film as a function of the (a) applied \gls{rf} power and (b) frequency. (c) and (d) are the enlarged views of the framed areas in (a) and (b), respectively. Lines are the guide to the eye.}
    \label{fig:coercivityreduction}
\end{figure}

Figure~\ref{fig:coercivityreduction}a shows the hysteresis loops of the thin films for different applied \gls{rf} power from 17.5 to 22.5 dBm (at 93.35 MHz), and also without a \gls{SAW}. 
The sharp switching of the magnetisation indicates a strong \gls{PMA} of the Co layer. 
The presence of the curved corners is because a higher (than coercivity) field is required to squeeze out the homochiral \glspl{DW} to saturate the thin film~\cite{shape_of_hysteresis_loop}.\par
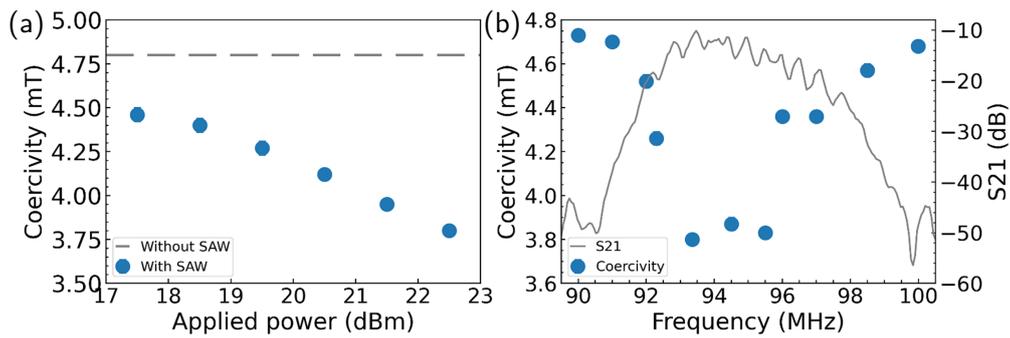
\begin{figure}
    \centering
    \includegraphics[width=.9\textwidth]{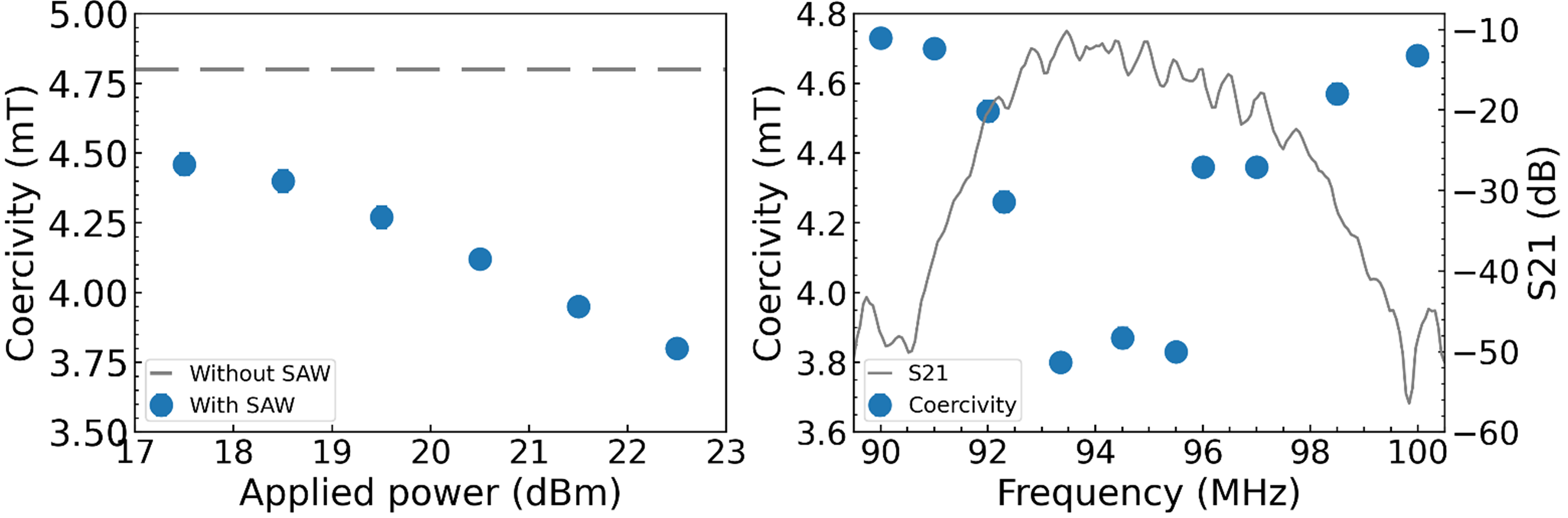}
\begin{picture}(0,0)
    \put(-380,115) {\small (a)}
    \put(-202,115) {\small (b)}
\end{picture}
    \caption[Coercivity of the Ta/Pt/Co/Ir/Ta thin film.]{Coercivity of the \revision{Ta/Pt/Co/Ir/Ta} thin film as a function of the (a) applied \gls{rf} power and (b) frequency. S21 is also displayed in (b). Error bars are smaller than the data points.}
    \label{fig:coercivityreduction_2}
\end{figure}

Significant coercivity reduction is observed in the presence of the standing \gls{SAW}: The coercivity decreases with increasing applied \gls{rf} power from 4.46 $\pm$ 0.04 mT at 17.5 dBm to 3.80 $\pm$ 0.02 mT at 22.5 dBm (Figure~\ref{fig:coercivityreduction}b and c). As shown in Figure~\ref{fig:coercivityreduction_2}, the coercivity is thus up to $\sim$21\% reduced compared to that measured without \gls{SAW} (4.80 $\pm$ 0.03 mT). The coercivity reduction can be explained by the strain-induced changes of the \gls{PMA}~\cite{shepley2015modification}. 
The \glspl{SAW} act as time-varying strain waves, which can locally change the energy landscape of the thin films, periodically raising and lowering the anisotropy of the thin film. The magnetisation reverses when the anisotropy at low values. Owing to the nature of the magnetisation reversal process, the magnetisation is not \revision{reversible} even when the anisotropy is at the highest value. The anisotropy changes increase with the increasing applied \gls{rf} powers.\par

%
\subsection{Enhanced magnetisation reversal speed}

\begin{figure}[ht!]
    \centering
    \includegraphics[width=1\textwidth]{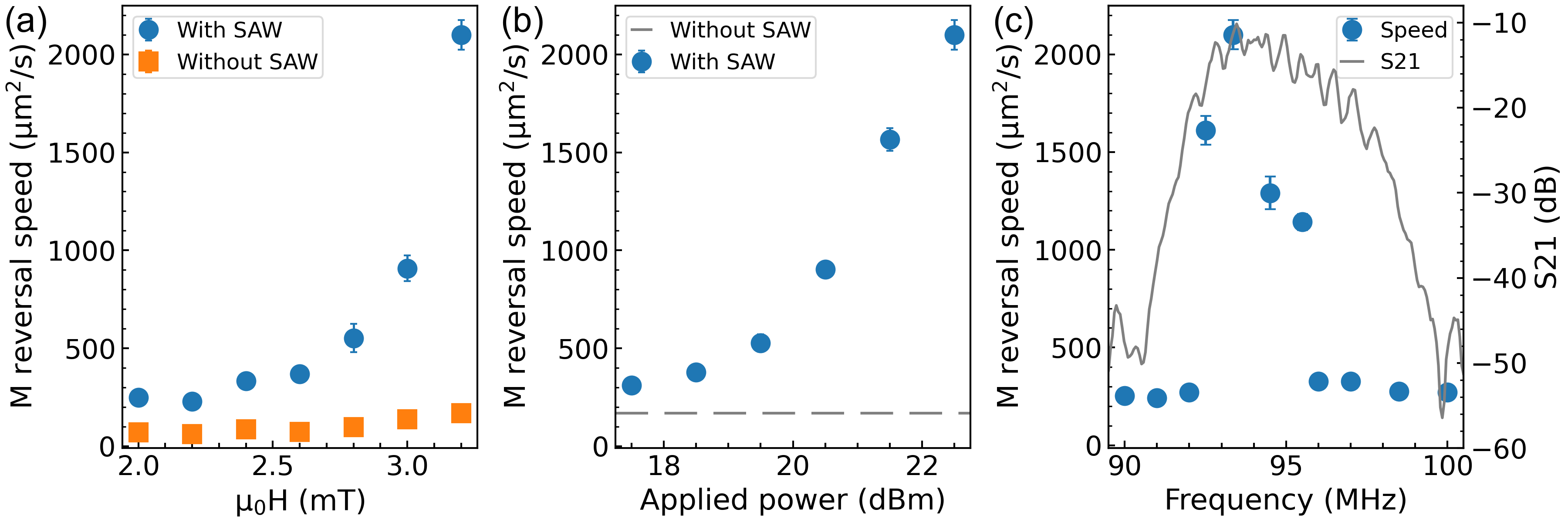}
    \caption[Magnetisation reversal speed of the Ta/Pt/Co/Ir/Ta thin film.]{Magnetisation reversal speed \revision{of the Ta/Pt/Co/Ir/Ta thin film} as a function of the (a) external magnetic field, (b) applied \gls{rf} power and (c) frequency of the standing \gls{SAW}. The applied \gls{rf} power and frequency of \gls{SAW} in (a) are 22.5 dBm and 93.35 MHz, respectively. The external field and frequency of \gls{SAW} in (b) are 3.2 mT and 93.35 MHz, respectively. The external field and applied \gls{rf} power of \gls{SAW} in (c) are 3.2 mT and 22.5 dBm, respectively.}
    \label{fig:magnetisation reversal speed}
\end{figure}

Figure~\ref{fig:magnetisation reversal speed}a to c show the influence of the external field, applied \gls{rf} power and frequency of the \gls{SAW} on the magnetisation reversal speed, respectively. 
As shown in Figure~\ref{fig:magnetisation reversal speed}a, with the external magnetic field increasing from 2.0 to 3.2 mT, the change in the magnetisation reversal speed without \gls{SAW} is very limited (from 71 $\pm$ 5 \textmu m\textsuperscript{2}/s to 168 $\pm$ 3 \textmu m\textsuperscript{2}/s), whereas in the presence of the standing \gls{SAW}, the magnetisation reversal speed increases from \revision{250 $\pm$ 20} \textmu m\textsuperscript{2}/s at 2.0 mT to 2100 $\pm$ 80 \textmu m\textsuperscript{2}/s at 3.2 mT, which is a $\sim$740\% increase. The magnetisation reversal speed is increased eleven-fold at 3.2 mT in the presence of \gls{SAW} compared to that with magnetic field only.
The implication is that the standing \gls{SAW} can significantly lower the required switching field and accelerate the magnetisation reversal.
At 93.35 MHz, the magnetisation reversal speed gradually increases with the increasing applied \gls{rf} power from 310 $\pm$ 30 \textmu m\textsuperscript{2}/s at 17.5 dBm to 2100 $\pm$ 80 \textmu m\textsuperscript{2}/s at 22.5 dBm with a 3.2 mT field (see Figure~\ref{fig:magnetisation reversal speed}b).\par

Similar to the effect of frequency on the coercivity reduction, the magnetisation reversal speed increases when the frequency approaches the centre frequency, achieving its maximum value near the centre frequency (as shown in Figure~\ref{fig:magnetisation reversal speed}c). We note the S21 response is asymmetric and biased towards better transduction at lower frequencies, resulting in the biased magnetisation reversal speed responding to the source frequency.
The change of the magnetisation reversal speed is caused by two main factors. Firstly, the presence of the standing \gls{SAW} significantly reduces the domain nucleation field owing to the magnetoelastic coupling effect (see Figure~\ref{fig:coercivityreduction}a). \revision{The \gls{DW} energy density can be expressed as: \revision{$\gamma_{\tiny\RM{DW}}=4\sqrt{A_\RM{exch}K_\RM{eff}}$}, where $A_\RM{exch}$ is the exchange stiffness and $K_\RM{eff}$ is the effective anisotropy. Meanwhile, the \gls{SAW} smooths the energy landscape of the thin film, accelerating the \gls{DW} velocity.} \revision{\glspl{SAW} locally and periodically modifying the anisotropy and the energy potential of the thin film, which benefits the \gls{DW} nucleation and motion}~\cite{shepley2015modification}. \par

%
\subsection{Standing SAW-induced stripe domain patterns}
\begin{figure}[!ht]
    \centering
    \includegraphics[width=.85\textwidth]{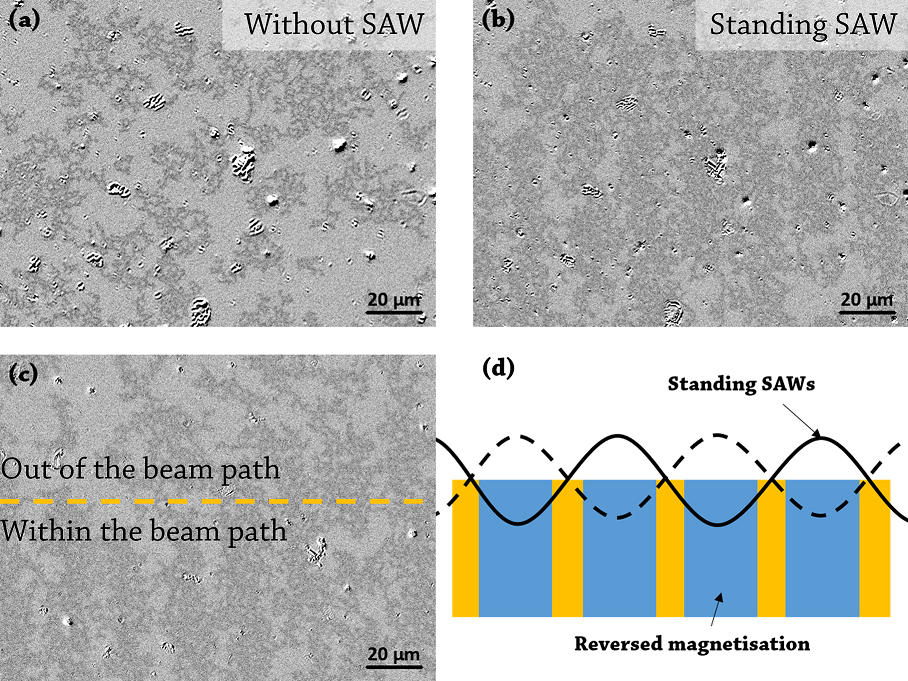}
    \caption[Domain patterns in the Ta/Pt/Co/Ir/Ta thin film.]{\revision{Domain patterns in the Ta/Pt/Co/Ir/Ta thin film.} (a) Domain patterns randomly distributed obtained with only magnetic field, (b) stripe domain patterns with both magnetic field and standing \gls{SAW}, (c) domain patterns at the boundary of the beam path, and (d) schematic of the standing \gls{SAW}-assisted magnetisation reversal.}
    \label{fig:domain pattern}
\end{figure}

\begin{figure}[h]
    \centering
    \includegraphics[width=.55\textwidth]{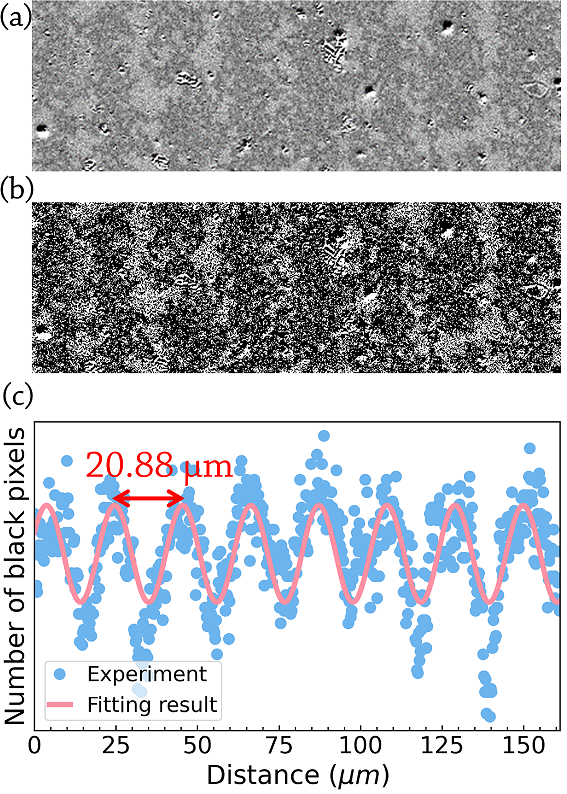}
    \caption[Analysis of the strip domain pattern in the Ta/Pt/Co/Ir/Ta thin film upon application of standing surface acoustic waves.]{\revision{Analysis of the strip domain pattern in the Ta/Pt/Co/Ir/Ta thin film upon application of standing \glspl{SAW}.} (a) Cropped image from Figure~\ref{fig:domain pattern}b, (b) is (a) in binary value, (c) the number of black pixels in (b) and its sinusoidal fitting. The fitting results show a wavelength of $\sim$20.88 \textmu m, which is very close to the half-wavelength of the standing \gls{SAW} ($\sim$21 \textmu m).}
    \label{fig:domain pattern_2}
\end{figure}

Figure~\ref{fig:domain pattern}a and b show domain patterns with both \gls{SAW} off and \gls{SAW} on obtained using Kerr microscopy, respectively. 
A 50 mT out-of-plane magnetic field was first applied to the sample to saturate the magnetisation in the ``up'' direction. 
A 4.6 mT field was then applied in the opposite direction. A large number of small branch domains nucleated and propagated randomly in the thin film without the \gls{SAW} (see Figure~\ref{fig:domain pattern}a). 
With application of standing \glspl{SAW}, the required domain nucleation field was much lower (around 3.6 mT for an applied \gls{rf} power of 22.5 dBm at the centre frequency of 93.35 MHz). 
The domains tended to nucleate in certain parts of the thin film and lined up forming a clear stripe pattern (as shown in Figure~\ref{fig:domain pattern}b). 
The alignment of the domain patterns only occurred within the \gls{SAW} beam path, with a random distribution of domains observed outside of the beam path (Figure~\ref{fig:domain pattern}c). This result itself indicates very strongly that the standing \glspl{SAW} are responsible for the domain patterns. Figure~\ref{fig:domain pattern_2}e is cropped from Figure~\ref{fig:domain pattern}b. \par

By converting Figure~\ref{fig:domain pattern_2}a into binary values (Figure~\ref{fig:domain pattern_2}b) and counting the number of the black pixels, we plotted the distribution of the domains (see Figure~\ref{fig:domain pattern_2}c). The experimental data was fitted by a sinusoidal function. The fit shows that the spacing between the lines is 20.88 $\pm$ 0.06 \textmu m, corresponding well to the nominal half-wavelength ($\sim$21.33 \textmu m) spacing expected of the standing \gls{SAW}. 
This is because the magnetoelastic coupling effect is the strongest at anti-nodes of the standing \gls{SAW}, where the surface deflection is largest, and where the local lowering the anisotropy is expected to be the greatest. The effect is expected to be the weakest at the nodes of the standing \gls{SAW} (as depicted in Figure~\ref{fig:domain pattern}d). \par

%
\subsection{Magnetisation dynamics under standing SAWs}

\begin{figure}[!ht]
    \centering
    \includegraphics[width=0.6\textwidth]{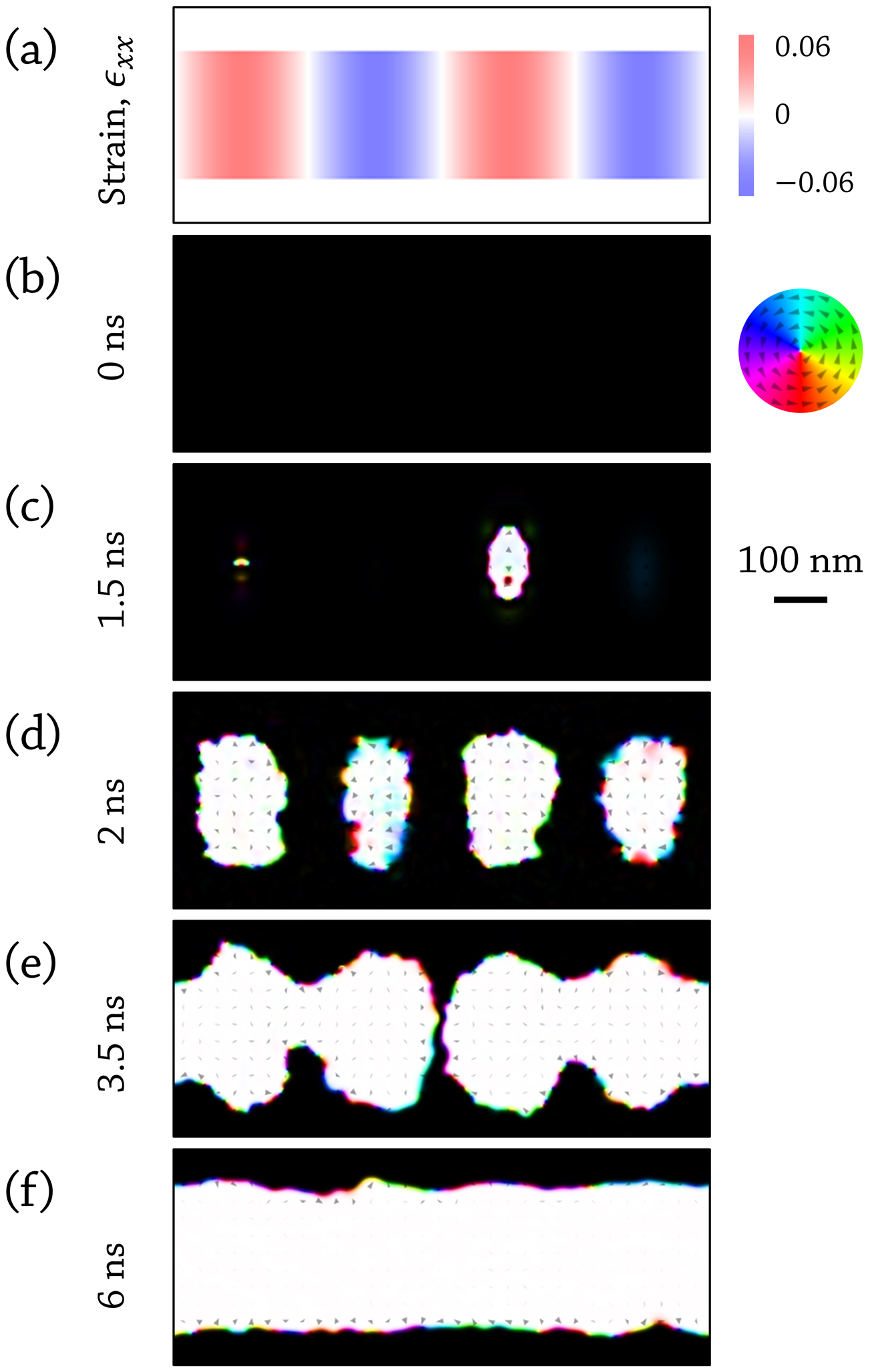}
    \caption[Simulation of magnetisation reversal process.]{\revision{Simulation of magnetisation reversal process.} (a) Spatial strain profile, (b) to (f) magnetisation at 0, 1.5, 2, 3.5 and 6 ns after applying field and standing \gls{SAW}. The black and white colours in (b) to (f) represent magnetisation pointing ``down'' and ``up'', respectively. The colour code for the in-plane magnetisation component is shown by a colour wheel.}
    \label{fig:mumax}
\end{figure}

To further understand the effect of the nodes and anti-nodes of the standing \gls{SAW} on the magnetisation reversal process, we performed micromagnetic simulations using Mumax3 with a built-in strain tensor~\cite{Vansteenkiste2014}.
As shown in Figure~\ref{fig:mumax}a, the red, white and blue indicate strains with maximum, zero, and minimum values, respectively. The system is firstly relaxed with magnetisation pointing ``down'' (see Figure~\ref{fig:mumax}b). Standing \glspl{SAW} ($\epsilon_{\RM{max}}$ = 0.06) together with external magnetic field (300 mT) pointing ``up'' are then applied. As shown in Figure~\ref{fig:mumax}c, at 1.5 ns, domains begin nucleating at the anti-nodes of the standing \gls{SAW}. Then, the domains continue nucleating and propagating from anti-nodes to nodes until the domains merge with each other (Figure~\ref{fig:mumax}d and e), followed by domain propagation out of the beam path (Figure~\ref{fig:mumax}f). The simulation results agree qualitatively with our experiments and in particular our assumption that the anti-nodes of the standing \gls{SAW} favour the nucleation of the domains and the propagation of the \glspl{DW}. The simulation and magnetisation reversal speed results also reveal that standing \glspl{SAW} lower the coercivity of the thin films by lowering local domain nucleation field and accelerating \gls{DW} propagation.\par

\clearpage

\section{Summary}
To summarise, in this chapter, a strong coercivity reduction (up to $\sim$21\%) was observed for the Ta/Pt/Co/Ir/Ta thin film with \gls{PMA} upon application of 93.35 MHz standing \glspl{SAW}. 
Owing to the \gls{SAW}-induced magnetoelastic coupling effect, the magnetisation reversal speed was significantly enhanced (by a factor of $\sim$11).
The experiments and Mumax simulations together indicated that, owing to the strain distribution difference, standing \glspl{SAW} were able to lower the domain nucleation field locally at the anti-nodes of the standing \gls{SAW}. Domains tended to nucleate and propagate from anti-nodes to nodes of the standing \gls{SAW}, forming striped domain patterns with spacing the same as the half-wavelength of the standing \gls{SAW}. \par
\cleardoublepagewithnumberheader
\chapter{Separation of heating and strain effects}
\label{Chapter5_heating_effect}

\section{Introduction} 
The motion of the magnetic \glspl{DW} in thin films with \gls{PMA} shows potential for technological applications in spintronics, mediated by the magnetoelastic coupling effect. The dynamic strain waves that \glspl{SAW} introduce create a dynamic energy landscape that can trigger the magnetisation precession~\cite{thevenard2013irreversible,thevenard2016precessional,camara2019field}, assist the magnetisation switching~\cite{shuai2022local,li2014acoustically}, and enhance the \gls{DW} motion~\cite{adhikari2021surface,edrington2018saw,vilkov2022magnetic,wei2020surface,cao2021surface}. 
However, the applied \gls{rf} power and the propagation of \glspl{SAW} through a structure can also result in a global or local temperature increase in devices owing to either the dissipation of the \gls{rf} power or by direct acoustothermal heating from the \gls{SAW}~\cite{zheng2018role,utko2006heating,han2021thermal,shilton2015rapid}.
These temperature increases can also contribute to the \gls{DW} motion. The acoustothermal heating effect is often-discussed in the context of the interaction of \glspl{SAW} with microfluidic systems~\cite{shilton2015rapid,ha2015acoustothermal,kondoh2009development,das2019acoustothermal,park2015acoustothermal,kulkarni2009surface,wang2021rapid,weser2022three}. For example, Wang \et investigated \gls{SAW} as a rapid and controllable acoustothermal microheater for both the sessile droplet and liquid within a polydimethylsiloxane microchamber~\cite{wang2021rapid}. On the other hand, Utko \et found that \gls{rf} heating effect \revision{(Joule heating)} plays a more important role than \gls{SAW} itself in a \gls{SAW}-driven single-electron pumps~\cite{utko2006heating}.\par

Evaluation of the different possible heating mechanisms in \gls{SAW}-magnetic film systems is therefore crucial in order to distinguish between heating- and \gls{SAW}-induced effects. In this chapter, we therefore examine heating within a \gls{SAW} device that comprises of a \revision{Ta/Pt/Co/Ta} thin film with \gls{PMA} in the presence of standing or travelling \glspl{SAW} within a frequency range of 46 to 51 MHz and at a power of 21 dBm. \revision{In contrast to the Ta/Pt/Co/Ir/Ta thin films discussed in the previous chapter, the Ta/Pt/Co/Ta thin film used in this chapter demonstrated a smoother \gls{DW} profile, attributable to the reduced \gls{DMI} at the Co/Ta interface. This feature enabled more precise measurements of \gls{DW} velocity under a range of experimental conditions.} The heating was measured \textit{in situ} using Pt thin film thermometers placed at several locations on the same chip, including within the \gls{SAW} beam path. \gls{DW} velocity was also separately determined at different temperatures firstly with no \gls{SAW} applied, and secondly in the presence of standing or travelling \glspl{SAW}, allowing us to discuss and investigate fully the origin and impact of heating, and to distinguish it from the \gls{SAW} driven effects. Results obtained in this chapter were published as \textit{``Separation of heating and magnetoelastic coupling effects in surface-acoustic-wave-enhanced creep of magnetic domain walls''}, \href{https://doi.org/10.1103/PhysRevApplied.20.014002}{Physical Review Applied, 20, 014002 (2023)}~\cite{shuai2023separation}.

\section{Methods}
\subsection{Magnetic thin film and SAW device}

\begin{figure}
\centering
\includegraphics[width=1\textwidth]{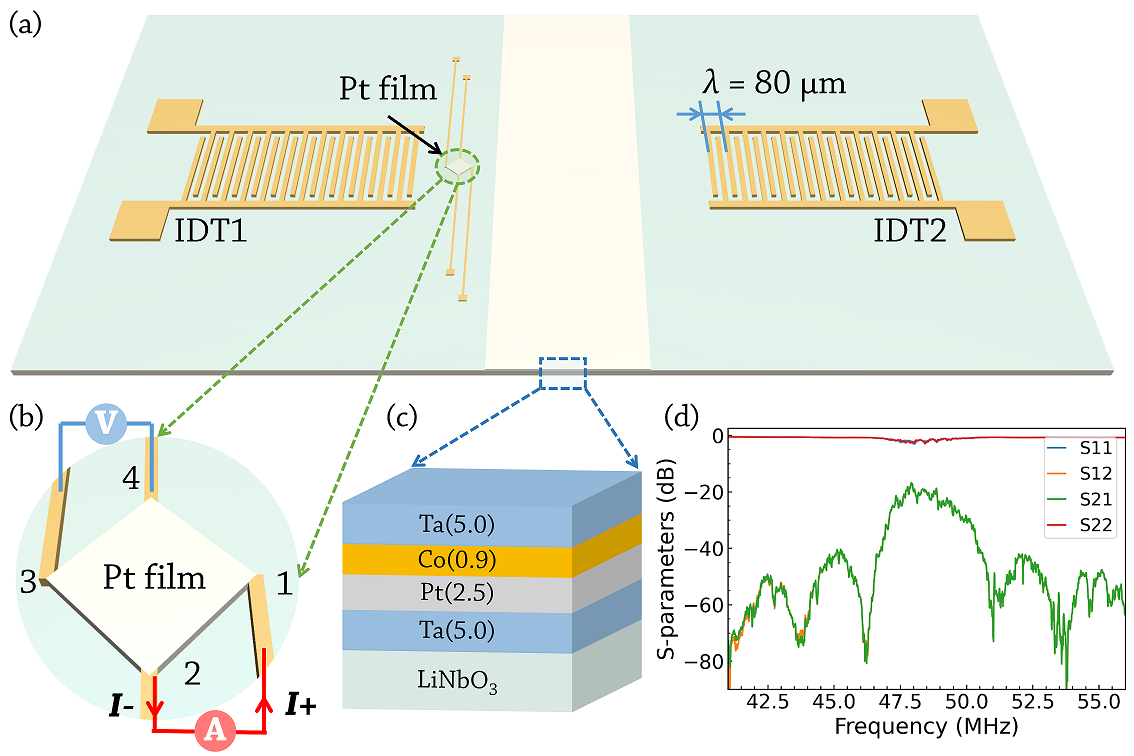}
\caption[Schematic diagram of the surface acoustic wave device, electrical transport measurement, and thin film structures.]{\label{fig:config}
(a) Schematic diagram of the \gls{SAW} device (not to scale). A 2-mm wide magnetic thin film is deposited by \gls{dcMS} onto a \LNO\ substrate. A pair of \glspl{IDT}, designed to launch \glspl{SAW} with a wavelength ($\lambda$) of 80 \textmu m, are positioned at the opposite ends of the magnetic thin film. In between the magnetic thin film and the \glspl{IDT}, and within the path of the \gls{SAW} beam, a Pt film is patterned. (b) Schematic representation of the electrical transport measurements using the Pt film. Current passes through Ports 1 and 2, while voltage is simultaneously measured at Ports 3 and 4. (c) The structure of the Ta(5.0 nm)/Pt(2.5 nm)/Co(0.9 nm)/Ta(5.0 nm) thin film. (d) \glspl{S-parameter} for the \glspl{IDT} used to launch \glspl{SAW}. The delay line, which includes both \glspl{IDT} and the substrate, exhibits a centre frequency of 48 MHz.}
\end{figure}

Figure~\ref{fig:config}a shows a schematic of the \gls{SAW} device used in this chapter.
A 2-mm wide Ta(5.0 nm)/Pt(2.5 nm)/Co(0.9 nm)/Ta(5.0 nm) (see Figure~\ref{fig:config}c) thin film with \gls{PMA} was prepared at the centre of an 128\textdegree\ Y-cut \LNO\ substrate with dimensions of 10 $\times$ 12 $\times$ 0.5 mm\textsuperscript{3} by \gls{dcMS}.
One pair of Ti(10 nm)/Au(90 nm) \glspl{IDT} was patterned at either end of the magnetic thin film using optical lithography, followed by metal evaporation and lift-off. Each \gls{IDT} consisted of 20 pairs of electrodes. The aperture of the \glspl{IDT} and the distance between the two \glspl{IDT} were set at 500 \textmu m and 3 mm, respectively. Both the electrode width and pitch were designed to be 20 \textmu m, resulting in a \gls{SAW} wavelength ($\lambda$) of $\sim$80 \textmu m.
A Pt film with dimensions of 200 $\times$ 200 $\times$ 0.075 \textmu m\textsuperscript{3} was deposited by \gls{dcMS} to determine temperature changes. This film was situated between the magnetic thin film and the \glspl{IDT} within the \gls{SAW} beam path. The distance from the centre of the thermometer (located within the \gls{SAW} beam path) to the edge of the magnetic thin film was 250 \textmu m. Figure~\ref{fig:config}b provides a schematic representation of the Pt film. Four-terminal electrical transport measurements were conducted using a combined Keithley 6221-2182A current source (connected to Ports 1 and 2 in Figure~\ref{fig:config}b) and a nano-voltmeter (connected to Ports 3 and 4 in Figure~\ref{fig:config}b).

\subsection{SAW circuit and S-parameters}
\glspl{S-parameter} were measured using the method discussed in Section~\ref{Chapter3_S-parameters_measurement}, allowing the determination of the peak value of transmission as a function of frequency, at which point \glspl{SAW} were generated most strongly (with a 19.67 dB transmission loss at a centre frequency of 48 MHz for the device used in the chapter, see Figure~\ref{fig:config}d).\par

\begin{figure}
    \centering
    \includegraphics[width=.76\textwidth]{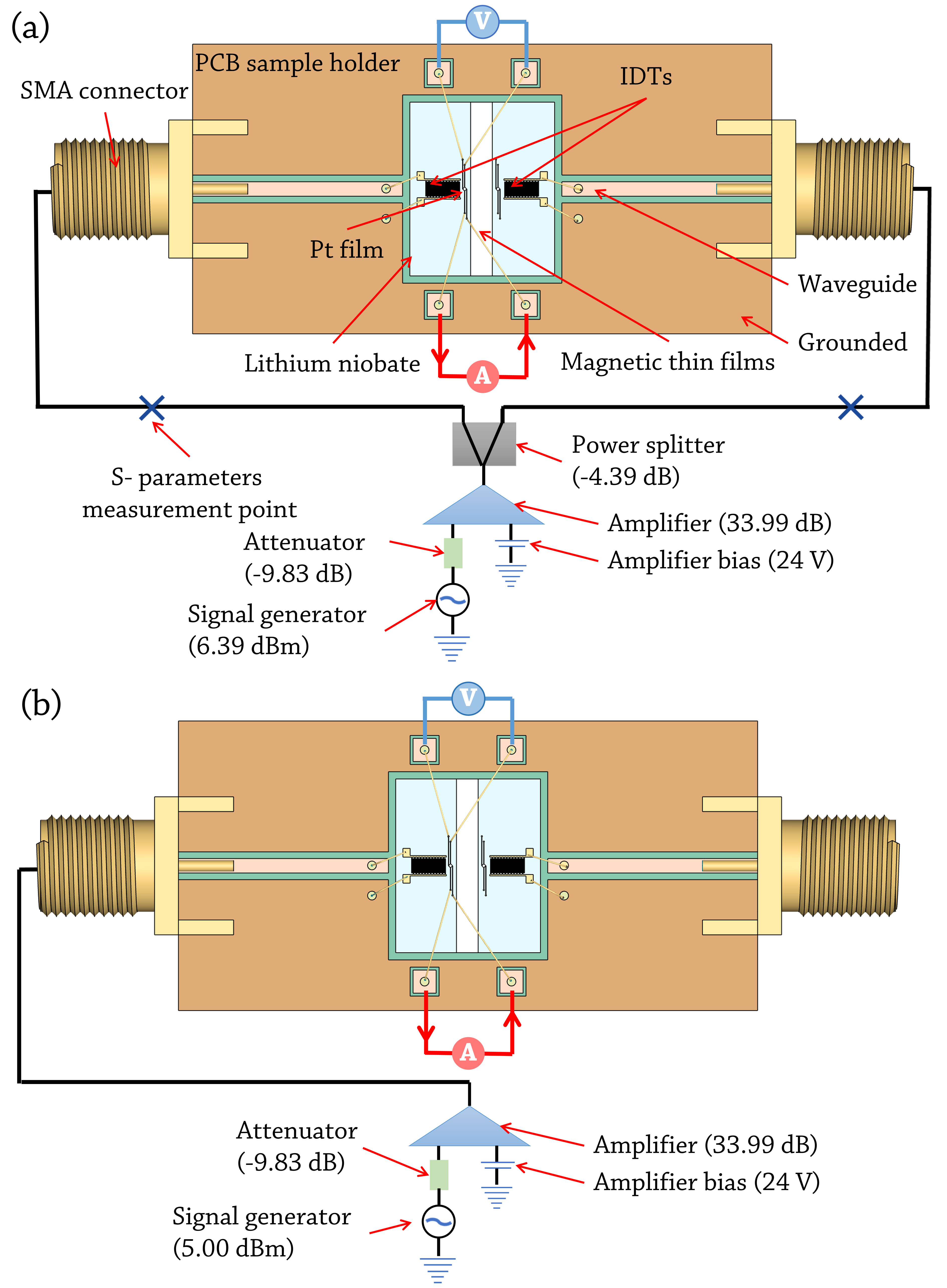}
    \caption[Schematic diagram of radio-frequency circuits used to generate standing and travelling surface acoustic waves.]{Schematic diagram of \gls{rf} circuits used to generate (a) standing \glspl{SAW} and (b) travelling \glspl{SAW}. The signal generator, attenuator, amplifier and power splitter were removed and a \gls{VNA} was connected between the ``X'' points for \gls{S-parameter} measurements. Electrical transport measurements were performed using a conventional four-point-probe measurement to determine the temperature changes while applying \glspl{SAW}.}
    \label{fig:SAW_circuit}
\end{figure}

Figure~\ref{fig:SAW_circuit}a and b show schematic diagrams of the \gls{rf} circuit for launching standing \glspl{SAW} and travelling \glspl{SAW}, respectively. For both circuits, the signal generator (Agilent E5062A) sent out \gls{rf} signals to an attenuator (9.83 dB, Mini-Circuit BW-S10W2+) before being amplified (with a gain of 33.99 dB at 48 MHz at 24 V \gls{dc} voltage, Mini-Circuits ZHL-2-8-S+) to ensure that the applied power did not exceed the linear regime of the amplifier.

To form standing \glspl{SAW}, a power splitter (Mini-Circuit ZAPD-30-S+) was used to divide the \gls{rf} signal into two, with each resulting signal then sent to one \gls{IDT} (Figure~\ref{fig:SAW_circuit}a). As for the \gls{rf} circuit for launching travelling \glspl{SAW}, \gls{rf} signals were sent directly to either one of the \glspl{IDT} (Figure~\ref{fig:SAW_circuit}b).
The total power loss from the power splitter was 4.39 dB, comprising 3 dB of split power loss and 1.39 dB of insertion loss for each branch. This loss was compensated for by the power from the signal generator (6.40 dBm for standing \glspl{SAW} and 5.00 dBm for travelling \glspl{SAW}) to ensure that the power level was the same for both standing \glspl{SAW} and travelling \glspl{SAW}. 
Thus, we can state that the total loss, associated with three factors: 
(i) cabling, 
(ii) transduction at the \gls{IDT} (expected to be the largest factor given no impedance matching was used), and 
(iii) \gls{SAW} propagation to the centre of the device, together introduces 19.67 dB/2 = 9.84 dB total loss from each port to the centre of the (symmetric) device. 
Therefore, the power sent to the \gls{IDT} for both standing \glspl{SAW} and travelling \glspl{SAW} was the same ($\sim$21 dBm), which can qualitatively compare the heating and strain (induced by standing \glspl{SAW} and travelling \glspl{SAW}) effects on \gls{DW} motion.
Figure~\ref{fig:config}d shows the reflection (S11 and S22) and transmission (S21 and S12) characteristics of the \gls{SAW} transducers and substrate, exhibiting a centre frequency of 48 MHz.\par

\subsection{Determination of temperature change}

\begin{figure}
    \centering
    \includegraphics[width=.7\textwidth]{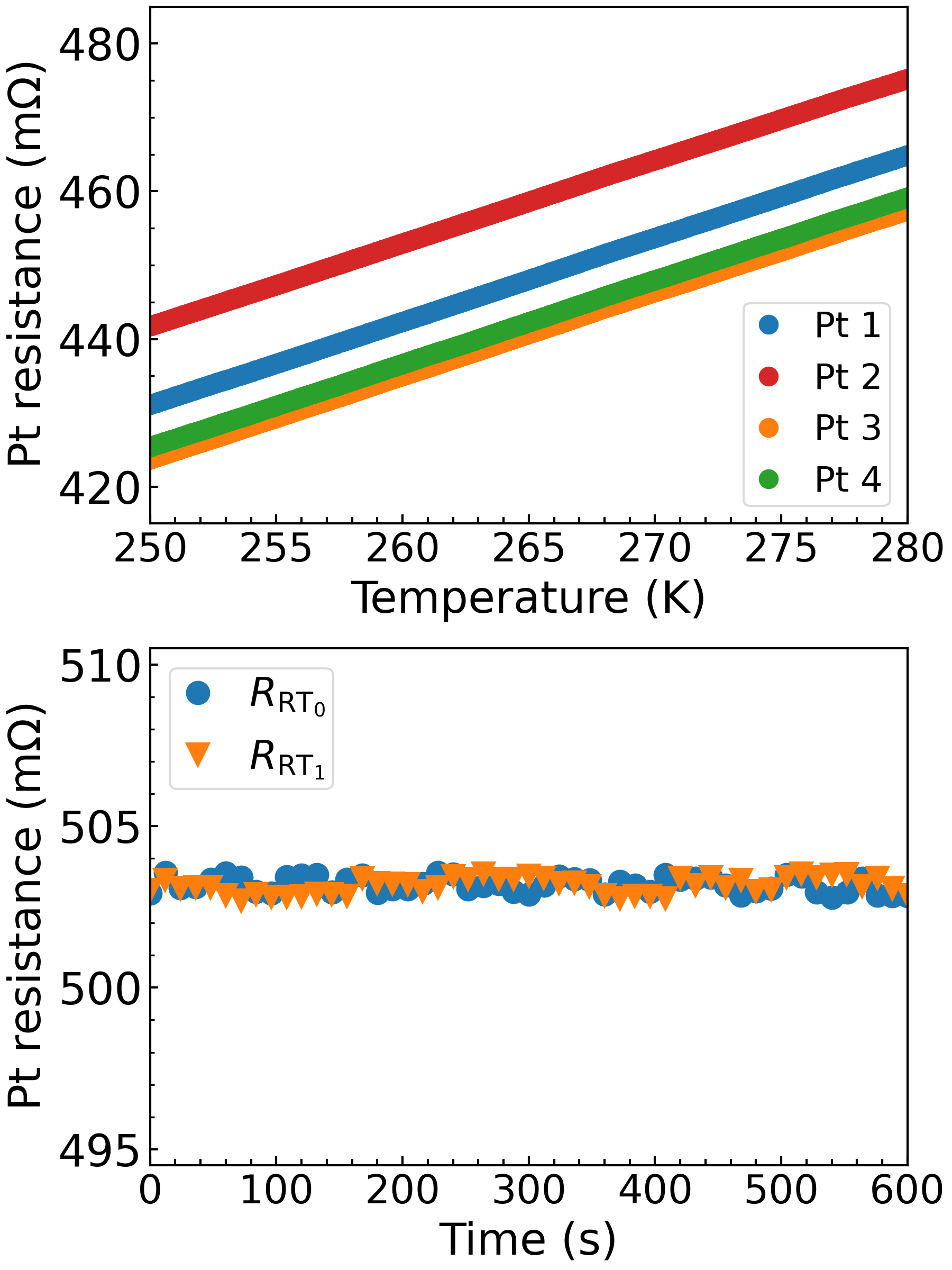}

\begin{picture}(0,0)
    \put(-170,390) {(a)}
    \put(-170,190) {(b)}
\end{picture}

    \caption[Pt resistance against temperature and against time at room temperature.]{(a) Pt resistance against temperature. Four different Pt thin films (Pt 1 to Pt 4) were measured within the temperature range of 250 to 280 K, and the mean of the four measurement was used to determine the resistance changing rate against temperature. (b) Pt resistance against time for continuous 10 minutes at room temperature before ($R_\RM{RT_0}$) and after ($R_\RM{RT_1}$) the application of \glspl{SAW}. The average of the Pt resistance was used as reference resistance at room temperature.}
    \label{fig:temperature}
\end{figure}
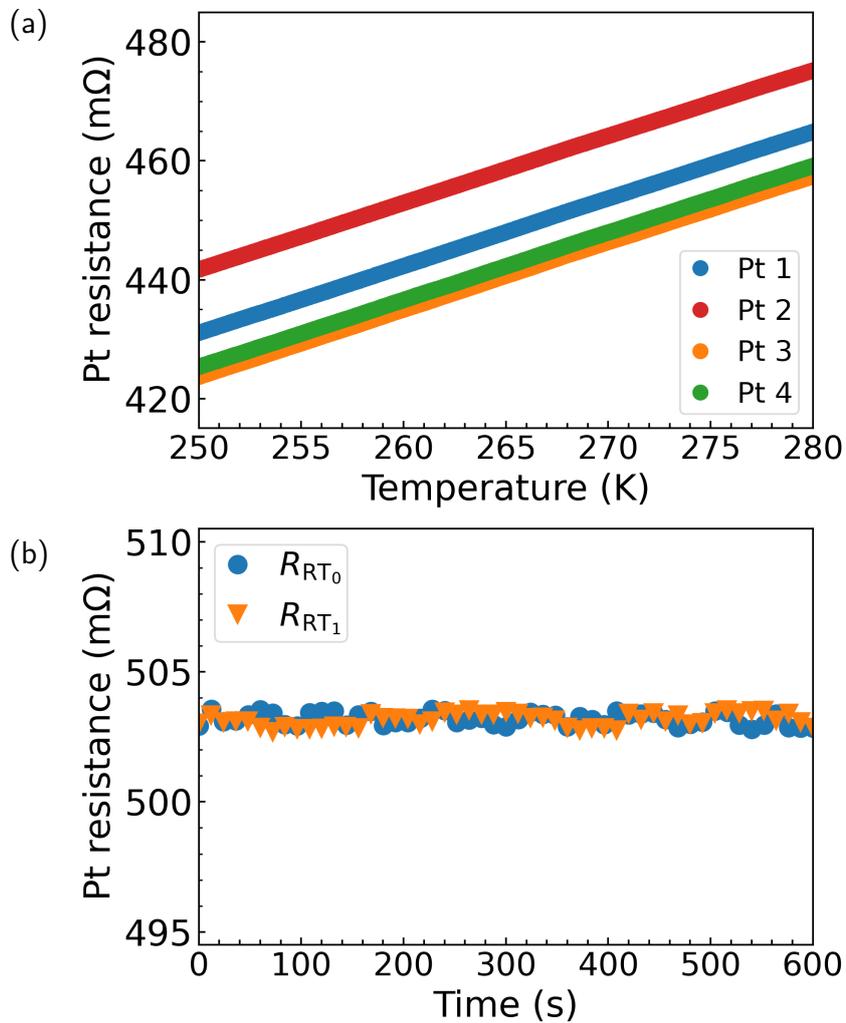

The temperature change ($\Delta\RM{T}$) at the Pt thermometer was obtained using the equation
\begin{equation}
\Delta \RM{T}=\alpha_0(R-R_\RM{RT}),
\end{equation}
where $\alpha_0$ is the temperature change of the Pt per unit resistance (K/m$\Omega$), $R$ is the measured Pt resistance, and $R_\RM{RT}$ is the Pt resistance at room temperature. The resistance of Pt linearly increases with increasing temperature (see Figure~\ref{fig:temperature}a), and the value of $\alpha_0$ is 0.888 $\pm$ 0.004 K/m$\Omega$, obtained by finding and averaging the reciprocal of the gradient of the Pt resistance-temperature curve for four different Pt thin films with the same dimensions (Pt 1 to Pt 4 in Figure~\ref{fig:temperature}a) across a temperature range from 250 to 280 K. The $R_\RM{RT}$ was measured at the beginning and the end of each measurement (Figure~\ref{fig:temperature}b).\par

\subsection{Determination of domain wall velocity}

\begin{figure}[ht!]
    \centering
    \includegraphics[width=1\textwidth]{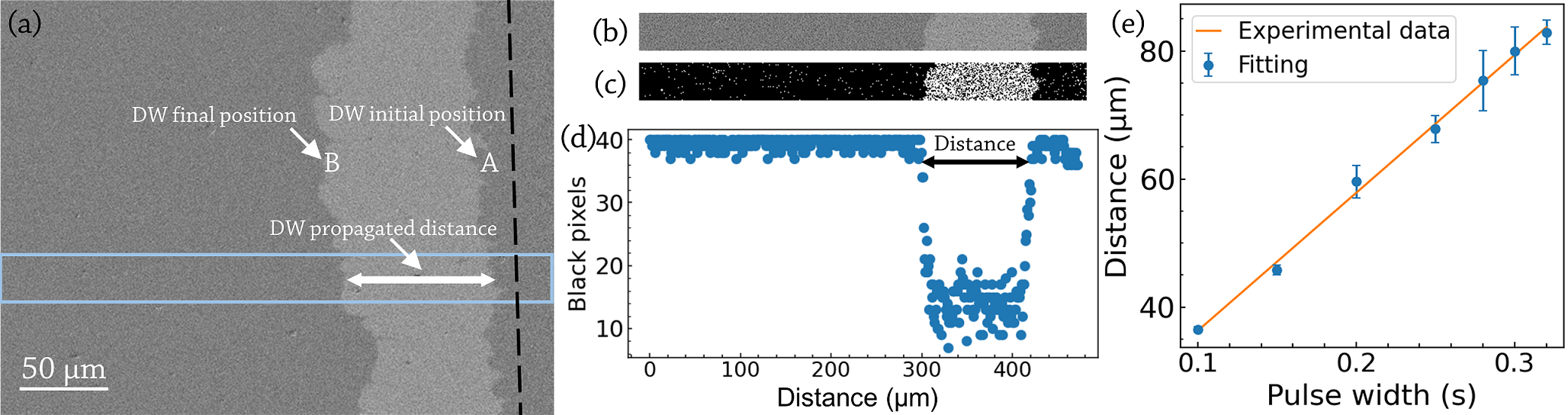}
    \caption[Determination of the domain wall velocity in the Ta/Pt/Co/Ta thin film.]{Determination of the \gls{DW} velocity \revision{in the Ta/Pt/Pt/Ta thin film}. (a) An example image used to determine the \gls{DW} velocity. The dash line is the boundary of the thin film (left) and substrate (right). Arrow A and B indicate initial and final positions the \gls{DW}, respectively. The distance between A and B is the \gls{DW} propagating distance. (b) The cropped image from the framed area in (a). (c) The binary value of the image (b). (d) The averaged number of black pixels in (c) against the distance. (e) The \gls{DW} propagating distance against pulse widths of the magnetic field. Data points and the straight line in the graph are experimental results and its linear fit, respectively.}
    \label{fig:DetermineDWV}
\end{figure}

The \gls{DW} velocity was measured using wide-field Kerr microscopy. Figure~\ref{fig:DetermineDWV}a shows an example image used for determining the \gls{DW} velocity. The magnetic thin film was first saturated using a large magnetic field of --300 Oe (the coercivity of the thin film is approximately 60 Oe). A short pulse (0.2 s) of a magnetic field with the opposite direction (50 Oe) was applied to nucleate a domain at the edge of the thin film (arrow A in Figure~\ref{fig:DetermineDWV}a). The first image was recorded, followed by the application of another magnetic field pulse to move the \gls{DW} (arrow B in Figure~\ref{fig:DetermineDWV}a). The second image was then taken, and the difference between the two images (the distance between A and B) was used to extract the distance that the \gls{DW} travels. The framed area in Figure~\ref{fig:DetermineDWV}a was cropped and converted into binary values, as shown in Figures~\ref{fig:DetermineDWV}b and~\ref{fig:DetermineDWV}c, respectively. \revisiontwo{There were few scratches formed on the substrate during the fabrication process, which created artificial pining sites affecting the \gls{DW} motion. The chosen area is free from such scratches.} The number of black pixels in Figure~\ref{fig:DetermineDWV}c was plotted against the distance, as shown in Figure~\ref{fig:DetermineDWV}d.

A series of pulses was applied to the thin film, and the velocity was then obtained by finding the gradient of the distance against the pulse width (see Figure~\ref{fig:DetermineDWV}e). Seven different pulse widths were applied to the thin film, and each pulse width was repeated three times. Images for all measurements were taken at the same position to ensure that the \gls{DW} velocity was comparable for all studied cases. \gls{DW} velocity was firstly measured from 19\degreeC (room temperature, RT) to 49\degreeC ($\Delta\RM{T}$ = 30 K) without \glspl{SAW}. The sample was heated by the hot side of a Peltier device placed underneath the sample. \gls{DW} velocity was also measured in the presence of standing \glspl{SAW} and then travelling \glspl{SAW} at the frequency of 48 MHz and a power of 21 dBm with no heating applied from the Peltier device.\par

\section{Results and discussion}

\subsection{Temperature change in device}

\begin{figure}
\centering
\includegraphics[width=0.7\textwidth]{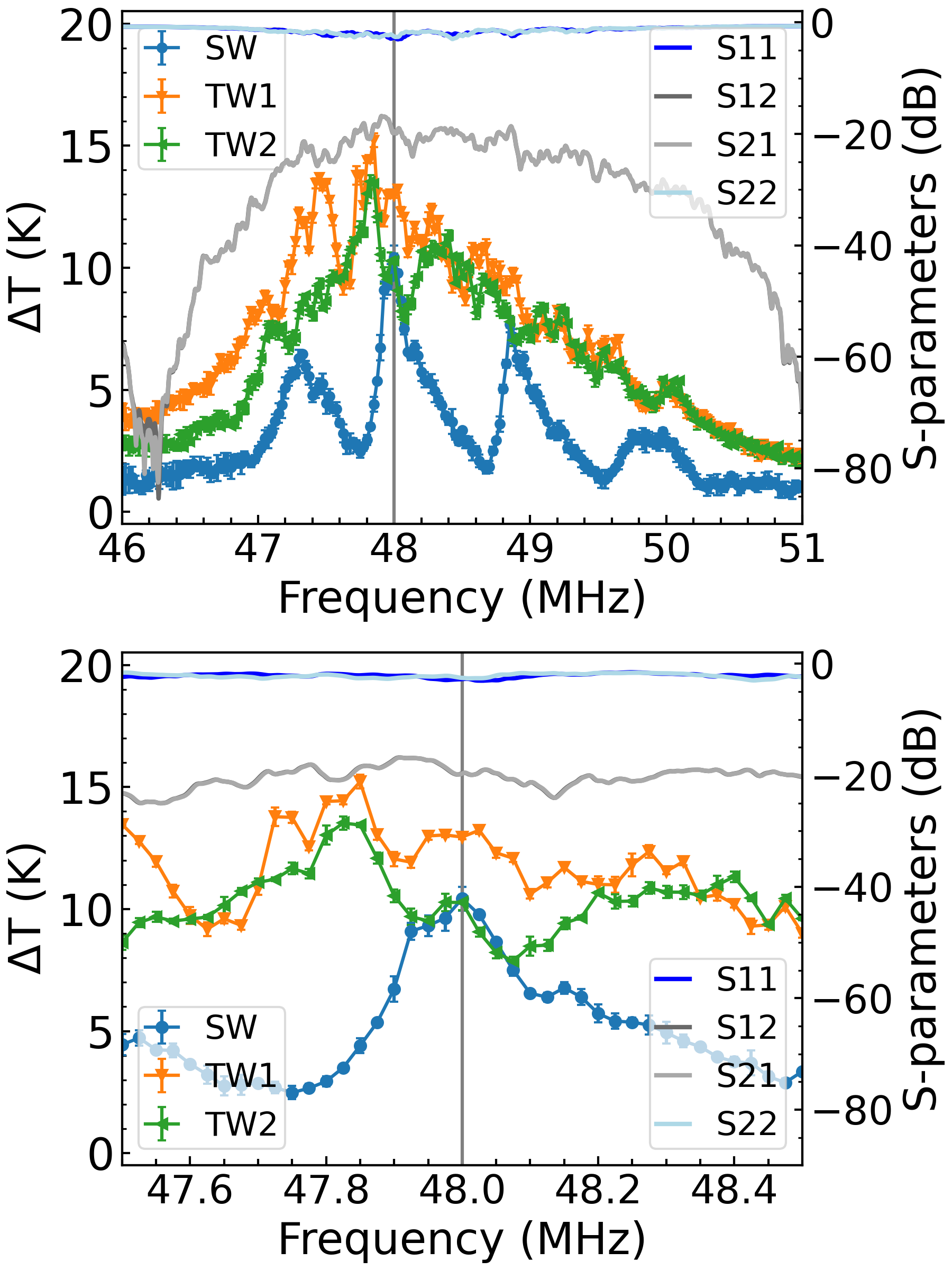}

\begin{picture}(0,0)
    \put(-170,390) {(a)}
    \put(-170,190) {(b)}
\end{picture}

\caption[Temperature changes of the Pt within the surface acoustic wave beam path.]{\label{fig:dTvf}
(a) Temperature changes ($\Delta\RM{T}$) of the Pt within the \gls{SAW} beam path as a function of source frequency from 46 to 51 MHz with \gls{rf} power of 21 dBm. (b) Zoomed-in temperature changes of the device against \gls{SAW} frequencies from 47.5 to 48.5 MHz. SW, TW1, and TW2 in the legend denote standing \glspl{SAW}, travelling \glspl{SAW} launched from \gls{IDT}1, and travelling \glspl{SAW} launched from \gls{IDT}2, respectively. Lines are the guide to the eye.}
\end{figure}
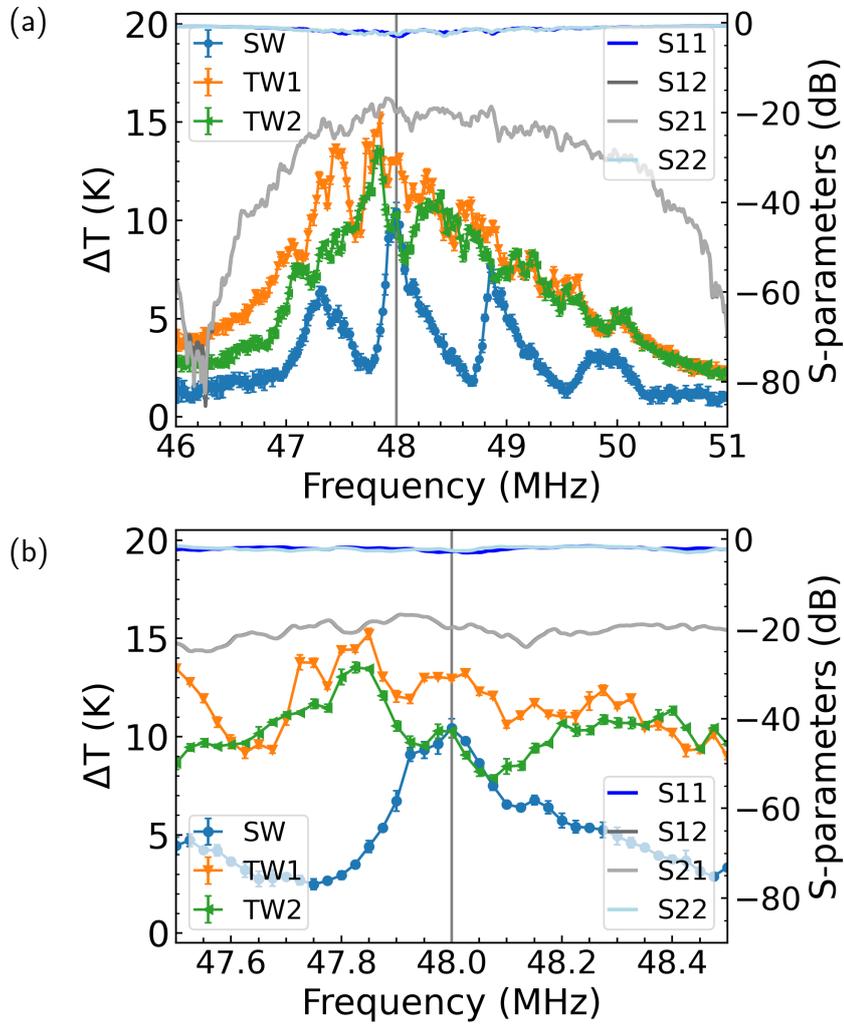

Figure~\ref{fig:dTvf}a and b show the temperature changes ($\Delta\RM{T}$) of the thermometer within the \gls{SAW} beam path against source frequency with a total \gls{rf} power of 21 dBm. We focus on the temperature changes within the \gls{SAW} bandwidth (from 46 MHz to 51 MHz, corresponding to the centre peak in Figure~\ref{fig:config}d). \revisiontwo{In the presence of standing \glspl{SAW}, four temperature peaks ($\sim$47.38 MHz, $\sim$48.00 MHz, $\sim$48.85 MHz, and $\sim$49.75 MHz) can be observed, with the largest $\Delta\RM{T}$ of 10.4 $\pm$ 0.5 K being at the centre frequency of 48 MHz (SW in Figure~\ref{fig:dTvf}a). The formation of the four resonant peaks can be explained by the constructive and destructive interference at certain frequencies causing different mechanical reflections at the \glspl{IDT}.}\par

Upon application of travelling \glspl{SAW} (TW1 and TW2 in Figure~\ref{fig:dTvf}a), the temperature of the Pt increased as the centre frequency was approached, reaching a maximum temperature near the centre frequency with a few local minimum temperatures due to \gls{rf} reflections from the \gls{IDT} and sample edges. \revision{This trend generally corresponds to the reflection and transmission coefficients (overlaid on the $\Delta\RM{T}$ in Figure~\ref{fig:dTvf}a) across the bandwidth.}
TW1 causes a slightly higher temperature increase at lower frequencies compared to TW2. For example, at 47 MHz, the temperature increases are 8.1 $\pm$ 0.2 K and 5.5 $\pm$ 0.3 K for TW1 and TW2, respectively. This is because this thermometer is closer to \gls{IDT}1, implying that dissipation of \gls{rf} power at the transducer or its bonded wire contributes to the observed change in temperature. The highest temperature increases for TW1 and TW2 occur at 47.85 MHz and 47.83 MHz, respectively, with values of 15.2 $\pm$ 0.3 K and 13.5 $\pm$ 0.3 K, as listed in Table~\ref{tb:dT_table}. \revision{Detailed temperature changes around the centre frequency can be found in Figure~\ref{fig:dTvf}b.} \revisiontwo{Within a specific narrow range, there is a discrepancy between the behavior of temperature and the \gls{S-parameter}. This difference might be from intricate interference generated by reflections occurring either at the IDTs or within the thin film.} At 48 MHz, the temperature increases are 12.9 $\pm$ 0.1 K and 10.3 $\pm$ 0.3 K for TW1 and TW2, respectively.\par

\renewcommand{\arraystretch}{1.5}
\begin{table}[htbp]
\centering
\caption[Device temperature changes in the presence of surface aocustic waves.]{Maximum temperature changes ($\Delta\RM{T}$$_\RM{max}$) of the device and corresponding frequencies, and temperature changes at centre frequency ($\Delta\RM{T}_\RM{48 MHz}$) in the presence of SAWs.}
\label{tb:dT_table}
\resizebox{0.8\textwidth}{!}{\begin{tabular}[t]{cccc}
\midrule[0.3mm]
\gls{SAW} type&$\Delta\RM{T}$$_\RM{max}$ (K)&Corresponding frequency (MHz)&$\Delta\RM{T}_\RM{48 MHz}$ (K)\\
\midrule[0.1mm]
SW & 10.4 $\pm$ 0.5 & 48.00 & 10.4 $\pm$ 0.5\\
TW1 & 15.2 $\pm$ 0.3 & 47.85 & 12.9 $\pm$ 0.1\\
TW2 & 13.5 $\pm$ 0.3 & 47.83 & 10.3 $\pm$ 0.3\\
\midrule[0.3mm]
\end{tabular}
}
\end{table}


\subsection{Heating dynamics model}

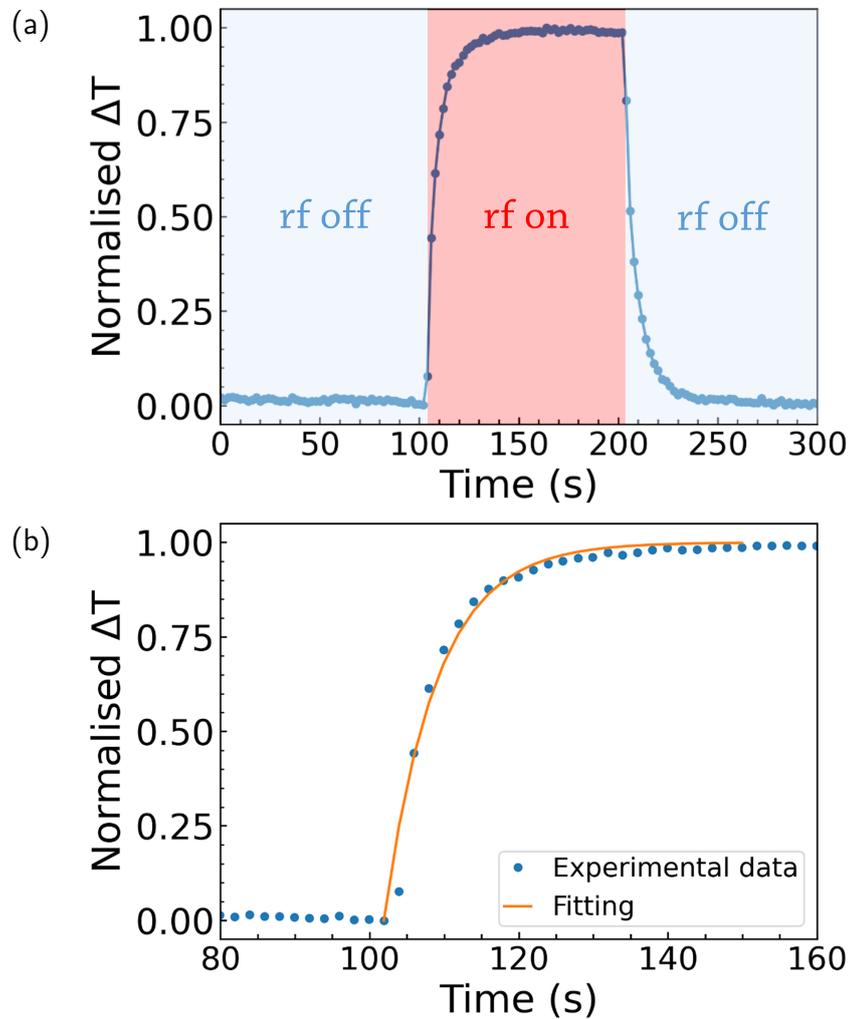
\begin{figure}
    \centering
    \includegraphics[width=.7\textwidth]{{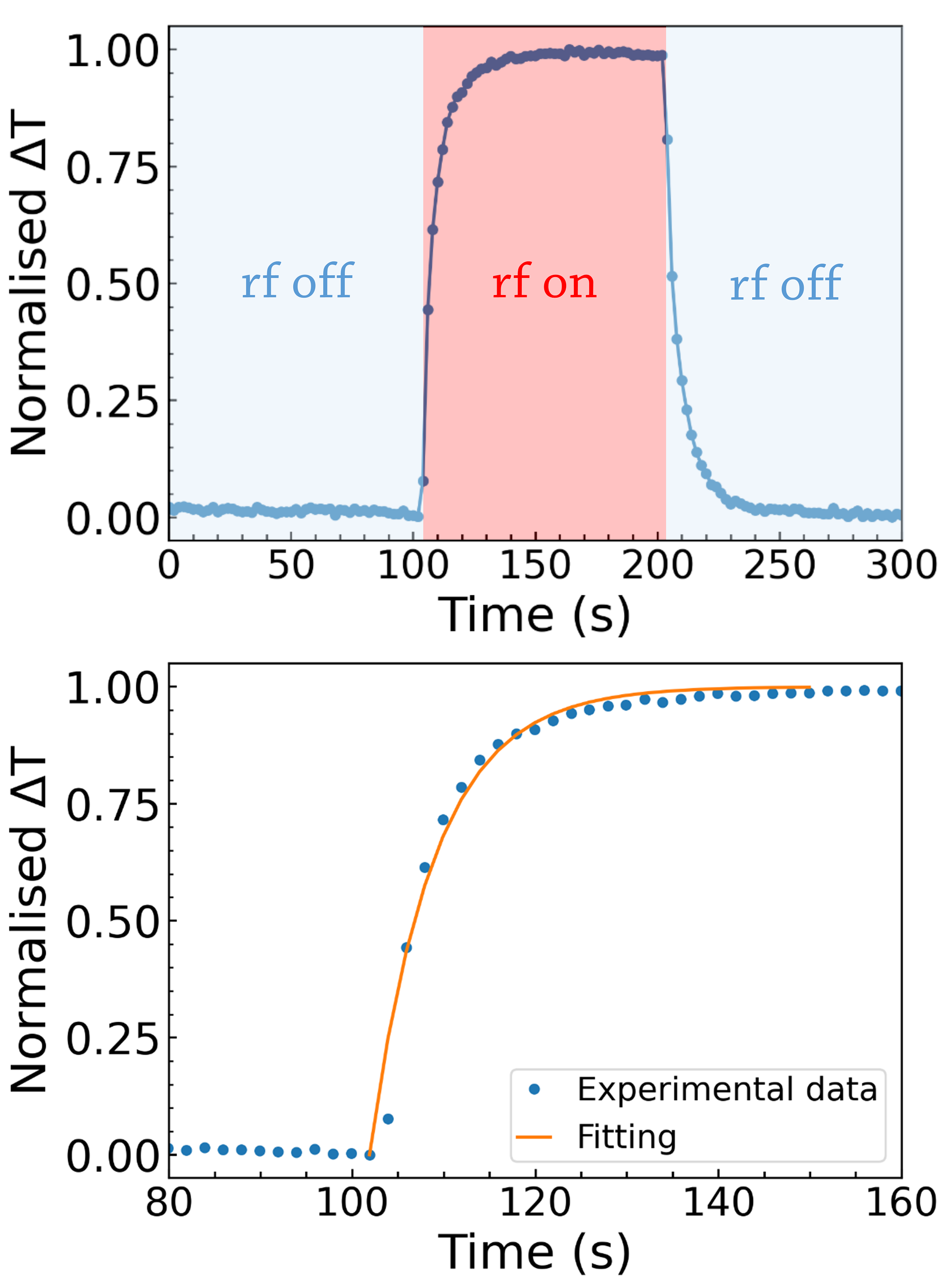}}
\begin{picture}(0,0)
    \put(-320,375) {(a)}
    \put(-320,180) {(b)}
\end{picture}
    \caption[Temperature progression sequence: from radio-frequency power-off state, through power-on state where temperature stabilises at equilibrium, back to power-off.]{(a) Normalised $\Delta\RM{T}$ progression: Temperature is first measured without \gls{rf} power (0--100 s), followed by measurement with \gls{rf} power applied (100--200 s), and concludes with a 100 s duration after the \gls{rf} power is turned off. (b) Normalised temperature variations ($\Delta\RM{T}$) plotted over time (80--160 s), capturing the rise in thermometer temperature from room conditions to an equilibrium state during \gls{SAW} presence. The solid line in the primary graph represents the experimental data fitted using Fourier's law.}
    \label{fig:heating_rate_fit}
\end{figure}

The source of heating in a \gls{SAW}-thin film system could be either \gls{rf} power dissipation due to Joule heating or by an acoustothermal effect introduced by the \gls{SAW} itself~\cite{zheng2018role,han2021thermal}. However, since \glspl{IDT} operate with the highest efficiency at their resonant frequency, and both \gls{rf} power dissipation and \gls{SAW} amplitude depend strongly on the \gls{rf} frequency, it is not possible to simply apply off-resonant \gls{rf} signals to distinguish between the two mechanisms.
To differentiate between these possible mechanisms, we investigate the detailed thermal behaviour of the collection of thermometers. The temperature of the thermometer rapidly rises after the \gls{rf} power is applied and reaches a steady state after about 30 s, as shown in Figure~\ref{fig:heating_rate_fit}.

We propose a simple model of the observed heating as follows:
Let $\RM{T}_{\RM{s}}$ be the temperature of the source, and $\RM{T}$ be the temperature of the thermometer. Based on the Fourier's law, the heat influx rate is estimated as
\begin{equation}
\dot{q}_{\RM{in}}(t) = \frac{k_0 \mathcal{A}}{r_\RM{s}} (\RM{T} - \RM{T}_{\RM{s}}),
\end{equation}
where $r_\RM{s}$ is the distance from the source of heating, $\mathcal{A}$ is the surface area of heat exchange and $k_0$ is thermal conductivity of the material. We denote $C_1 = {k_0 \mathcal{A}}/{r_\RM{s}}$ to be the thermal conductance of the material. To account for heat loss, we assume that it occurs via conduction and convection, which can be expressed as  
\begin{equation}
\dot{q}_{\RM{out}} = C_2 (\RM{T} - \RM{T}_{\RM{env}}),
\end{equation}
where $\RM{T}_{\RM{env}}$ is the temperature of the surroundings and $C_2$ is the corresponding thermal conductance. The equilibrium temperature $\RM{T}_\RM{{eq}}$ can be expressed as 
\begin{equation}
\RM{T}_\RM{{eq}} = \frac{C_1 \RM{T}_{\RM{s}} + C_2 \RM{T}_{\RM{env}}}{C_1 + C_2}.
\end{equation}
Before the equilibrium is achieved we use 
\begin{equation}\label{eqn: energy conservation}
    \dot{q}_{\RM{t}} = \dot{q}_{\RM{in}} - \dot{q}_{\RM{out}},
\end{equation}
where $\dot{q}_{\RM{t}}$ is the heat absorbed by the thermometer. If $c$ is the heat capacity of the thermometer, then Equation~\ref{eqn: energy conservation} can be re-written as 
\begin{equation}
\dot{\RM{T}} = \frac{C_1 + C_2}{c} \left( - \RM{T} + \RM{T}_\RM{{eq}}\right).  \label{eqn: temperature equation}
\end{equation}
Solving Equation~\ref{eqn: temperature equation} gives the temperature change over time
\begin{equation}\label{eqn:normalise}
    \RM{T} = \RM{T}_{\RM{eq}} - (\RM{T}_{\RM{eq}} - \RM{T}_{0}) \exp\left( -\frac{C_1 + C_2}{c} t \right).
\end{equation}
From the experimental data, we fit $\RM{T}$ vs $t$ to the $\RM{T} = \RM{T}_{\RM{eq}} -(\RM{T}_{\RM{eq}} - \RM{T}_{\RM{env}}) \exp(-\tilde{\gamma} t)$. The experimentally obtained $\tilde{\gamma}$ is expected to correspond to $({C_1 + C_2})/{c}$. Substituting $C_1 =  k_0{\mathcal{A}}/{r_\RM{s}}$ we obtain 
\begin{equation}
    \tilde{\gamma}  = \left( \frac{k_0 \mathcal{A}}{c} \right)  \frac{1}{r_\RM{s}} + \frac{C_2}{c}.
\end{equation}
\par

We choose to study the $\tilde{\gamma}$ value because it represents the ``heating rate'' and provides information about the entire heating process, enabling us to better understand its dynamics and variations at different locations. In contrast, temperature only represents the final result of the heating process and does not provide information about its dynamics.

\subsection{Source of heating}

\begin{figure}
    \centering
    \includegraphics[width=1\textwidth]{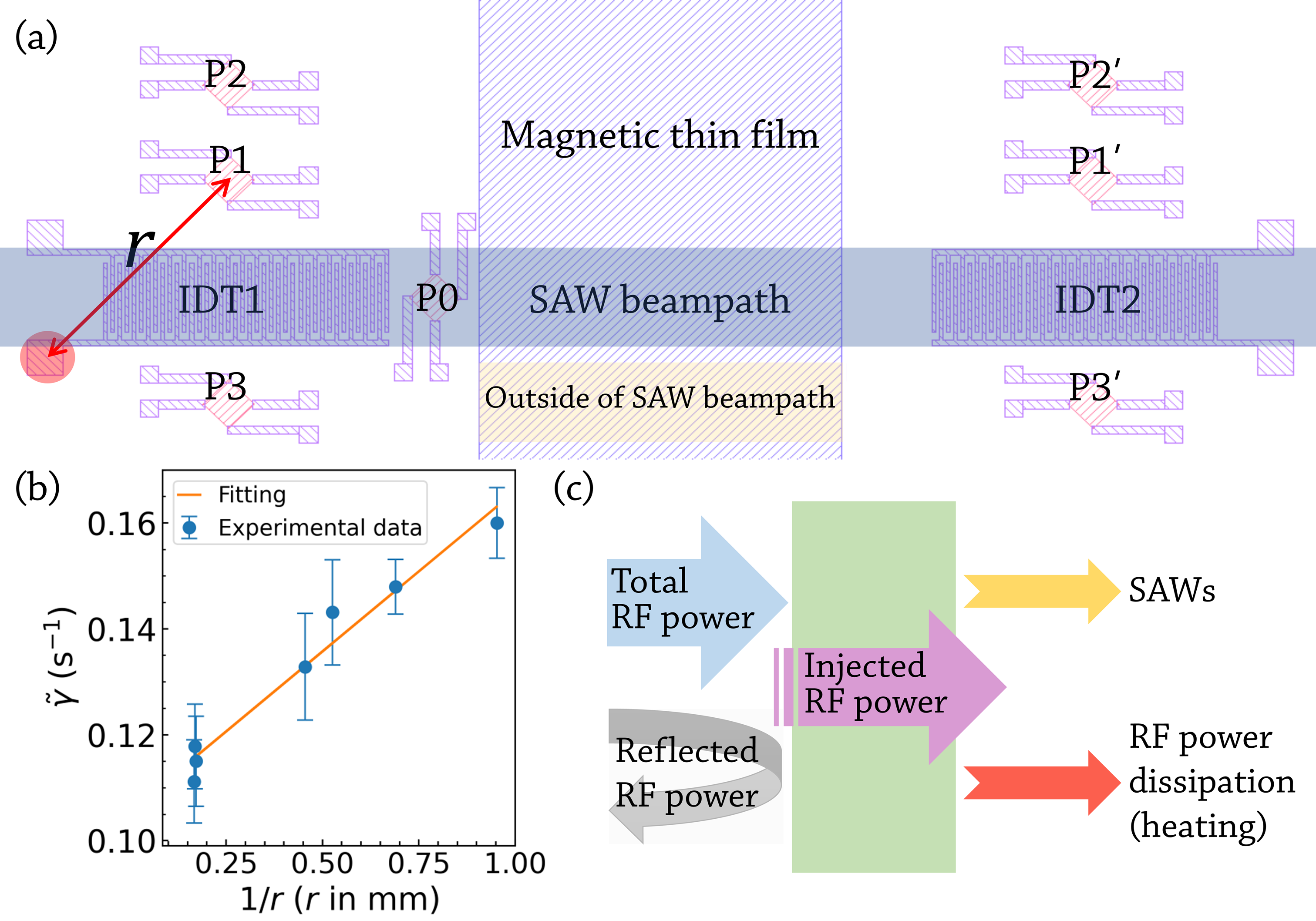}
    \caption[Analysis on the origin of the heating.]{(a) Schematic layout of the \gls{SAW} device showing seven thermometer positions: P0, located within the \gls{SAW} beam path, and P1 (P1$^{\prime}$), P2 (P2$^{\prime}$), and P3 (P3$^{\prime}$) located outside of the \gls{SAW} beam path, at different distances from \gls{IDT}1 (\gls{IDT}2). \gls{DW} velocity is also measured outside of the \gls{SAW} beam path (labelled as ``Outside of \gls{SAW} beam path'') to confirm the magnetoelastic coupling effect on \gls{DW} motion. (b) Plot of the $\tilde{\gamma}$ against 1/$r$ in the presence of a travelling \gls{SAW} launched from \gls{IDT}1, with a frequency of 48 MHz and power of 21 dBm. $r$ is defined as the distance from the \gls{rf} bond pad (red circle) to the centre of the thermometer. The solid line represents the linear fitting of $\tilde{\gamma}$ against 1/$r$. (c) Diagram of the \gls{rf} power breakdown.}
    \label{fig:heating_rate_results}
\end{figure}

Figure~\ref{fig:heating_rate_results}a shows a schematic layout of the \gls{SAW} device. Temperature measurements were conducted both inside and outside of the \gls{SAW} beam path in the presence of the travelling \gls{SAW} launched from \gls{IDT}1. The \gls{rf} power and frequency were set as 21 dBm and 48 MHz, respectively.
The thermometer P0 in Figure~\ref{fig:heating_rate_results}a is the same one used in Figure~\ref{fig:config}a, situated within the \gls{SAW} beam path and expected to experience the \glspl{SAW} passing through. P1, P1$^{\prime}$, P2, P2$^{\prime}$, and P3, P3$^{\prime}$ are thermometers situated outside the \gls{SAW} beam path where the \gls{SAW} amplitude is negligible~\cite{winkler2017compact}.
P1 and P1$^{\prime}$, P2 and P2$^{\prime}$, as well as P3 and P3$^{\prime}$ were equidistant from the \gls{SAW} beam path, which were 650 \textmu m, 1250 \textmu m above \glspl{IDT}, and 650 \textmu m below \glspl{IDT}, respectively.
This arrangement allowed us to determine whether heating was from the \gls{SAW} itself (acoustothermal heating) or from \gls{rf} power dissipation (Joule heating). We consider the two possible sources of heating mentioned above: acoustothermal heating effect and \gls{rf} power dissipation. If the heating is predominantly caused by the acoustothermal effect, the highest $\tilde{\gamma}$ value would be expected within the beam path, while if it is due to \gls{rf} power dissipation, the highest $\tilde{\gamma}$ should occur at the thermometer closest to the \gls{rf} bond pad.\par

Figure~\ref{fig:heating_rate_results}b presents the heating rate $\tilde{\gamma}$ obtained at various locations in response to the travelling \glspl{SAW} generated from \gls{IDT}1. The heating rate at P0 is 0.130 $\pm$ 0.010 s$^{-\text{1}}$, which is lower than the values measured at P1 (0.148 $\pm$ 0.005 s$^{-\text{1}}$) and P3 (0.160 $\pm$ 0.007 s$^{-\text{1}}$). This indicates that P1 and P3 are located closer to the heating source compared to P0. Furthermore, under the same conditions, the heating rate at P1$^{\prime}$ (0.118 $\pm$ 0.008 s$^{-\text{1}}$) and P3$^{\prime}$ (0.115 $\pm$ 0.008 s$^{-\text{1}}$) is lower than those measured at P1 and P3, respectively, suggesting that the heating source is situated close to \gls{IDT}1 where the \gls{rf} power is applied. The origin of this heating is therefore likely to be power dissipated at the \gls{IDT} in the \gls{rf} to \gls{SAW} transduction process. Notably, we observe a good linear relationship when plotting the $\tilde{\gamma}$ values against the reciprocal of the distance between the bond pad (where the \gls{rf} power is applied, red circle shown in Figure~\ref{fig:heating_rate_results}a) and the centre of the thermometer. Furthermore, the heating rates $\tilde{\gamma}$ are much higher at P1 and P3 than that at P0, where \glspl{SAW} are excited. Based on these results, we conclude that the heating observed in our \gls{SAW}-magnetic thin film system is predominantly due to \gls{rf} power dissipation \revision{(introduced by Joule heating)} rather than the acoustothermal heating effect introduced by the \gls{SAW} itself. \par

Figure~\ref{fig:heating_rate_results}c presents a breakdown of the \gls{rf} power flow in our device. When \gls{rf} signals are applied to the \glspl{IDT}, a portion of the power is reflected back to the source, while the rest is injected into the \glspl{IDT}. The proportion of the \gls{rf} power converted accordingly is influenced by several factors: (i) the reflection at the connection between the \gls{SMA}cable and bond pads, (ii) the reflection at the bond pads to \glspl{IDT}, and (iii) the reflection coefficient of the \glspl{IDT} (S11 or S22), all of which are sensitive to the \gls{rf} frequency. Among these factors, the greatest reflection occurs from (iii) given that no impedance matching was used. We calculate the injected power by subtracting the reflected \gls{rf} power from the total \gls{rf} power. A portion of the injected power is converted into surface acoustic waves, which is dependent on the electromechanical coefficient of the \LNO, while the remaining injected power is dissipated in the form of heating. Therefore, the \glspl{S-parameter} and other reflections influence both the \gls{rf} heating and \gls{SAW} amplitude, with a higher heating effect corresponds to a higher \gls{SAW} amplitude.\par


\begin{figure}
\centering
\includegraphics[width=0.7\textwidth]{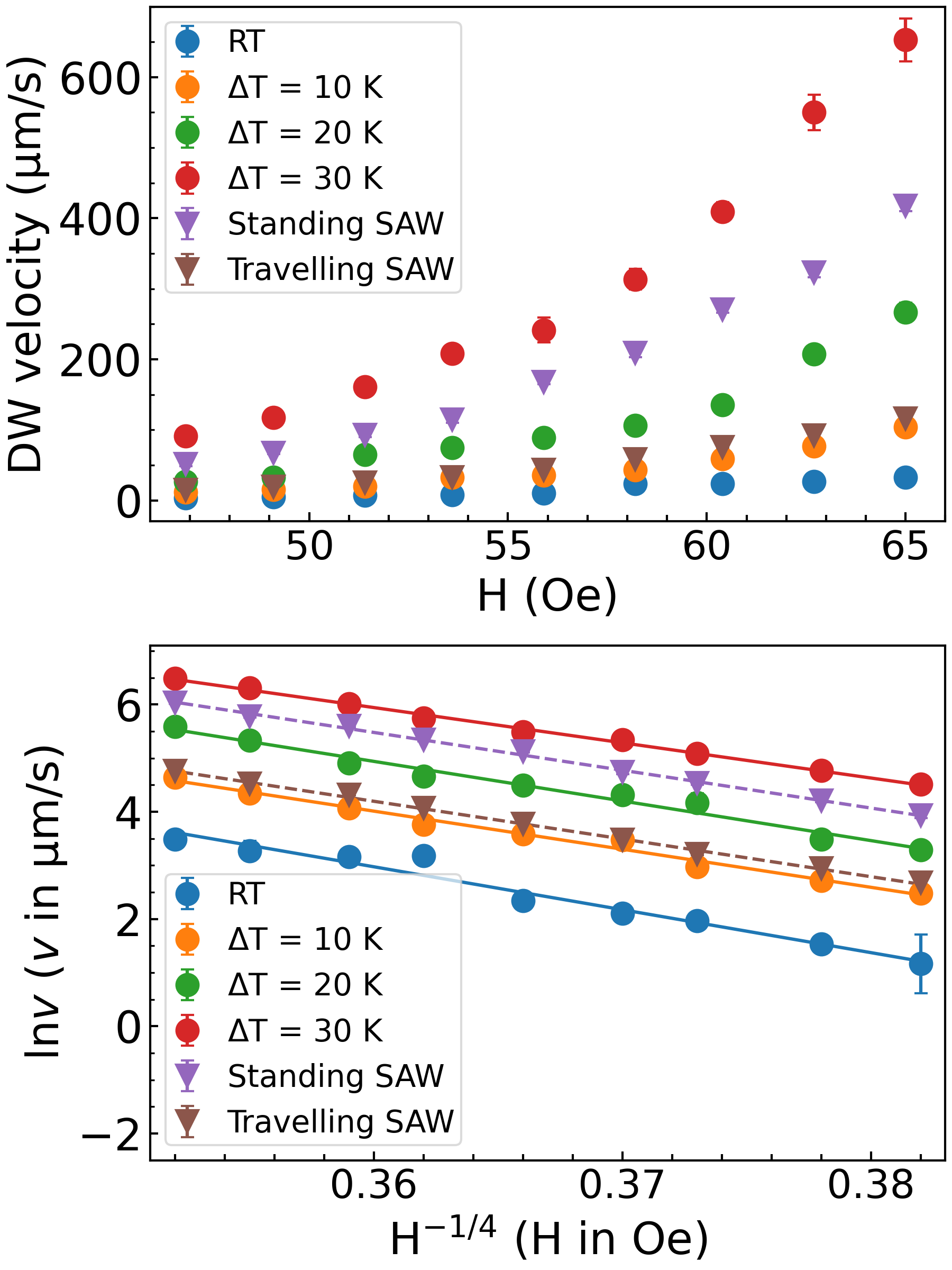}
\begin{picture}(0,0)
    \put(-310,375) {(a)}
    \put(-310,180) {(b)}
\end{picture}
\caption[Domain wall velocity in the Ta/Pt/Co/Ta thin film as a function of applied magnetic field.]{\label{fig:DWVvdT}
\revision{(a) \gls{DW} velocity \revision{in the Ta/Pt/Co/Ta thin film} as a function of the applied field (H). (b) Natural logarithm of \gls{DW} velocity ($v$) as a function of $\RM{H}^{-\text{1/4}}$. \gls{DW} velocity is measured at different temperatures from room temperature (RT) up to $\Delta\RM{T}$ = 30 K without \glspl{SAW} (circles) and in the presence of standing \glspl{SAW} and travelling \glspl{SAW} at centre frequency of 48 MHz and \gls{rf} power of 21 dBm without additional heating (triangles). The solid and dashed lines are linear fitting of the \gls{DW} velocity experimental data at different temperatures and in the presence of the \glspl{SAW}, respectively.}}
\end{figure}
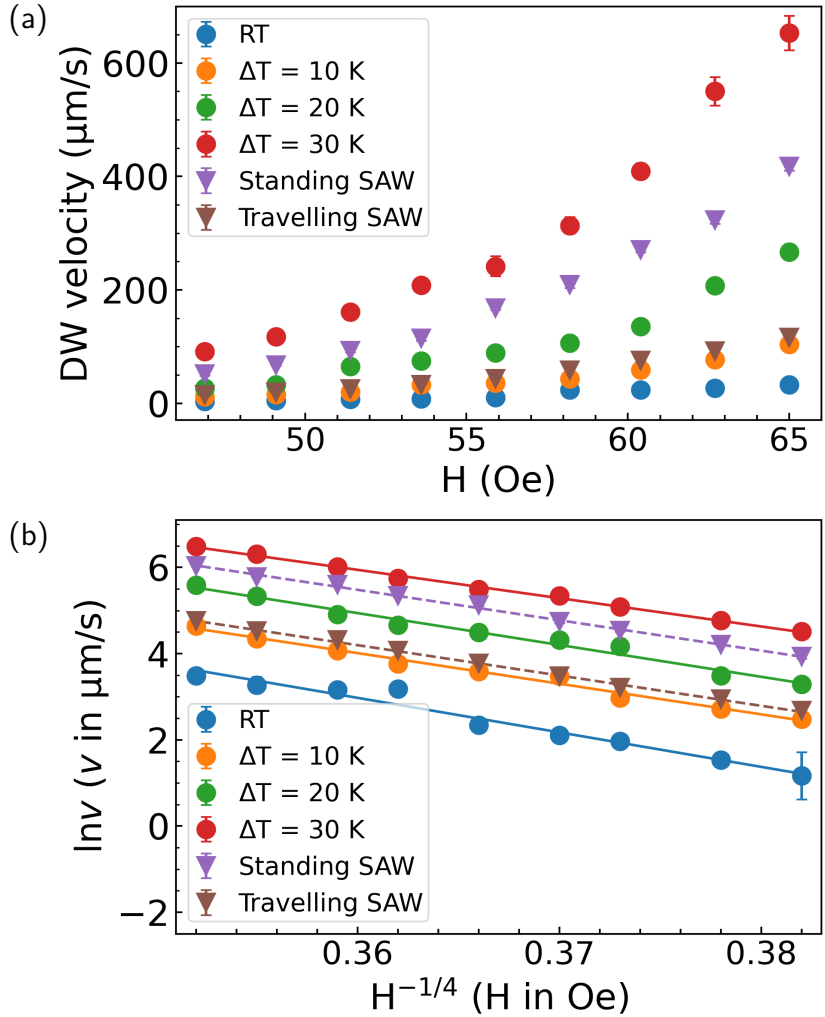

\subsection{Heating and strain effects on domain wall velocity}

We now examine the relationship between temperature and \gls{DW} velocity. Figure~\ref{fig:DWVvdT}a shows the obtained \gls{DW} velocity within the \gls{SAW} beam path plotted against the applied magnetic field. The \gls{DW} velocity was initially measured at room temperature ($\sim$19\degreeC). As the field increases from 47 to 65 Oe, the \gls{DW} velocity was found to increase from 3 $\pm$ 2 to 33 $\pm$ 3 \textmu m/s. As depicted in Figure~\ref{fig:DWVvdT}b, a plot of $\ln v$ against $\RM{H}^{-\text{1/4}}$ shows a linear dependence using Equation~\ref{eqn:creep_law}. This indicates that the \gls{DW} motion is in the creep regime, where thermal energy enables \glspl{DW} to overcome the pinning barriers. The \gls{DW} creep motion can be enhanced by increasing the temperature. We study the temperature dependence of the \gls{DW} velocity by heating our device by 10, 20, and 30 K using a Peltier device. As shown in Figure~\ref{fig:DWVvdT}a, \gls{DW} motion is significantly enhanced as the temperature increases. For instance, under a field of 65 Oe, the \gls{DW} velocity increases from 33 $\pm$ 3 \textmu m/s at room temperature to 650 $\pm$ 30 \textmu m/s when the temperature is increased by 30 K. According to creep law (Equation~\ref{eqn:creep_law}), the \gls{DW} motion still remains in the creep regime as the $\ln v$ shows a linear dependence on $\RM{H}^{-\text{1/4}}$ (see Figure~\ref{fig:DWVvdT}b).\par

\begin{figure}[ht!]
\centering
\includegraphics[width=1\textwidth]{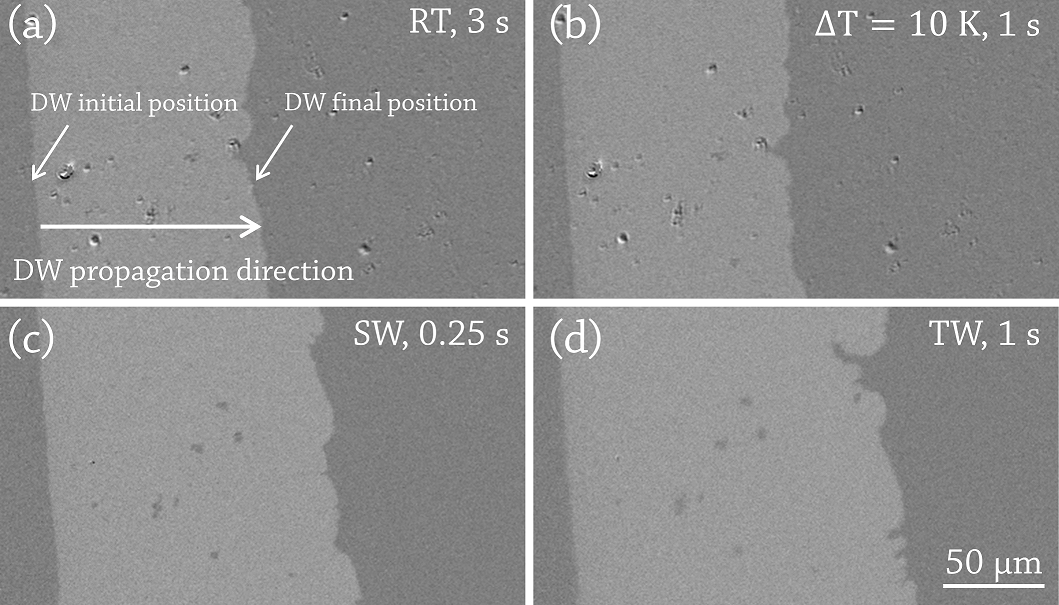}
\caption[Domain wall profile in the Ta/Pt/Co/Ta thin film under different experimental conditions imaged using a Kerr microscope]{\label{fig:DWProfile}
\gls{DW} profile \revision{in the Ta/Pt/Co/Ta thin film} under different experimental conditions imaged using a Kerr microscope. (a) Magnetic field applied at room temperature for 3 s, (b) Magnetic field applied with a temperature increase of 10 K for 1 s, (c) Standing \glspl{SAW} with magnetic field applied for 0.25 s, and (d) Travelling \glspl{SAW} with magnetic field applied for 1 s. The magnetic field strength used is 65 Oe, and the frequency and power of the \glspl{SAW} are 48 MHz and 21 dBm, respectively. The arrow in (a) indicates the direction of the \gls{DW} propagation.}
\end{figure}

\gls{DW} velocity was measured in the presence of \glspl{SAW} at the same position within the \gls{SAW} beam path. The frequency and power of both standing and travelling \glspl{SAW} were set at 48 MHz and 21 dBm, respectively. \gls{DW} motion was found to be enhanced in the presence of both types of \glspl{SAW} (see Figure~\ref{fig:DWVvdT}a). With application of travelling \glspl{SAW}, the \gls{DW} velocity is 116 $\pm$ 3 \textmu m/s at 65 Oe, representing a $\sim$2.5-fold increase compared to the measurement at room temperature. However, an even greater enhancement of \gls{DW} motion can be observed in the presence of standing \glspl{SAW}, with a velocity of 418 $\pm$ 8 \textmu m/s, which is a $\sim$11.7-fold increase from the room temperature measurement. \gls{DW} motion still remains within the creep regime in the presence of both travelling and standing \glspl{SAW} over the measured field range (see Figure~\ref{fig:DWVvdT}b).
The Kerr microscope images of the \gls{DW} profile under different experimental conditions are presented in Figure~\ref{fig:DWProfile}. The \glspl{DW} are nucleated at the left-hand side (as indicated as \gls{DW} initial position in Figure~\ref{fig:DWProfile}a). The \glspl{DW} move towards the right-hand side (as indicated as \gls{DW} final position in Figure~\ref{fig:DWProfile}a) driven by the magnetic field with/without \glspl{SAW}. The \gls{DW} exhibits a mostly smooth profile when subjected to a magnetic field of 65 Oe, as shown in Figure~\ref{fig:DWProfile}a. Increasing the temperature by 10 K (Figure~\ref{fig:DWProfile}b) or introducing standing \glspl{SAW} (Figure~\ref{fig:DWProfile}c) and travelling \glspl{SAW} (Figure~\ref{fig:DWProfile}d) does not cause any significant changes in the \gls{DW} profile.\par

The enhancement of \gls{DW} motion observed in the presence of \glspl{SAW} is likely caused by a combination of both the magnetoelastic coupling effect and the heating effect induced by \gls{rf} power dissipation. The thermometer experiences a temperature increase of $\sim$10 K at 48 MHz (as shown in Table~\ref{tb:dT_table}). In Figure~\ref{fig:DWVvdT}a, the \gls{DW} velocity curve with a 10 K temperature increase is similar to the travelling \gls{SAW}-assisted \gls{DW} velocity curve. However, the \gls{DW} motion enhancement observed with standing \glspl{SAW} is much greater than that achieved by increasing the temperature by 10 K alone. These results suggest that heating plays a major role in promoting \gls{DW} motion in the presence of travelling \glspl{SAW}, while the magnetoelastic coupling effect is the primary factor for significant \gls{DW} motion with the application of standing \glspl{SAW}.\par

Our study highlights the significance of heating in \gls{SAW}-thin film systems, particularly with higher power \glspl{SAW}. We perform \gls{DW} velocity measurements at a location 100 \textmu m away from the \gls{SAW} beam path (referred to as ``Outside of \gls{SAW} beam path'' in Figure~\ref{fig:heating_rate_results}a). This location was chosen because it still experiences some heating effect, albeit to a lesser degree, yet the \gls{SAW} amplitude was negligible in that area~\cite{han2021thermal, winkler2017compact}. With an applied magnetic field of 65 Oe, the \gls{DW} velocity increased from 33 $\pm$ 2 \textmu m/s at room temperature to 75 $\pm$ 3 \textmu m/s and 78 $\pm$ 4 \textmu m/s in the presence of standing and travelling \glspl{SAW}, respectively. This \gls{DW} velocity increase outside of the \gls{SAW} beam path confirms the role of Joule heating in enhancing \gls{DW} motion.\par

Furthermore, strain can modify the magnetic anisotropy of thin films through the magnetoelastic coupling effect~\cite{shepley2015modification,ba2021electric,feng2021field,bhattacharya2020creation}. The resulting changes in magnetic anisotropy create a dynamic energy landscape that reduces the energy requirement for \gls{DW} motion. Specifically, the dynamic strain waves associated with the \glspl{SAW} cause the magnetic anisotropy to periodically change, enhancing the possibility of \gls{DW} depinning from pinning sites and facilitating faster \gls{DW} motion. \revision{However, the distinct impacts of standing versus travelling \glspl{SAW} on \gls{DW} velocity remain to be fully understood and require further investigation. Notably, observations from samples with varying magnetic layer thicknesses (not included in this thesis) have revealed a divergence in \gls{DW} velocity when subjected to standing and travelling \glspl{SAW} with different frequencies. This phenomenon suggests a complex underlying mechanism and indicates that considerable investigative effort is required to elucidate these dynamics.}

\section{Summary}
In this chapter, we explored the impact of both heating and \glspl{SAW} in a \gls{SAW}-magnetic film system, measuring the impact of both heating and \glspl{SAW} on \gls{DW} velocity. Heating of approximately 10 K was observed within the \gls{SAW} beam path when \gls{rf} power was applied. \gls{DW} velocity was measured at various temperatures both with and without \glspl{SAW}. The \gls{DW} velocity increased from 33 $\pm$ 3 to 650 $\pm$ 30 \textmu m/s as the temperature rose from 19\degreeC to 49\degreeC at 65 Oe. The \gls{DW} velocity enhancement by travelling \glspl{SAW} (116 $\pm$ 3 \textmu m/s) was found to be slightly higher than that obtained a 10-K temperature increase (104 $\pm$ 8 \textmu m/s), suggesting that heating played a major role in promoting \gls{DW} motion. On the other hand, \gls{DW} motion was significantly enhanced in the presence of standing \glspl{SAW} (418 $\pm$ 8 \textmu m/s) due to the dominant effect of magnetoelastic coupling. Our study underscores the importance of considering heating in \gls{SAW} devices, especially those using high \gls{rf} power, and presents a straightforward way of measuring heating in \gls{SAW} devices and interpreting such effects.
\cleardoublepagewithnumberheader
\chapter{SAW effect on domain wall dynamics}
\label{Chapter6_DW_dynamics}

\section{Introduction} 
\glspl{SAW} are capable of introducing dynamic strain waves into magnetic thin films~\cite{yang2021acoustic}. Through the magnetoelastic coupling effect, these dynamic strain waves generate a dynamic energy landscape, thereby triggering magnetisation precession~\cite{thevenard2013irreversible, thevenard2016precessional, camara2019field}, facilitating magnetisation switching~\cite{shuai2022local, li2014acoustically}, and enhancing the \gls{DW} motion~\cite{dean2015sound, edrington2018saw, vilkov2022magnetic, wei2020surface, cao2021surface, shuai2023separation}. The utilisation of \glspl{SAW} has demonstrated significant promise for achieving energy-efficient control of \gls{DW} motion. However, a comprehensive understanding of the detailed mechanisms and effects underlying the interaction between \glspl{SAW} and \glspl{DW} is currently lacking. Moreover, the influence of defects and disorders in thin films on \gls{SAW}-assisted \gls{DW} motion remains largely unexplored. Thus, there is a knowledge gap regarding the systematic investigation of the impact of \gls{SAW} frequency on \gls{DW} dynamics in films with varying levels of disorders. Addressing this gap is crucial for advancing our understanding of \gls{SAW}-assisted \gls{DW} motion and unlocking its full potential for energy-efficient \gls{DW} control. \par

In this chapter, we investigate the influence of \gls{SAW} frequency (50, 100, and 200 MHz) on \gls{DW} motion within thin films with different levels of anisotropy disorder (1\% and 3\%) at low fields (up to 1 mT) using micromagnetic simulations. This chapter also delves into the \gls{DW} depinning process under different conditions, with a particular focus on \glspl{VBL}. By conducting these investigations, we aim to gain insights into the intricate dynamics of \gls{SAW}-assisted \gls{DW} motion and its interaction with disorders, ultimately contributing to the development of energy-efficient \gls{DW} control techniques. Results obtained in this chapter were accepted for publication as \textit{``Surface acoustic wave effect on magnetic domain wall dynamics''}, \href{https://doi.org/10.1103/PhysRevB.108.104420}{Physical Review B, 108, 104420 (2023)}~\cite{shuai2023surface}.\par

\section{Model and computational details}
\subsection{Proposed device and simulation configuration}

\begin{figure}[!ht]
    \centering
    \includegraphics[width=0.7\textwidth]{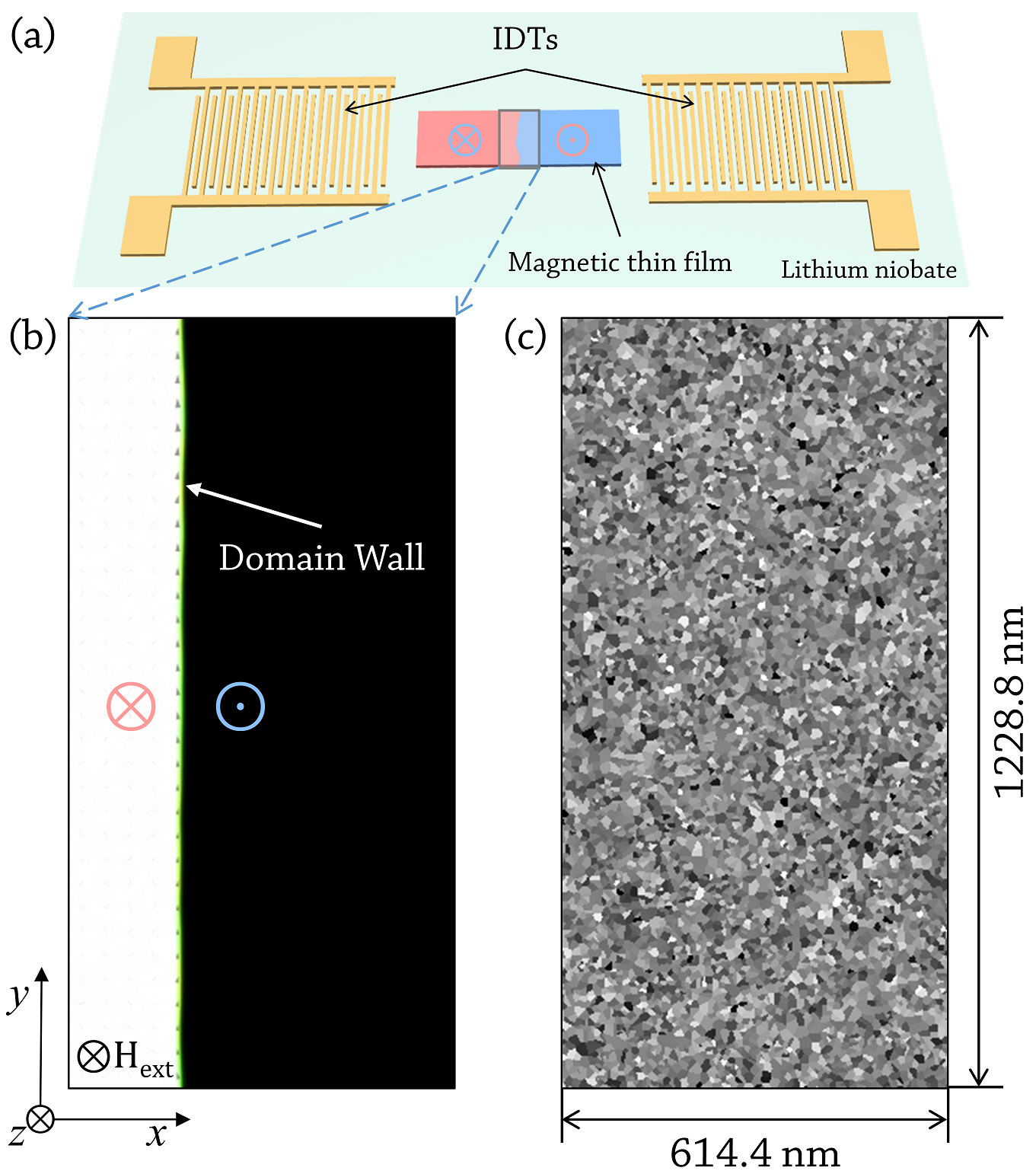}
    \caption[Configuration for domain wall dynamics simulations.]{(a) Schematic diagram of the proposed device (not to scale) consisting of a magnetic micro-wire with \gls{PMA} deposited onto a \LNO\ substrate and a pair of \glspl{IDT} used to launch \glspl{SAW}. (b) Initial magnetisation of the micro-wire with a Bloch \gls{DW} separating left and right domains. The external magnetic field favors the expansion of the left domain, and the travelling \glspl{SAW} propagate from left to the right along the $x$-axis. (c) A typical grain distribution where the grey level is proportional to the value of $K_\RM{u}$ for each grain. \revision{Figure (b) and (c) have the same dimensions.}}
    \label{fig:sim_config}
\end{figure}

Figure~\ref{fig:sim_config}a depicts a schematic diagram of the proposed device that can be implemented by depositing a Pt/Co/Pt micro-wire with \gls{PMA} onto a 128\textdegree\ Y-cut \LNO. At both ends of the magnetic thin film, a pair of \glspl{IDT} are patterned. By adjusting the width and pitch of the \gls{IDT} electrodes, \glspl{SAW} with various frequencies can be generated. Figure~\ref{fig:sim_config}b shows the simulated system employed in this study. The simulations in this work were performed using Mumax3, a GPU-accelerated micromagnetic simulation program~\cite{Vansteenkiste2014}. Mumax3 integrates numerically the \gls{LLG} equation (Equation~\ref{eqn:LLG}).
A Bloch \gls{DW} separating left and right domains was initialised at a distance of one-fourth from the left edge. The system then relaxed to equilibrium before an external magnetic field ($\B H_{\RM{ext}}$) pointing ``in'' was applied favouring the \gls{DW} motion towards the right.
\revision{The material parameters used in the simulations were chosen to match those of the Pt/Co/Pt thin films as referenced in the literature~\cite{yokouchi2020creation}:} 
saturation magnetisation $M_\RM{s}=\text{6}\times\text{10}^\text{5}$ A/m, 
exchange constant $A_\RM{exch}=\text{1.0}\times\text{10}^\text{{--11}}$ J/m, 
anisotropy constant $K_\RM{u}=\text{8}\times\text{10}^\text{5}$ J/m$^\text{3}$, 
and Gilbert damping constant $\alpha=\text{0.01}$. \revision{A low Gilbert damping parameter was selected in the simulations to enhance the \gls{DW} motion, thereby decreasing the required computational time to observe significant \gls{DW} displacement.}
The grid size and cell size of the simulation were set at 256 $\times$ 512 $\times$ 1 nm$^\text{3}$ and 2.4 $\times$ 2.4 $\times$ 1.0 nm$^\text{3}$, respectively, which resulted in a computational region of 614.4 $\times$ 1228.8 $\times$ 1 nm$^\text{3}$. \revision{Simulations were intentionally conducted at 0 K to distinctly elucidate the influences induced by \glspl{SAW} in the absence of thermal activity. Temperature is a crucial factor in thermally-assisted \gls{DW} motion within the creep regime; nonetheless, incorporating temperature into the simulation adds a layer of complexity that can obscure the specific effects of \glspl{SAW}. To clarify the impact of temperature, a comparative analysis of simulations with and without thermal influence is presented towards the end of the chapter. This approach allows for a focused examination of the \gls{SAW} effects while also acknowledging the role of temperature in experimental conditions.}\par

\subsection{Incorporating disorder and SAW in simulations}

To introduce disorder in our simulations, we utilised the built-in function of Mumax3, which defines grain-like regions using Voronoi tessellation~\cite{leliaert2014current}. A typical grain distribution with an average diameter $d_\RM{grain}=\text{10}$ nm, \revision{consistent with experimental observation,} is shown in Fig~\ref{fig:sim_config}c~\cite{leliaert2014influence,voto2016effects}. We assigned anisotropy constants ($K_\RM{u}$) to each grain, randomly distributed according to a normal distribution centered around their nominal value of $\text{8}\times\text{10}^\text{5}$ J/m$^\text{3}$. The standard deviation ($\sigma$) was utilised to control the level of disorder. \revision{Specifically, we consider two typical values of $\sigma$: 1\% and 3\% used in simulations~\cite{kuszewski2018resonant,voto2016effects,kim2017current}. The reason for selecting lower anisotropy disorder values than those observed experimentally is related to the computational timescales; simulating larger anisotropy disorder would necessitate a significantly long computational time to observe \gls{DW} depinning.}
A higher $\sigma$ indicates a greater variation in the anisotropy constants within the thin film, resulting in a higher pinning energy and enhanced pinning effects. By manipulating the $\sigma$ value, we can effectively modulate the degree of disorder and its impact on the magnetic behavior of the system during simulations.\par

The magnetoelastic energy density in the system can be described by Equation~\ref{eqn:E_me}. It should be noted that the film is considered acoustically thin, meaning that it is sufficiently thin and rigid for the chosen \gls{SAW} frequency~\cite{campbell2012surface}. As a result, only the in-plane strain component ($\epsilon_{xx}$) needs to be taken into account~\cite{yokouchi2020creation,dean2015sound,nepal2018magnetic}. In our simulations, we set magnetoelastic coupling coefficient $B_1$ to be $\text{1.5}\times\text{10}^\text{7}$ J/m$^\text{3}$, and $B_2=0$. Travelling \glspl{SAW} ($\epsilon_{xx}$) were implemented using Equation~\ref{eqn:travellingSAW}. The amplitude of the \gls{SAW} was set as 0.006, \revision{a value selected for its and alignment with parameters achievable in experimental settings.} The velocity of \glspl{SAW} propagating in \LNO\ is reported as $\sim$4000 m/s~\cite{paskauskas1995velocity}.\par

\subsection{Determination of domain wall velocity}

The velocity of the \gls{DW}, denoted as $v_\RM{DW}$, was determined by the change in the average normalised magnetisation along the $z$-axis ($\Delta \langle m_z \rangle$) and can be expressed as
\begin{equation}
v_\RM{DW}=\frac{L_x\Delta \langle m_z\rangle}{2 \Delta t_\RM{s}} ,
\end{equation}
where $L_x = \text{614.4}$ nm represents the length of the computational region, and $\Delta t_\RM{s}$ is the simulation time.
To investigate the effect of \gls{SAW} frequency on \gls{DW} dynamics, we set the frequencies of the \glspl{SAW} as either 50, 100, and 200 MHz. 
The dynamics of the \gls{DW} was simulated for 100 ns. \revisiontwo{This timescale accommodates five complete cycles of \glspl{SAW} at the lowest frequency employed in these simulations (50 MHz). This simulation time was optimised so that allows SAWs to promote \gls{DW} motion (as evidenced in Figure~\ref{fig:DW_motion}).}
For each value of the applied field the magnetisation was set to initial magnetisation. To account for the variation in depinning energy among different samples, we generated 10 stochastic realisations for each frequency and disorder level, using the same material parameters but different grain distributions. The data points presented in the results are the ensemble averages of these 10 stochastic realisations, along with corresponding error bars, which indicate the ensemble spread of the 10 simulations.\par


\section{Results and discussion}
\subsection{Enhanced domain wall velocity}
\begin{figure}[ht!]
    \centering
    \includegraphics[width=0.6\textwidth]{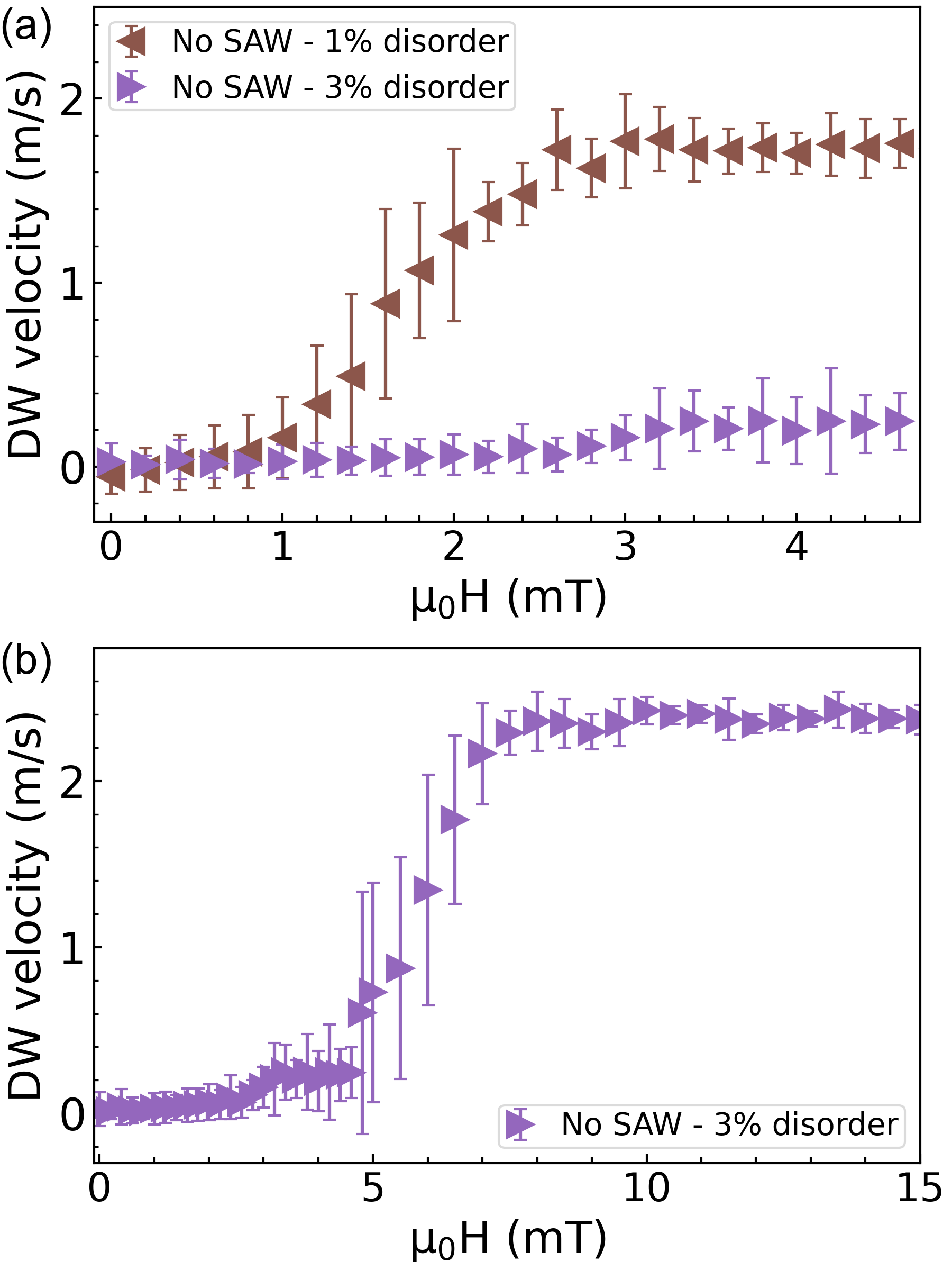}
    \caption[Domain wall velocity as a function of the applied external magnetic field without surface acoustic waves.]{(a) \gls{DW} velocity as a function of the applied external magnetic field from 0 to 4.6 mT without \glspl{SAW} in thin film with 1\% and 3\% anisotropy disorder. (b) shows the \gls{DW} velocity ($y$-axis, m/s) of the thin film with 3\% anisotropy disorder against the applied field ($x$-axis, mT) from 0 to 15 mT. Error bars represent the ensemble spread of 10 simulations.}
    \label{fig:DW_velocity_noSAW}
\end{figure}

Figure~\ref{fig:DW_velocity_noSAW} shows the \gls{DW} velocity as a function of the external magnetic field. 
For the thin film with 1\% anisotropy disorder, \gls{DW} velocity gradually increases with external magnetic field (see Figure~\ref{fig:DW_velocity_noSAW}a), which can be attributed to the relatively low level of anisotropy disorder. Following this, the \gls{DW} velocity experiences a rapid increase as the field strength rises from 1 to 3 mT. Eventually, a plateau is reached where the \gls{DW} velocity depends very little on the external magnetic field. The emergence of the plateau in thin film with disorder has been investigated in previous studies~\cite{voto2016effects,nakatani2003faster,martinez2007thermal}. 
For thin films with 3\% anisotropy disorder, the \gls{DW} velocity is almost zero until a higher depinning field of 3 mT is applied. It enters a plateau at approximately 7 mT (as depicted in Figure~\ref{fig:DW_velocity_noSAW}b).\par

\begin{figure}[ht!]
    \centering
    \includegraphics[width=0.6\textwidth]{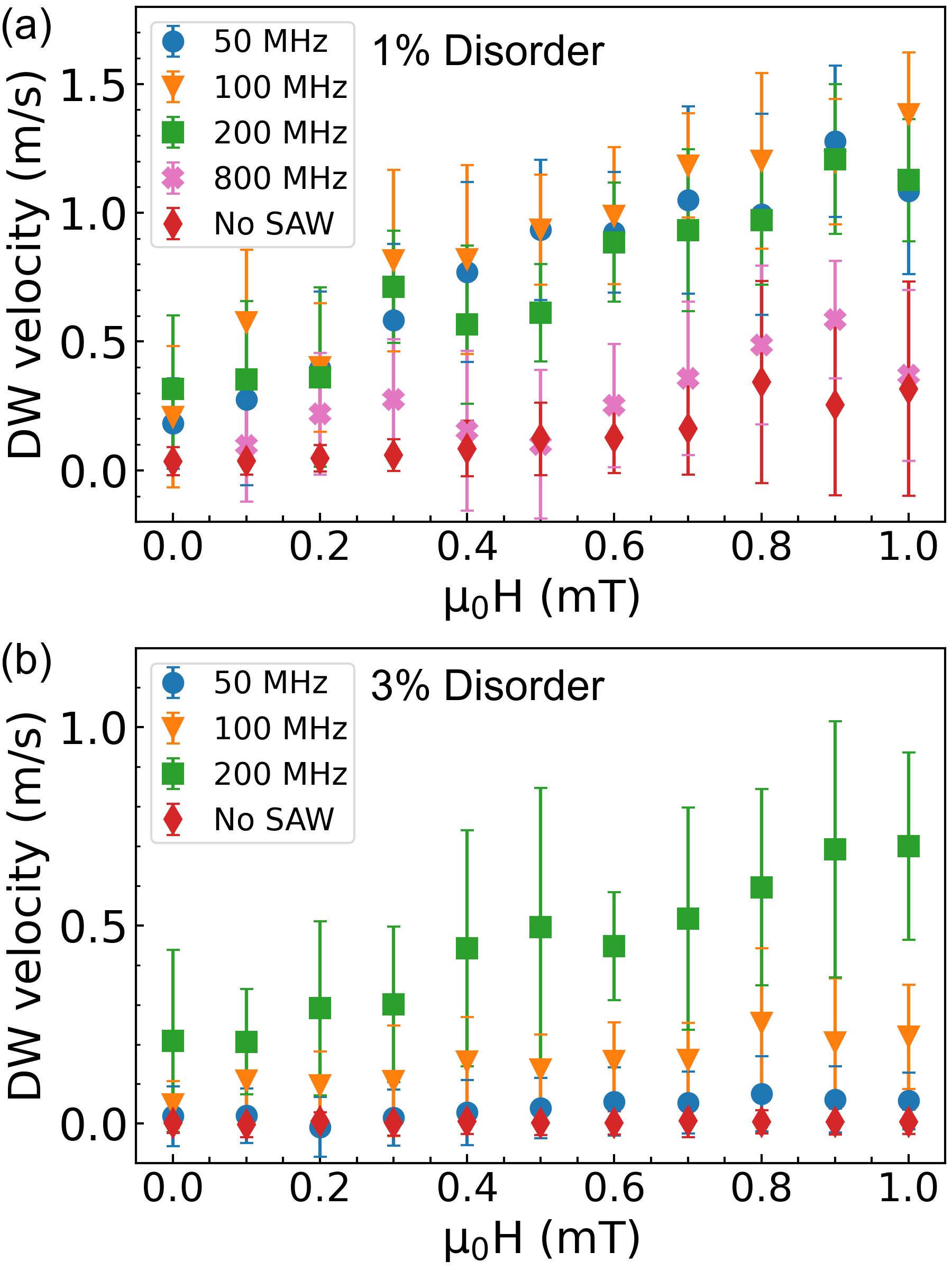}
    \caption[Domain wall velocity as a function of the applied external magnetic field with and without surface acoustic waves.]{\gls{DW} velocity as a function of the applied external magnetic field from 0 to 1 mT without \glspl{SAW} and in the presence of \glspl{SAW} with frequencies of 50, 100, and 200 MHz in thin film with (a) 1\% and (b) 3\% anisotropy disorder. Error bars represent the ensemble spread of 10 simulations.}
    \label{fig:DW_velocity_SAW}
\end{figure}

This chapter primarily focuses on the motion of magnetic \glspl{DW} under low magnetic fields (up to 1 mT). At weak applied fields, particularly around the depinning field, the probability of the \gls{DW} escaping from pinning sites is contingent upon the local energy potential associated with those sites. The \revision{shape} of this energy potential is influenced by material defects, with a specific emphasis on anisotropy disorder in this work. Notably, the pinning energy exhibits significant variation across different samples. To account for this variation, we performed simulations using 10 stochastic samples, resulting in a broad ensemble spread.
In the thin film with 1\% anisotropy disorder, and with no \glspl{SAW} applied, the \gls{DW} velocity steadily increases from 0.04 $\pm$ 0.05 m/s to 0.3 $\pm$ 0.4 m/s as the magnetic field rises from 0 to 1 mT (see Figure~\ref{fig:DW_velocity_SAW}a).
However, the presence of \glspl{SAW} substantially enhances the \gls{DW} motion across all studied frequencies. For instance, compared to the case without \glspl{SAW}, the \gls{DW} velocity is amplified by a factor of 2.7, reaching 1.1 $\pm$ 0.3 m/s with an applied field of 1 mT and 50 MHz \glspl{SAW}.
It is worth noting that the \gls{DW} velocity gradually decreases with increasing \gls{SAW} frequency. For example, in the presence of 200 MHz \glspl{SAW}, the overall trend of the \gls{DW} velocity curve lies below that observed with 50 MHz \glspl{SAW}. To further illustrate this effect, we performed simulations with 800 MHz \glspl{SAW}, and the results show that the \gls{DW} velocity is significantly slower compared to the velocities observed with \gls{SAW} frequencies ranging from 50 to 200 MHz.\par

In the case of the thin film with 3\% anisotropy disorder (see Figure~\ref{fig:DW_velocity_SAW}b), the \gls{DW} velocity within the studied field range is notably slower compared to the thin film with 1\% anisotropy disorder. This difference can be attributed to the higher pinning energy associated with the increased disorder level~\cite{chauve2000creep,voto2016effects}.
Within the applied field range of 0--1 mT, the \gls{DW} velocity shows minimal change due to the depinning field being at 3 mT (as shown in Figure~\ref{fig:DW_velocity_noSAW}a). However, the introduction of \glspl{SAW} leads to a significant enhancement in the \gls{DW} velocity, particularly with the presence of 200 MHz \glspl{SAW}. Comparing the \gls{DW} velocity with 1 mT alone to the \gls{DW} velocity in the presence of 200 MHz \glspl{SAW}, an increase from 0.004 $\pm$ 0.03 m/s to 0.7 $\pm$ 0.2 m/s is observed.
Interestingly, the \gls{DW} velocity increases with \gls{SAW} frequency in the thin film with 3\% anisotropy disorder, which is contrary to the observations made in the thin film with 1\% anisotropy disorder.\par

\subsection{Domain wall depinning}
\begin{figure}[!ht]
    \centering
    \includegraphics[width=1\textwidth]{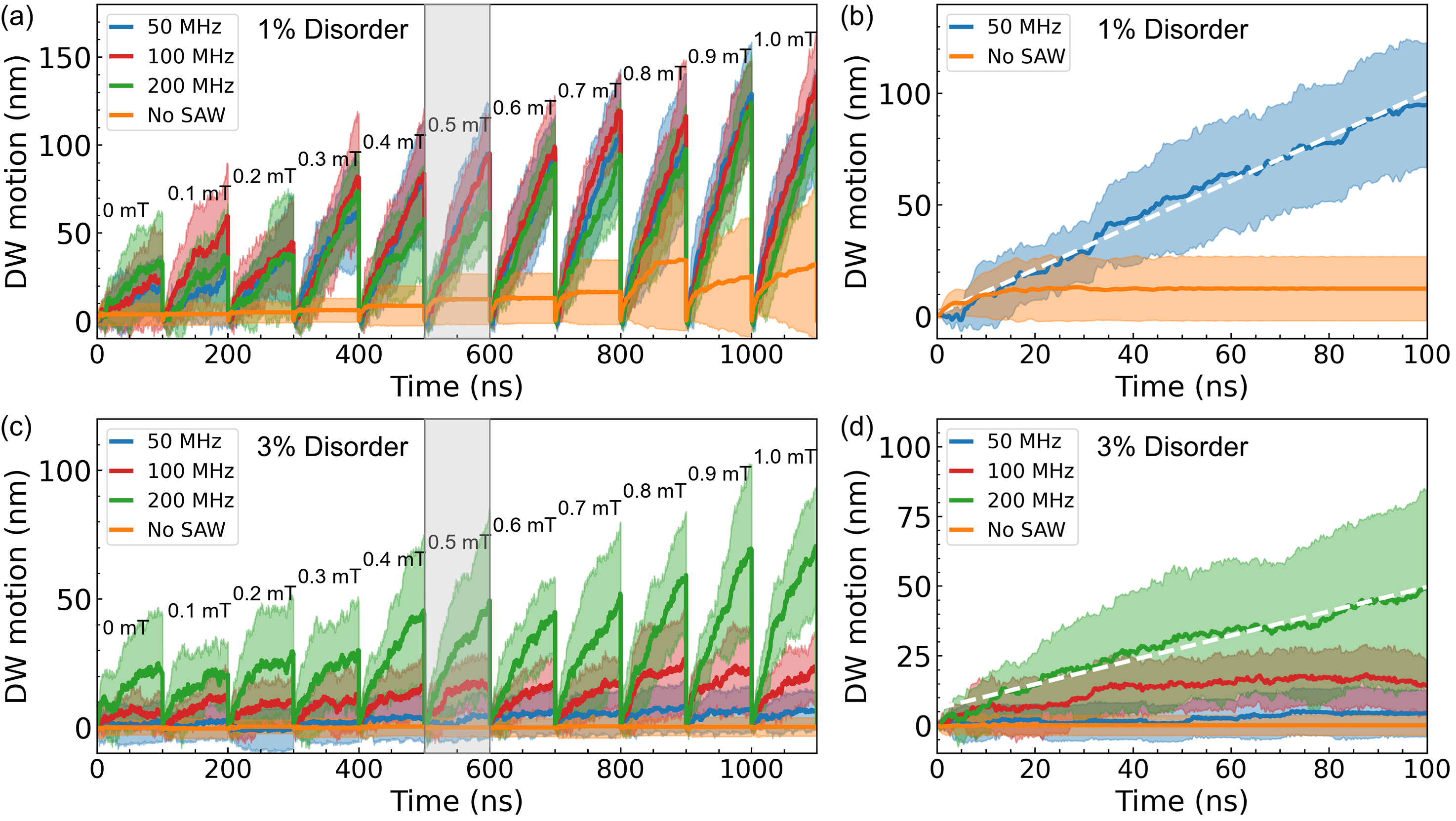}
    \caption[Time evolution of domain wall motion in thin films with anisotropy disorder.]{Time evolution of \gls{DW} motion in thin films with (a) 1\% and (c) 3\% anisotropy disorder. \gls{DW} motion at each applied field was performed for 100 ns, followed by returning to the initial magnetisation for the next simulation. Shaded areas represents the ensemble spread of 10 simulations. The applied field for each simulations is labeled on the Figure (b) and (d) are zoomed-in views of the enclosed areas in (a) and (c), respectively. The dash lines in (b) and (d) are the linear fits of the corresponding curves. \revisiontwo{SAWs are able to promote the \gls{DW} motion within the simulation timescale.}}
    \label{fig:DW_motion}
\end{figure}

To explore the influence of \gls{SAW} frequency on the dynamics of \glspl{DW} in thin films with varying levels of disorder, we examined the \gls{DW} motion as a function of simulation time, as shown in Figure~\ref{fig:DW_motion}. For each value of applied field, simulation was conducted for a duration of 100 ns, followed by resetting the initial magnetisation before the start of the subsequent simulation.
We begin by discussing the thin film with 1\% anisotropy disorder in the absence of \glspl{SAW}. From Figure~\ref{fig:DW_motion}a, it is evident that the depinning field of the \gls{DW} is approximately 0.4 mT in the absence of \glspl{SAW}. Below this field, no noticeable \gls{DW} motion is observed. In Figure~\ref{fig:DW_motion}b, a more detailed view of \gls{DW} motion at 0.5 mT is presented. Initially, in the presence of only the magnetic field, the \gls{DW} moves towards the right for approximately the first 20 ns, followed by a pinning event that persists until the end of the simulation. 
As the field strength increases, the \gls{DW} advances more until the \gls{DW} is pinned at a position for which the depinning field is larger than the applied field. For instance, at external magnetic fields of 0.6, 0.7, and 0.8 mT, the pinning occurs after approximately 30, 35, and 50 ns, respectively.\par

We further explore the impact of \glspl{SAW} on \gls{DW} motion within the same thin film. Even at 0 mT applied field, a significant displacement of the \gls{DW} is observed when \glspl{SAW} with frequencies of 50, 100, and 200 MHz are applied. This observation suggests that the introduced \glspl{SAW} effectively facilitate \gls{DW} depinning from pinning sites. Moreover, regardless of the \gls{SAW} frequency, the \gls{DW} exhibits continuous motion at a consistent velocity, as indicated by the dashed line overlaying the \gls{DW} motion curve in Figure~\ref{fig:DW_motion}b. The pinning events observed in the absence of \glspl{SAW} are absent when \glspl{SAW} are applied, as depicted in Figure~\ref{fig:DW_motion}a. Instead, steeper slopes in the \gls{DW} displacement over time are observed with increasing external magnetic field, indicative of higher \gls{DW} velocities, as illustrated in Figure~\ref{fig:DW_velocity_SAW}a.\par

Next, we examine the behavior of the \gls{DW} in the thin film with 3\% anisotropy disorder. Figure~\ref{fig:DW_motion}c illustrates that when only the magnetic field is applied, the pinning events are significantly more prominent compared to the thin film with 1\% anisotropy disorder under the same conditions. In fact, the \gls{DW} remains pinned even with a field strength of 1 mT, indicating a higher depinning field in thin films with increased anisotropy disorder.
The \gls{DW} motion exhibits distinct characteristics depending on the frequency of the applied \glspl{SAW}. As shown in Figure~\ref{fig:DW_motion}c, compared to the case without \glspl{SAW}, the overall trend of the \gls{DW} motion is enhanced in the presence of \glspl{SAW} within the studied frequency range. However, the frequency of the \gls{SAW} significantly influences the \gls{DW} motion.
Figure~\ref{fig:DW_motion}d illustrates the \gls{DW} displacement in the thin film with 3\% anisotropy disorder at 0.5 mT with and without \glspl{SAW}. Without \glspl{SAW}, the \gls{DW} hardly moves. In contrast, the presence of 50 MHz \glspl{SAW} leads to an observable \gls{DW} motion with a few pinning events. For example, the \gls{DW} becomes pinned during the initial 10 ns, followed by motion for a short period until the next pinning event at 20 ns. With an increase in \gls{SAW} frequency to 100 MHz, more significant \gls{DW} motion is observed, accompanied by additional pinning events. Notably, when the \gls{SAW} frequency reaches 200 MHz, continuous \gls{DW} motion with minimal pinning effects occurs (as indicated by the dashed line in Figure~\ref{fig:DW_motion}d). In fact, even at 0 mT, 200 MHz \glspl{SAW} facilitate the \gls{DW} depinning process, while 50 MHz \glspl{SAW} exhibit limited effects on \gls{DW} motion, even at 1 mT. These observations suggest that \glspl{SAW} with higher frequencies exert a stronger influence on \gls{DW} motion in thin films with higher levels of anisotropy disorder.
It is worth noting that the \gls{DW} velocity, even in the presence of \glspl{SAW}, is slower compared to the thin film with lower anisotropy disorder. This is attributed to the higher energy required to depin the \gls{DW} in thin films with increased anisotropy disorder.
\revisiontwo{Note that within the first five complete cycles (100 ns for 50 MHz, 50 ns for 100 MHz, and 20 ns for 200 MHz), \gls{DW} depins from the pinning sites and shows a continuous motion in the presence of the SAWs with 100 and 200 MHz. This observation confirms that SAWs with five complete cycles offer sufficient opportunities for depinning, thereby validating the chosen simulation duration.}\par


\begin{figure}
    \centering
    \includegraphics[width=1\textwidth]{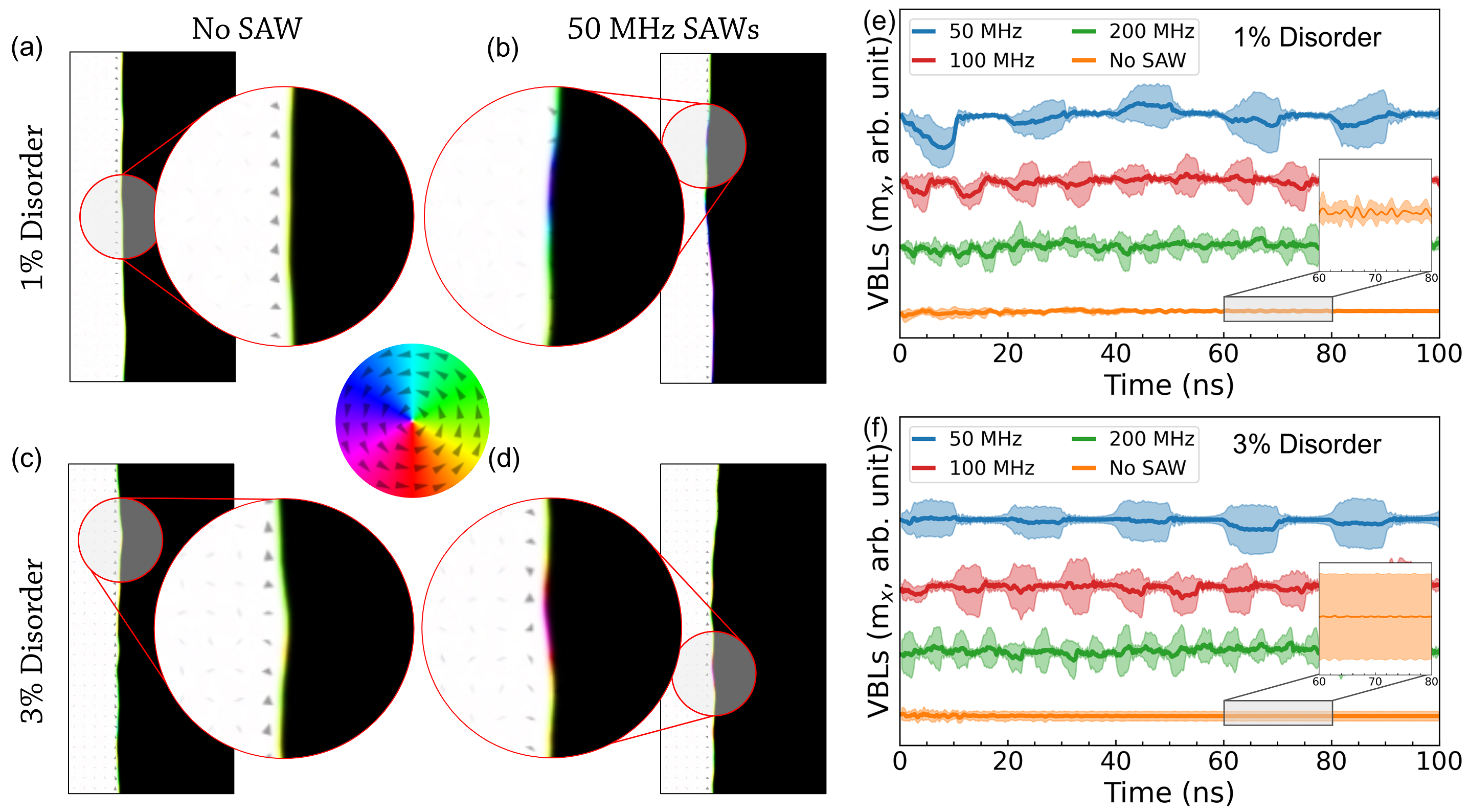}
    \caption[Snapshots of the magnetisation configuration without surface acoustic waves and in the presence of surface acoustic waves.]{Snapshots of the magnetisation configuration under an applied field of 0.5 mT at 10 ns in thin films with (a) 1\% disorder without \gls{SAW}, (b) 1\% disorder with 50 MHz \glspl{SAW}, (c) 3\% disorder without \gls{SAW}, and (d) 3\% disorder with \glspl{SAW}. Inset graphs in (a)--(d) represent the enlarged magnetisation configuration of the corresponding enclosed area. The colour code for the in-plane magnetisation component is shown by a colour wheel. The number of \glspl{VBL} against the simulation time in thin film with (e) 1\% and (f) 3\% anisotropy disorder under an applied field of 0.5 mT without \glspl{SAW} and in the presence of \glspl{SAW} with frequencies of 50, 100, and 200 MHz. Inset graphs in (e) and (f) are zoomed-in view of \glspl{VBL} number against simulation time (60--80 ns) without \glspl{SAW} (two graphs share the same scale). Shaded areas indicate the ensemble spread of 10 simulations. The number of \glspl{VBL} ($y$-axis) is shifted for clarity of changes.}
    \label{fig:vbls}
\end{figure}

\subsection{Formation of vertical Bloch lines}

To gain insight into the \gls{DW} depinning process, we examine the magnetisation configuration under the influence of the magnetic field alone and in the presence of \glspl{SAW}.
At 10 ns into the simulation, the magnetisation configuration is captured and presented in Figure~\ref{fig:vbls}a--d. In the thin film with 1\% anisotropy disorder and without \glspl{SAW} at 0.5 mT, the \gls{DW} moves steadily and coherently while maintaining its structure. A smooth \gls{DW} with no significant curvature is observed, depicted as a light green line separating the left and right domains in Figure~\ref{fig:vbls}a.
However, in the presence of \glspl{SAW}, the \gls{DW} exhibits curvatures and Bloch segments with opposite orientation (up/down) separated by \glspl{VBL} as shown in Figure~\ref{fig:vbls}b. The different coloured sections indicate the introduction of additional spin configurations in the wall structure induced by the \glspl{SAW}~\cite{voto2016effects,yamada2015excitation,yoshimura2016soliton}.\par

In our simulations, the \gls{DW} exhibits a Bloch wall structure, where the magnetisation is primarily oriented in the ``up'' or ``down'' direction (along the $y$-axis). The \glspl{VBL} in our system, on the other hand, are oriented in the ``left'' or ``right'' direction (along the $x$-axis). Thus, the additional spin structures introduced by the \glspl{VBL} can be represented by the magnetisation component along the $x$-axis ($m_x$). It is important to note that the ``number of \glspl{VBL}'' in this study is not an absolute quantity; rather, we utilised $m_x$ as a representation of the presence and density of \glspl{VBL} within the system.
The number of \glspl{VBL} is plotted against simulation time in Figure~\ref{fig:vbls}e and f. In the thin film with 1\% anisotropy disorder and the application of the magnetic field alone, the number of \glspl{VBL} shows minimal change, indicating rare occurrence of spin rotations (inset graph in Figure~\ref{fig:vbls}e).
However, when \glspl{SAW} are applied, the number of \glspl{VBL} exhibits periodic fluctuations, as shown in Figure~\ref{fig:vbls}e. The period of these fluctuations corresponds to the frequencies of the applied \glspl{SAW}. For instance, \glspl{SAW} with frequencies of 50, 100, and 200 MHz exhibit periodicities of approximately 20, 10, and 5 ns, respectively. The periodic variation in the number of \glspl{VBL} with the \gls{SAW} frequency arises from the magnetoelastic coupling effect. When \glspl{SAW} are applied, strain is introduced to the magnetic thin film, resulting in the emergence of an effective magnetic field. This effective magnetic field plays a crucial role in both the nucleation and annihilation of \glspl{VBL}, as well as influencing the motion of the \gls{DW} and the \glspl{VBL} within the \gls{DW}, contributing to their dynamics.\par

In the magnetisation configuration of the thin film with 3\% anisotropy disorder at 0.5 mT, the \gls{DW} exhibits greater curvatures (see Figure~\ref{fig:vbls}c) and a larger number of \glspl{VBL} (see Figure~\ref{fig:vbls}f) compared to the thin film with 1\% anisotropy disorder under the same conditions.
The inset graphs in Figure~\ref{fig:vbls}e and f (same scale) provide a closer look at the number of \glspl{VBL} when only the magnetic field is applied within the range of 60 to 80 ns.
The ensemble spread of \glspl{VBL} in the thin film with 3\% anisotropy disorder is approximately 2.5 times wider than that in the thin film with 1\% anisotropy disorder, indicating a greater occurrence of spin rotations within the \gls{DW} in the thin film with 3\% anisotropy disorder.
This can be attributed to the fact that the magnetic field alone is strong enough to induce \gls{DW} motion and maintain \gls{DW} structure in the thin film with 1\% anisotropy disorder, thereby avoiding the significant formation of \glspl{VBL}. However, in the thin film with 3\% anisotropy disorder, the magnetic field is insufficient to enable the \gls{DW} to overcome the local energy potential~\cite{yoshimura2016soliton}. Consequently, the magnetic field introduces spin rotations within the \gls{DW}, contributing to the broader ensemble spread of \glspl{VBL}.
Similar to the \gls{DW} in the thin film with 1\% anisotropy disorder, the number of \glspl{VBL} in the thin film with 3\% anisotropy disorder also undergoes periodic changes with the \gls{SAW} frequency. The fluctuation behavior of \glspl{VBL} in relation to the \gls{SAW} frequencies is observed consistently across thin films and various applied fields.\par


\begin{figure}
    \centering
    \includegraphics[width=0.9\textwidth]{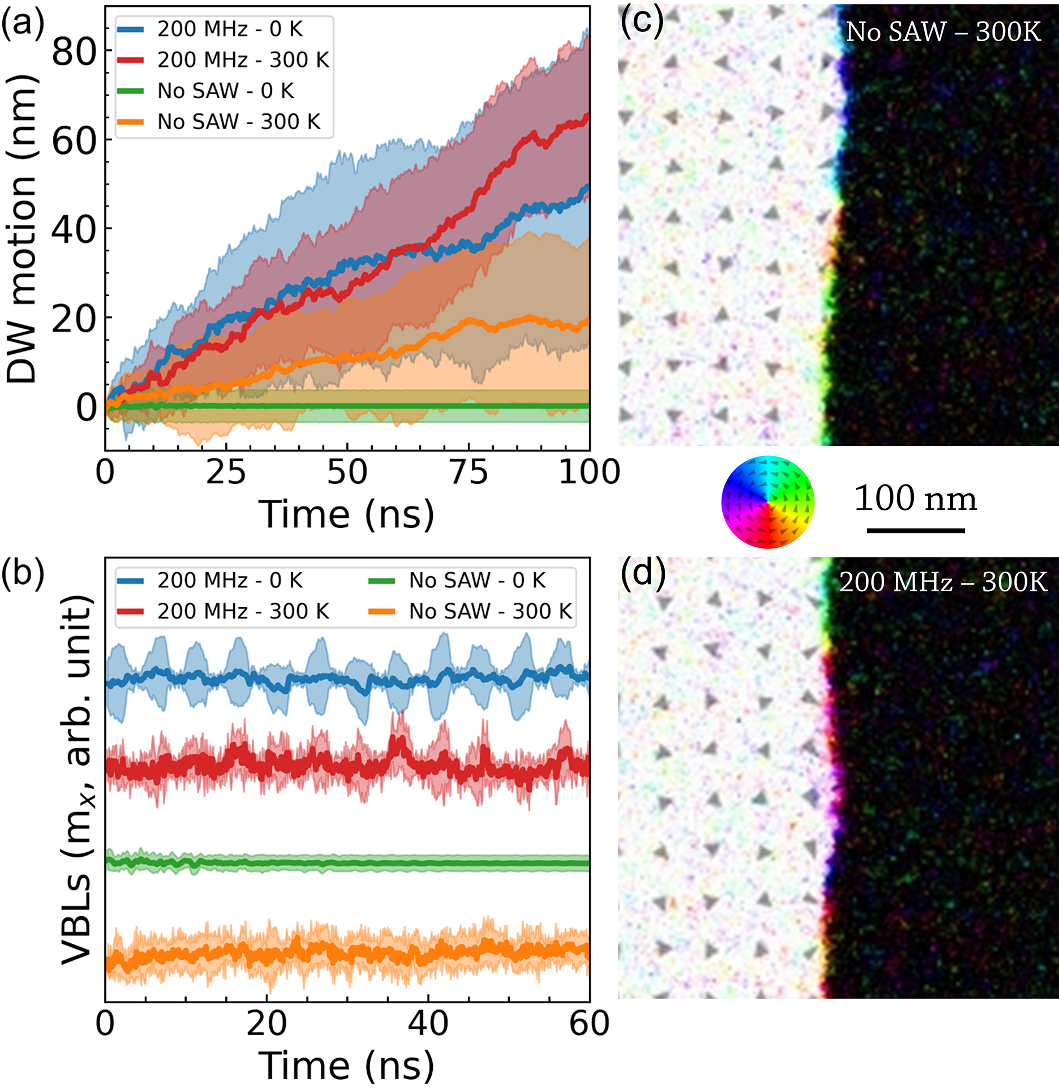}
    \caption[Domain wall motion and vertical Bloch lines plotted against simulation time, and snapshots of the magnetisation configuration both without surface acoustic waves and in the presence of surface acoustic waves at 0 and 300 K.]{(a) \gls{DW} motion and (b) \glspl{VBL} plotted against simulation time both without \glspl{SAW} and with 200 MHz \glspl{SAW} at 0 and 300 K. Snapshots of the magnetisation configuration at 300 K at 10 ns (c) without \glspl{SAW} and (d) with 200 MHz \glspl{SAW}. The simulations are conducted in the thin film with 3\% disorder under an applied field of 0.6 mT. The colour code for the in-plane magnetisation component is shown by a colour wheel. Shaded areas indicate the ensemble spread of 10 simulations. The number of \glspl{VBL} ($y$-axis) is shifted for clarity of changes.}
    \label{fig:temperature_effect}
\end{figure}

The introduction of \glspl{SAW} induces spin rotation in the \gls{DW}, resulting in a smoothing of the energy landscape within the thin film. This phenomenon leads to enhanced \gls{DW} motion. It is instructive to compare the effect of \glspl{SAW} with the effect of thermal fluctuations on \gls{DW} motion. To do this, we separately examined the effects of \glspl{SAW} and temperature on \gls{DW} motion~\cite{martinez2007thermal,leliaert2015thermal}.
Figure~\ref{fig:temperature_effect}a displays the time evolution of the \gls{DW} at both 0 K and 300 K in a thin film with 3\% anisotropy disorder. In the absence of \glspl{SAW}, the presence of thermal fluctuations at 300 K significantly enhances \gls{DW} motion compared to the motion observed at 0 K. This enhancement is also evident through the presence of multiple \glspl{VBL} at 300 K, as depicted in Figure~\ref{fig:temperature_effect}c, and the variation in the number of \glspl{VBL} over time, illustrated in Figure~\ref{fig:temperature_effect}b. The number of \glspl{VBL} is notably higher than in simulations conducted at 0 K without \glspl{SAW}, although the changes in the number of \glspl{VBL} occur randomly over time.
In the presence of 200 MHz \glspl{SAW} at 300 K, the \gls{DW} also contains multiple \glspl{VBL}, as shown in Figure~\ref{fig:temperature_effect}d. However, in this case, the changes in the number of \glspl{VBL} over time have a periodicity that aligns with the \gls{SAW} frequency. Moreover, there is an additional level of randomness introduced by thermal fluctuations when compared to the scenario with 200 MHz \glspl{SAW} at 0 K, as depicted in Figure~\ref{fig:temperature_effect}c.
The introduction of \glspl{SAW} in magnetic thin films creates an effective field with the same periodicity as the \gls{SAW} frequency. This effective field induces spin rotation, leading to the formation of \glspl{VBL} within the \gls{DW}. The spin rotation caused by \glspl{SAW} enhances the possibility of \gls{DW} depinning by smoothing the energy landscape of the thin films, similar to the effect of thermal fluctuation~\cite{martinez2007thermal}.
It is important to note that the spin rotation induced by \glspl{SAW} exhibits the same frequency as the \gls{SAW}, in contrast to the randomly distributed nature of thermal fluctuation. This coherent spin rotation induced by \glspl{SAW} introduces a more controlled and synchronised influence on the \gls{DW} dynamics, offering potential advantages for precise manipulation and control of \gls{DW} motion in magnetic thin films.\par

\begin{figure}
    \centering
    \includegraphics[width=0.8\textwidth]{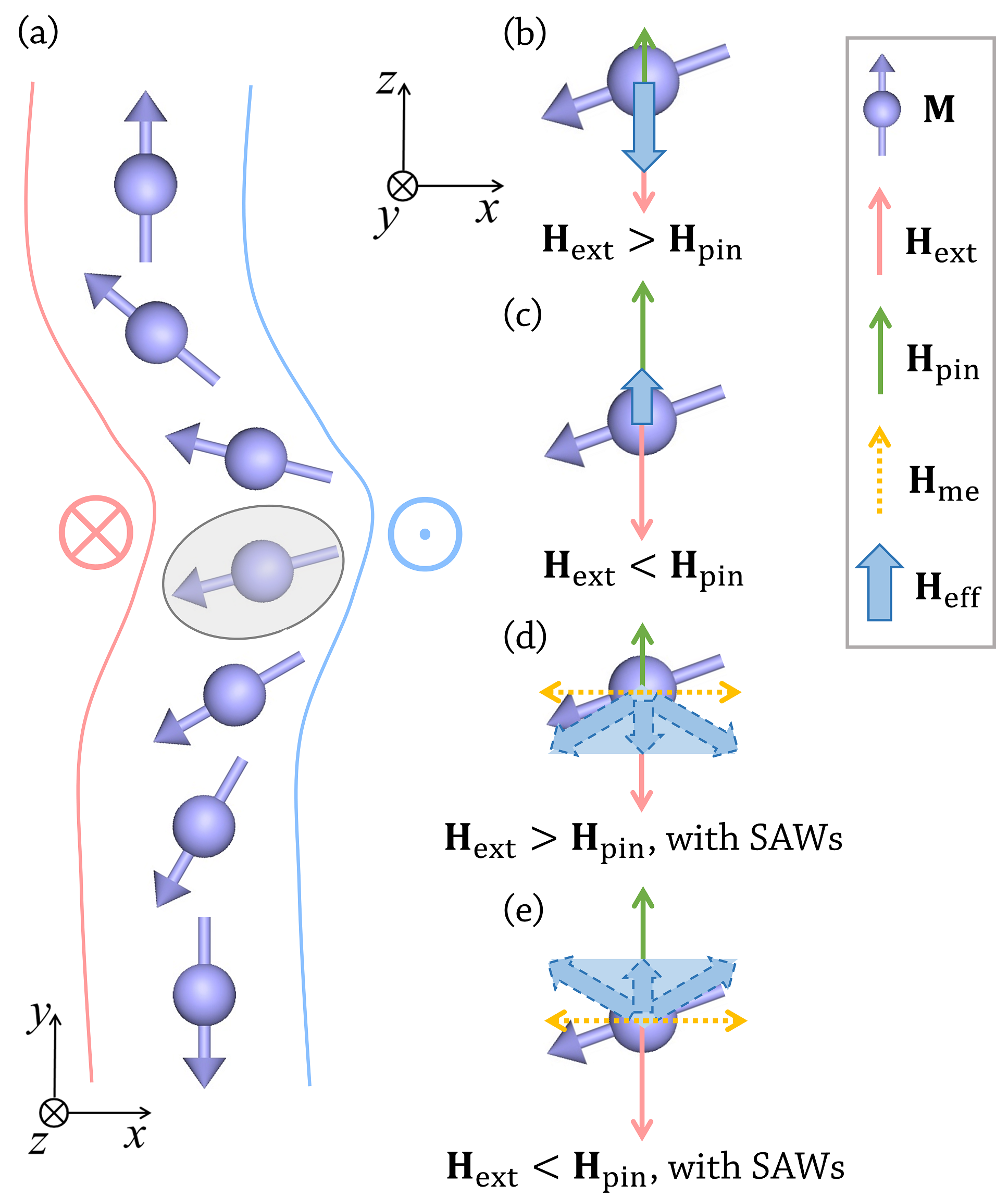}
    \caption[Diagram of spin dynamics under different conditions.]{(a) The left and right domains are pointing ``in'' and ``out'' of the paper, respectively, separated by a \gls{DW} with \glspl{VBL}. The circled spin is used to analyse the effective fields under different conditions in (b)--(e). (b) $\B{H}_{\RM{ext}}>\B{H}_{\RM{pin}}$ without \gls{SAW}. (c) $\B{H}_{\RM{ext}}<\B{H}_{\RM{pin}}$ without \gls{SAW}. (d) $\B{H}_{\RM{ext}}>\B{H}_{\RM{pin}}$ with \glspl{SAW}. (e) $\B{H}_{\RM{ext}}>\B{H}_{\RM{pin}}$ with \glspl{SAW}.}
    \label{fig:sketch}
\end{figure}

The \gls{SAW}-induced effective field and spin rotation, are critical in modulating the local energy landscape. 
Fig.~\ref{fig:sketch}a shows a schematic of a \gls{DW} containing a \gls{VBL} with a curvature separating the left and right domain. Note that only external magnetic field  ($\B{H}_{\RM{ext}}$) and effective fields generated by pinning potential ($\B{H}_{\RM{pin}}$) and \glspl{SAW} ($\B{H}_{\RM{me}}$) are demonstrated in the sketch.
When only external magnetic field applied is applied and no \gls{SAW}, the magnetic moment mainly experiences two torques generated by external magnetic field and local effective pinning field, respectively. In the simulation, the pinning is introduced by the magnetic anisotropy in out-of-plane direction ($K_{\RM{u}}$). Therefore, only pinning in the $z$-axis is taken into consideration.  
When $\B{H}_{\RM{ext}}$ is larger than $\B{H}_{\RM{pin}}$ (as shown in Fig.~\ref{fig:sketch}b), the magnetic moment aligns to the net effective field resulting in the \gls{DW} motion towards the right. 
However, when $\B{H}_{\RM{ext}}$ is smaller than $\B{H}_{\RM{pin}}$ (as shown in Fig.~\ref{fig:sketch}c), the \gls{DW} is pinned at the local pinning site since the net effective field is smaller than the local effective pinning field. 
Owing to the magnetoelastic coupling effect, the dynamic strain induced by \glspl{SAW} generates a dynamic effective field (as shown in Fig.~\ref{fig:sketch}d and e).
This effective field exerts an extra dynamic torque on the magnetic moment within the \gls{DW} (in our simulation, only in-plane strain is taken into consideration).
The \gls{SAW}-induced dynamic effective field can promote diverse spin configurations within the \gls{DW} (as evidenced by the formation of \glspl{VBL}).
Under certain configurations together with other effective fields (such as exchange interaction), the \gls{DW} can overcome these local energy barriers.
Hence, \glspl{SAW} not only generate an effective field for \gls{DW} motion but also induce spin rotations and \glspl{VBL}, facilitating the most conducive spin configurations for \gls{DW} depinning.\par

The \gls{SAW}-induced spin rotation has two \revision{possible} distinct effects on \gls{DW} motion:
(i) The spin rotation increases the likelihood of \gls{DW} depinning from pinning sites, thereby promoting \gls{DW} motion.
\revision{(ii) The presence of the \glspl{SAW} drives the \gls{DW} motion into a non-linear regime, where the spin rotation leads to enhanced energy dissipation at the \gls{DW} due to magnetic damping.}
These two effects have contrasting consequences on \gls{DW} motion. On one hand, they can increase the \gls{DW} velocity by enhancing the probability of depinning. \revision{On the other hand, the energy dissipation can limit the energy available to enhance \gls{DW} motion.}
Higher \gls{SAW} frequencies correspond to higher spin rotation frequencies, which, in turn, lead to increased energy dissipation \revision{at the \gls{DW}}. In the thin film with 3\% disorder, the effect (i) is more significant, resulting in enhanced \gls{DW} velocity and depinning with increasing \gls{SAW} frequency. In contrast, the thin film with 1\% disorder experiences a greater impact from effect (ii), leading to a smaller \gls{DW} velocity in the presence of \glspl{SAW} with higher frequencies.
At low magnetic fields, material defects or grain boundaries act as potential wells, serving as pinning sites for the motion of the \gls{DW}~\cite{voto2016effects,chauve2000creep,metaxas2007creep}. \glspl{SAW} in our thin films introduce energy fluctuations with the same frequency as the \glspl{SAW}, which promote the depinning of the \gls{DW} from these pinning sites~\cite{leliaert2014current, PhysRevB.84.075469, leliaert2015thermal}. \revision{The implication of effect (ii) is a hypothesis inferred from our simulation data; however, it necessitates additional examination to confirm its validity.}\par


\section{Summary}
In summary, we investigated the impact of \gls{SAW} frequency (50, 100, and 200 MHz) on the \gls{DW} motion in thin films with different levels of anisotropy disorder through micromagnetic simulations.
The results demonstrated that \glspl{SAW} enhance \gls{DW} velocity by promoting the depinning of \glspl{DW} from pinning sites through \gls{SAW}-induced spin rotation, which was consistently observed across all cases studied. The spin rotation not only increased the likelihood of \gls{DW} depinning but also contributed to energy dissipation at the \gls{DW}.
In the thin film with 3\% anisotropy disorder, the \gls{DW} velocity increased with the \gls{SAW} frequency, primarily due to the amplifying effect of spin rotation, which enhanced the probability of \gls{DW} depinning. Conversely, in the thin film with 1\% anisotropy disorder, the \gls{DW} velocity decreased with increasing \gls{SAW} frequency due to the significant \gls{SAW}-induced spin rotation, leading to pronounced energy dissipation at the \glspl{DW} as the primary factor influencing \gls{DW} motion.
These findings provide valuable insights into the intricate interplay between \glspl{SAW}, spin rotation, and \gls{DW} dynamics, highlighting the crucial role of anisotropy disorder in governing the response of \glspl{DW} to \glspl{SAW}.
\cleardoublepagewithnumberheader
\chapter{SAW effect on skyrmion dynamics}
\label{Chapter7_Skyrmion_motion}

\section{Introduction}
Magnetic skyrmions, which are topologically protected particle-like magnetic structures, show significant potential in applications including in data storage and processing devices~\cite{nagaosa2013topological,fert2013skyrmions}. 
Skyrmions in thin films can be manipulated by spin-polarised current via \gls{STT} or \gls{SOT} owing to the large \gls{SOC} with heavy metals~\cite{torrejon2014interface,litzius2020role,woo2016observation,caretta2018fast,woo2018current}. 
However, these methods require a high current density, which can cause Joule heating thereby wasting energy and affecting the stability of skyrmions. 
In addition, skyrmions typically show both longitudinal and transverse motion owing to the skyrmion Hall effect, which can cause the annihilation of skyrmions at device edges, thus complicating device realisations~\cite{nagaosa2013topological,juge2019current,woo2018current}.\par

Typically, the stability of skyrmions in thin films with \gls{PMA} is a result of the balance between the magnetic anisotropy and the interfacial \gls{DMI} induced by the broken interfacial inversion symmetry~\cite{bhattacharya2020creation,dzyaloshinsky1958thermodynamic,moriya1960anisotropic,barker2023breathing}. 
To avoid Joule heating, one can modify the anisotropy of thin films using strain to control magnetisation~\cite{shepley2015modification,matsukura2015control,radaelli2014electric,shirahata2015electric}. 
For instance, Wang \et created skyrmions in Pt/Co/Ta multilayer nano-dot systems using an electrical field-induced strain~\cite{wang2020electric}, while Ba \et demonstrated the creation, reversible deformation, and annihilation of skyrmions in a Pt/Co/Ta multilayer thin film using strain by applying an electrical field to a PMN-PT substrate~\cite{ba2021electric}. 
The dynamic strain induced by \glspl{SAW} has also been suggested as an attractive approach to control thin film magnetisation~\cite{shuai2022local,dean2015sound,thevenard2016strong,adhikari2021surface,edrington2018saw,ryu2013chiral,cao2021surface,li2014acoustically,shuai2023precise}. 
\gls{SAW} control of skyrmions has a number of potential advantages. \glspl{SAW} can be exerted by voltage instead of current making it attractive from the energy-efficiency perspective, while \glspl{SAW} can propagate over distances of several millimetres with very little power loss. This allows one pair of electrodes to control multiple devices. Besides, pinning sites can be created by electrodes remotely, which potentially allows one to control skyrmions precisely without complex design.\par
\revisiontwo{Nepal \et theoretically demonstrated the dynamics of skyrmion motion under the influence of standing \glspl{SAW} as reported in their study~\cite{nepal2018magnetic}. Their findings indicate that standing \glspl{SAW} are capable of creating pinning sites for skyrmions, with the driving force being proportional to the strain gradient. In the initial section of this chapter, we replicate their investigation on the formation of pinning sites induced by standing \glspl{SAW}. We then explore the influence of travelling \glspl{SAW} and orthogonal \glspl{SAW} -- created by the transverse standing and longitudinal travelling \glspl{SAW} -- on skyrmion trajectories.}

\section{Model and computational details}
\begin{figure}
\centering
\includegraphics[width=0.75\textwidth]{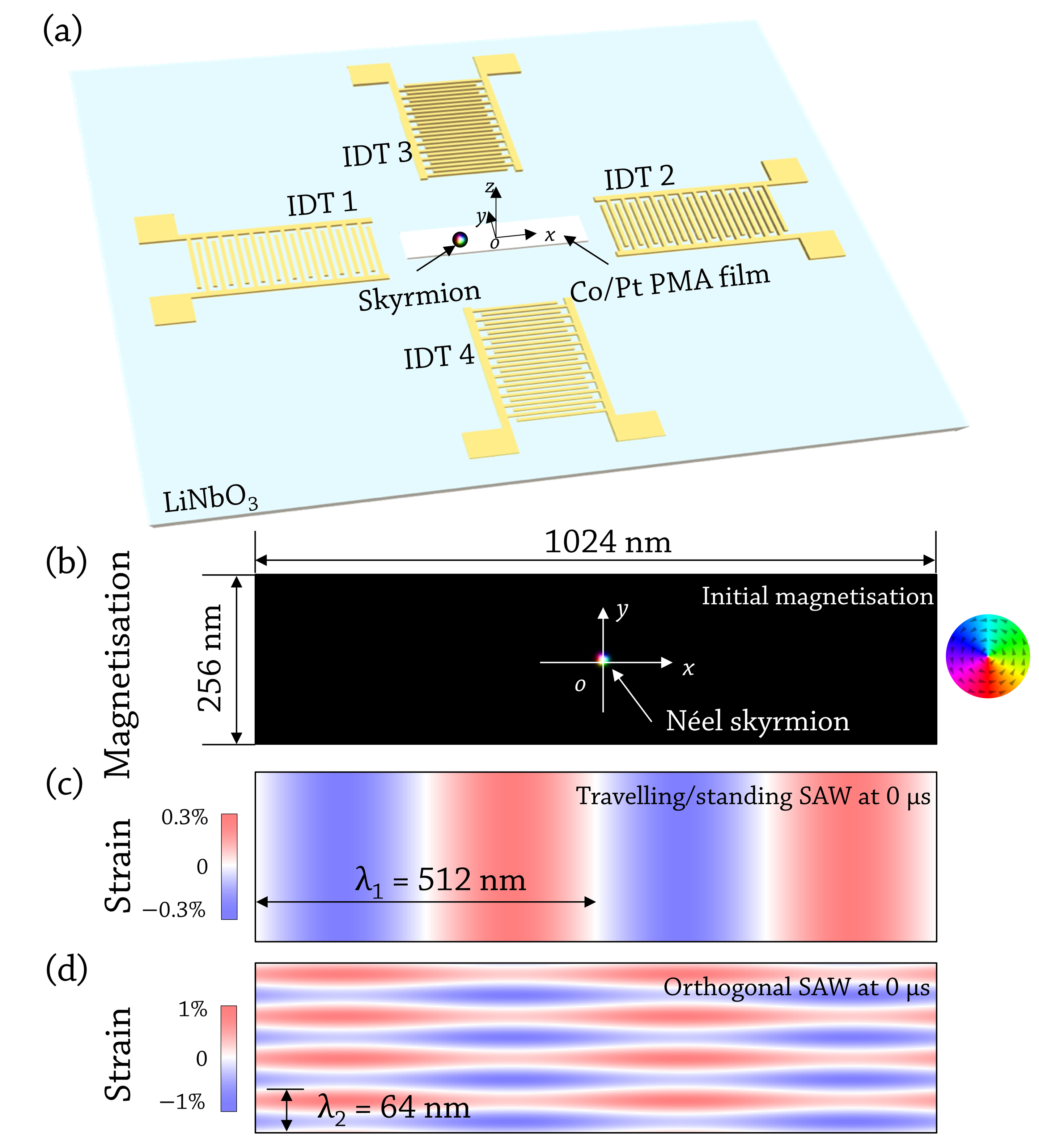}
\caption[Diagram of the proposed device and its initial magnetisation, strain spatial profile of the surface acoustic wave.]{\label{fig:config_sim}
(a) Diagram of the proposed device. A N\'eel skyrmion is initialised in a \gls{PMA} thin film grown onto a \LNO\ substrate. The thin film is surrounded by two pairs of \glspl{IDT} (\gls{IDT}1 to \gls{IDT}4) used to generate \glspl{SAW} in different modalities. (b) An example of the initial magnetisation. A N\'eel skyrmion is initialised and relaxed in the thin film. (c) Strain spatial profile of travelling \glspl{SAW} and standing \glspl{SAW} at 0 \textmu s. (d) Spatial strain profile of orthogonal \glspl{SAW} (longitudinal travelling \glspl{SAW} and transverse standing \glspl{SAW}) at 0 \textmu s. \revision{The colour code for the in-plane magnetisation component is shown by a colour wheel.}}
\end{figure}

Figure~\ref{fig:config_sim}a shows the schematic diagram of the proposed device that can be implemented by preparing a Co/Pt \gls{PMA} thin film (with dimensions of 1024 $\times$ 256 $\times$ 1 nm\textsuperscript{3} employing repeated boundary conditions) onto a 128\textdegree\ Y-cut \LNO\ substrate surrounded by two pairs of \glspl{IDT} (labelled as \gls{IDT}1 to \gls{IDT}4). 
As shown in Figure~\ref{fig:config_sim}b, the origin of the thin film geometry is located at its centre, with $x$ and $y$ correspond to the longitudinal and transverse directions of the thin film, respectively. 
The electrode spacing of the first (\gls{IDT}1 and \gls{IDT}2) and second (\gls{IDT}3 and \gls{IDT}4) pair of \glspl{IDT} was set as 128 nm and 16 nm, respectively, producing \glspl{SAW} with wavelength $\lambda_1$ (see Figure~\ref{fig:config_sim}c) and $\lambda_2$ (see Figure~\ref{fig:config_sim}d) of 512 nm and 64 nm, respectively.\par

\glspl{SAW} with different propagation modes can be achieved by applying \gls{rf} signals to one or more \glspl{IDT}: travelling \glspl{SAW} propagating in $x$ direction can be generated by applying \gls{rf} signals to \gls{IDT}1; standing \glspl{SAW} can be formed by applying \gls{rf} signals to \gls{IDT}1 and \gls{IDT}2 simultaneously; while orthogonal \glspl{SAW} consisting of horizontal travelling \glspl{SAW} and transverse standing \glspl{SAW}, can be formed by applying \gls{rf} signals to \gls{IDT}1, \gls{IDT}3 and \gls{IDT}4 at the same time. 
\revision{N\'eel skyrmions, which are common type of skyrmions found in thin films with \gls{PMA}, were initialised and relaxed at different positions in the thin film with \gls{PMA} in order to study the \gls{SAW} effect on their motion (Figure~\ref{fig:config_sim}b).} The strain amplitude of standing \glspl{SAW} and travelling \glspl{SAW} was set as 0.3\%. As for orthogonal \glspl{SAW}, amplitudes of the longitudinal travelling \glspl{SAW} and transverse standing \glspl{SAW} were 0.3\% and 0.7\%, respectively. 
Figure~\ref{fig:config_sim}c and d show the spatial strain profile of travelling \glspl{SAW}, standing \glspl{SAW} and orthogonal \gls{SAW} at 0 \textmu s, respectively. \par
Micromagnetic simulations were performed using Mumax3~\cite{Vansteenkiste2014,Exl2014,Mulkers2017}, based on the \gls{LLG} equation (Equation~\ref{eqn:LLG}).
The magnetoelastic energy density can be expressed as Equation~\ref{eqn:E_me}.
The material parameters are set corresponding to Co/Pt thin film as follows~\cite{sampaio2013nucleation,gutjahr2000magnetoelastic}: 
saturation magnetisation $M_\RM{s}=\text{5.8}\times\text{10}^\text{5}$ A/m, 
exchange constant $A_\RM{exch}=\text{1.5}\times\text{10}^\text{{--11}}$ J/m, 
anisotropy constant $K_\RM{u}=\text{8}\times\text{10}^\text{5}$ J/m$^\text{3}$, 
interfacial \gls{DMI} strength $D = \text{3}\times\text{10}^\text{--3}$ J/m$^\text{2}$,
magnetoelastic coupling coefficient $B_1 = \text{2}\times\text{10}^\text{7}$ J/m$^\text{3}$.
\revisiontwo{The Gilbert damping constant of Co/Pt multilayer is reported in the range of 0.02--0.1~\cite{fujita2008damping,barman2007ultrafast}. To resolve high frequency magnetisation dynamics and to enhance the skyrmion motion, $\alpha$ value of 0.01 was chosen~\cite{yokouchi2020creation,barker2023breathing}.} The standing \glspl{SAW} and travelling \glspl{SAW} were implemented using Equation~\ref{eqn:standingSAW} and Equation~\ref{eqn:travellingSAW}, respectively.

\section{Results and discussions}
\subsection{Impact of standing SAWs on skyrmion motion}
\label{Chapter7_Section:standingSAWs}
\begin{figure}[ht!]
\centering
\includegraphics[width=1\textwidth]{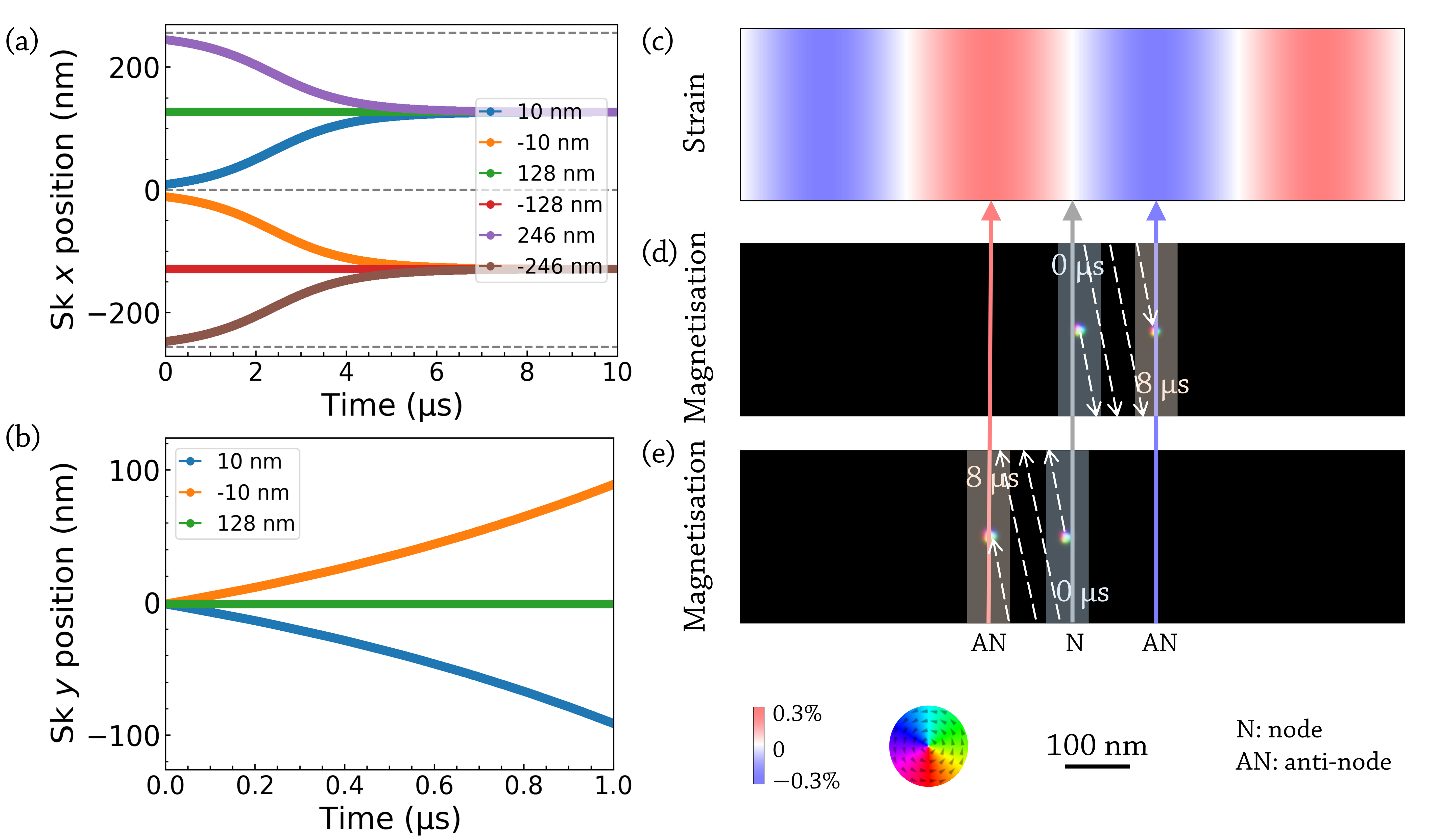}
\caption[Skyrmion motion under the standing surface acoustic waves.]{\label{fig:standingSAW} 
Translation of skyrmion in (a) $x$ and (b) $y$ directions driven by standing \glspl{SAW}. Numbers in the legend indicate initial $x$ positions of skyrmions relative to the origin. (c) Spatial strain profile of standing \glspl{SAW} at 0 \textmu s. Examples of the skyrmion initial and final positions driven by standing \glspl{SAW}: (d) skyrmion initialised at $-$10 nm and (e) 10 nm along $x$ direction \revision{relative to the origin}. ``N'' and ``AN'' demote the node and anti-node, respectively. \revision{The colour code for the in-plane magnetisation component is shown by a colour wheel.}}
\end{figure}

We firstly studied the skyrmion motion driven by standing \glspl{SAW}, and in particular the effect of nodes and anti-nodes of standing \glspl{SAW} on skyrmion motion. The dynamic strain waves form a strain gradient, which periodically changes between nodes and anti-nodes. Owing to the magnetoelastic coupling effect, the strain gradient provides a driving force for skyrmion motion in the $x$ direction. However, this strain gradient vanishes at the anti-nodes of standing \glspl{SAW}. The induced strain gradient pushes the skyrmion moving towards the anti-nodes, with pinning therefore occurring at the anti-nodes. This behaviour is shown in Figure~\ref{fig:standingSAW}a. Skyrmions were initialised at different $x$ positions along the thin film (relative to the origin). We observed skyrmions moving towards the nearest anti-nodes from their initial position. However, skyrmions were pinned without any motion in both $x$ and $y$ directions if they were initialised at anti-nodes. Figure~\ref{fig:standingSAW}d and e show examples of skyrmion motion from initial positions (10 nm and --10 nm in $x$ direction) to finial positions at 8 \textmu s. We also observed skyrmion Hall-like motion in the $y$ direction since skyrmions are rotationally symmetric magnetisation textures (Figure~\ref{fig:standingSAW}b).\par

\subsection{Impact of travelling SAWs on skyrmion motion}

\begin{figure}[ht!]
\centering
\includegraphics[width=0.7\textwidth]{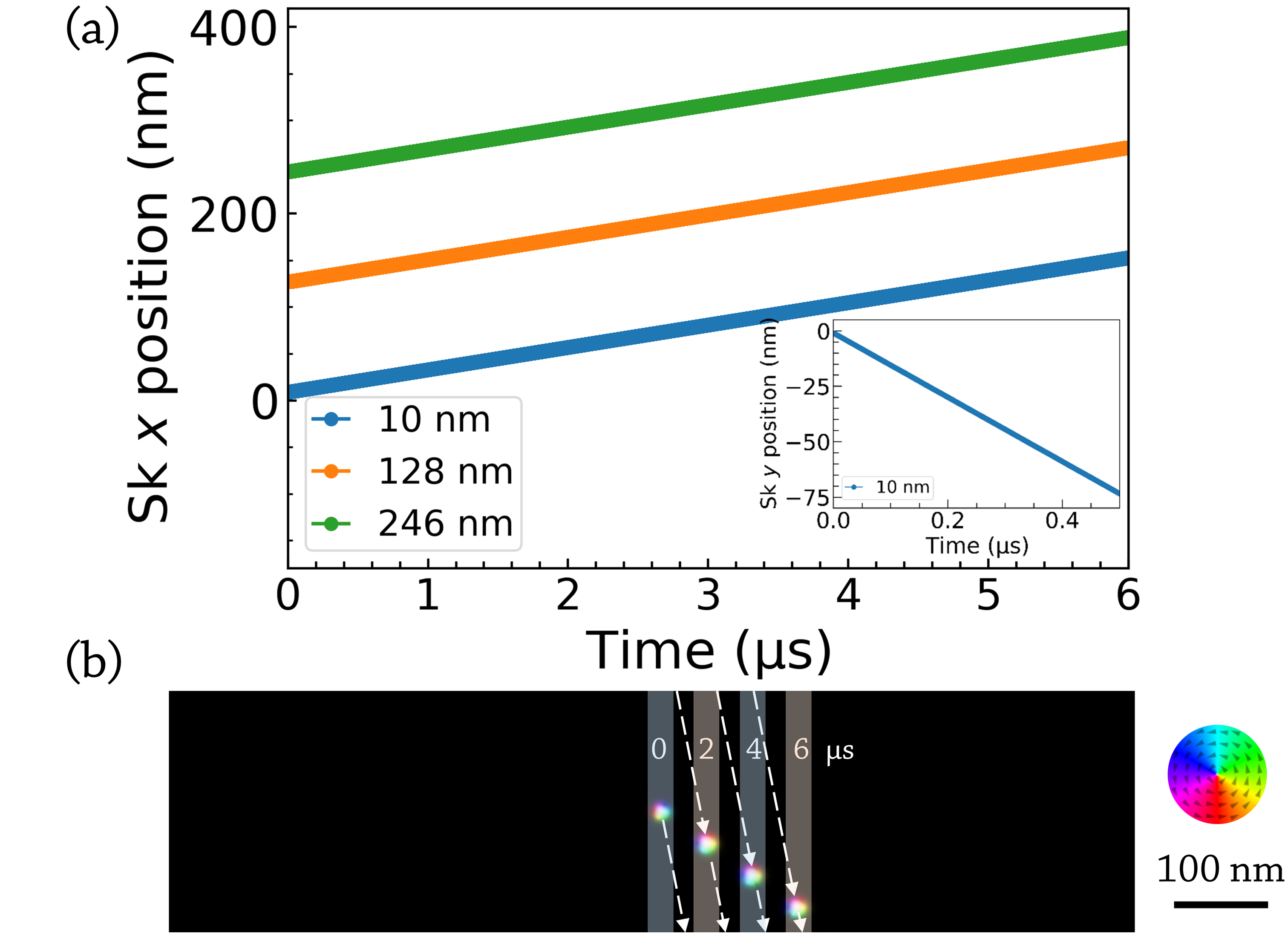}
\caption[Skyrmion motion under the travelling surface acoustic waves.]{\label{fig:travellingSAW} 
(a) Translation of skyrmion in the $x$ direction driven by travelling \glspl{SAW}. Numbers in the legend indicate the initial $x$ positions of skyrmions relative to the origin. The insert graph shows the skyrmion motion in the $y$ direction during the first 0.5 \textmu s. (b) Snapshots of skyrmion motion from 0 to 6.0 \textmu s. The arrows indicate the skyrmion moving direction. \revision{The colour code for the in-plane magnetisation component is shown by a colour wheel.}}
\end{figure}

Secondly, we studied the skyrmion motion driven by travelling \glspl{SAW}. Figure~\ref{fig:travellingSAW}a shows the translation of skyrmions in the $x$ direction driven by travelling \glspl{SAW}. Unlike the skyrmion motion driven by standing \glspl{SAW}, regardless of initial positions of skyrmions (10 nm, 128 nm, and 246 nm along the $x$ direction), skyrmions move continuously in the $x$ direction at a speed of 2.40 cm/s without any pinning with the application of travelling \glspl{SAW} (Figure~\ref{fig:travellingSAW}a). This is because skyrmions experience nodes and anti-nodes at all positions with the application of travelling \glspl{SAW}, which provide a dynamic but continuous strain gradient (driving force). Skyrmions also exhibit Hall-like motion in the $y$ direction (see the insert graph in Figure~\ref{fig:travellingSAW}a). Figure~\ref{fig:travellingSAW}b show an example of skyrmion position changes with time driven by travelling \glspl{SAW}. \revision{The velocity at which skyrmions are driven by travelling \glspl{SAW} is substantially less than the speed of the \glspl{SAW} themselves, which is around 4000 m/s. This implies that multiple wave periods are required to induce skyrmion motion.} \par

\subsection{Impact of orthogonal SAWs on skyrmion motion}

\begin{figure}[ht!]
\centering
\includegraphics[width=0.7\textwidth]{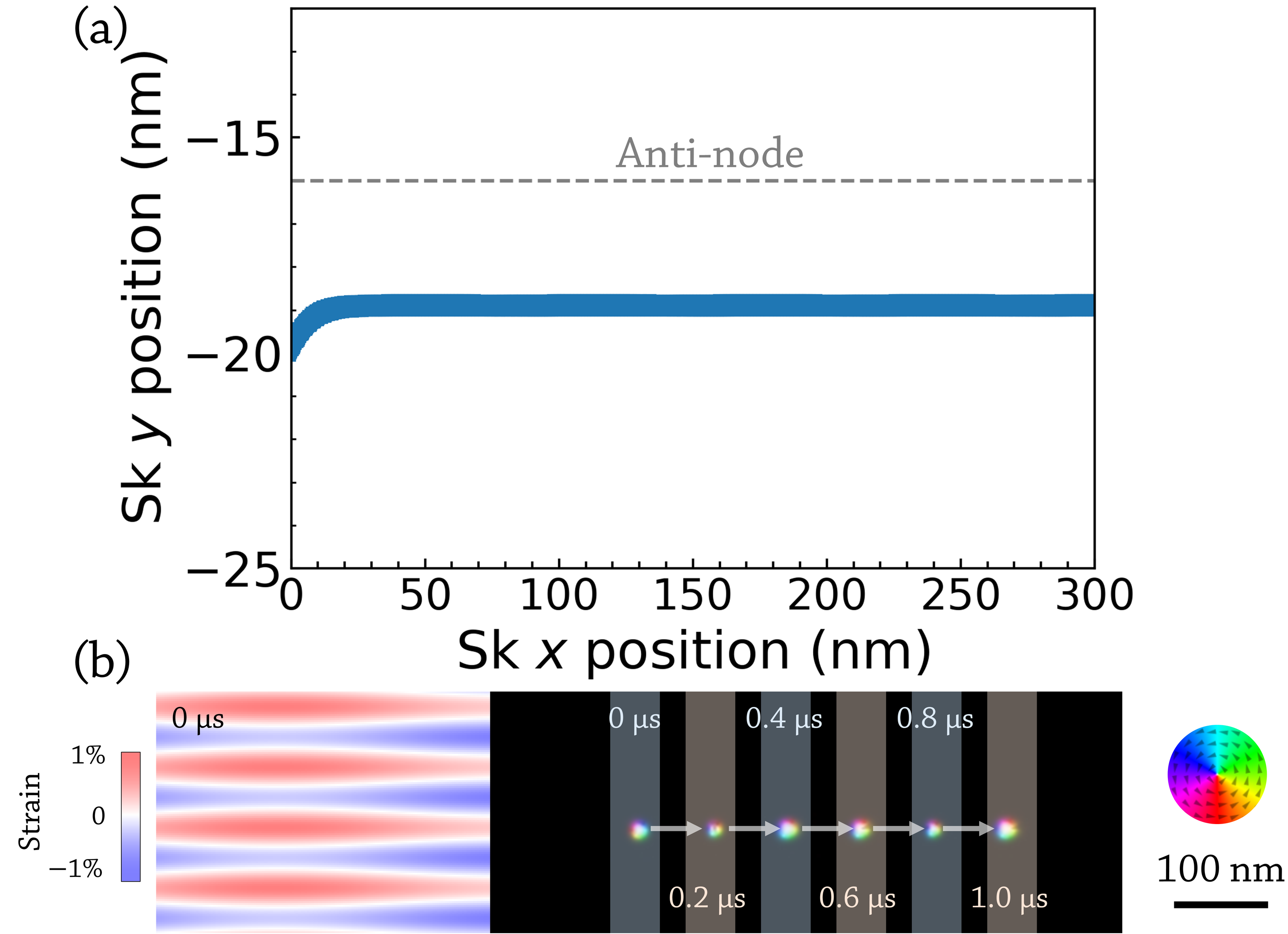}
\caption[Skyrmion motion under the orthogonal surface acoustic waves.]{\label{fig:orthogonalSAW} 
(a) Translation of skyrmion in $x$ and $y$ directions driven by orthogonal \glspl{SAW}. (b) Left-hand side of the \revision{figure} shows the spatial stain profile of orthogonal \glspl{SAW} at 0 \textmu s. Right-hand side of the \revision{figure} shows the snapshots of the skyrmion motion from 0 to 1.0 \textmu s. The arrows indicate the direction of the skyrmion moving direction, which is in line with the anti-node of standing \glspl{SAW} in orthogonal \glspl{SAW}.  \revision{The colour code for the in-plane magnetisation component is shown by a colour wheel.}}
\end{figure}

We then demonstrate the precise transport of a single skyrmion using orthogonal \glspl{SAW}, which are formed by combining longitudinal travelling \glspl{SAW} and transverse standing \glspl{SAW}. With this configuration, we are able to move skyrmions continuously in the $x$ direction using travelling \glspl{SAW} and also to create pinning ``channels'' using the anti-nodes of the standing \glspl{SAW} along the $y$ direction to suppress the skyrmion transverse motion. In this simulation, the wavelengths and strain amplitudes of travelling \glspl{SAW} and standing \glspl{SAW} were set as 512 nm and 0.3\%, and 64 nm and 0.7\%, respectively (see Figure~\ref{fig:config_sim}d). Figure~\ref{fig:orthogonalSAW}a shows the translation of a skyrmion. Skyrmion moves continuously in the $x$ direction with a very limited motion of $\sim$1 nm in the $y$ direction at the beginning, when the skyrmion tries to move from its initial position towards the pinning channel that standing \glspl{SAW} create. Note that the skyrmion velocity significantly increases to 38.64 cm/s compared to that of travelling \gls{SAW}-induced skyrmion motion (2.40 cm/s). \revision{This is because the standing \glspl{SAW} induced force contributes to the longitudinal motion~\cite{nepal2018magnetic}.}\par

\begin{figure*}[ht!]
\includegraphics[width=1\textwidth]{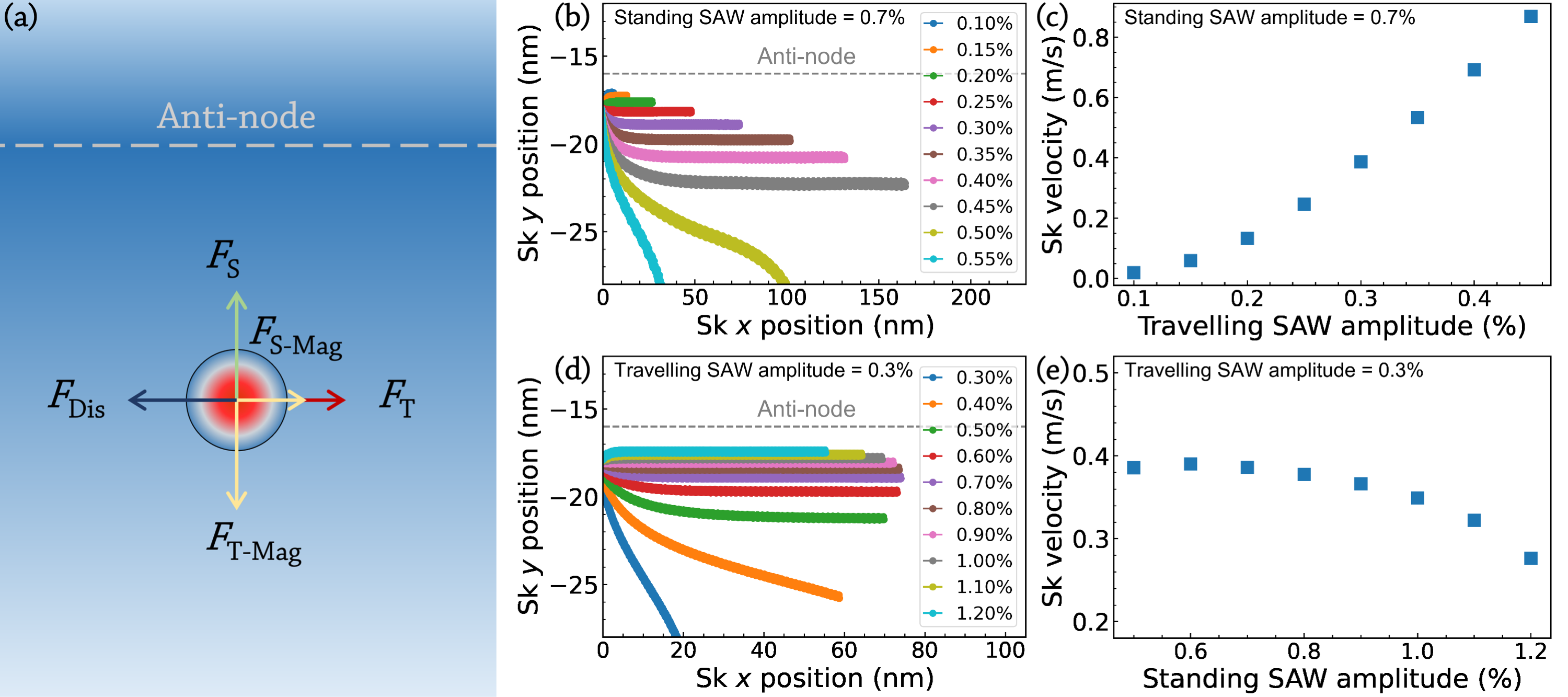}
\caption[Illustration of forces acting on a skyrmion induced by orthogonal surface acoustic waves, and skyrmion motion under orthogonal surface acoustic waves with different properties.]{\label{fig:sk_velocity} 
\revision{(a) Illustration of forces acting on a skyrmion induced by orthogonal \glspl{SAW}. The forces include the driving force ($F_\RM{T}$), Magnus force ($F_{\RM{T}\text{-}\RM{Mag}}$) from the travelling \gls{SAW}, as well as the driving force ($F_\RM{S}$), Magnus force ($F_{\RM{S}\text{-}\RM{Mag}}$) from the standing \gls{SAW}, and dissipative force ($F_\RM{Dis}$) due to damping. 
(b) Skyrmion trajectories under a fixed standing \gls{SAW} amplitude of 0.7\%, with varying travelling \gls{SAW} amplitudes from 0.1\% to 0.55\%.
(c) Corresponding skyrmion velocities for varying travelling \gls{SAW} amplitudes.
(d) Skyrmion trajectories under a fixed travelling \gls{SAW} amplitude of 0.3\%, with varying standing \gls{SAW} amplitudes from 0.3\% to 1.2\%.
(e) Corresponding skyrmion velocities for varying standing \gls{SAW} amplitudes.}}
\end{figure*}

\revision{To understand the influence of the amplitude of standing and travelling \glspl{SAW} on skyrmion velocity, we simulated skyrmion motion driven by orthogonal \glspl{SAW} with varying configurations. The amplitude of the standing \gls{SAW} was maintained at 0.7\%, while the amplitude of the travelling \gls{SAW} was varied from 0.1\% to 0.55\%. It was observed that the skyrmion is pinned in the transverse direction when the travelling \gls{SAW} amplitude is at most 0.45\%. The skyrmion velocity significantly increases from 0.01 m/s to 0.86 m/s as the travelling \gls{SAW} amplitude rises from 0.1\% to 0.45\% (see Figure~\ref{fig:sk_velocity}c).
Furthermore, we set the travelling \gls{SAW} amplitude to 0.3\% and altered the standing \gls{SAW} amplitude from 0.3\% to 1.2\%. The transverse motion of the skyrmion is effectively restrained when the standing \gls{SAW} amplitude exceeds 0.5\% (see Figure~\ref{fig:sk_velocity}d). A slight decrease in skyrmion velocity is observed with an increase in standing \gls{SAW} amplitude (see Figure~\ref{fig:sk_velocity}e).
In other words, the strain amplitude of the standing \gls{SAW} must be greater than that of the travelling \glspl{SAW} to provide sufficient pinning energy that confines the skyrmion motion in the $x$ direction.}
\revision{Skyrmion motion trajectory can be attributed to the balance of forces acting on the skyrmion (as depicted in Figure~\ref{fig:sk_velocity}).
In the following analysis, we will use $-x$, $+x$, $-y$, and $+y$ to denote the directions as left, right, down, and up relative to the origin (centre of the simulation region), respectively. 
The travelling \gls{SAW} generates a force $F_\RM{T}$ that moves the skyrmion longitudinally in the same direction of the travelling \gls{SAW} ($+x$). 
At the same time, the travelling \gls{SAW} also induces a Magnus force $F_{\RM{T}\text{-}\RM{Mag}}$ perpendicular to the $F_\RM{T}$ ($-y$) due to the skyrmion Hall effect. 
Similarly, the standing \gls{SAW} exerts a driving force $F_\RM{S}$ along the $+y$ axis and a Magnus force $F_{\RM{S}\text{-}\RM{Mag}}$ in $-x$ axis, orthogonal to each other. 
Therefore, skyrmion velocity decreases with increasing amplitude of the standing \gls{SAW}, as shown in Figure~\ref{fig:sk_velocity}e.
There is also dissipative force $F_\RM{Dis}$ at the $-x$ direction due to damping. The the net force of the above mentioned forces determines the motion of the skyrmion. To suppress the skyrmion motion in the transverse direction, $F_\RM{S}$ needs to be equal to the $F_{\RM{T}\text{-}\RM{Mag}}$. }\par

\subsection{Multichannel skyrmion racetrack}

\begin{figure}[ht!]
\centering
\includegraphics[width=1\textwidth]{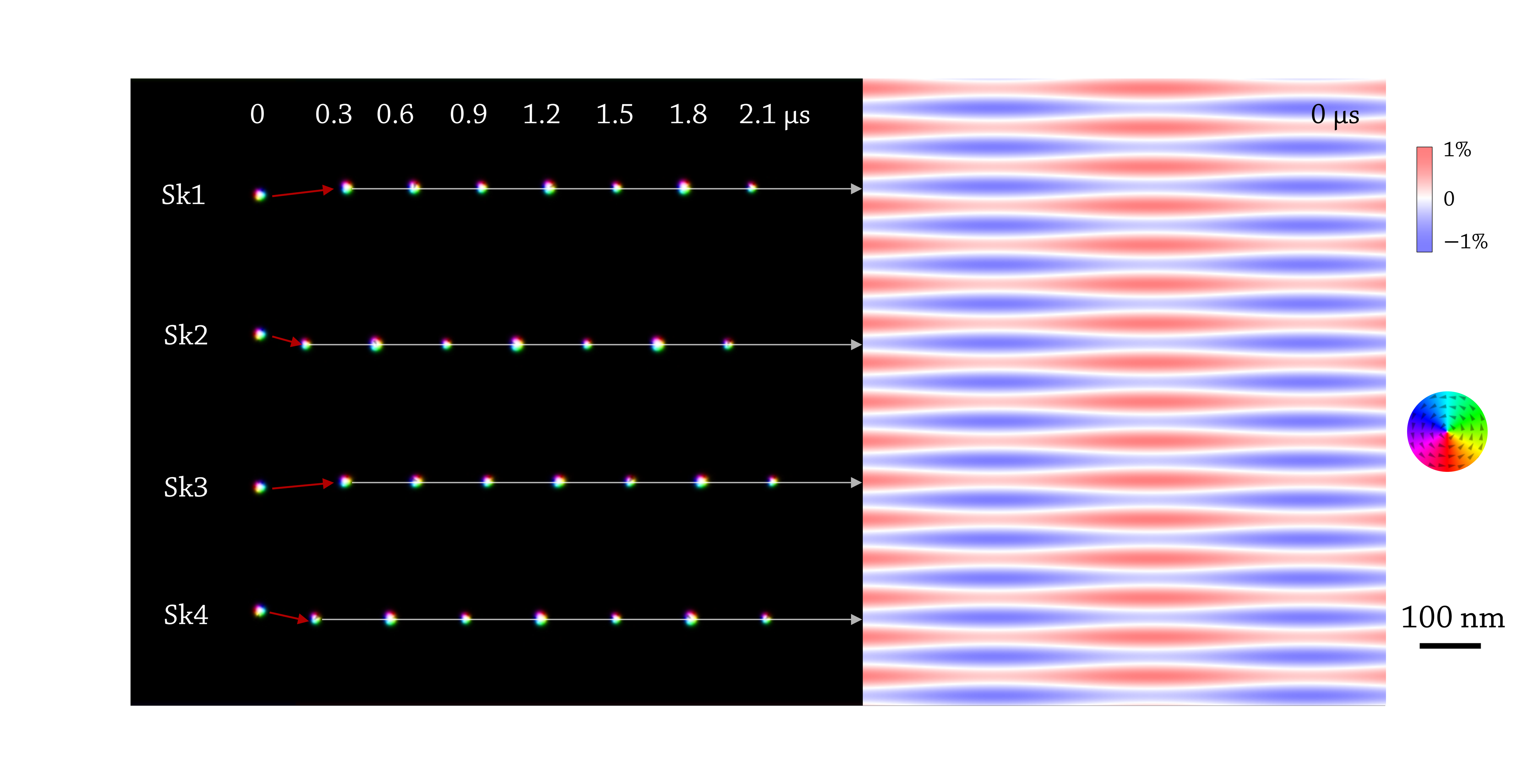}
\caption[Skyrmion trajectories in a multichannel racetrack with orthogonal surface acoustic waves.]{\label{fig:ProposedDevice} 
\revision{Skyrmion trajectories in a multichannel racetrack with orthogonal \glspl{SAW}. The left side displays the motion of four skyrmions (Sk1 to Sk4) in a multichannel racetrack, influenced by orthogonal \glspl{SAW}. The skyrmions are initially positioned randomly along the $y$-axis but at an identical $x$-coordinate. The time elapsed is denoted above the graph in \textmu s, and the arrows indicate the direction of skyrmion motion, aligning with the anti-nodes of the standing \gls{SAW}. On the right side, the spatial strain profile is shown at time 0 \textmu s, and the color wheel represents the color code for the in-plane magnetisation component.}}
\end{figure}

Finally, we proposed a multi-channel skyrmion racetrack obtained using orthogonal \glspl{SAW}. In this simulation, we use the same orthogonal \gls{SAW} property as above (i.e. the wavelengths and strain amplitudes of standing \glspl{SAW} and travelling \glspl{SAW}, which together form orthogonal \glspl{SAW}, are 512 nm and 0.3\%, and 64 nm and 0.7\%, respectively) but with a larger space comprising 2048 $\times$ 1024 $\times$ 1 nm\textsuperscript{3}. We initialised four skyrmions (Sk1 to Sk4 in Figure~\ref{fig:ProposedDevice}) randomly along transverse direction but with the same longitudinal position. As shown in Figure~\ref{fig:ProposedDevice}, skyrmions are transported different distances during the first 0.3 \textmu s. This is owing to the fact that standing \glspl{SAW} push skyrmions towards their nearest anti-nodes. This means that standing \glspl{SAW} provide a driving force with the same/opposite direction as/to travelling \glspl{SAW} depending their positions relative to the anti-nodes (see Figure~\ref{fig:standingSAW}a). From 0.3 to 2.1 \textmu s, the velocity of all skyrmions remains the same. This is because the driving force is the same for all skyrmions after they arrive the pinning channel. With this design, one can create multiple channels to transport skyrmions without skyrmion interaction to increase the transporting data density. The width and density of the channel can be determined by the wavelength of standing \glspl{SAW} that forms orthogonal \glspl{SAW}.

\revision{The results presented above provide a proof-of-concept for purely \gls{SAW}-driven skyrmion motion. However, several questions remain to be explored. Notably, some material parameters employed in this study exceed those achievable in experimental settings. Additionally, the SAW parameters, including wavelength and amplitude, that facilitate skyrmion motion require adjustment to align with experimentally viable scales for practical device realisation. Furthermore, the simulations were conducted under ideal conditions, devoid of temperature fluctuations and disorder, raising the need to investigate how skyrmions respond to these external variables.}

\section{Summary}
To summarise, we demonstrated the skyrmion motion by \glspl{SAW} with different modalities. Skyrmions were moved by standing \glspl{SAW} and travelling \glspl{SAW} in both longitudinal and transverse directions due to the \gls{SAW}-induced strain gradient. Standing \glspl{SAW} created pinning sites at their anti-nodes, whereas travelling \glspl{SAW} provided a constant driving force to move skyrmions continuously. By combining longitudinal travelling \glspl{SAW} and transverse standing \glspl{SAW}, we demonstrated the precise transport of skyrmions in the longitudinal direction without transverse motion by orthogonal \glspl{SAW}. These results suggest the possibility of multi-channel skyrmion racetrack memory/logic devices using \glspl{SAW}.\par
\cleardoublepagewithnumberheader
\chapter{Conclusions and outlook}
\label{chapter8_conclusion_and_outlook}
Magnetic spin structures, such as \acrfull{DW} and skyrmion, in thin films with \acrfull{PMA} hold significant promise for technological advancements in spintronics, particularly in applications like magnetic racetrack memory and logic devices. The utilisation of \acrfullpl{SAW} presents a novel and energy-efficient approach to control magnetic spin structures through the magnetoelastic coupling effect.
This thesis offers a comprehensive investigation of the interactions between \glspl{SAW} and magnetic thin films with \gls{PMA}, employing both experimental and micromagnetic techniques. 
The primary objective is to gain an in-depth understanding of the effects of \glspl{SAW} on magnetisation dynamics, with a specific focus on magnetic spin structures like \glspl{DW} and skyrmions.

\section{Conclusions}
Chapter~\ref{Chapter4_anisotropy_control} provided insights into the effect of the standing \glspl{SAW} on the coercivity, magnetisation switching, and domain patterns in \revision{Ta/Pt/Co/Ir/Ta} thin films with \gls{PMA}. The results indicated that the coercivity of the thin film decreased by up to 21\% in the presence of the standing \glspl{SAW} at 93 MHz with a power of 22.5 dBm. The magnetisation reversal speed also significantly increased by a factor of 11. Due to the difference in strain distribution, standing \glspl{SAW} locally lowered the domain nucleation field at the anti-nodes of the standing \gls{SAW}. Domains tended to nucleate and propagate from anti-nodes to nodes of the standing \gls{SAW}, resulting in striped domain patterns with spacing identical to the half-wavelength of the standing \gls{SAW}.

Chapter~\ref{Chapter5_heating_effect} delved into the impact of both heating and \glspl{SAW} on \gls{DW} motion in a \revision{Ta/Pt/Co/Ta} thin film with \gls{PMA}. A heating of approximately 10 K was observed within the \gls{SAW} beam path when \gls{rf} power was applied, as measured \textit{in situ} using an on-chip Pt thermometer. \gls{DW} velocity was assessed at various temperatures both with and without \glspl{SAW}. The increase in \gls{DW} velocity caused by travelling \glspl{SAW} was slightly more pronounced than that resulting from a 10-K temperature increase, suggesting that heating was the primary factor promoting \gls{DW} motion. However, \gls{DW} motion was substantially enhanced in the presence of standing \glspl{SAW} due to the dominant effect of magnetoelastic coupling. This chapter emphasised the importance of considering heating in \gls{SAW} devices, particularly with high \gls{rf} power, and presented a straightforward method for measuring heating in \gls{SAW} devices.

Chapter~\ref{Chapter6_DW_dynamics} studied the impact of \gls{SAW} frequency on \gls{DW} motion in thin films with varying levels of anisotropy disorder using micromagnetic simulations. The results showed that \glspl{SAW} boosted \gls{DW} velocity by facilitating the depinning of \glspl{DW} from pinning sites through \gls{SAW}-induced spin rotation. This not only increased the likelihood of \gls{DW} depinning but also contributed to energy dissipation at the \gls{DW}. In the thin film with 3\% anisotropy disorder, \gls{DW} velocity rose with increasing \gls{SAW} frequency due to the amplified effect of spin rotation. In contrast, in the thin film with 1\% anisotropy disorder, \gls{DW} velocity decreased with rising \gls{SAW} frequency because of pronounced \gls{SAW}-induced spin rotation. This chapter shed light on the intricate relationship between \glspl{SAW}, spin rotation, and \gls{DW} dynamics, emphasising the important role of anisotropy disorder in determining the \gls{DW} response to \glspl{SAW}.

Chapter~\ref{Chapter7_Skyrmion_motion} illustrated the proof-of-concept of the motion of skyrmions in thin films with \gls{PMA} driven by different modes of \glspl{SAW} using micromagnetic simulations. Skyrmions were moved by both standing and travelling \glspl{SAW} in longitudinal and transverse directions due to the introduced strain gradient. Standing \glspl{SAW} formed pinning sites at their anti-nodes, while travelling \glspl{SAW} provided a consistent force that moved skyrmions continuously. By integrating longitudinal travelling \glspl{SAW} with transverse standing \glspl{SAW}, the study showcased the precise transport of skyrmions longitudinally without transverse motion using orthogonal \glspl{SAW}. This chapter pointed towards the potential of multi-channel skyrmion racetrack memory/logic devices driven by \glspl{SAW}.

In conclusion, these chapters have contributed to a deeper understanding of the interactions between \glspl{SAW} and magnetic thin films with \gls{PMA}. This thesis not only offered valuable theoretical and experimental insights but also proposed potential energy-efficient methods for controlling magnetic spin structures. By elucidating the complex mechanisms through which \glspl{SAW} impact magnetisation dynamics, this research has paved the way for future studies and technological advancements in the energy-efficient manipulation of magnetic structures using \glspl{SAW}.

\section{Outlook}
The findings of this thesis illuminate the potential for utilising \glspl{SAW} as an innovative tool for manipulating magnetic structures in an energy-efficient manner. However, the limited time frame of this research inevitably leaves some questions unanswered and challenges unaddressed. The following paths for future work and development could continue to build upon and extend the findings of this study:

\subsubsection*{Further exploration of magnetisation dynamics introduced by \glspl{SAW}} 
The interactions between \glspl{SAW} and magnetic spin structures, such as \glspl{DW} and skyrmions, are very complicated and need additional theoretical and experimental investigation. For example, extending the work in Chapter~\ref{Chapter6_DW_dynamics} may involve the preparation and examination of thin films with varying levels of anisotropy disorder and the fabrications of \glspl{IDT} with different geometrical configurations. These explorations could yield deeper insights into the effects of \gls{SAW} frequencies on \gls{DW} dynamics, thus enriching the understanding of magnetisation control mechanisms.

\subsubsection*{Addressing heating challenges} 
The heating effects caused by \gls{rf} power dissipation can significantly influence the magnetic properties of thin films. Addressing these challenges necessitates a focus on innovative device design that can mitigate or better manage the thermal impact.

\subsubsection*{Development of novel devices}
The ability to precisely control skyrmion motion, as demonstrated in Chapter~\ref{Chapter7_Skyrmion_motion}, opens exciting possibilities for the development of multi-channel skyrmion racetrack memory/logic devices utilising \glspl{SAW}. As a continuation of the work presented, future research efforts could be directed towards the development of skyrmions with low depinning energy and experimental validation of skyrmion motion using \glspl{SAW}. Such advancements could mark a significant step towards practical applications, bridging the gap between theoretical exploration and technological innovation.\par

\cleardoublepagewithnumberheader

\bibliographystyle{unsrt} 
\renewcommand{\bibname}{References} 
\bibliography{References/library} 
\addcontentsline{toc}{chapter}{References} 
\cleardoublepagewithnumberheader

\thispagestyle{empty}
\mbox{}
\newpage
\thispagestyle{empty}
\mbox{}
\newpage
\end{document}